\documentclass[12pt,hidelinks]{report}
\usepackage{import} % better way to handle subdirectories
\usepackage{charter}
\usepackage{easyReview}
\usepackage[many]{tcolorbox}
\usepackage{geometry}                % See geometry.pdf to learn the layout options. There are lots.close
\usepackage{graphicx}
\graphicspath{figures/}  %% Please place figures in the subdirectory https://www.overleaf.com/project/63696f74ce40f19dc9226cd3

\usepackage{hyperref}    % Hyperlinks in references
\usepackage[all]{hypcap} % Internal hyperlinks to floats.
\usepackage{multirow}
\usepackage{amssymb}
\usepackage{epstopdf}
\usepackage{amssymb}
\usepackage{siunitx}
\usepackage[inline]{enumitem} 
\setlist[itemize]{noitemsep}
\setlist[enumerate]{noitemsep}
\usepackage{booktabs} 
\usepackage{rotating}
\usepackage{lscape}
\usepackage[modulo,pagewise]{lineno}
\usepackage{subcaption}
\usepackage{tabularx}
\usepackage{bm}% bold math
\usepackage{adjustbox}
\usepackage{wrapfig}

\usepackage{tikz-imagelabels}
\usepackage{circuitikz}
\usepackage{venndiagram}
\usepackage{ascmac}

\usepackage[version=4]{mhchem}
\usepackage[backend=biber, style=numeric, sorting=none]{biblatex}
\usepackage{titlesec}
\titleformat{\paragraph}
{\normalfont\normalsize\bfseries}{\theparagraph}{1em}{}
\titlespacing*{\paragraph}
{0pt}{3.25ex plus 1ex minus .2ex}{1.5ex plus .2ex}
\nolinenumbers
\newcommand{\editor}[1]{\relax}    % Remove all editor indications
\usepackage[official]{eurosym}

\newcommand{\cdrauthor}[2]{  \author{#1\,\orcidlink{#2}}  }
\usepackage{authblk}
\usepackage{orcidlink}
\usepackage[symbol]{footmisc}  % footnotes with symbols instead of numbers

\input{belle2-uwg-sym}

\begin{document}
	
%
% front matter: title page and authors
%

%
% redefine title commands without
%
\makeatletter         
\newcommand\mymaketitle{
	{\raggedright % Note the extra {
			\begin{center}
				{\Huge \bfseries  \@title }\\[4ex] 
	\end{center}}} % Note the extra }

\newcommand\mymakeauthors{
	{\raggedright % Note the extra {
			\begin{center}
				{\@author}\\[4ex] 
	\end{center}}} % Note the extra }

\makeatother

\title{The \belletwo\ Detector Upgrades \\ 
	Framework Conceptual Design Report}

\cdrauthor{H.~Aihara}{0000-0002-1907-5964}
\cdrauthor{A.~Aloisio}{0000-0002-3883-6693}
\cdrauthor{D.~P.~Auguste}{0009-0007-7522-848X}
\cdrauthor{M.~Aversano}{0000-0001-9980-0953}
\cdrauthor{M.~Babeluk}{0000-0001-7951-452X}
\cdrauthor{S.~Bahinipati}{0000-0002-3744-5332}
\cdrauthor{Sw.~Banerjee}{0000-0001-8852-2409}
\cdrauthor{M.~Barbero}{0000-0002-7824-3358}
\cdrauthor{J.~Baudot}{0000-0001-5585-0991}
\cdrauthor{A.~Beaubien}{0000-0001-9438-089X}
\cdrauthor{F.~Becherer}{0000-0003-0562-4616}
\cdrauthor{T.~Bergauer}{0000-0002-5786-0293}
\cdrauthor{F.U.~Bernlochner.}{0000-0001-8153-2719}
\cdrauthor{V.~Bertacchi}{0000-0001-9971-1176}
\cdrauthor{G.~Bertolone}{0000-0001-9544-7486}
\cdrauthor{C.~Bespin}{0000-0002-5413-1730}
\cdrauthor{M.~Bessner}{0000-0003-1776-0439}
\cdrauthor{S.~Bettarini}{0000-0001-7742-2998}
\cdrauthor{A.~J.~Bevan}{0000-0002-4105-9629}
\cdrauthor{B.~Bhuyan}{0000-0001-6254-3594}
\cdrauthor{M.~Bona}{0000-0002-9660-580X}
\cdrauthor{J.~F.~Bonis}{0009-0006-1453-6019}
\cdrauthor{J.~Borah}{0000-0003-2990-1913}
\cdrauthor{F.~Bosi}{0000-0002-1922-0092}
\cdrauthor{R.~Boudagga}{0009-0007-1198-2651}
\cdrauthor{A.~Bozek}{0000-0002-5915-1319}
\cdrauthor{M.~Bra\v cko}{0000-0002-2495-0524}
\cdrauthor{P. Branchini}{0000-0002-2270-9673}
\cdrauthor{P.~Breugnon}{0000-0002-1508-218X}
\cdrauthor{T.~E.~Browder}{0000-0001-7357-9007}
\cdrauthor{Y.~Buch}{0000-0002-8050-4000}
\cdrauthor{A.~Budano}{0000-0002-0856-1131}
\cdrauthor{M.~Campajola}{0000-0003-2518-7134}
\cdrauthor{G.~Casarosa}{0000-0003-4137-938X}
\cdrauthor{C. ~Cecchi}{0000-0002-2192-8233}
\cdrauthor{C.~Chen}{0000-0003-1589-9955}
\cdrauthor{S.~Choudhury}{0000-0001-9841-0216}
\cdrauthor{L.~Corona}{0000-0002-2577-9909}
\cdrauthor{G.~de~Marino}{0000-0002-6509-7793}
\cdrauthor{G.~De~Nardo}{0000-0002-2047-9675}
\cdrauthor{G.~De~Pietro}{0000-0001-8442-107X}
\cdrauthor{R.~de~Sangro}{0000-0002-3808-5455}
\cdrauthor{S.~Dey}{0000-0003-2997-3829}
\cdrauthor{J.~C.~Dingfelder}{0000-0001-5767-2121}
\cdrauthor{T.~V.~Dong}{0000-0003-3043-1939}
\cdrauthor{A.~Dorokhov}{0000-0001-5809-524X}
\cdrauthor{G.~Dujany}{0000-0002-1345-8163}
\cdrauthor{D.~Epifanov}{0000-0001-8656-2693}
\cdrauthor{L.~Federici}{0000-0002-3401-9522}
\cdrauthor{T.~Ferber}{0000-0002-6849-0427}
\cdrauthor{T.~Fillinger}{0000-0001-9795-7412}
\cdrauthor{Ch.~Finck}{0000-0002-5068-5453}
\cdrauthor{G.~Finocchiaro}{0000-0002-3936-2151}
\cdrauthor{F.~Forti}{0000-0001-6535-7965}
\cdrauthor{A.~Frey}{0000-0001-7470-3874}
\cdrauthor{M.~Friedl}{0000-0002-7420-2559}
\cdrauthor{A.~Gabrielli}{0000-0001-7695-0537}
\cdrauthor{L.~Gaioni}{0000-0001-5499-7916}
\cdrauthor{Y.~Gao}{0000-0001-6326-4773}
\cdrauthor{G.~Gaudino}{0000-0001-5983-1552}
\cdrauthor{V.~Gaur}{0000-0002-8880-6134}
\cdrauthor{A.~Gaz}{0000-0001-6754-3315}
\cdrauthor{R.~Giordano}{0000-0002-5496-7247}
\cdrauthor{S.~Giroletti}{0009-0003-8954-7773}
\cdrauthor{B.~Gobbo}{0000-0002-3147-4562}
\cdrauthor{R.~Godang}{0000-0002-8317-0579}
\cdrauthor{I.~Haide}{0000-0003-0962-6344}
\cdrauthor{Y.~Han}{0000-0001-6775-5932}
\cdrauthor{K.~Hara}{0000-0002-5361-1871}
\cdrauthor{K.~Hayasaka}{0000-0002-6347-433X}
\cdrauthor{C.~Hearty}{0000-0001-6568-0252}
\cdrauthor{A.~Heidelbach}{0000-0002-6663-5469}
\cdrauthor{T.~Higuchi}{0000-0002-7761-3505}
\cdrauthor{A.~Himmi }{0009-0003-6971-233X}
\cdrauthor{M.~Hoferichter}{0000-0003-1113-9377}
\cdrauthor{D.~A.~Howgill}{0009-0008-9414-8968}
\cdrauthor{C.~Hu-Guo}{0000-0001-9626-4673}
\cdrauthor{T.~Iijima}{0000-0002-4271-711X}
\cdrauthor{K.~Inami}{0000-0003-2765-7072}
\cdrauthor{C.~Irmler}{0009-0008-8290-8472}
\cdrauthor{A.~Ishikawa}{0000-0002-3561-5633}
\cdrauthor{R.~Itoh}{0000-0003-1590-0266}
\cdrauthor{D.~Iyer}{0009-0005-2141-8806}
\cdrauthor{W.~W.~Jacobs}{0000 0002 9996 6336}
\cdrauthor{D.~E.~Jaffe}{0000-0003-3122-4384}
\cdrauthor{Y.~Jin}{0000-0002-7323-0830}
\cdrauthor{T.~Junginger }{0000-0002-2228-2809}
\cdrauthor{J.~Kandra}{0000-0001-5635-1000}
\cdrauthor{K.~Kojima}{0000-0002-3638-0266}
\cdrauthor{T.~Koga}{0000-0002-1644-2001}
\cdrauthor{A.~A.~Korobov}{0000-0001-5959-8172}
\cdrauthor{S.~Korpar}{0000-0003-0971-0968}
\cdrauthor{P.~Kri\v zan}{0000-0002-4967-7675}
\cdrauthor{H.~Kr\"uger}{0000-0001-8287-3961}
\cdrauthor{T.~Kuhr}{0000-0001-6251-8049}
\cdrauthor{A.~Kumar}{0000-0002-4346-7335}
\cdrauthor{R~.Kumar}{0000-0002-6277-2626}
\cdrauthor{A.~Kuzmin}{0000-0002-7011-5044}
\cdrauthor{Y.-J.~Kwon}{0000-0001-9448-5691}
\cdrauthor{S.~Lacaprara}{0000-0002-0551-7696}
\cdrauthor{C.~Lacasta}{0000-0002-2623-6252}
\cdrauthor{Y.~-T.~Lai}{0000-0001-9553-3421}
\cdrauthor{K.~Lalwani}{0000-0002-7294-396X}
\cdrauthor{T.~Lam}{0000-0001-9128-6806}
\cdrauthor{L.~Lanceri}{0000-0001-8220-3095}
\cdrauthor{M.~J.~Lee}{0000-0003-4528-4601}
\cdrauthor{C.~Leonidopoulos}{0000-0002-7241-2114}
\cdrauthor{D.~Levit}{0000-0001-5789-6205}
\cdrauthor{P.~M.~Lewis}{0000-0002-5991-622X}
\cdrauthor{J.~F.~Libby}{0000-0002-1219-3247}
\cdrauthor{Q.~Y.~Liu}{0000-0002-7684-0415}
\cdrauthor{Z.~Y.~Liu}{0009-0000-6360-2867}
\cdrauthor{D.~Liventsev}{0000-0003-3416-0056}
\cdrauthor{S.~Longo}{0000-0002-8124-8969}
\cdrauthor{G.~Mancinelli}{0000-0003-1144-3678}
\cdrauthor{M.~Manghisoni}{0000-0001-5559-0894}
\cdrauthor{E.~Manoni}{0000-0002-9826-7947}
\cdrauthor{C.~Marinas}{0000-0003-1903-3251}
\cdrauthor{C.~Martellini}{0000-0002-7189-8343}
\cdrauthor{A.~Martens}{0000-0003-1544-4053}
\cdrauthor{M.~Massa}{0000-0001-6207-7511}
\cdrauthor{L.~Massaccesi}{0000-0003-1762-4699}
\cdrauthor{F.~Mawas}{0000-0002-7176-4732}
\cdrauthor{J.~Mazorra}{0000-0003-0525-2736}
\cdrauthor{M.~Merola}{0000-0002-7082-8108}
\cdrauthor{C.~Miller}{0000-0003-2631-1790}
\cdrauthor{M.~Minuti}{0000-0001-9577-2588}
\cdrauthor{R.~Mizuk}{0000-0002-2209-6969}
\cdrauthor{A.~Modak}{0000-0003-1198-1441}
\cdrauthor{A.~Moggi}{0000-0002-2323-8017}
\cdrauthor{G.~B.~Mohanty}{0000-0001-6850-7666}
\cdrauthor{S.~Moneta}{0000-0003-2184-7510}
\cdrauthor{Th.~Muller}{0000-0003-4337-0098}
\cdrauthor{I.~Na}{0000-0002-2281-8968}
\cdrauthor{K.~R.~Nakamura}{0000-0001-9987-6288}
\cdrauthor{M.~Nakao}{0000-0001-8424-7075}
\cdrauthor{A.~Natochii}{0000-0002-1076-814X}
\cdrauthor{C.~Niebuhr}{0000-0002-4375-9741}
\cdrauthor{S.~Nishida}{0000-0001-6373-2346}
\cdrauthor{A.~Novosel}{0000-0002-7308-8950}
\cdrauthor{P.~Pangaud}{0000-0003-1148-9182}
\cdrauthor{B.~Parker}{0000-0002-7054-0833}
\cdrauthor{A.~Passeri}{0000-0003-4864-3411}
\cdrauthor{A.~Passeri}{0000-0003-4864-3411}
\cdrauthor{T.~K.~Pedlar}{0000-0001-9839-7373}
\cdrauthor{Y.~Peinaud}{0009-0008-8611-0468}
\cdrauthor{Y.~Peng}{0009-0008-5109-5958}
\cdrauthor{R.~Peschke}{0000-0002-2529-8515}
\cdrauthor{R.~Pestotnik}{0000-0003-1804-9470}
\cdrauthor{T.~H.~Pham}{0009-0007-9814-6284}
\cdrauthor{M. Piccolo}{0000-0001-9750-0551}
\cdrauthor{L.~E.~Piilonen}{0000-0001-6836-0748}
\cdrauthor{S.~Prell}{0000-0002-0195-8005}
\cdrauthor{M.~V.~Purohit}{0000-0002-8381-8689}
\cdrauthor{L.~Ratti}{0000-0003-1906-1076}
\cdrauthor{V.~Re}{0000-0003-0697-3420}
\cdrauthor{L.~Reuter}{0000-0002-5930-6237}
\cdrauthor{E.~Riceputi}{0000-0002-0542-553X}
\cdrauthor{I.~Ripp-Baudot}{0000-0002-1897-8272}
\cdrauthor{G.~Rizzo}{0000-0003-1788-2866}
\cdrauthor{J.~M.~Roney}{0000-0001-7802-4617}
\cdrauthor{A.~Russo}{0009-0009-5212-8704}
\cdrauthor{S.~Sandilya}{0000-0002-4199-4369}
\cdrauthor{L.~Santelj}{0000-0003-3904-2956}
\cdrauthor{V.~Savinov}{0000-0002-9184-2830}
\cdrauthor{B.~Scavino}{0000-0003-1771-9161}
\cdrauthor{L.~Schall}{0000-0003-0002-3790}
\cdrauthor{G.~Schnell}{0000-0002-7336-3246}
\cdrauthor{C.~Schwanda}{0000-0003-4844-5028}
\cdrauthor{A.~J.~Schwartz}{0000-0002-7310-1983}
\cdrauthor{B.~Schwenker}{0000-0002-7120-3732}
\cdrauthor{M.~Schwickardi}{0000-0003-2033-6700}
\cdrauthor{A.~Seljak}{0000-0002-7950-9277}
\cdrauthor{J.~Serrano}{0000-0003-2489-7812}
\cdrauthor{J.-G.~Shiu}{0000-0002-8478-5639}
\cdrauthor{B.~Shwartz}{0000-0002-1456-1496}
\cdrauthor{F.~Simon}{0000-0002-5978-0289}
\cdrauthor{A.~Soffer}{0000-0002-0749-2146}
\cdrauthor{W.~M.~Song}{0000-0003-1376-2293}
\cdrauthor{M.~Stari\v c}{0000-0001-8751-5944}
\cdrauthor{P.~Stavroulakis}{0000-0001-9914-7261}
\cdrauthor{S.~Stefkova}{0000-0003-2628-530X}
\cdrauthor{R.~Stroili}{0000-0002-3453-142X}
\cdrauthor{S.~Tanaka}{0000-0002-6029-6216}
\cdrauthor{N.~Taniguchi}{0000-0002-1462-0564}
\cdrauthor{V.~Teotia}{0000-0002-8463-9535}
\cdrauthor{N.~Tessema}{0009-0004-2929-5532}
\cdrauthor{R.~Thalmeier}{0009-0003-4480-0990}
\cdrauthor{E.~Torassa}{0000-0003-2321-0599}
\cdrauthor{K.~Trabelsi}{0000-0001-6567-3036}
\cdrauthor{F.~F.~Trantou}{0000-0003-0517-9129}
\cdrauthor{G.~Traversi}{0000-0003-3977-6976}
\cdrauthor{P.~Urquijo}{0000-0002-0887-7953}
\cdrauthor{S.~E.~Vahsen}{0000-0003-1685-9824}
\cdrauthor{I.~Valin}{0009-0005-6517-9755}
\cdrauthor{G.~S.~Varner\footnote[2]{Deceased}}{0000-0002-0302-8151}
\cdrauthor{K.~E.~Varvell}{0000-0003-1017-1295}
\cdrauthor{L.~Vitale}{0000-0003-3354-2300}
\cdrauthor{V.~Vobbilisetti}{0000-0002-4399-5082}
\cdrauthor{X.~L.~Wang}{0000-0002-4617-2006}
\cdrauthor{C.~Wessel}{0000-0003-0959-4784}
\cdrauthor{H.U.~Wienands}{0009-0007-8146-995X}
\cdrauthor{E.~Won}{0000-0002-4245-7442}
\cdrauthor{D. ~Xu}{0009-0002-9893-7813}
\cdrauthor{S.~Yamada}{0009-0004-2929-5532}
\cdrauthor{J.~H.~Yin}{0000-0002-1479-9349}
\cdrauthor{K.~Yoshihara}{0000-0002-3656-2326}
\cdrauthor{C.~Z.~Yuan}{0000-0002-1652-6686}
\cdrauthor{L.~Zani}{0000-0003-4957-805X}
\cdrauthor{Z.~Zong}{0000-0001-9259-4296}
\cdrauthor{S.~Zou}{0000-0003-3377-7222}
\affil{Belle II Upgrades Working Group}

\begin{titlepage}

\vspace*{-3\baselineskip}
\resizebox{!}{3cm}{\includegraphics{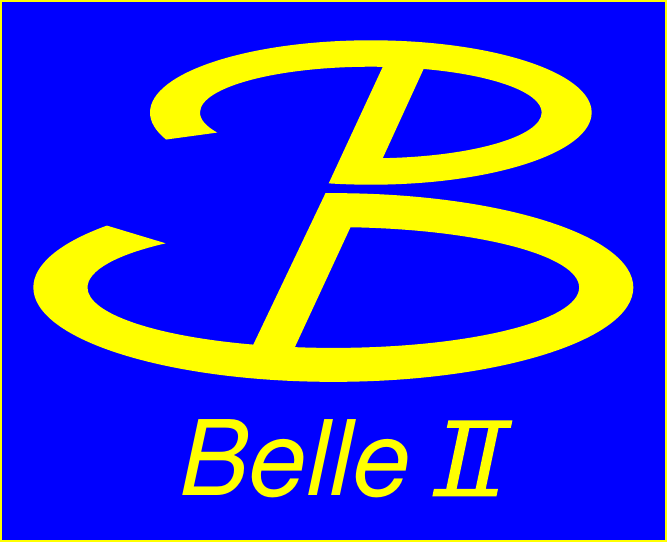}}

\vspace*{-5\baselineskip}
\begin{flushright}
	BELLE2-REPORT-2024-042\\
	KEK-REPORT-2024-1\\
        4 July 2024
\end{flushright}

\vspace{3cm}

\mymaketitle
		
\begingroup
\let\newpage\relax% Void the actions of \newpage
%		\maketitle
		
\begin{abstract}
			We describe the planned near-term and potential longer-term upgrades of the Belle II detector at the  SuperKEKB electron-positron collider operating at the KEK laboratory in Tsukuba, Japan. These upgrades will allow increasingly sensitive  searches for possible new physics beyond the Standard Model in flavor, tau, electroweak and dark sector physics that are both complementary to and competitive with the LHC and other experiments. 
\end{abstract}
\endgroup

\end{titlepage}

\newpage
%
% KEK Related information
%

\vspace*{\fill}

\textbf{© High Energy Accelerator Research Organization (KEK), 2024}

\includegraphics[width=2cm]{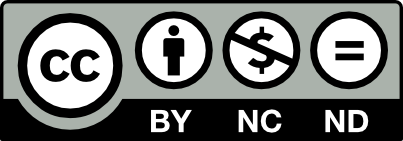}

\textbf{
CC BY-NC-ND 4.0 }\href{https://creativecommons.org/licenses/by-nc-nd/4.0/}{\textbf{https://creativecommons.org/licenses/by-nc-nd/4.0/}}\textbf{ }
\vspace*{0.5cm}

KEK Reports are available from:
\vspace*{0.5cm}

\phantom{Some text}High Energy Accelerator Research Organization (KEK)

\phantom{Some text}1-1 Oho, Tsukuba-shi

\phantom{Some text}Ibaraki-ken, 305-0801

\phantom{Some text}JAPAN
\vspace*{0.5cm}

Phone: +81-29-864-5137

Fax: +81-29-864-4604

E-mail: irdpub@mail.kek.jp

Internet: https://www.kek.jp/en/

\newpage
\mymakeauthors

\chapter*{Executive Summary}
\label{sec:ExecSummary}
The SuperKEKB accelerator and the \belletwo\ detector started operation in 2019 with a steady increase in luminosity, reaching a peak instantaneous luminosity of $\mathcal{L}_{\rm peak}$ = \SI{4.65 e 34}{cm^{-2}s^{-1}} in June 2022 and collecting an integrated luminosity of \qty{424}{\fb^{-1}}.
It has become clear that reaching the target peak luminosity of \SI{6 e 35}{cm^{-2}s^{-1}} and the integrated luminosity \qty{50}{ab^{-1}} will require an upgrade of the accelerator complex, including potentially a redesign of the interaction region (IR).  

A new IR would require a replacement of the vertex detector, and would imply the need for a long shutdown (named LS2, since LS1 extended from 6/2022 to 12/2023), providing also a window of opportunity for the installation of  improvements in many detector subsystems. The timescale for LS2 is placed after 2027--2028 but not well defined yet, as it depends on the precise definition of the accelerator upgrade plans. It should become clearer after the restart of the operation in 2024 and the full evaluation of the LS1 improvements. It should be added that the extrapolation of backgrounds to full luminosity is affected by large uncertainties which may erode the \belletwo\ safety margins and reduce the overall operational efficiency.

The \belletwo\ upgrade program is therefore motivated by the desire to improve the overall physics performance of the experiment by: (i) improving the robustness against backgrounds and the radiation resistance, providing larger safety factors to run at high luminosity; (ii) developing the technology to cope with different future paths, improving the detector performance, and also providing a safety net in case anything goes wrong with the current detector.

This Framework Conceptual Design Report (FCDR) presents a number of possible detector improvements  that can be implemented in the short or medium term, along with other ideas for longer term upgrades, that go beyond the currently planned program. The various upgrades are largely independent of each other and can be selected individually on the basis of actual need, physics performance, technical readiness, and funding availability. In this sense they do not form a single coherent picture of the future Belle II, but rather a menu of possibilities.
A different kind of opportunity is explored by the beam polarization option, which requires a SuperKEKB upgrade but no detector modification, opening a new and unique program of precision physics. Only a brief summary is presented in this FCDR.

\subsection*{BPAC Review}
In December 2023 the FCDR was submitted to the B-Factory Programme Advisory Committee who reviewed it during the February 2024 meeting. In the report~\cite{bpac-reports}, the committee observes: \emph{The BPAC has reviewed the CDR, and recognises the large amount of excellent work that it contains. However, given the structure of the document explained above, it does not represent a coherent proposal for a coordinated upgrade of the experiment and collider on a common timescale, and as such the committee cannot make a recommendation concerning the document as a whole. Instead, comments from the BPAC review are provided here to the individual sections of the CDR, following the order of their presentation in the document.} 

As a response, the FCDR has been modified to show that the specific comments and recommendations have been received and that the  work required to address them is being planned. 

The  recommendations from BPAC are listed below:
\begin{itemize}
    \item The R\&D required for defining the machine upgrade should proceed in a timely manner, to avoid any unnecessary delay to the schedule for the upgrade, and the timescale for LS2 should be adapted as required.
    \item Studies are well advanced for the new vertex detector system (VTX) that is envisioned to replace the current vertex detector. Work on this detector should advance towards the engineering level studies, including the aspects of system integration and operation. This will enable progress to be made quickly on the final design once the machine upgrade has been defined.
    \item Further R\&D work should also continue for the other proposed detector upgrades, but the BPAC considers that a new document providing a coherent and integrated description of the accelerator and detector upgrade plan, with a quantitative demonstration of expected physics performance, will be required once the overall scope has been defined, for a decision to be made on the project. Such a document should also demonstrate the tolerance against background rates at a luminosity of 
    \SI{6 e 35}{cm^{-2}s^{-1}}  and radiation damage corresponding to an integrated luminosity of  \qty{50}{ab^{-1}}.
    \item It is important to make the progression from qualitative to quantitative statements on the expected physics impact of the upgrade (for instance the KLM upgrade) and to demonstrate the overall performance of the upgraded detector for the main physics channels.
    \item The committee appreciates that the proposed test of production of polarised elec- trons in the injector and measuring the polarisation decay time in the SuperKEKB ring will provide essential information for making further decisions on the idea to collect data with a longitudinally polarised beam. However, it should be noted that the highest priority of the coming period must remain to achieve stable machine operation with a luminosity above \SI{e35}{cm^{-2}s^{-1}} for data taking.
   
\end{itemize}

The various options are the result of an internal selection process over the course of the past 2-3 years and are clearly at different levels of readiness. This FCDR documents the current state so that R\&D and engineering activities can be properly organized. The next step will be the preparation of Technical Design Reports for those options that are worth pursuing, possibly on different timescales, as well as a document providing a coherent and integrated description of the accelerator and detector upgrade plan, as specifically requested by the BPAC.

\tableofcontents

\clearpage

%\vspace{0.4in}

%%%%%%%  ========= BIBLIOGRAPHIES  ============== %%%%%%%
%
% Insert all the references in the belle2upgrades.bib file
% use \cite in your section to cite them

\chapter{Introduction}
\label{sec:Introduction}
%%%%%%%%%%%%%%%%%%%%%%%%%%%
\editor{F.Forti}
\section{\belletwo: status and perspective}

% What is Belle II 
% brief history of Belle II commissioning years.
%  Current status and perspectives
% The summary of LS1 work will go in 1.2

% cite TDR and detector paper
The first generation asymmetric B-Factories experiments, \babar\ at PEP-II and Belle at KEKB, discovered large CP violating effects in the $B$ meson sector and provided experimental verification of the Cabibbo-Kobayashi-Maskawa mechanism for quark mixing and CP-violating phase in the Standard Model, along with a wealth of other measurements~\cite{BaBar:2014omp}.

\begin{figure}[htb]
	\centering
	\includegraphics[width=0.95\linewidth]{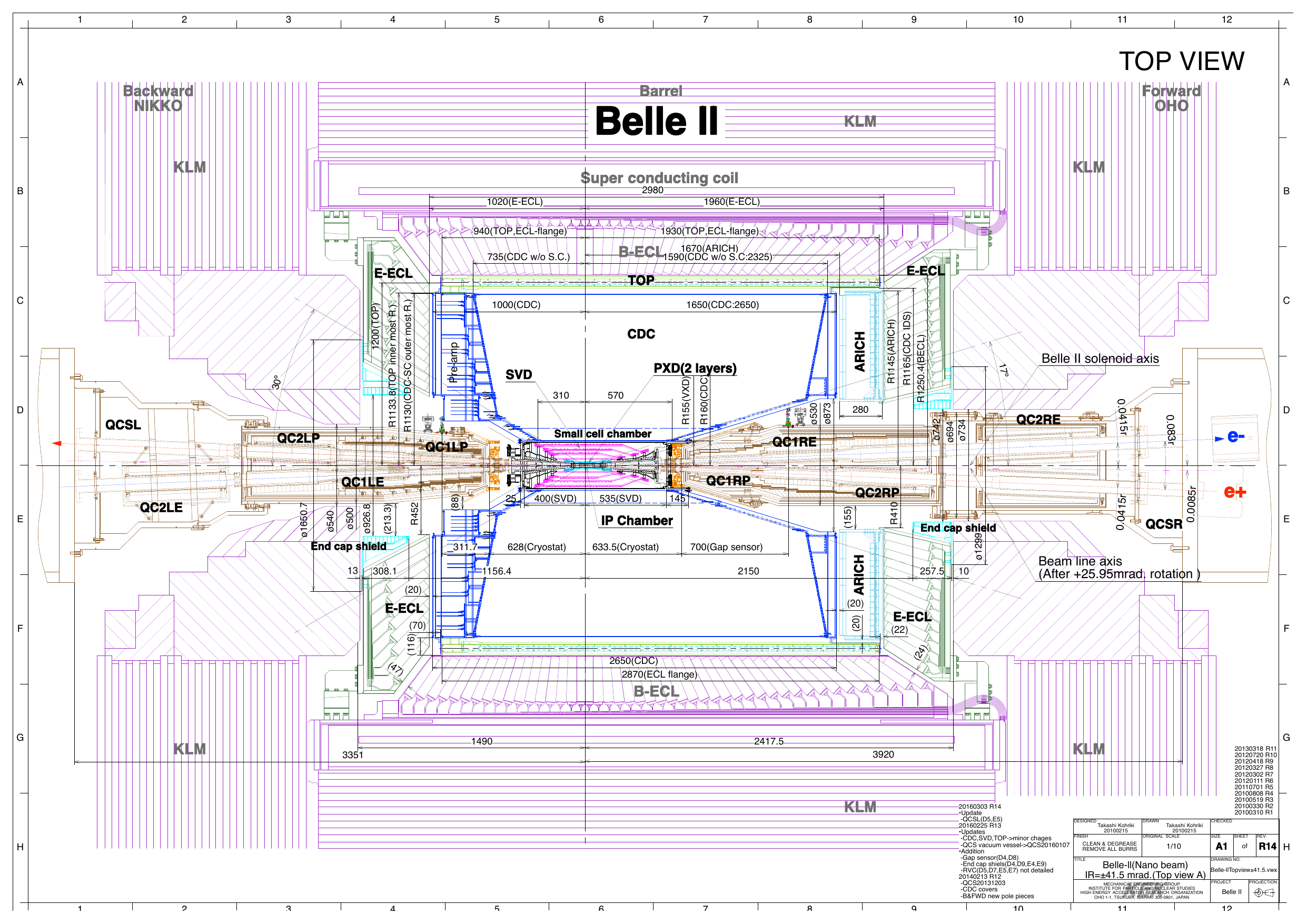}
	\caption{\belletwo\ detector top view. }
	\label{fig:c1belle-lltopview41}
\end{figure}

The SuperKEKB accelerator~\cite{ref:skb} and the  \belletwo\ detector~\cite{Abe:2010gxa, Adachi:2018qme}, shown in figure~\ref{fig:c1belle-lltopview41}, are a major upgrade of the Belle/KEKB complex. 
The nomenclature of the \belletwo\  subdetectors is the following:  Pixel Detector (PXD), Silicon Vertex Detector (SVD), forming together the VerteX Detector (VXD), Central Drift Chamber (CDC),  Time of Propagation Counter (TOP), Aerogel Ring Imaging Cherenkov Counter (ARICH), Electromagnetic Calorimeter (ECL),  K-Long Muon System (KLM), Trigger and Data acquisition (TRG/DAQ), and High Level Trigger (HLT). 
A summary of the subdetector technology is presented in table~\ref{tab:detector_summary}~\cite{Belle-II:2018jsg, Adachi:2018qme}

%\begin{landscape}
	\begin{sidewaystable}
	\centering
	%\small
	\caption{Summary of the detector components\cite{Adachi:2018qme}.}
	\label{tab:detector_summary}
	
	\begin{tabular}{lcp{3cm}p{7cm}p{3cm}p{3cm}} 
		\hline
		Purpose & Name & Component & Configuration & Readout channels & $\theta$ coverage \\%& Performance\\
		\hline
		Beam pipe & Beryllium& & Cylindrical, inner radius 10 mm, 10 $\mu$m Au, 0.6 mm Be,
		1 mm paraffin, 0.4 mm Be \\
		\hline
		Tracking 	& PXD& Silicon Pixel (DEPFET)& Sensor size: 15$\times$(L1 136, L2 170) mm$^2$, Pixel size: 50$\times$(L1a 50, L1b 60, L2a 75, L2b 85) $\mu$m$^2$; two layers at radii: 14, 22 mm & 10M & [17$^\circ$;150$^\circ$]\\ 
		& SVD& Silicon Strip& Rectangular and trapezoidal, strip pitch: 50(p)/160(n) - 75(p)/240(n) $\mu$m, with one floating intermediate strip; four layers at radii: 38, 80, 115, 140 mm  & 245k& [17$^\circ$;150$^\circ$]\\  
		& CDC& Drift Chamber with He-C$_2$H$_6$ gas& 14336 wires in 56 layers, inner radius of 160mm outer radius of 1130 mm & 14k & [17$^\circ$;150$^\circ$]\\ 
		\hline
		Particle ID & TOP& RICH with quartz radiator &  16 segments in $\phi$ at $r\sim120$ cm, 275 cm long, 2cm thick quartz bars with 4$\times$4 channel MCP PMTs& 8k& [31$^\circ$;128$^\circ$] \\
		& ARICH& RICH with aerogel radiator & $2\times$2 cm thick focusing radiators with different $n$, HAPD photodetectors & 78k & [14$^\circ$;30$^\circ$]\\
		\hline
		Calorimetry& ECL&  CsI(Tl)& Barrel: $r=125-162$cm, end-cap: $z=-102-+196$cm & 6624~(Barrel), 1152~(FWD), 960~(BWD) & [12.4$^\circ$;31.4$^\circ$], [32.2$^\circ$;128.7$^\circ$], [130.7$^\circ$;155.1$^\circ$]\\% & $\frac{\sigma E}{E}=\frac{0.2\%}{E}\oplus\frac{1.6\%}{\sqrt[4]{E}}\oplus1.2\%\sim 1.7\%$\\
		\hline
		Muon ID & KLM& barrel:RPCs and scintillator strips 	& 2 layers with scintillator strips  and 12 layers with 2 RPCs& $\theta$ 16k, $\phi$ 16k & [40$^\circ$;129$^\circ$] \\
		               & KLM& end-cap: scintillator strips 	& 12 layers of (7-10)$\times$40 mm$^2$ strips & 17k & [25$^\circ$;40$^\circ$], [129$^\circ$;155$^\circ$]\\		
		%\hline
		%Trigger \\
		\hline
	\end{tabular}
\end{sidewaystable}
%\end{landscape}

SuperKEKB and \belletwo\ were initially commissioned with colliding beams in 2018 without the Vertex Detector VXD (Phase 2 operation). After establishing that the level of machine background was acceptable,  at the end of 2018 the VXD was installed, but only with 2 out of 12 ladders in the PXD second layer, because of delays in the construction of the ladders. 

In spring 2019 accelerator and detector operation started in earnest (Phase 3 operation), leading  SuperKEKB to surpass the highest recorded instantaneous luminosities of the previoius B factories and of the LHC, obtaining a peak instantaneous luminosity of $\mathcal{L}_{\rm peak}$ = \SI{4.65 e 34}{cm^{-2}s^{-1}} in June 2022. 

\begin{figure}[htb]
	\centering
	\includegraphics[width=0.9\linewidth]{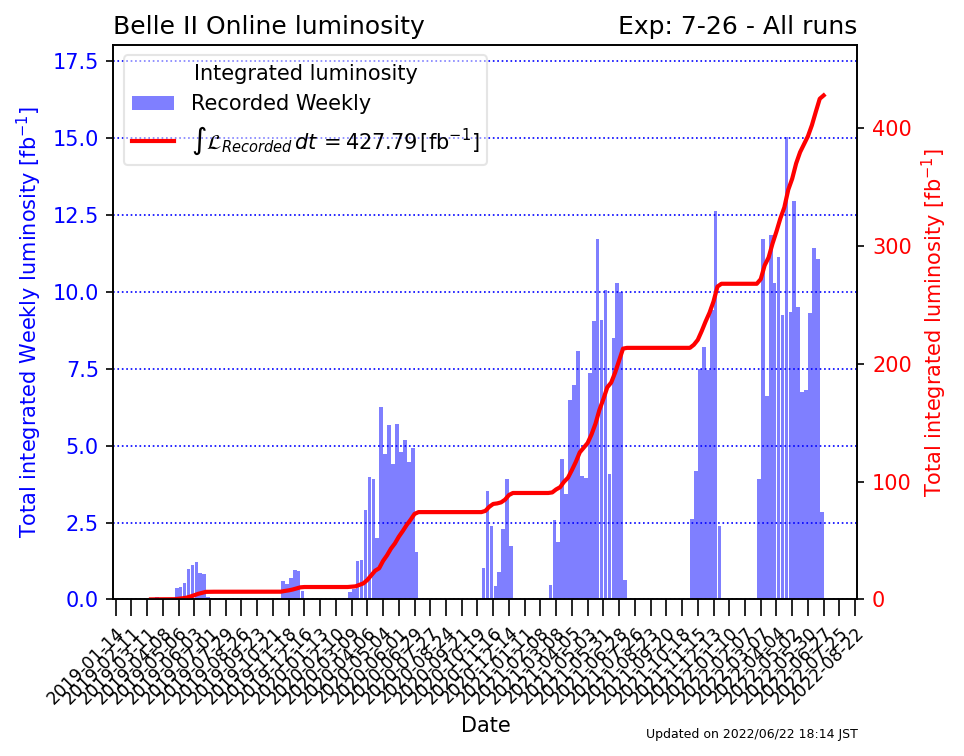}
	\caption{Belle II integrated luminosity in the first run period 2019 - 2022.}
	\label{fig:lumionlineweeklyexp7-26}
\end{figure}

In summer 2022 the accelerator and detector operations were suspended for the Long Shutdown 1 (LS1), used to perform a large number of technical improvements, both on the accelerator side and the detector side, including the installation of a complete PXD detector. As shown in figure~\ref{fig:lumionlineweeklyexp7-26}, in the first run period 2019--2022 a total integrated luminosity of \qty{427}{\fb^{-1}} was collected, of which \qty{424}{\fb^{-1}} in good runs, allowing the publication of numerous measurements~\cite{b2-publications}. 
\belletwo\ is an international collaboration of 1000 members at more than 100 institutions in 28 countries/regions. 

The current program foresees the accumulation of an integrated luminosity of  $\mathcal{L}_{\rm int}$ = \SI{50}{ab^{-1}} by the first years of the next decade. 

\begin{figure}[htb]
	\centering
	\includegraphics[width=0.95\linewidth]{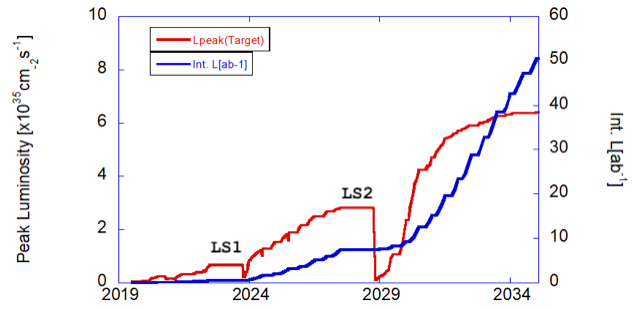}
	\caption{Projected luminosity for SuperKEKB.}
	\label{fig:lumiproject220220610}
\end{figure}

SuperKEKB is expected to be able to reach 
$\mathcal{L}_{\rm peak}$ of \SI{2 e 35}{cm^{-2}s^{-1}} 
with the existing accelerator complex. 
However, in order to reach the target luminosity of \SI{6 e 35}{cm^{-2}s^{-1}}, an upgrade of the interaction region 
and the QCS superconducting final focus will  be required, as discussed in section~\ref{sec:SKB}.
 A second Long Shutdown (LS2) in a time frame after 2027--2028 will be required to install  accelerator and detector upgrades, although it is not clear at this time whether  the  interaction region can be redesigned in that time scale. A notional projection of the integrated luminosity for the coming years is shown in figure~\ref{fig:lumiproject220220610}. 
An International Task Force has been setup to provide advice to SuperKEKB on the possible technical solutions~\cite{ref:SKB-ITF}.  

The LS2 and the possibility of the SuperKEKB interaction region redesign provide both a window of opportunity for the installation of detector improvements in many subsystems, and the requirement for a vertex detector replacement, since even small changes in the interaction region would make it impossible to reuse the current VXD. Even without an interaction region redesign, upgrading the VXD would provide better physics performance as well as creating a safety net in case the current detector gets damaged.

In section~\ref{sec:upgrades}  the detector upgrade options are briefly summarized, while the rest of the document provides a detailed description of the proposed upgrades and their status of development.  It should be recognized that at this time significant R\&D is still required to arrive at a final design, and that there are still large uncertainties on the time frame and exact path of the upgrade project. 

The detailed design of the detector upgrade will be prepared and documented in a future Technical Design Reports, to be prepared once the plans for the SuperKEKB upgrade are clarified.
%\clearpage
\section{\belletwo\ upgrades overview}
\label{sec:upgrades}

Soon after the construction of the \belletwo\ detector was completed (at the end of 2018) an Upgrade Working Group (UWG) was established by the collaboration, with the charge of examining the limits of the current detector and develop upgrade possibilities. 
Both short and long term options were to be considered, in order to ensure successful completion of the existing physics program, and longer term options to cope with possible higher luminosity upgrades.

As \belletwo\ started full operation in 2019, it became clear, not surprisingly, that machine background plays an important role in the safety, efficiency, and performance of data taking. SuperKEKB is a new and challenging accelerator with a relatively narrow operating margin between high luminosity and high backgrounds. It was realized that reaching the design \SI{8 e 35}{cm^{-2}s^{-1}}  luminosity was probably too ambitious and an intermediate target of  \SI{6 e 35}{cm^{-2}s^{-1}} was set, that in any case required some machine improvement program with the possibility of an interaction region redesign. 

In summer 2020 a MEXT roadmap 2020 was defined in order to reach the luminosity target of \SI{50}{ab^{-1}}, calling for two long shutdowns in the lifetime of the experiment: LS1 in 2022 for the replacement of PXD, and LS2 in 2026 or later for the upgrade of the machine elements and the interaction region. LS2 provides a window of opportunity for a significant detector upgrade, possibly improving robustness against machine backgrounds and physics performance.

The UWG has requested the submission of Expression of Interests from the various detector groups documenting upgrade ideas, mainly focused on the LS2 time scale, but also considering longer time scales and future ugprade paths.
These internal EOIs were collected in February 2021 and reviewed by an internal Upgrade Advisory Committee formed at the end of 2020 to provide recommendations on the upgrade process.  As a result of this internal review the initial EOIs were adjusted and somewhat reduced in scope, resulting in  the submission of a 
\belletwo\ Upgrades White Paper to the Snowmass process~\cite{Forti:2022mti}. %Snowmass whitepaper on Belle II upgrades

\belletwo\ is designed to operate efficiently even with the background levels expected at design  luminosity, but safety margins are not large. 
It should be noted that the extrapolation of the level of backgrounds to full luminosity is affected by large uncertainties due to many elements, 
such as: limited accuracy of the present parameterization; lack of modeling for the injection background; lack of a specific design for the interaction region upgrade.
The expected evolution of these backgrounds with luminosity is described in~\cite{Natochii:2022vcs} and~\cite{NATOCHII2023168550}. 
The P5 report~\cite{ref:p5report} considers the Belle II and SuperKEKB upgrades as key elements to pursue the quantum imprints of new phenomena and recommends supporting the program.

The upgrade program is therefore motivated by a number of considerations:
\begin{itemize}
	\item Improve detector robustness against backgrounds
	\item  Provide larger safety factors for running at higher luminosity 
	\item Increase longer term subdetector radiation resistance
	\item Develop the technology to cope with different future paths, for instance if a major interaction region (IR) redesign is required to reach the target luminosity
	\item Improve overall physics performance
\end{itemize}

The  upgrade activities can be classified over three different time scales:
\begin{enumerate}
\item {\bf short term}:
Long Shutdown 1 (LS1) was started in July 2022, and concluded at the end of 2023, with the main purpose of installing a complete pixel detector (PXD2). Many other minimal improvements on other subsystems were also  implemented. 

\item {\bf medium term}:
Long Shutdown 2 (LS2), in a time frame after 2027--2028, will probably be needed for the upgrade of the interaction region to reach $\mathcal{L}_{\rm peak}$ = \SI{6 e 35}{cm^{-2}s^{-1}}. A new vertex detector might be required to accommodate the new IR design, and other sub-detector upgrades are possible, with various options being evaluated. 

\item {\bf long term}: 
Studies have started to explore upgrades beyond the currently planned program, such as beam polarization and ultra-high luminosity, possibly $\mathcal{L}_{\rm peak}$ in excess of \SI{1 e 36}{cm^{-2}s^{-1}}. While the beam polarization has a concrete proposal, for ultra-high luminosity studies have just started. 
\end{enumerate}

LS1 was completed in December 2023, with a number of activities carried out: 
 
\begin{itemize}
	\item{\textbf{VXD}:} Extraction and re-installation of the VXD, with  the complete PXD2 detector.
	\item {\textbf{CDC}:} Improvement in gas circulation, monitoring and modification of the high voltage resistor chain.
	\item {\textbf{TOP}:} MCP-PMT replacement with ALD life extended PMTs
	\item {\textbf{ECL}:} Improvement in pedestal correction.  Gain adjustment in the shaper DSP
	\item {\textbf{KLM}:} Efficiency recovery of some damaged chambers. Reinforcement of monitoring system
	\item {\textbf{TRG}:} Optimization of trigger veto. 
	\item {\textbf{DAQ}:} New readout board PCIe40 long-term stability test with realistic high-occupancy data.
	\item {\textbf{MDI}}: Installation of additional loss monitors and speed-up of abort signal, as well as of additional shields against neutrons and electromagnetic showers. 
\end{itemize}

Table \ref{tab:upgrades} shows the current ideas for the medium and long term Belle II detector upgrade, that evolved from the Snowmass White paper, reducing the number of options and focusing on a concrete path as much as possible.

In the following sections a more detailed description of the various options is provided. Some approximate cost estimates are also indicated. 
The simulations reported in the text, if not specified otherwise, have used the Belle II Analysis Software Framework basf2~\cite{Kuhr:2018lps,basf2-zenodo}.

%\begin{landscape}
\begin{sidewaystable}
	\centering
	\caption{Known short and medium-term Belle II subdetector upgrade plans, sorted by time scale.  MDI is the Machine-Detector-Interface, while RMBA is Radiation Monitoring and Beam Abort system. Moving from inner to outer radius, the current Belle II sub-detectors are: Silicon Pixel Detector (PXD), Silicon Strip Detector (SVD), forming the VerteX Detector (VXD), Central Drift Chamber (CDC),  Time of Propagation Counter (TOP), Aerogel Rich Counter (ARICH), Electromagnetic Calorimeter (ECL),  K-Long Muon System (KLM), Trigger and Data aquistion (TRG/DAQ), including the High Level Trigger (HLT).}
	\label{tab:upgrades}
	\begin{tabular} {llll}
	\toprule
	Subdetector   &	Function 		& upgrade  activity	        & time scale	 \\
	\hline
	MDI & RMBA & Faster and more performant electronics & medium-term \\
	\hline
   VXD	&	Vertex Detector	& all-pixels DMAPS CMOS sensors (VTX) & medium-term \\
%		    	&					& all-pixels: SOI sensors option & medium-term  \\ 
	\hline
	CDC		&	Tracking		& upgrade front end electronics     & short/medium-term \\
	\hline
	TOP   	& PID, barrel		& Replace not-life-extended ALD MCP-PMTs & medium-term \\
			   	&   	& Front end electronics upgrade  & medium-term \\
                    &       & Replace PMTs with SiPMs & long-term \\
	\hline
	KLM		& $K_{L}$, $\mu$ ID	& replace 13 barrel layers of legacy RPCs with scintillators & medium/long-term\\
			&					& upgrade of electronics readout and proportional mode RPC readout & medium/long-term\\
			&					& timing upgrade for K-long momentum measurement & medium/long-term\\
	\hline
	Trigger	& 					& hardware  and firmware improvements & continuous\\
	\hline
	DAQ		&					& add 1300-1900 cores to HLT& short/medium-term\\
	\hline
	ARICH  	& PID, forward	&  replace HAPD with Silicon PhotoMultipliers & long-term \\
	&				&  replace HAPD with Large Area Picosecond Photodetectors & long-term\\
	\hline
	ECL 	&  $\gamma$, $e$ ID	& Add pre-shower detector in front of ECL & long-term\\
	& 					& Complement ECL PiN diodes with APDs or SiPM & long-term\\
	& 					& Replace CsI(Tl) with pure CsI crystals & long-term\\
	\bottomrule
	\end{tabular}
\end{sidewaystable}
%\end{landscape}

\chapter{SuperKEKB}
\label{sec:SKB}
%%%
\section{SuperKEKB status}
\editor{Y.~Ohnishi, M.~Masuzawa}

\begin{figure}
    \centering
    \includegraphics[width=0.6\linewidth]{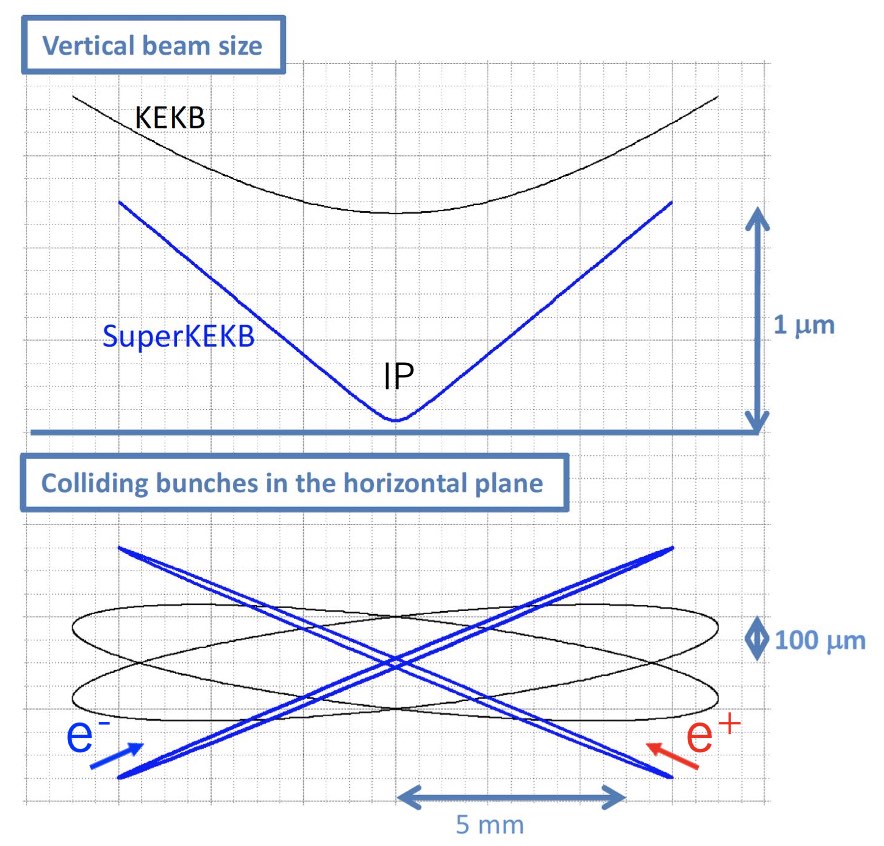}
    \caption{Schematic view of the nanobeam collision scheme.}
    \label{fig:NanoBeamScheme}
\end{figure}

SuperKEKB is an energy asymmetry electron-positron collider aiming for a peak luminosity of 
%6 $\times$ 10$^{35}$\,cm$^{-2}$s$^{-1}$, 
\SI{6E35}{\per\square\cm \per\second}, 
the highest luminosity ever achieved by a collider, and an integral luminosity of \SI{50}{\per\atto\barn}. The adoption of a collision method based on the nano-beam scheme\cite{ref:cw1} is expected to open up unexplored areas of luminosity. In the nano-beam scheme, sufficiently thin bunches collide at a large horizontal crossing angle. The length of the region where bunches overlap in the direction of travel ($\sigma_{z,eff}$) can be sufficiently shortened compared to the bunch length, as schematically shown in Fig.\,\ref{fig:NanoBeamScheme}. This makes it possible to reduce the vertical beta function at the interaction point (IP),  $\beta_y^*$, to a value as small as $\sigma_{z,eff}$, and a luminosity increase is expected. In SuperKEKB, $\beta_y^*$ is squeezed down to about \SI{0.3}{mm}, equivalent to about 1/20 of that of KEKB, to achieve the target luminosity.  

\begin{figure}
    \centering
    \includegraphics[width=0.95\linewidth]{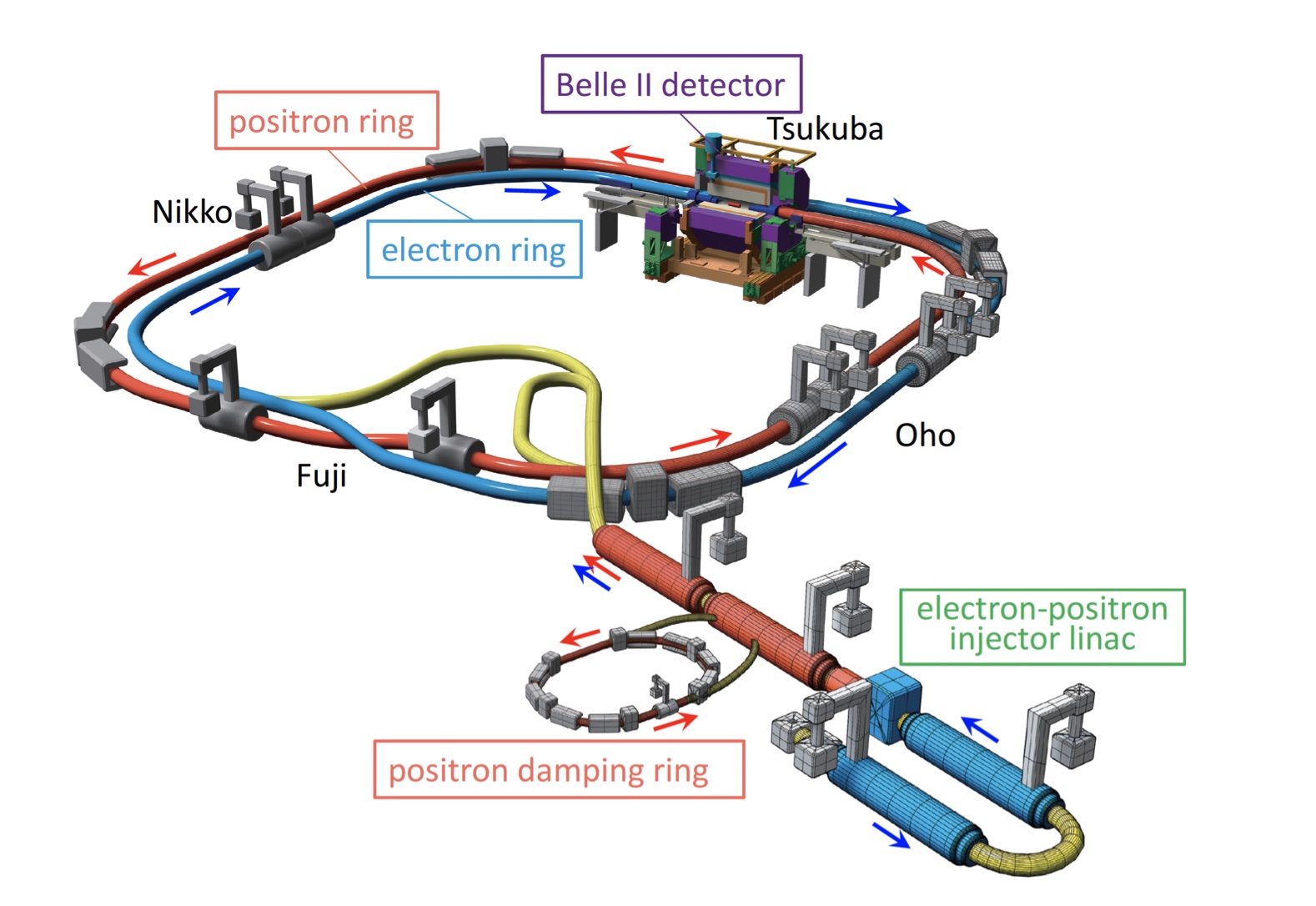}
    \caption{Schematic view of SuperKEKB Main Ring. The electron and positron rings have four straight sections named Tsukuba, Oho, Fuji, and Nikko. The electron and positron beams collide at the interaction point in the Tsukuba straight section.}
    \label{fig:superKEKB}
\end{figure}

The SuperKEKB collider complex consists of a 7-GeV electron ring (the high-energy ring, HER), a 4-GeV positron ring (the low-energy ring, LER), and an injector linear accelerator (LINAC) with a 1.1-GeV positron damping ring (DR), as shown in  Fig.\,\ref{fig:superKEKB}. The extremely high luminosity of SuperKEKB required significant upgrades to the HER, LER, and final-focus system of KEKB. The injector linac also required significant upgrades for injection beams with high current and low emittance. A new RF-gun was developed and installed, and a new DR was designed and constructed to realize low-emittance electron and positron injection to the SuperKEKB Main Ring (MR).

\begin{figure}[ht!]
	\centering
	\includegraphics*[width=0.98\columnwidth]{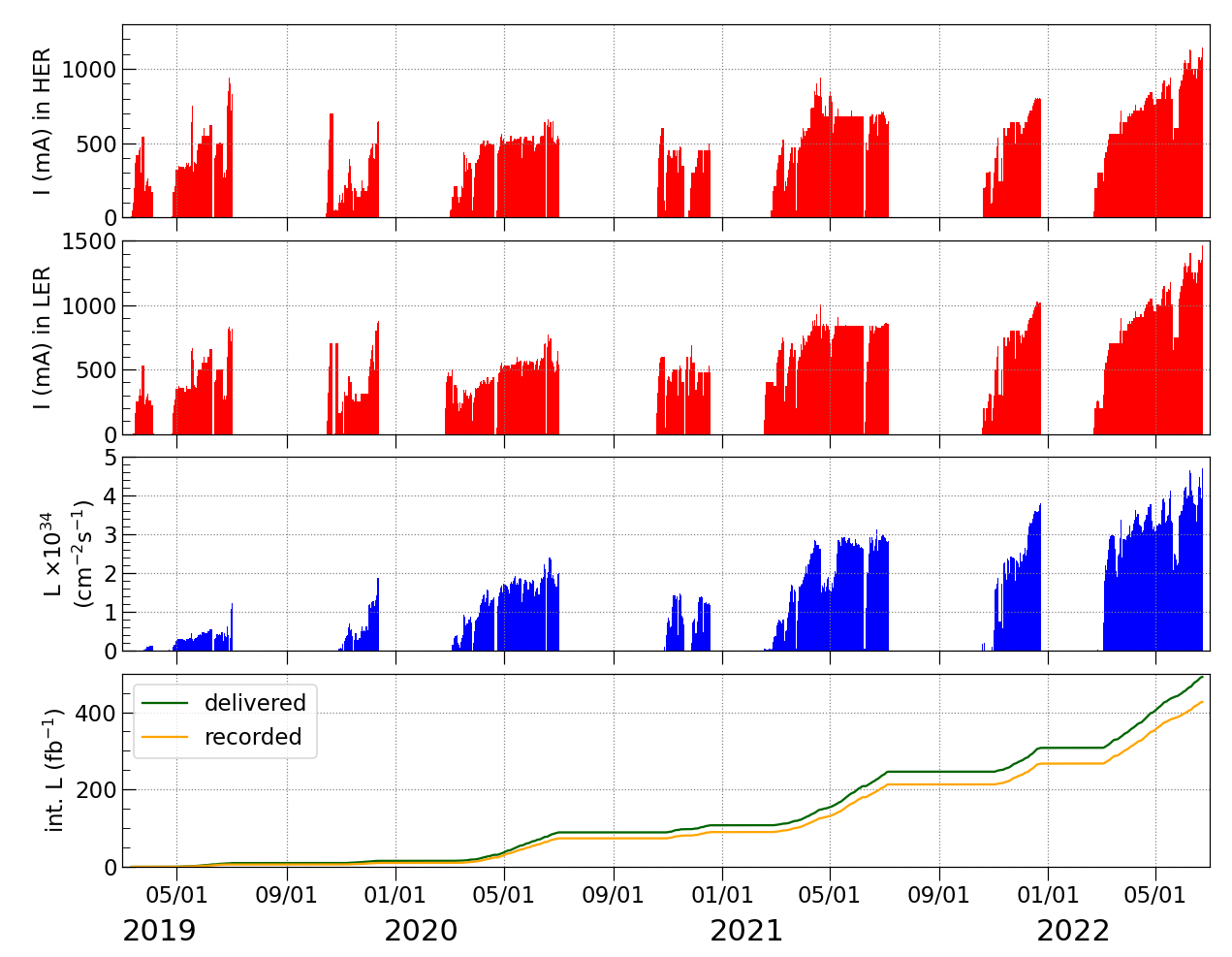}
	\caption{
	Operation history from 2019 to 2022: beam current in the HER, LER, daily luminosity, and integrated luminosity. 
	}
	\label{fig:historyNew}
\end{figure}

Phase 1 commissioning was carried out in 2016 without beam collisions. After that, the interaction region (IR) was completed by installing the final focusing superconducting quadrupole magnet system and the Belle II detector (without VXD), so that the Phase 2 commissioning could start in March 2018 with collision tuning. The first hadron event was observed by Belle II on April 26 of the same year. Phase 3 commissioning started in March 2019 with the fully instrumented Belle II detector.  A summary of the operation since 2019 is shown in Fig.\,\ref{fig:historyNew}.

\begin{figure}
    \centering
    \includegraphics[width=0.98\linewidth]{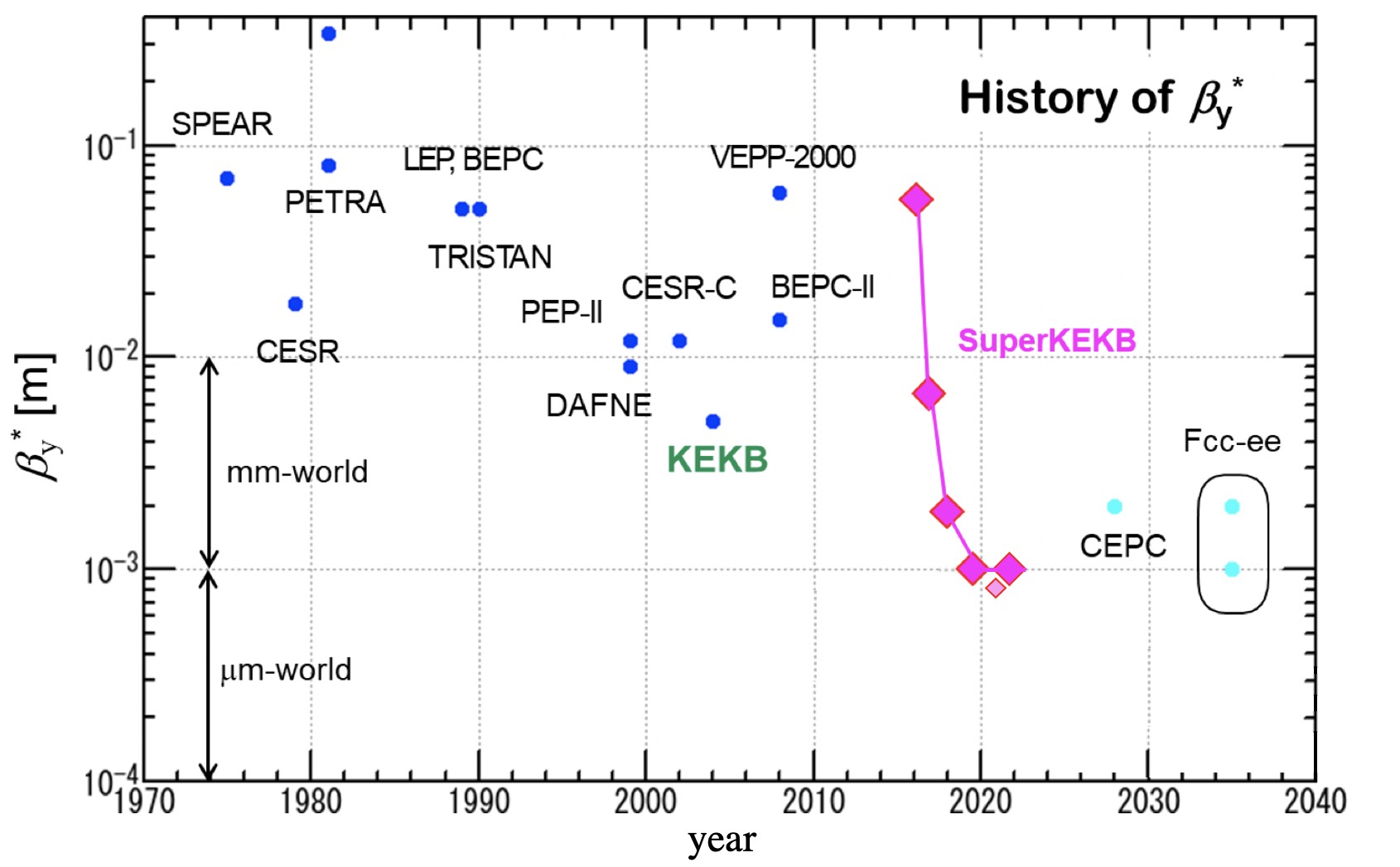}
    \caption{History of  $\beta_y^*$}
    \label{fig:betay}
\end{figure}

Figure~\ref{fig:betay}  shows the evolution of $\beta_y^*$ together with reference to values at other accelerators. By carefully looking at the relationship between the injection efficiency and the background in the detector and adjusting the collimator opening, we were able to perform the first collision experiment at $\beta_y^*=\SI{1.0}{mm}$ during the 2019 run. A test of squeezing $\beta_y^*$ down to \SI{0.8}{mm} was carried out for approximately one week in 2020 and 2022.

The crab waist (CW) scheme \cite{ref:cw1,ref:cw2} was adopted for both rings since the 2020 spring run.
 The CW ratio, the ratio to the magnetic field required to align the waist of the electron (positron) beam on the positron (electron) orbit, is at this point 80\% in the LER and 40\% in the HER. 

A peak luminosity of 
%4.65 $\times$ 10$^{34}$\,cm$^{-2}$s$^{-1}$ 
\SI{4.65E34}{\per\square\cm \per\second} 
was achieved 
with $\beta_y^*=\SI{1.0}{mm}$ in the physics run, 
and 
%4.71 $\times$ 10$^{34}$\,cm$^{-2}$s$^{-1}$ 
\SI{4.71E34}{\per\square\cm \per\second} 
is the highest luminosity obtained without data acquisition by the Belle II detector up to now. This peak luminosity is a world record in collider history.

The machine parameters that accomplished the peak luminosity are presented in Table\,\ref{tab:AchievedParams}
together with those valid for $\beta_y^*=\SI{0.8}{mm}$.
\begin{table}[ht!]
	\centering
	\caption{Machine Parameters achieved at SuperKEKB}
	\begin{tabular}{lcccc}
		\toprule
		& \multicolumn{2}{c}{June 8, 2022} & \multicolumn{2}{c}{May 22, 2022} \\
		\textbf{Parameters} & \textbf{LER} & \textbf{HER} & \textbf{LER} & \textbf{HER} \\
		\midrule
			$I$ (\si{A}) & \num{1.321} & \num{1.099} & \num{0.744} & \num{0.600}        \\
			$n_b$   & \multicolumn{2}{c}{\num{2249}} & \multicolumn{2}{c}{\num{1565}} \\
			$I_b$ (\si{mA}) & \num{0.587} & \num{0.489} & \num{0.475} & \num{0.383} \\
			$\beta_x^*$ (\si{mm}) & \num{80} & \num{60} & \num{80} & \num{60} \\
			$\beta_y^*$ (\si{mm})   & \multicolumn{2}{c}{\num{1.0}} & \multicolumn{2}{c}{\num{0.8}} \\
			$\xi_y$   & \num{4.07E-2} & \num{2.79E-2} & \num{3.09E-2} & \num{2.19E-2} \\
			$\varepsilon_y$ (\si{pm}) & \multicolumn{2}{c}{\num{31.7}} & \multicolumn{2}{c}{\num{29.6}} \\
			$\Sigma_y^*$ (\si{\um}) & \multicolumn{2}{c}{\num{0.252}} & \multicolumn{2}{c}{\num{0.218}} \\
			$\sigma_z$ (\si{mm}) & \num{5.69} & \num{6.02} & \num{5.40} & \num{5.78} \\
			CW ratio (\%) & \num{80} & \num{40} & \num{80} & \num{40} \\
			$L$ (\si{\per\square\cm \per\second}) & \multicolumn{2}{c}{\num{4.65E34}} & \multicolumn{2}{c}{\num{2.49E34}} \\
			\bottomrule
	\end{tabular}
	\label{tab:AchievedParams}
 \end{table}
The luminosity formula in the nano-beam scheme is expressed as
\begin{eqnarray}\label{eq:lum}
	L=\frac{N_+N_-n_bf_0}{2\pi\phi_x\Sigma_z\Sigma_y^*},
\end{eqnarray}
where $\pm$ indicates positrons (+) or electrons (-), $N$ is the number of particles in a bunch, $n_b$ is the number of bunches, $f_0$ is the revolution frequency, and $\phi_x$ is the half crossing angle. 
$\Sigma_z$ and $\Sigma_y^*$ are defined by 
\begin{eqnarray}\label{eq:capsigma}
	\Sigma_z = \sqrt{\sigma_{z+}^2+\sigma_{z-}^2}~~~~~~~
	\Sigma_y^* = \sqrt{\sigma_{y+}^{*2}+\sigma_{y-}^{*2}},
\end{eqnarray}
where $\sigma_{y+}$ and $\sigma_{y-}$ indicate the vertical beam sizes of the positron and electron beams at the IP, respectively.
In the case of the nano-beam scheme, a geometrical luminosity loss due to an hourglass effect \cite{ref:hourglass} is sufficiently small, even if the beta function at the IP is squeezed down to the bunch length divided by the Piwinski angle, and is omitted here. 
The vertical beam--beam parameter is expressed by 
\begin{eqnarray}\label{eq:beam-beam}
	\xi_{y\pm} \simeq \frac{r_e N_\mp}{2\pi\gamma_\pm}\frac{1}{\phi_x\sigma_z}
	\sqrt{\frac{\beta_y^*}{\varepsilon_y}},
\end{eqnarray}
where $r_e$ is the classical electron radius, $\gamma$ is the Lorentz factor, and $\varepsilon_y$ is the vertical emittance.
Here, the beam spot-size at the IP between the LER and HER is assumed to be equal. Alternatively, the vertical beam--beam parameter is expressed 
using Eq.\,\eqref{eq:lum} and Eq.\,\eqref{eq:beam-beam} as
\begin{eqnarray}\label{eq:beam-beamL}
	\xi_{y\pm}=2er_e\frac{L\beta_y^*}{\gamma_\pm I_\pm},	
\end{eqnarray}
where $e$ is the elementary charge, $I_{+}$ and $I_{-}$ are LER beam current and HER beam current, respectively.
Therefore luminosity is proportional to $\xi_{y\pm}$ and $I_\pm$ and inversely proportional to $\beta_y^*$, as is shown by Eq.\,\eqref{eq:lumP}.
\begin{eqnarray}\label{eq:lumP}
    L \propto \frac{I_\pm\xi_{y\pm}}{\beta_y^*} 	
\end{eqnarray}

As the beam-beam parameter increases, the collision probability increases, but so does the instability. How high this value can be raised is the key to increasing luminosity.

\begin{figure}[htb]
	\centering
	\includegraphics*[width=0.98\columnwidth]{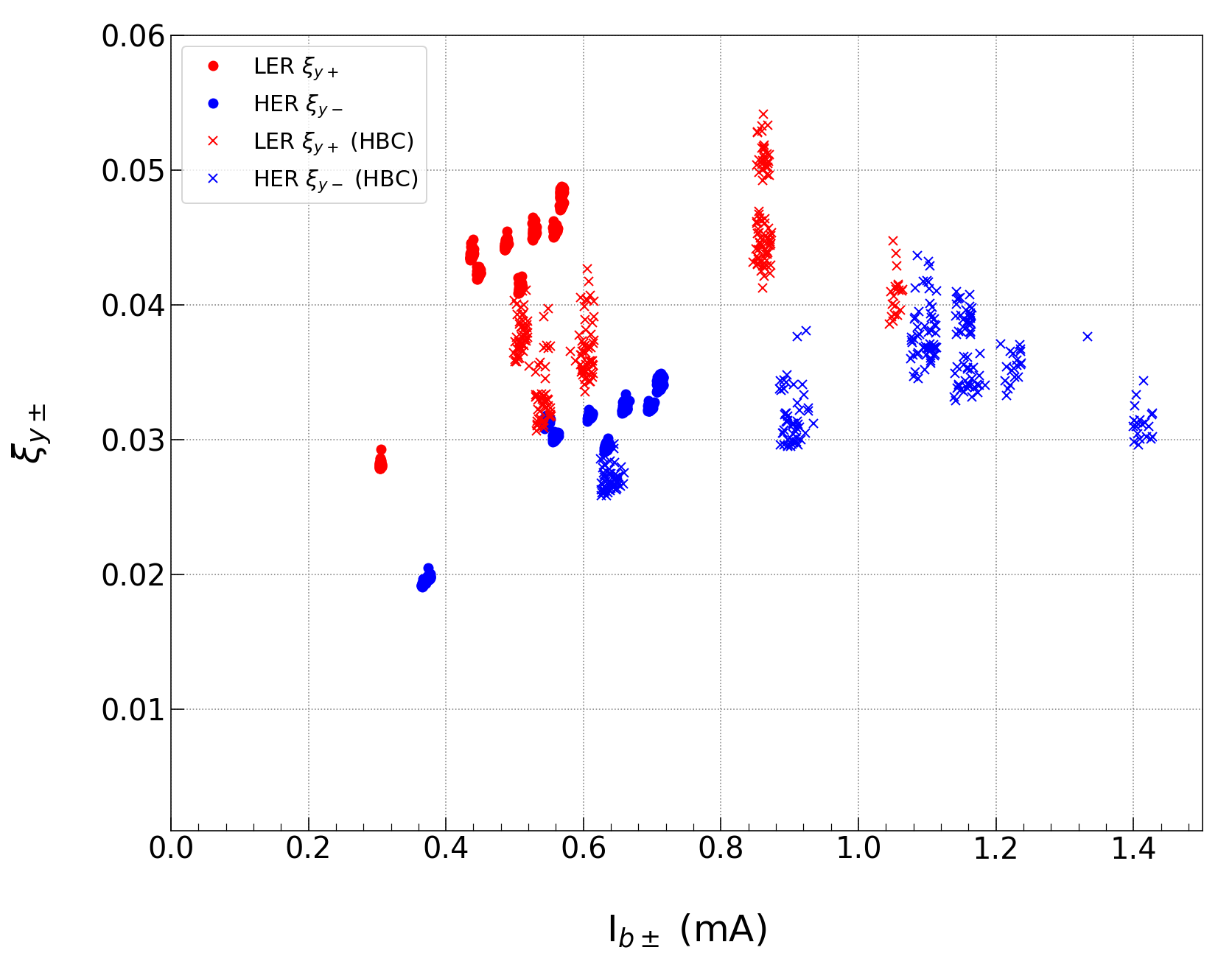}
	\caption{Beam--beam parameters as a function of the bunch current. 
	The positrons in the LER are indicated by red, whereas blue is used for the electrons in the HER. 
	The crosses are data from the high bunch current (HBC) test.}
	\label{fig:beam-beam}
\end{figure}

Figure\,\ref{fig:beam-beam} shows the vertical beam--beam parameter 
defined by Eq.\,\eqref{eq:beam-beamL} for $\beta_y^*=\SI{1.0}{mm}$.
When the ratio of beam currents in the LER to that of the HER is 7 to 4,
it becomes the energy transparent condition, $\gamma_+I_+=\gamma_-I_-$. 
However, the ratio of bunch current in the LER to that of the HER is 5 to 4 
in the physics run, which is the optimized value based on the operational experiences.
The maximum beam--beam parameter of 0.0565 (LER)/0.0434 (HER) 
at the bunch current of \SI{1.1}{mA} in the LER was achieved 
by the ratio of 4 to 3 in the machine study called high bunch current (HBC) study.
The specific luminosity is defined as the total luminosity divided by the number of bunches and by the bunch current product as Eq.\,\eqref{eq:splum}.

\begin{eqnarray}\label{eq:splum}
	L_{sp}=\frac{L}{I_{b+}I_{b-}n_b}\propto\frac{1}{\Sigma_z\Sigma_y^*},
\end{eqnarray}
where $I_b$ is the bunch current.

\begin{figure}[ht!]
	\centering
	\includegraphics*[width=0.98\columnwidth]{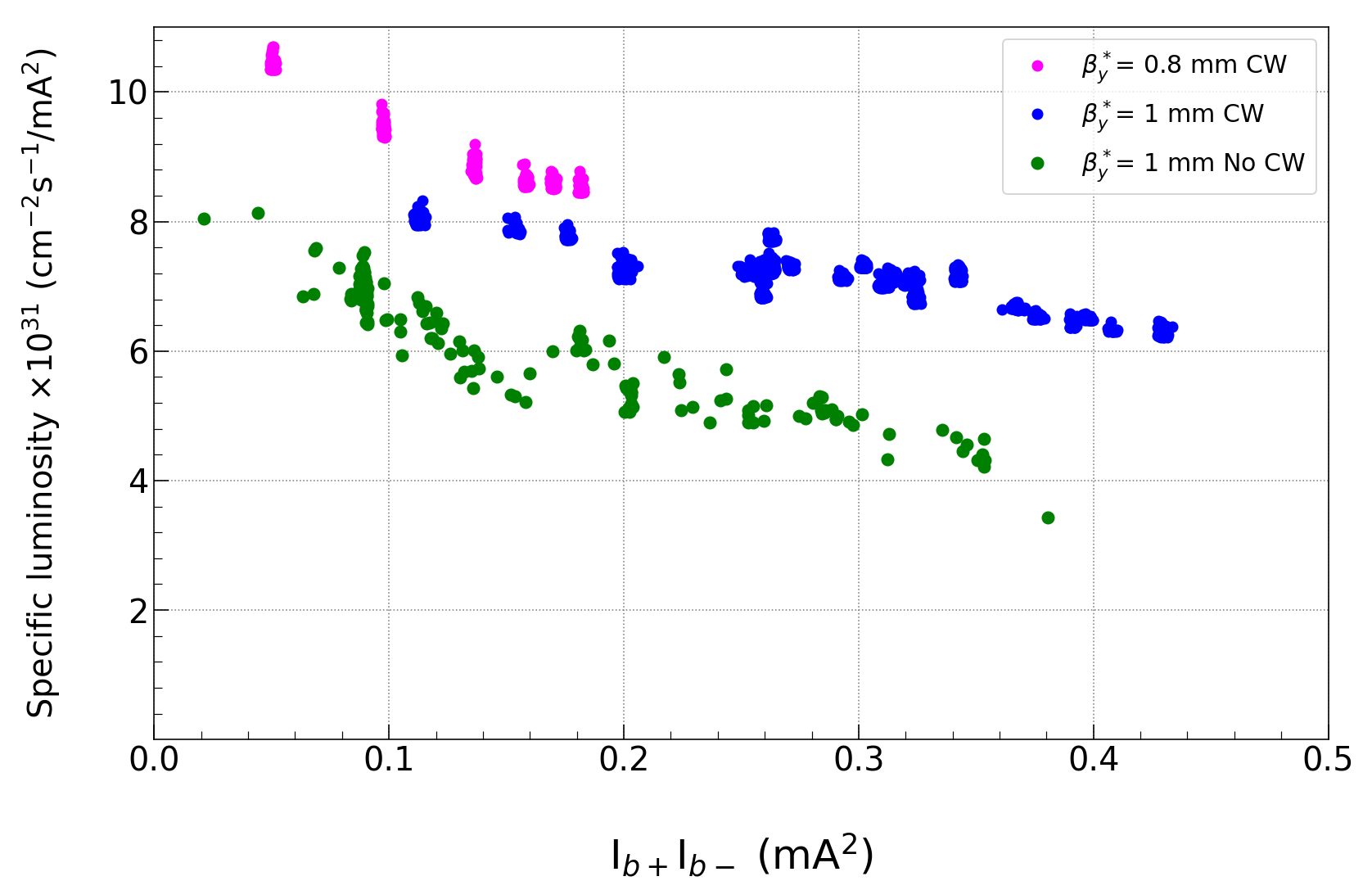}
	\caption{Specific luminosity for $\beta_y^*=\SI{0.8}{mm}$ and \SI{1}{mm}. 
	Comparison between the CW and no CW for $\beta_y^*=\SI{1}{mm}$.}
	\label{fig:splum}
\end{figure}

Figure \ref{fig:splum} shows the specific luminosity achieved during the 2022 physics run.
The specific luminosity decreased as the bunch current product increased, 
which implies that the beam--beam blowup was observed \cite{ref:dzhou}. 
The comparison between the CW and no CW for $\beta_y^*=\SI{1.0}{mm}$ is also shown in the figure.
The specific luminosity was improved by utilizing the CW scheme \cite{ref:zobov}.

The specific luminosity with $\beta_y^*=\SI{0.8}{mm}$ was higher than with $\beta_y^*=\SI{1}{mm}$  by 12\%
at \SI{0.18}{\square \mA}. 
The specific luminosity is by more than 20\% higher at a lower bunch current product of less than \SI{0.1}{mA^2}.
 The reason that the specific luminosity decreases quickly can be the chromatic X--Y couplings at the IP for $\beta_y^*=\SI{0.8}{mm}$.

In the case of $\beta_y^*=\SI{1.0}{mm}$, two of the chromatic X--Y couplings 
($\partial r_1^*/\partial\delta$ and $\partial r_2^*/\partial\delta$) 
in the LER are corrected by utilizing 24 rotating sextupole magnets \cite{ref:nakamura}.
The sextupole magnet is mounted on a rotating table to induce a skew sextupole field.
Prior to the correction of chromatic X--Y couplings, the machine error at the IP is locally corrected using IP tuning knobs such as X--Y couplings ($r_1,r_2,r_3,r_4$) 
and vertical dispersions ($\eta_y^*$, $\eta_{py}^*$) so as to maximize luminosity. 

\subsection{Impedance}

\begin{figure}[ht!]
    \centering
    \includegraphics[width=0.8\linewidth]{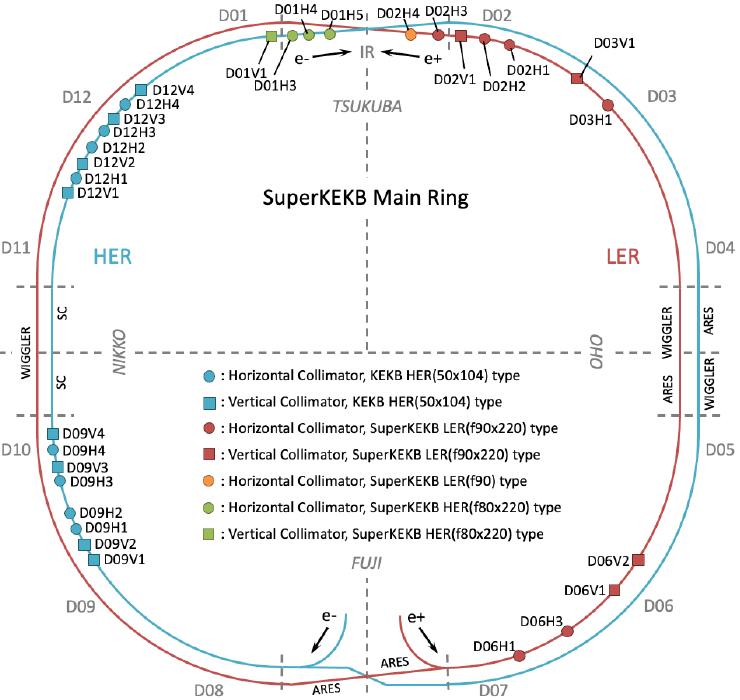}
    \caption{Collimator Map in 2022.}
    \label{fig:CollimatorMap}
\end{figure}

The collimators are important to reduce beam related backgrounds at SuperKEKB.
Movable collimators for the horizontal and vertical plane are installed in the LER and HER, respectively, as shown in Fig.\,\ref{fig:CollimatorMap} \cite{ref:ishibashi}. 

A kick factor from each collimator becomes large because the vertical half aperture of the collimator is very small, typically of approximately 
\SIrange[range-phrase = {-}]{1}{2}{mm}.
%1--2\,mm.
 Therefore, transverse mode coupling instability (TMCI) might become serious.
 The TMCI, also known as strong head-tail instability, was first observed at PETRA \cite{ref:petra}.
 For a large collider with a small chamber height and collimator aperture, the TMCI limits the performance.
 At SuperKEKB, we observed a vertical beam blowup at the bunch current of approximately \SI{0.8}{mA},
 which is much smaller than the TMCI threshold $I_\text{th}$ expressed as Eq.\,\eqref{eq:thresh}\cite{ref:ishibashi}.

\begin{eqnarray}\label{eq:thresh}
	I_\text{th}=\frac{C{f_s}E}{{\Sigma_i}{\beta_{y,i}}{k_{y,i}(\sigma_z)}}.
\end{eqnarray}

Here, $C$, $f_s$, $E$, $\beta_{y,i}$, $k_{y,i}$, and $\sigma_z$ are constants, the synchrotron frequency, beam energy, vertical beta function at a component $i$, vertical kick factor at a component $i$, and longitudinal bunch length, respectively. The expected TMCI threshold from Eq.\,\eqref{eq:thresh} is about \SI{2}{mA}, while the observed value is \SI{0.8}{mA}. We call this phenomenon ``-1 mode instability'' because the coherent-mode frequency corresponding to $\nu_y-\nu_s$ appears when the beam blowup occurs.
 Recently, it was found that tuning of the vertical bunch-by-bunch feedback system \cite{ref:bbFB} affects the -1 mode instability with an interplay of the vertical short-range wake field due to the collimators.
 Optimization of the bunch-by-bunch feedback system is necessary to suppress the -1 mode instability.
\subsection{Sudden beam loss}

A fast and large beam loss has been observed frequently 
when the bunch current increases over \SI{0.7}{mA} \cite{ref:ikeda}.
We call this fast beam loss ``sudden beam loss (SBL).''
More than half of the beam current was lost within a few turns (\SIrange[range-phrase = {-}]{20}{30}{\us}).
The SBL occurs in both rings, but it is more serious in the LER.
%(20--30\,$\mu$s).
%No large coherent and incoherent beam oscillations have been observed in the %horizontal and vertical direction before the beam loss.
When a SBL occurs, the vertical collimators become damaged and sometimes QCS quenches occur.

The mechanism of SBL is not understood yet.
%It was found that the events were located near the vertical collimators. 
The highest and earliest beam losses during the SBL event are usually detected near the narrowest vertical collimators in the LER (e.g., D06V1, see Fig.~\ref{fig:CollimatorMap}).
 A hypothesis of ``fireball'' has been proposed to explain the SBL, 
which is based on an RF cavity breakdown mechanism~\cite{ref:abe}.
In the fireball, isolated microparticles with a higher sublimation point, such as tungsten and tantalum, 
become fireballs heated by the beam-induced field.
 The fireball touches a material surface with a lower sublimation point, 
generating plasma around the fireball.
 The plasma grows and might lead to significant interactions with the beam.
 We will apply the copper-coated collimator head to mitigate the occurrence of fireballs during the operation after LS1.
\subsection{Dynamic aperture and Touschek lifetime}
The dynamic aperture (DA),  the stable region in phase space for a beam that can orbit stably, is estimated by a particle tracking simulation code such as SAD\cite{SAD2022} and compared to the measured beam lifetime.
The measured lifetime at $\beta_y^*=\SI{1.0}{mm}$ was 8\,minutes for $I_{b+}=\SI{0.62}{mA}$ with $\varepsilon_y=\SI{40}{pm}$ in the LER 
and 23\,minutes for $I_{b-}=\SI{0.5}{mA}$ with $\varepsilon_y=\SI{35}{pm}$ in the HER.
The simulation can almost reproduce the Touschek lifetime in the LER. However, the HER simulation overestimates the measured DA by a factor of 2. Consequently, a larger machine error is needed to understand the HER.
 The smaller DA will become a serious problem as one squeezes $\beta_y^*$ and increases the bunch current, as well as for beam injections~\cite{ref:iida}.

\section{SuperKEKB evolution and upgrade path}

The SuperKEKB physics run began in March 2019, with a fully instrumented Belle II detector. The crab waist scheme, which was not included in the original SuperKEKB design plan,  was implemented in April 2020 to enhance beam performance. In 2020, SuperKEKB surpassed the luminosity record of %2.11\,$\times$\,10$^{34}$\,cm$^{-2}$s$^{-1}$  
\SI{2.11E34}{\per\square\cm  \per\second}
achieved by its predecessor, KEKB, with vertical and horizontal beta functions at the IP set at \SI{1.0}{mm} and \SI{80}{mm} (LER)/\SI{60}{mm} (HER), respectively. Thanks to the nano-beam scheme, the peak luminosity record was surpassed with smaller beam currents than those used at KEKB. In June 2022, SuperKEKB achieved a new world record for a peak luminosity of %4.7\,$\times$\,10$^{34}$\,cm$^{-2}$s$^{-1}$  
\SI{4.7E34}{\per\square\cm  \per\second}
with a $\beta_y^*$ of \SI{1.0}{mm} at the collision point.

\begin{table}[!h]
	\centering
	\caption{Machine Parameters in the Two-stage Improvement}
	\begin{tabular}{lcccc}
{ } & \multicolumn{2}{c}{First stage}& \multicolumn{2}{c}{Second stage} \\
		\toprule
		\textbf{Parameters} & \textbf{LER} & \textbf{HER} & \textbf{LER} & \textbf{HER} \\
		\midrule
			I (\si{A}) & \num{2.08} & \num{1.48} & \num{2.75} & \num{2.20}        \\
			n$_b$   & \multicolumn{2}{c}{\num{2345}} & \multicolumn{2}{c}{\num{2345}} \\
			I$_b$ (\si{mA}) & \num{0.89} & \num{0.63} & \num{1.17} & \num{0.938} \\
			$\beta_y^*$ (\si{mm})   & \multicolumn{2}{c}{\num{0.8}} & \multicolumn{2}{c}{\num{0.6}} \\
			$\xi_y$   & \num{4.44E-2} & \num{3.56E-2} & \num{6.04E-2} & \num{4.31E-2} \\
			$\varepsilon_y$ (\si{pm}) & \multicolumn{2}{c}{\num{30}} & \multicolumn{2}{c}{\num{21}} \\
			$\Sigma_y^*$ (\si{\micro\meter}) & \multicolumn{2}{c}{\num{2.18E-1}} & \multicolumn{2}{c}{\num{1.60E-1}} \\
			$\sigma_z$ (\si{mm}) & \num{6.49} & \num{6.35} & \num{7.23} & \num{7.05} \\
			L (\si{\per\square\cm  \per\second}) & \multicolumn{2}{c}{\num{1E35}} & \multicolumn{2}{c}{\num{2.4E35}} \\
			\bottomrule
	\end{tabular}
	\label{tab:machineParams}
 \end{table}

 \begin{figure}
     \centering
     \includegraphics[width=0.98\linewidth]{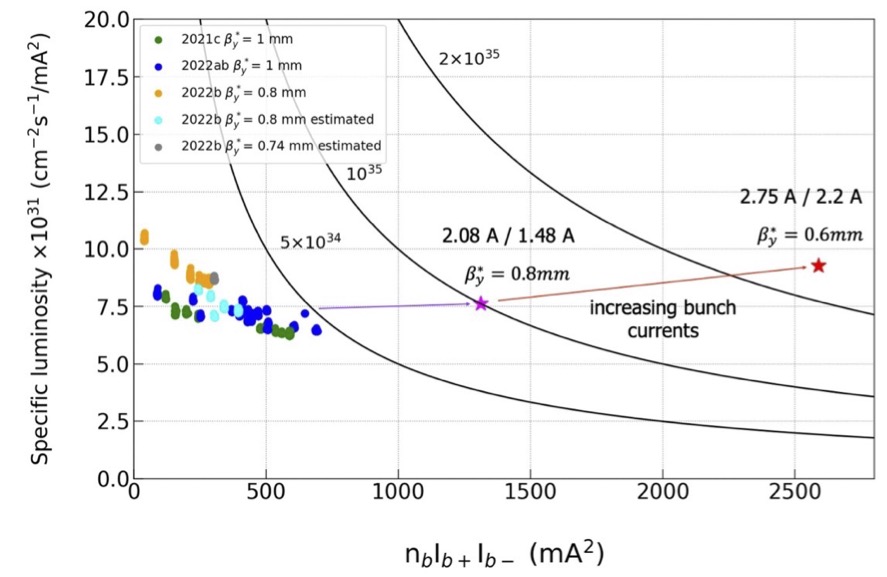}
     \caption{Achieved specific luminosity as a function of beam current product (circles) and its targeted development between LS1 and LS2.}
     \label{fig:Strategy1}
 \end{figure}
 
The operation of SuperKEKB was halted for one year and a half since the summer of 2022 to replace some components of the Belle II detector and upgrade the machine. The installation of a nonlinear collimation (NLC) scheme was one of many accelerator upgrades during LS1. The NLC is installed in the OHO section of the LER. A pair of skew sextupole magnets are installed to create an angular kick to deflect mainly halo particles. A regular vertical collimator placed between the two skew sextupole magnets can then stop deflected particles. 
The vertical collimator can then be opened much wider than typical vertical collimators. 
Since the betatron function at this collimator is small, the wakefield contribution of the NLC to the machine impedance and, hence, to the TMCI threshold is negligible. We can tighten this collimator and relax other vertical collimators to increase the TMCI limit while keeping the IR beam backgrounds at acceptable levels. 
%The ides of NLC was proposed originally in the 1990s %\cite{ref:NLC1},\cite{ref:NLC2}but SuperKEKB is the first %collider in the world to actually install NLC and test it.
 The idea of NLC was initially proposed 
in the 1990s~\cite{ref:NLC1, ref:NLC2}, but SuperKEKB is the world's first collider to install NLC and test it.

The initial target of the luminosity is 
%10$^{35}$\,cm$^{-2}$s$^{-1}$ 
\SI{1E35}{\per\square\cm \per\second}
after LS1.
By increasing beam currents and squeezing $\beta_y^*$ down to 0.6 mm while keeping the specific luminosity high, we plan to increase the luminosity up to \SI{2.4E35}{\per\square\cm \per\second} as shown in Fig.\,\ref{fig:Strategy1}
%2.4\,$\times$\,10$^{35}$\,cm$^{-2}$s$^{-1}$ 
by the second long shutdown (LS2), 
when further upgrade work is required to achieve more than %6\,$\times$\,10$^{35}$\,cm$^{-2}$s$^{-1}$. 
\SI{6E35}{\per\square\cm \per\second}.
Table\,\ref{tab:machineParams} lists the machine parameters for these milestones after the LS1.
We have achieved the designed number of bunches. 
Therefore, we attempt to increase the bunch current toward the target luminosity while suppressing beam blowup. 

 \section{Future plans and prospect}
\begin{figure}
    \centering
    \includegraphics[width=0.98\linewidth]{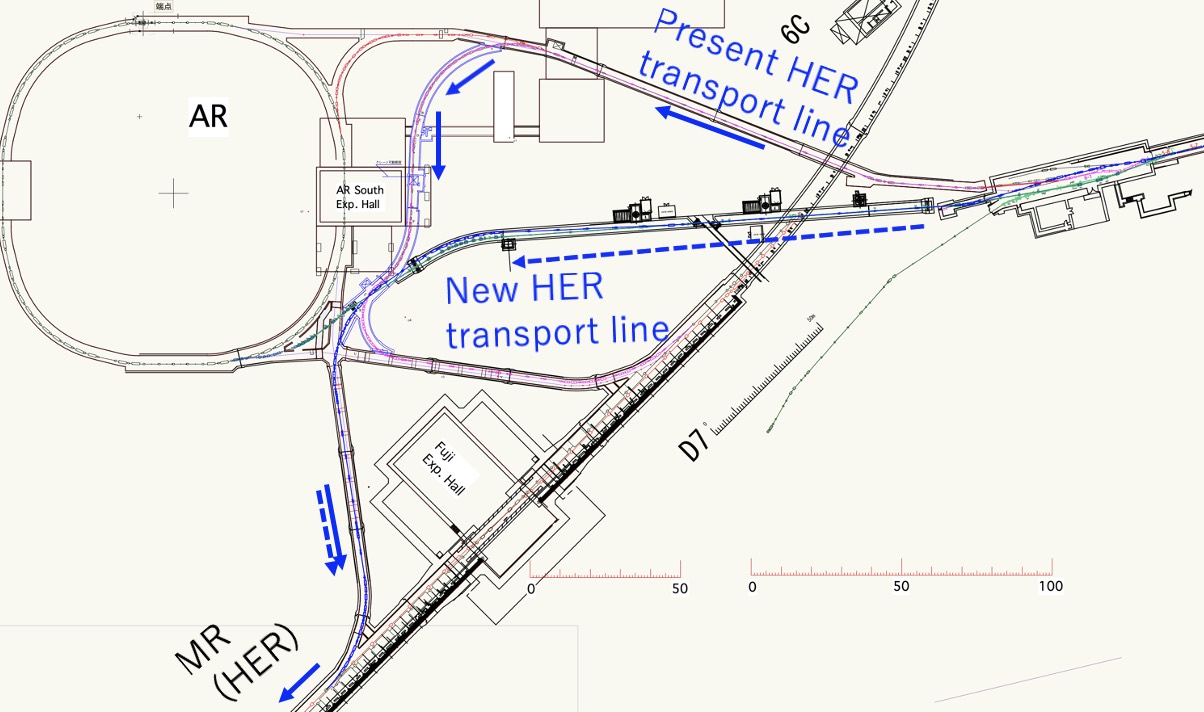}
    \caption{The new beam transport line and the present beam transport line for the HER.}
    \label{fig:NewBT}
\end{figure}

During the next long shutdown, several major modifications are considered to further improve the machine's performance. Modification of the IR, an upgrade of the injection complex, a new HER beam transport (BT) line (Fig.\,\ref{fig:NewBT}), an increase of the HER RF stations, and replacement of various aging components are among modification candidate items. A new BT line aims at keeping the emittance growth by coherent synchrotron radiation (CSR) and incoherent synchrotron radiation (ISR) under control. Additional RF stations are needed to increase the stored beam to the design current in the HER.

While discussions on when to start LS2 are underway, consideration has already begun regarding modifications during the LS2 period. The modifications must be effective enough that the integrated luminosity lost during LS2 is recovered quickly afterwards. Design and evaluation progress of the IR upgrade, which is one of the upgrade items, are described in the following subsection. 

\section{Interaction region design}
\label{sec:IRdesign}
\editor{M.~Masuzawa, N.~Ohuchi}

 It should be noted that this is not a completely new design from a blank slate. We have begun reconsidering the unfinished issues at the SuperKEKB IR design stage. We are currently considering items that could be renovated over several years.  Below is a status report about the new IR design.
 
\subsection{Introduction of the IR optics} 

The SuperKEKB interaction region (IR) is designed to achieve extremely small beta functions at the IP,
$\beta_y^*$ in the vertical direction, and $\beta_x^*$ in the horizontal direction.
 Superconducting magnets QC1s and QC2s provide the focusing magnetic field required to squeeze down
$\beta_y^*$ and $\beta_x^*$, respectively~\cite{Ohuchi:2022nima}.

The solenoid field of Belle II is fully compensated with the superconducting compensation solenoids
(anti-solenoid) on each side of the IP, therefore

\begin{align}
 \int_\text{IP}B_z(s) ds = 0,
\end{align} 
where $B_z$ is magnetic field along the solenoid axis and $s$ is the distance from the IP.
 The crossing angle between the Belle II solenoid axis and LER or HER beam axis is $\pm$41.5 mrad.  The X--Y coupling and the horizontal and vertical dispersions are corrected to zero at the IP by skew quadrupole magnets and horizontal and vertical dipole magnets on each IP side.  In the LER, the chromatic X--Y coupling arising from the solenoid field overlapping on QC1s is corrected by using rotatable sextupole magnets.

%The X--Y coupling and the horizontal and vertical dispersions arising from the coupling
%between the Belle II solenoid and the QCS anti-solenoid are corrected to be zero
%at the IP by the skew quadrupole magnets and sextupole magnet pairs at each side of the IP.
%The geometrical constraints are satisfied by adjusting the horizontal and vertical dipole magnets.

%Because the crossing angle, that is, the angle between the Belle~II solenoid axis and each beam %axis of the LER and HER is
%41.5~mrad,
%the X--Y axis rotates around the beam axis along the beam trajectory.
%&Vertical emittance $\epsilon_y$ is estimated approximately as follows:

%Vertical emittance $\epsilon_y$ is estimated approximately as follows:

%\begin{align}
%\epsilon_y\propto \left(\frac{p}{\rho}\right) ^2 \int H(s) ds \propto %B_x^4(s)
%\end{align}

The finite crossing angle between both beam orbits and the solenoid axis causes vertical dispersion.
 This results in a vertical emittance that is strongly dependent on the transverse component of the magnetic field along the orbit represented by $B_x(s)$,

\begin{align}
B_x(s)\cong -\frac{x}{2}B_z^\prime(s) = -\frac{s\phi}{2} B_z^\prime(s).
\end{align}

A half crossing angle $\phi$ or/and $B^\prime_z(s)$ should be decreased to reduce the vertical emittance.
 Because decreasing half crossing angle $\phi$  requires a complete change of the IR, we focus on modifying $B_z(s)$ such that its differential along beam trajectory $B_z^\prime(s)$ becomes smaller than that in the present system.

\subsection{Characteristics of the new interaction region} 

\begin{figure}[htb]
\centering
\includegraphics[width=\textwidth]{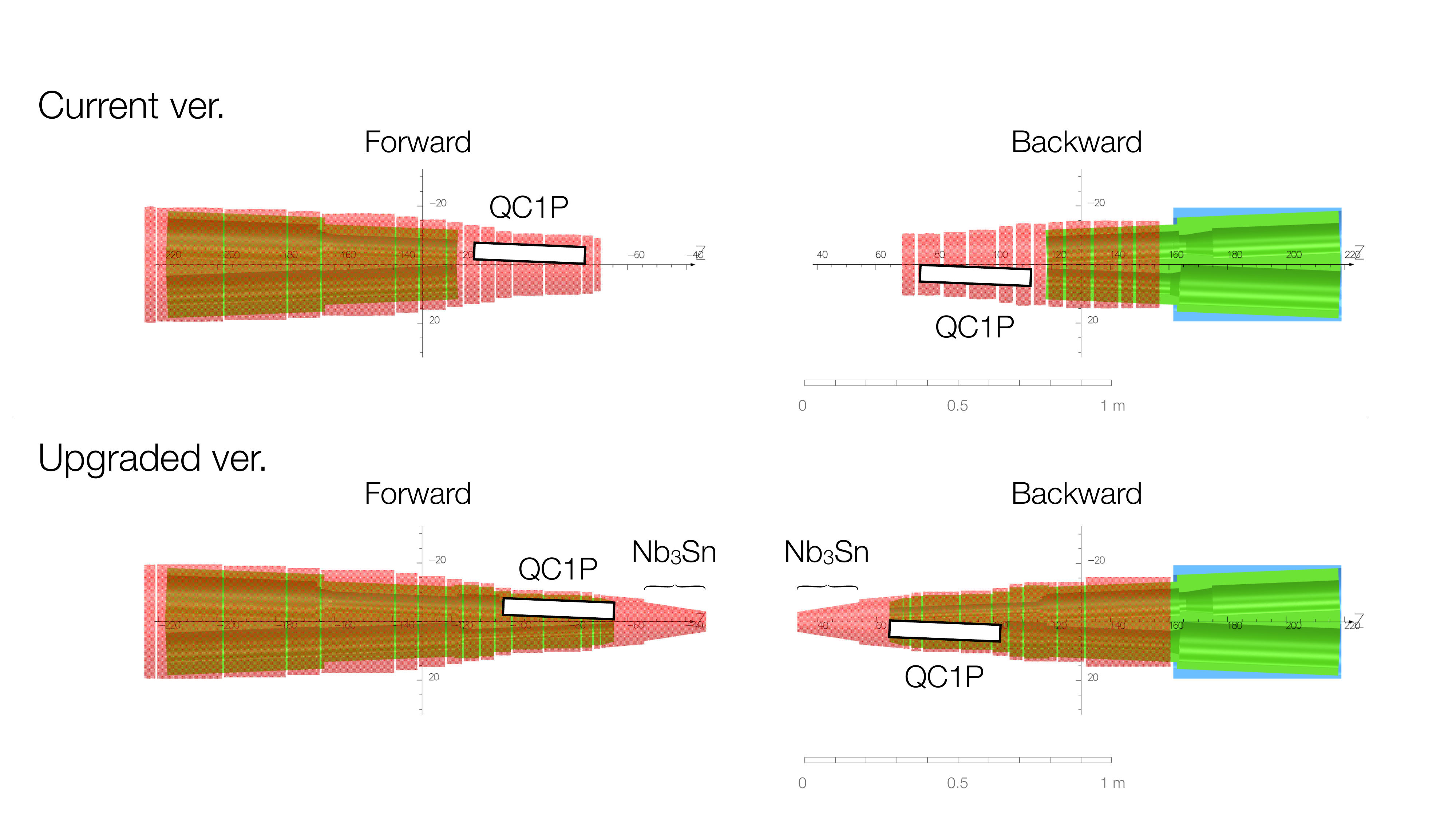}
\caption{Comparison between the current QCS configuration (top) and the new QCS configuration (bottom).  Anti-solenoid coils and iron (or permendur) shield are indicated in red and green, respectively.}
\label{fig:QC1Comp.pdf}
\end{figure}

The new design concept of the IR upgrade is to make the beam trajectory as parallel to the QC1 magnet axis as possible to
 cancel the X--Y coupling and chromaticity between the IP and
 QC1 as close to zero as possible and minimize $\epsilon_y$ by redesigning the $B_z(s)$  profile.
 The vertical offsets and the rotations of QC1P and QC2P and the horizontal offsets of QC1E and QC2E are expected to be considerably
 smaller than the present values with the new anti-solenoid field design and with incorporating the magnetic yokes on QC1P. The design concepts of the new IR are listed below and shown schematically in  Fig.~\ref{fig:QC1Comp.pdf}.
 
 \begin{itemize}
    \item Separation of the solenoid field and QC1P
    \item QC1P relocation, closer to the IP by \SI{100}{mm} (from the present $L^*=\SI{935}{mm}$ to \SI{835}{mm})
\end{itemize}

\begin{figure}[ht!]
\centering
\includegraphics[width=\textwidth]{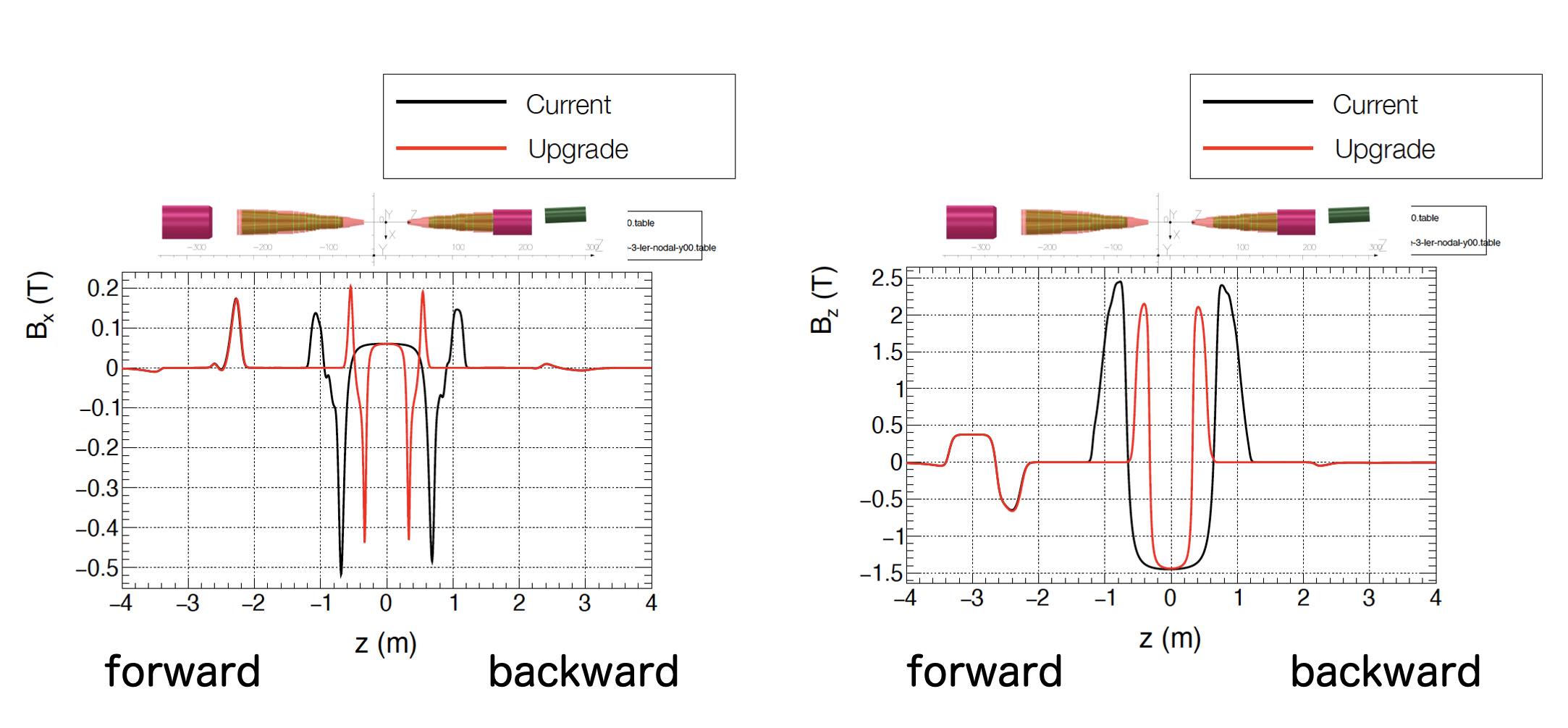}
\caption{Comparison between the current and the new anti-solenoid field profile of $B_x$ and $B_z$ for the magnet configurations in Fig.~\ref{fig:QC1Comp.pdf} for the LER.  The anti-solenoid coil cancels the magnetic field generated by the Belle II detector solenoid before QC1P.}
\label{fig:FieldComp.jpg}
\end{figure}

\begin{figure}
    \centering
    \includegraphics[trim=0 220 0 220, width=\textwidth, clip]{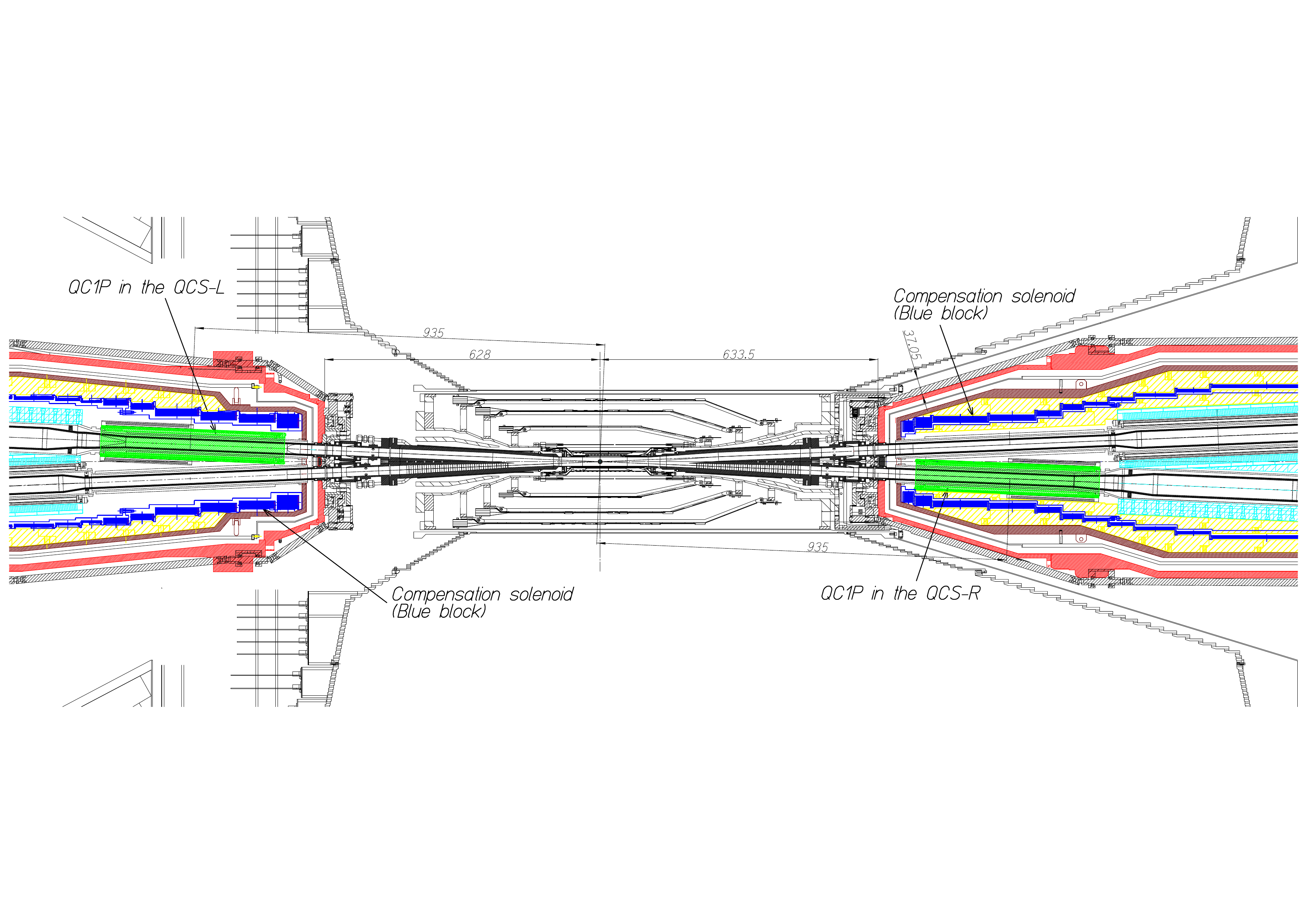}
    \caption{The current QCS system with cryostat. The red, brown, yellow, blue, cyan, and green shaded areas correspond to the vacuum vessels of the cryostat, the liquid helium vessels, the tungsten radiation shields, the anti-solenoid coils, the magnetic yokes/shields, 
and the QC1P magnets, respectively.  The left side corresponds to the backward side.}
    \label{fig:CurrentCryostat.pdf}
\end{figure}

\begin{figure}[ht!]
    \centering
       \includegraphics[trim=0 220 0 220, width=\textwidth, clip]{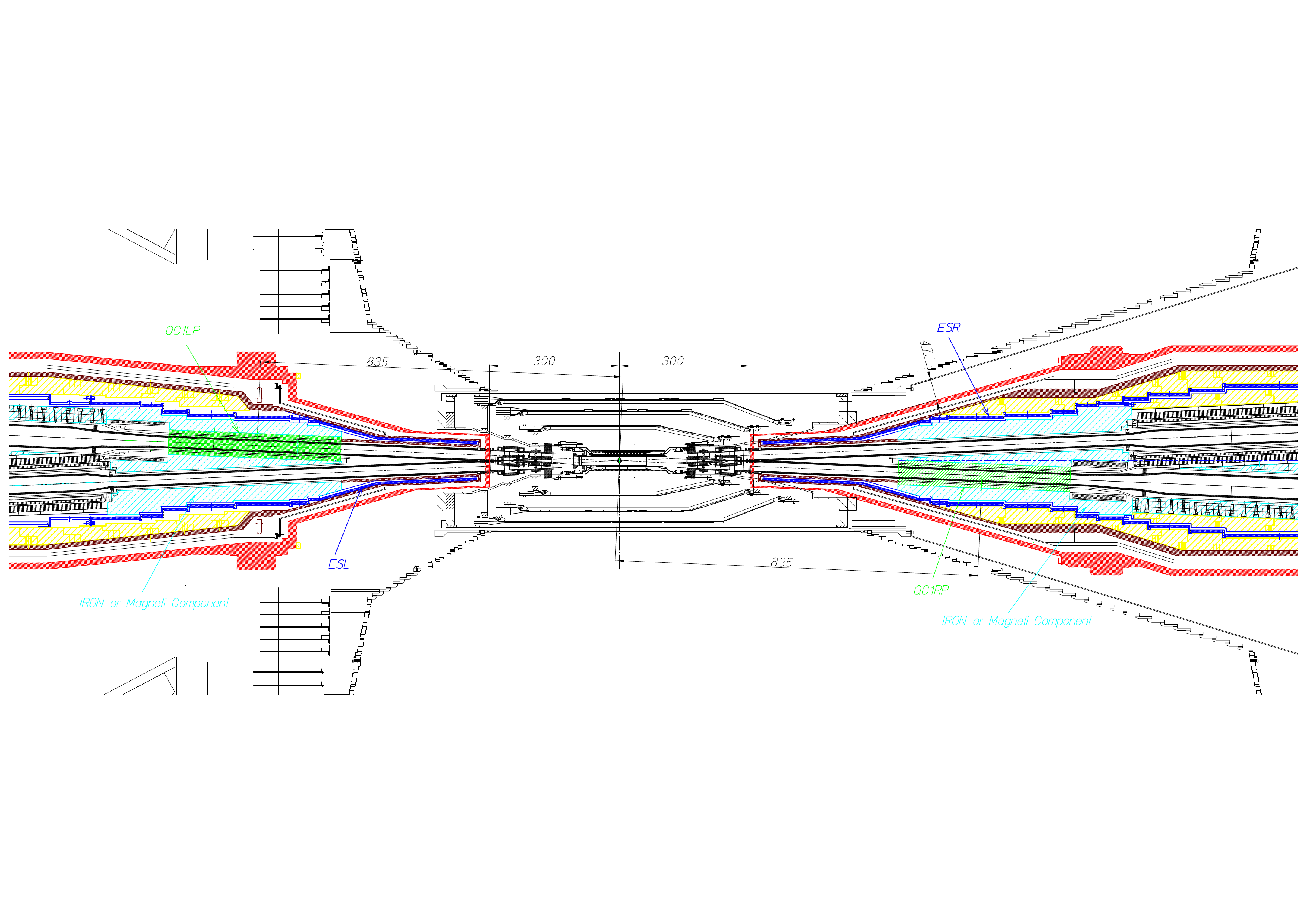}
     \caption{The new QCS system with cryostat. The red, brown, yellow, blue, cyan, and green shaded areas correspond to the vacuum vessels of the cryostat, the liquid helium vessels, the tungsten radiation shields, the anti-solenoid coils, the magnetic yokes/shields, 
and the QC1P magnets, respectively.  The left side corresponds to the backward side.}
     \label{fig:UpgradeCryostat.pdf}
 \end{figure}

 The magnetic field from the line current distribution was evaluated using Opera3D ~\cite{ref:OPERA3D}, where the magnetic field by the Belle II solenoid coil is included. As is shown in Fig.~\ref{fig:FieldComp.jpg}, the peak field becomes lower with the new IR, resulting in thinner coil.  The detector solenoid field is cancelled between IP and QC1P. The cryostats are included in Fig.~\ref{fig:CurrentCryostat.pdf} and Fig.~\ref{fig:UpgradeCryostat.pdf} for the current and new QCS systems, respectively. 

 Some parts of the cryostat and the Belle~II detector (e.g., the vertex detector support structure) interfere on the forward side. A different solution for connecting the vacuum systems has to be developed.
 The thinner solenoid coil creates more space in the radial direction between the detector and the cryostat, which makes cabling easier around the cryostat.  It should be noted, though, that these are very simplified preliminary drawings. Drawings with final dimensions cannot be produced until the magnet design and a detailed design of each part are finalized.

%%%%%%%%%%%%%%%%%%%%%%%%%%%%%%%%%%%%%%%%%%%%%%%%%%%%%%%%%%%%%%%%%%%%%%%%%%%
%%%%%%%%%%%%%%%%%%%%%%%%%%%%%%%%%%%%%%%%%%%%%%%%%%%%%%%%%%%%%%%%%%%%%%%%%%%
 
%%%%%%%%%%%%%%%%%%%%%%%%%%%%%%%%%%%%%%%%%%%%%%%%%%%%%%%%%%%%%%%%%%%%%%%%%%%
%%%%%%%%%%%%%%%%%%%%%%%%%%%%%%%%%%%%%%%%%%%%%%%%%%%%%%%%%%%%%%%%%%%%%%%%%%%
%%%%%%%%%%%%%%%%%%%%%%%%%%%%%%%%%%%%%%%%%%%%%%%%%%%%%%%%%%%%%%%%%%%%%%%%%%%
%%%%%%%%%%%%%%%%%%%%%%%%%%%%%%%%%%%%%%%%%%%%%%%%%%%%%%%%%%%%%%%%%%%%%%%%%%%
% 
%%%%%%%%%%%%%%%%%%%%%%%%%%%%%%%%%%%%%%%%%%%%%%%%%%%%%%%%%%%%%%%%%%%%%%%%%%%

 \clearpage
%%%%%%%%%%%%%%%%%%%%%%%%%%%%%%%%%%%%%%%%%%%%%%%%%%%%%%%%%%%%%%%%%%%%%%%%%%%
 
 \subsection{Optics evaluation} 

 Optics evaluation was carried out, focusing on the effects on the dynamic aperture and Touschek lifetime, chromatic coupling, and vertical emittance.  Neither Beam-beam effects nor octupole correction are included in this evaluation. The R-side backward compensation solenoid is not implemented in the discussed optical evaluation model. The upstream vertical orbit and vertical emittance are provisional values.

\begin{figure}[h]
    \centering
    \includegraphics[width=0.8\linewidth]{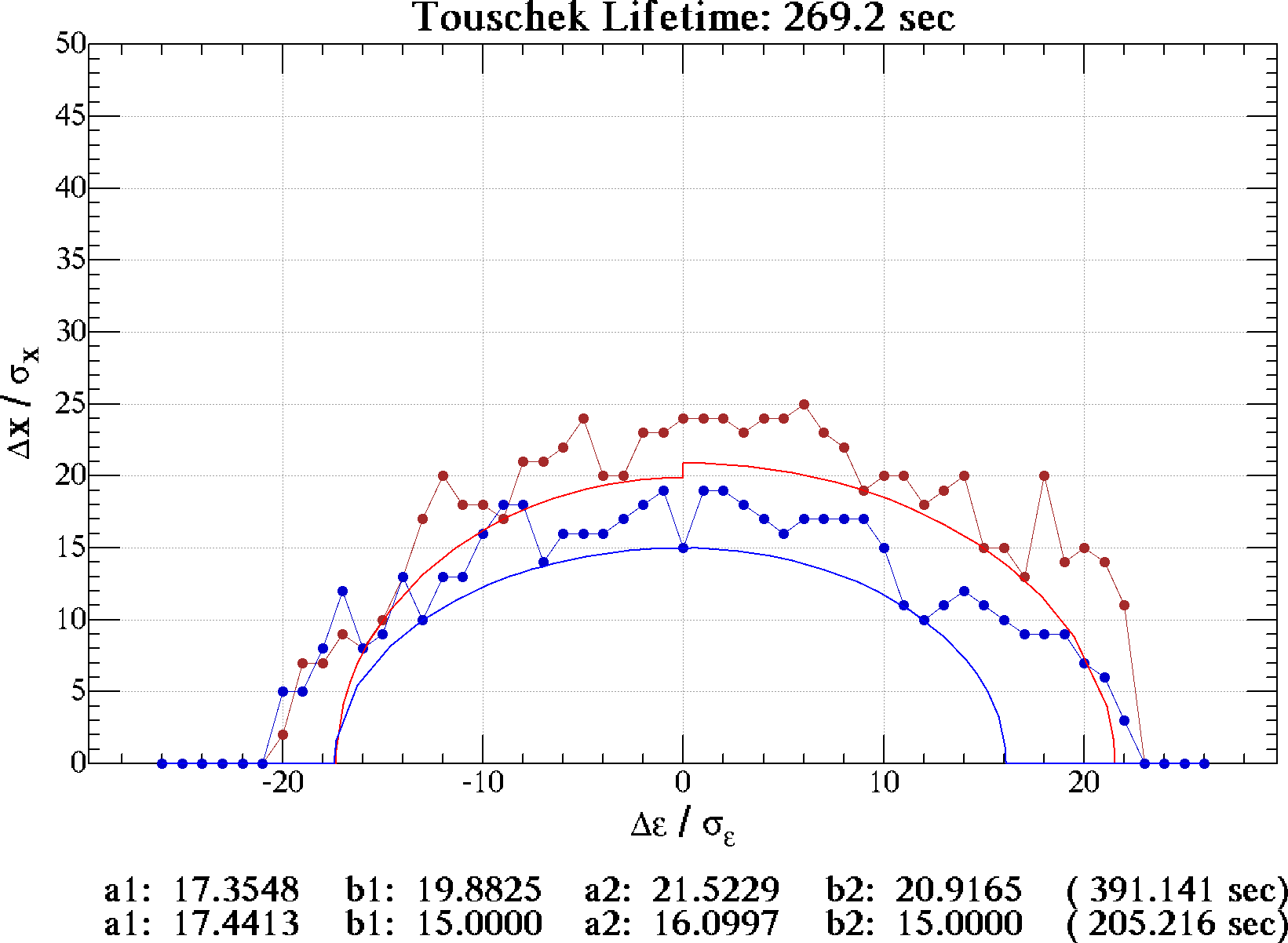}
    \caption{The present lattice ($\beta_y^*$  = \SI{0.27}{mm})}
    \label{fig:DAdesign}
%\end{figure}
%
%\begin{figure}
%    \centering

    \vspace{10mm}
    
    \includegraphics[width=0.8\linewidth]{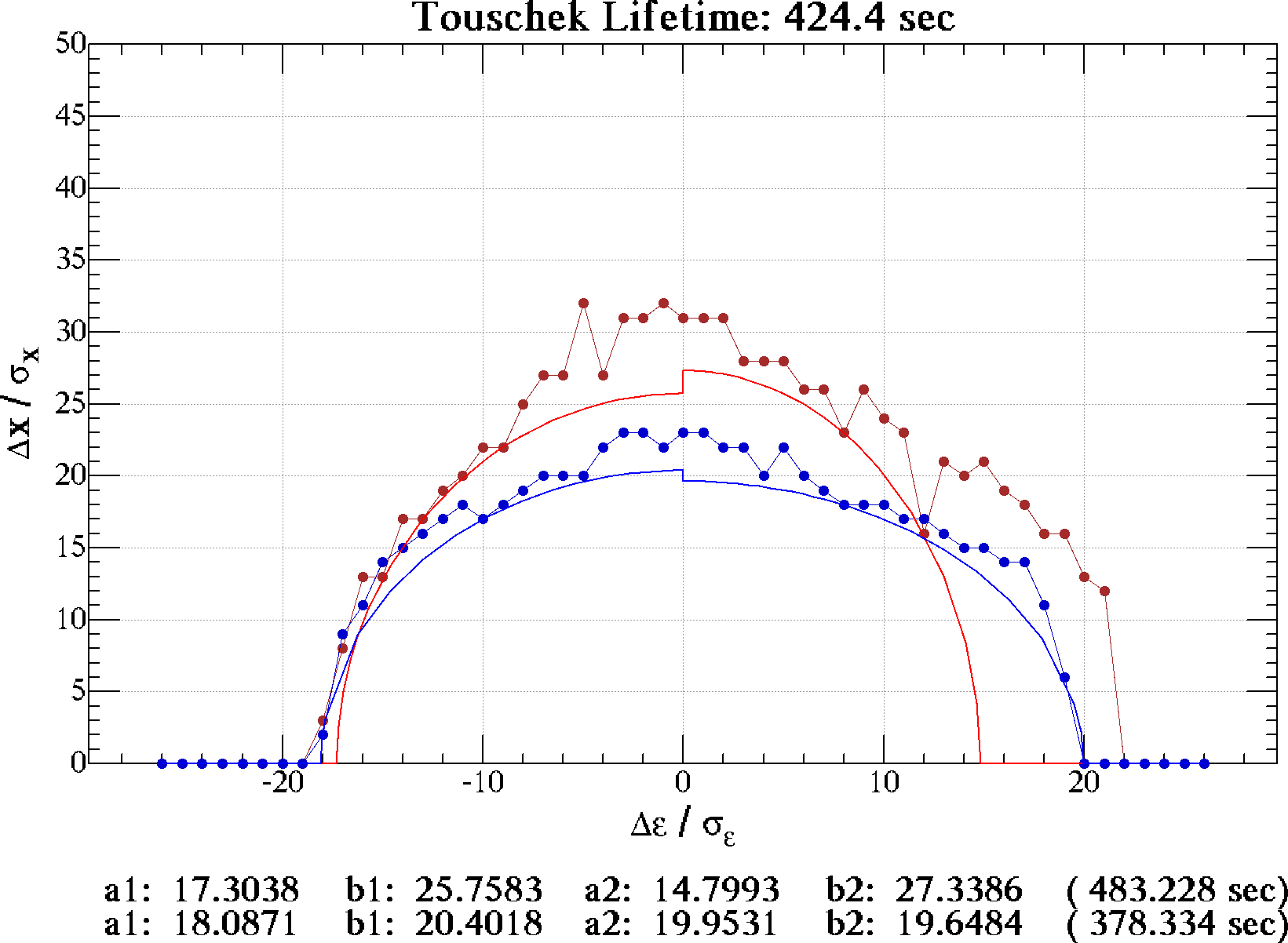}
    \caption{The new lattice ($\beta_y^*$  = \SI{0.27}{mm})}
    \label{fig:DAnew}
\end{figure}

 Fig.~\ref{fig:DAdesign} and Fig.~\ref{fig:DAnew} compare the dynamic aperture and Touschek lifetime of the present IR lattice (sler\_1704) and the new IR lattice (V21-r0g0) for LER, respectively.  The design parameters, such as \SI{0.27}{mm} for $\beta_y^*$, were used as input parameters for the simulation.  By relocating the QC1P \SI{100}{mm} closer to the IP, Touschek lifetime increases from 261 seconds to 424 seconds.

 %%%%%%%%%%%%%%%%%%%%%%%%%%%%%%%%%%%%%%%%%%%%%%%%%%%%%%%%%%%%%%%%%%%%%%%%%%%
%\begin{figure}
%    \begin{minipage}[b]{.5\linewidth}
%     \centering
%     \includegraphics[width=0.8\linewidth]{fig/DAdesign.jpeg}
%     \subcaption{The present lattice ($\beta_y^*$  = 0.27~mm)}
%     \label{fig:DAdesign.jpeg}
%     \end{minipage}
%     \begin{minipage}[b]{.5\linewidth}
%    \centering
%     \includegraphics[width=0.8\linewidth]{fig/DAnew.jpeg}
%     \subcaption{The new lattice ($\beta_y^*$  = 0.27~mm)}
%     \label{fig:DAnew.jpeg}
%     \end{minipage}
% \end{figure}

Table\, \ref{tab:ChromaticCoupling} summarizes chromatic X--Y coupling for the present and new IR lattice designs.
It is shown that the chromatic X--Y coupling parameters, $\partial R_1/\partial \delta$, $\partial R_2/\partial \delta$, $\partial R_3/\partial \delta$ and $\partial R_4/\partial \delta$, are few orders of magnitude smaller with the new IR with/without the sextupole correction on.  A clear improvement is seen with the new IR lattice.

\begin{table}[ht]
	\centering
	\caption{Chromatic Coupling comparison between the designed lattice and the new lattice with (ON) and without (OFF) chromatic coupling correction by rotatable sextupole magnets. }
	\begin{tabular}{lccccc}
		\toprule
%		& \multicolumn{2}{c}{June 8, 2022} & \multicolumn{2}{c}{May 22, 2022} \\
		\textbf{Lattice} & \textbf{Sext.cor} & \textbf{$\partial R_1/\partial \delta$} & \textbf{$\partial R_2/\partial \delta$} & \textbf{$\partial R_3/\partial \delta$} & \textbf{$\partial R_4/\partial \delta$}\\
		\midrule
			sler\_1704 & ON &  \num{-8.9E-3} & \num{+4.0E-3} & \num{-5.0E1}   & \num{+2.9}    \\	
            sler\_1704 & OFF &  \num{-2.3} & \num{-1.0E-2} & \num{-4.2E2}   & \num{-6.1E2}     \\
            V21-r0g0 & ON &  \num{+2.3E-5} & \num{-6.0E-6} & \num{-4.4E-2}   & \num{+5.5E-3}     \\	
             V21-r0g0 & OFF &  \num{+1.1E-1} & \num{+2.8E-4} & \num{+8.1}  & \num{+2.6E1}     \\
   \bottomrule
	\end{tabular}
	\label{tab:ChromaticCoupling}
 \end{table}

\begin{figure}
    \centering
    \includegraphics[width=0.8\linewidth]{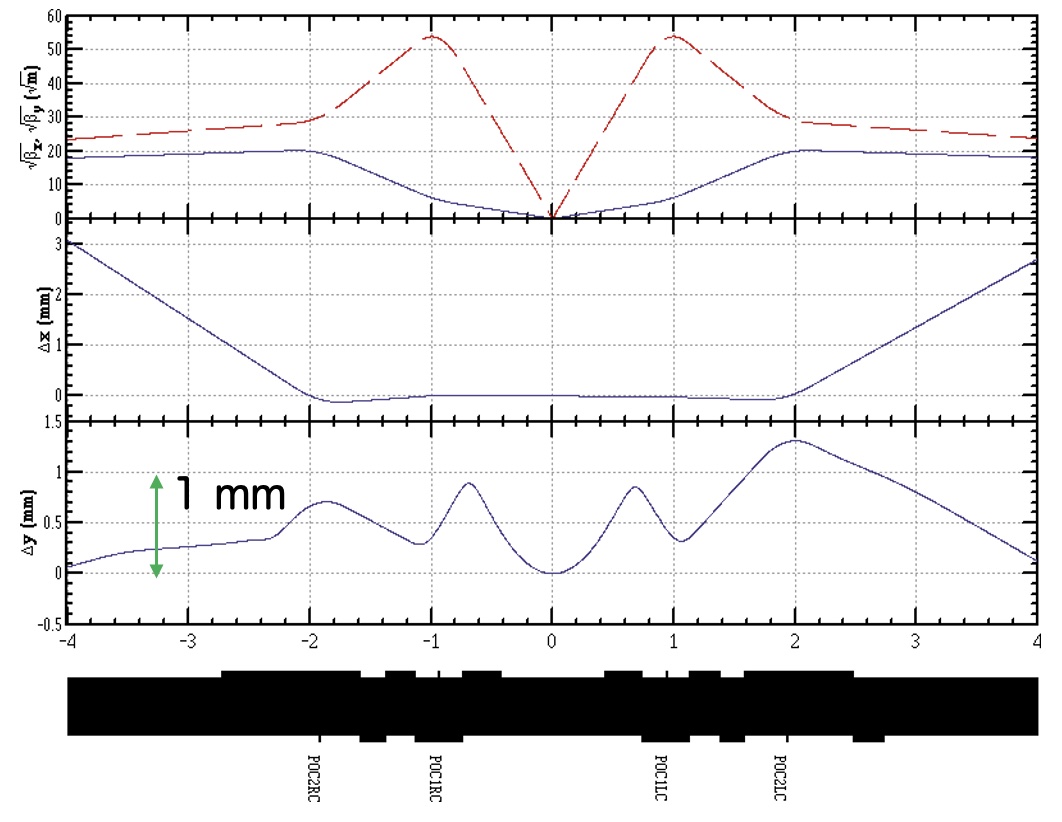}
    \caption{The present orbit and beta functions for $\beta_y^*$  = \SI{0.27}{mm}.The horizontal and vertical beta functions are indicated by red and blue lines, respectively.}
    \label{fig:PresentOrbitandBeta}
\end{figure}

\begin{figure}
    \centering
    \includegraphics[width=0.8\linewidth]{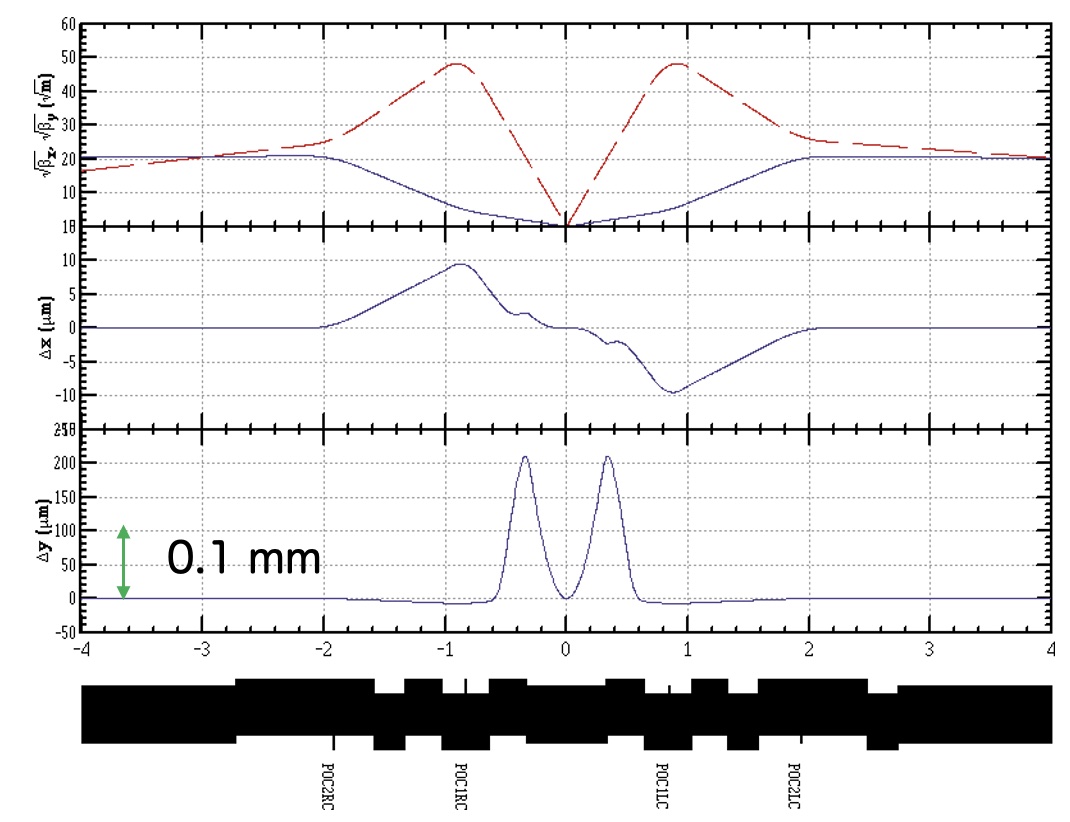}
    \caption{The new orbit and beta functions for $\beta_y^*$  = \SI{0.27}{mm}.The horizontal and vertical beta functions are indicated by red and blue lines, respectively.}
    \label{fig:NewOrbitandBeta}
\end{figure}

\begin{figure}
    \centering
    \includegraphics[width=0.8\linewidth]{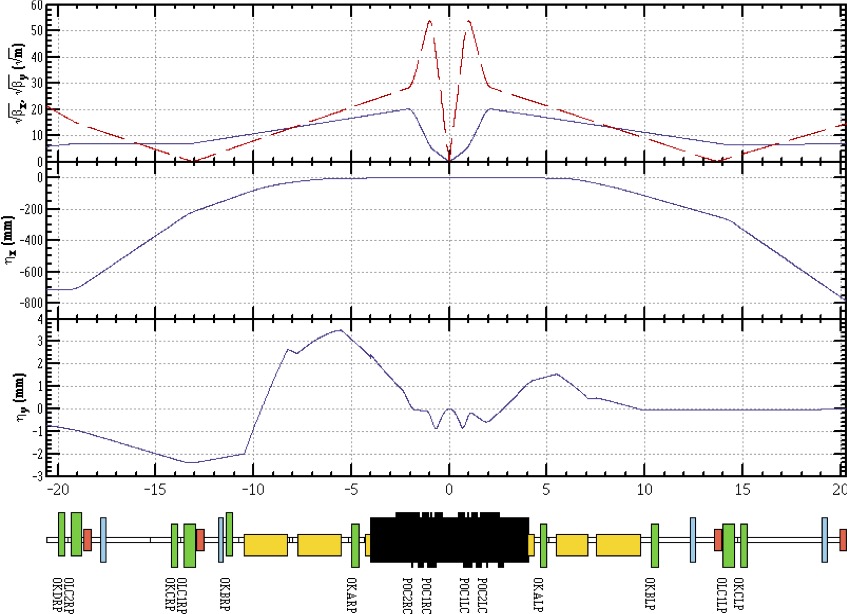}
    \caption{The present dispersion functions for $\beta_y^*$  = \SI{0.27}{mm}.The horizontal and vertical beta functions are indicated by red and blue lines, respectively.}
    \label{fig:PresentDispersion}
\end{figure}

\begin{figure}
    \centering
    \includegraphics[width=0.8\linewidth]{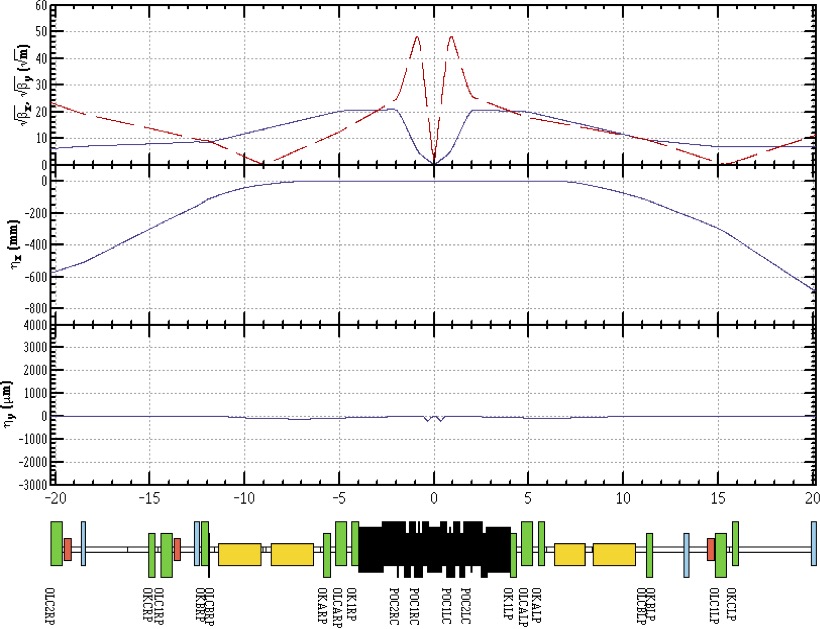}
    \caption{The new dispersion functions for $\beta_y^*$  = \SI{0.27}{mm}.The horizontal and vertical beta functions are indicated by red and blue lines, respectively.}
    \label{fig:NewDispersion}
\end{figure}

The present and new IR orbit and the beta functions are shown in Fig.~\ref{fig:PresentOrbitandBeta} and Fig.~\ref{fig:NewOrbitandBeta}, respectively.
The orbit displacement is $\sim$\SI{10}{\micro\meter}  at QC1, while it is about \SI{1}{mm} with the present lattice. 
The dispersion functions of the present and new lattice are shown in Fig.~\ref{fig:PresentDispersion} 
and Fig.~\ref{fig:NewDispersion}, respectively.
The vertical emittance from the IR is calculated to be \SI{14}{fm}, which is
 negligibly small.

%%%%%%%%%%%%%%%%%%%%%%%%%%%%%%%%%%%%%%%%%%%%%%%%%%%%%%%%%%%%%%%%%%%%%%%%%%%
%\begin{figure}
%    \begin{minipage}[b]{.5\linewidth}
%     \centering
%     \includegraphics[width=0.8\linewidth]{fig/NewOrbit.jpg}
%     \subcaption{New orbit and beta functions ($\beta_y^*$  = 0.27~mm)}
%     \label{fig:NewOrbit.jpg}
%     \end{minipage}
%     \begin{minipage}[b]{.5\linewidth}
%     \centering
%     \includegraphics[width=0.8\linewidth]{fig/NewDispersion.jpg}
%     \subcaption{New dispersion functions ($\beta_y^*$  = 0.27~mm)}
%     \label{fig:NewDispersion.jpg}
%     \end{minipage}
% \end{figure}
 %%%%%%%%%%%%%%%%%%%%%%%%%%%%%%%%%%%%%%%%%%%%%%%%%%%%%%%%%%%%%%%%%%%%%%%%%%%

  To summarize, 1)~dynamic aperture and Touschek lifetime improved by relocating QC1 closer to the IP, 2)~chromatic X--Y coupling improves significantly, 3)~the contribution to the vertical emittance from the IR is in order of several tens of femtometers.  The new IR removes the difficulties in luminosity tuning that we are facing now.  Luminosity tuning is expected to become easier with the new IR.

 \subsection{R\&D of Nb$_\text{3}$Sn cables and QC1P quadrupole}

 QC1P must be designed more compactly to fit into the smaller space when moved \SI{100}{mm} closer to the IP than its current position.
 It is, therefore, necessary to develop superconducting cables that can operate at higher current densities.  The design current density of the new QC1P superconducting cable is twice that of the current QC1P.
 
 Nb$_\text{3}$Sn is known to have a higher critical current density and critical temperature than NbTi, which is commonly used as an accelerator magnet material~\cite{Barzi:2019xdq}.  In the case of Nb$_\text{3}$Sn quadrupole magnet, the temperature margin for keeping 
the superconducting condition of the magnet can be doubled compared to the current quadrupole magnet~\cite{Devred:2004jt}. This effect is very important for stable beam operation at larger beam currents (twice larger than the present) after  LS2.  The development of superconducting magnets using Nb$_\text{3}$Sn cables to produce a required magnetic field with a quite high current density in the superconducting coil has been carried out, though there are many items that need to be considered and investigated to be used for QC1P. The items include issues such as countermeasures against quenches caused by magnetic flux jumps within a superconductor and time decay of the electromagnet magnetic field due to magnetic flux creep. Fabrication of a small coil using Nb$_\text{3}$Sn also needs to be confirmed.  Nb$_\text{3}$Sn magnet fabrication is quite difficult because of the material's brittleness compared to NbTi. To produce Nb$_\text{3}$Sn as an A15 compound, the coil needs to be heated over 650 degrees Celsius for tens of hours after completing the coil winding. The heat-treated coils are very brittle and need to be impregnated and assembled into the magnet very carefully; the higher multipole fields in the magnet must be at the level of a few 10$^{-4}$ of the main quadrupole field.

 At present, we have the following development schedule:
 \begin{quote}
 \begin{itemize}
  \item Study of the Nb$_\text{3}$Sn cable for the new QC1P magnet.
 \end{itemize}
\end{quote}
 \begin{quote}
 \begin{itemize}
  \item Production of the Nb$_\text{3}$Sn cable, and design of the prototype QC1P and the R\&D coil.
 \end{itemize}
\end{quote}
 \begin{quote}
 \begin{itemize}
  \item Construction of the first R\&D coil and a mirror magnet~\cite{Andreev:2009zza}, followed by a test of the coil in the mirror magnet configuration.
 \end{itemize}
\end{quote}
\begin{quote}
 \begin{itemize}
  \item Review of the proto-type QC1P magnet design based on the test results of the mirror magnet and confirmation of the design.
 \end{itemize}
\end{quote}
\begin{quote}
 \begin{itemize}
  \item Start and completion of the construction of the first  Nb$_\text{3}$Sn prototype QC1P magnet after fabrication of 4+1 coils. The magnet will be tested, and the magnetic field measurements will be performed.
 \end{itemize}
\end{quote}

 We will proceed with feasibility studies of Nb$_\text{3}$Sn cables for QC1P and possibly for the anti-solenoid coils. 
 
 %\subsection{Remaining tasks at the desk review stage} 
  \subsection{R\&D program for the new IR} 
 
A detailed machine optics study, including beam pipe geometry optimization, is essential to identify the narrowest machine aperture, where the majority of IR beam loss concentration is expected based on our operation experience. The new outcome of this study might require additional detector shielding installation around the final focusing magnets, which could potentially impact the vertex detector envelope. The remaining tasks related to beam optics and lattice design are listed below.

Regarding the second compensation solenoid:
\begin{quote}
 \begin{itemize}
  \item Optical modeling, evaluation of effects on trajectory and optics optimization of LER R side (upstream side of IP) trajectory with consideration of the effect on the detector background.
 \end{itemize}
\end{quote}

For reviewing and evaluating machine performance:
\begin{quote}
 \begin{itemize}
  \item Consistency checks of machine parameters, such as 
    \begin{itemize}
    \item  Are the assumed collision parameters, lifetime, stored beam current, and injector performance consistent?
     \item Is constructing an evaluation model for detector background estimate necessary?
       \begin{itemize}
        \item The difficulties of this scheme of large Piwinski angle collision at $\beta_y^*$ of  \SI{0.3}{mm} with Crab Waist are the narrow dynamic aperture and short Touschek lifetime.  The maximum stored beam current is determined by the balance between injection capacity and lifetime while controlling the background to the detector.
      \end{itemize}
       \end{itemize}
 \item Evaluation with beam-beam, wakefield, and machine errors, which is necessary to estimate actual performance.
  \begin{itemize}
    \item  We need to identify the causes of the current machine performance defects and establish reproduction techniques and evaluation models on simulation.
     \end{itemize}
 \end{itemize}
\end{quote}

Tasks required for construction:
\begin{quote}
 \begin{itemize}
  \item Completion of the HER IR optics
  \begin{itemize}
    \item  Review of horizontal orbit design and correction of optics due to introduction of magnetic shielding to QC1P.
%    \item There should be no problem since the design will be straightforward except that QC1E will be stronger due to convergence caused by QC1P leakage magnetic field.
     \end{itemize}
 \item Redesign of outside QCS region (IP \SI{+-4}{m}) for both LER and HER
  \begin{itemize}
    \item  The model being considered in the LER has not been tested to see if it will physically fit in the existing tunnel. This point of view also needs to be considered when constructing beamlines.
      \item Add and relocate elements needed for matching/adjusting luminosity tuning knobs and resolve interference with tunnel boundary and HER.
      \item HER also needs to reflect IR horizontal orbit change. 
     \end{itemize}
 \end{itemize}
\end{quote}

\chapter{Machine Detector Interface}
\label{sec:MDI}
\editor{H.~Nakayama, A.~Natochii}
%% MDI
\section{Beam backgrounds and countermeasures}
\label{sec:BeamBackgroundsAndCountermeasures}
\editor{H.Nakayama, A.Natochii}

Here, we briefly review the main sources of beam-induced backgrounds, usually simply referred to as beam backgrounds for short. Beam backgrounds can arise from particle losses from the beam during machine operation, leading to electromagnetic (EM) showers. Secondary particles from such showers then cause higher hit rates and other detrimental effects in Belle~II. While this is the typical example, any aspect of machine operation, even the intentional collision of beams, which results in higher Belle~II occupancy or radiation dose, can be considered a beam background. Since SuperKEKB is designed to operate at much higher beam currents and collision luminosities than its predecessor KEKB, the IR backgrounds are also expected to be much higher.

\subsection{Particle scattering} 
Beam particles circulating in SuperKEKB may undergo single-beam processes caused by i) scattering with residual gas molecules due to Bremsstrahlung and Coulomb interactions and ii) Coulomb scattering between particles in the same beam bunch, commonly referred to as the Touschek effect. These processes lead to the scattered particles falling out of the stable orbit and reaching the physical or dynamic aperture of the machine. This creates EM showers when these stray particles hit the beam pipe walls. The beam-gas background rate is proportional to the vacuum pressure and the beam current, while the Touschek background is proportional to the beam current squared and inversely proportional to the number of bunches, beam size, and the third power of the beam energy.

\subsection{Colliding beams}
While particle collisions at the IP are the goal of SuperKEKB, there are several undesirable collision processes that have very high cross sections but are of little interest for physics measurements. Examples are radiative Bhabha scattering ($e^+ e^- \rightarrow e^+ e^- \gamma$) and two-photon processes ($e^+ e^- \rightarrow e^+ e^- e^+ e^-$). These processes increase the Belle~II occupancy and radiation dose, and we refer to these increases as luminosity background. The rate of this background component is proportional to the luminosity.

\subsection{Synchrotron radiation (SR)}
X-rays are emitted when electrons or positrons pass through the strong magnetic field of the QCS or dipole magnets upstream and downstream of the IP. SR can potentially damage the inner layers of the vertex detector. Because SR power is proportional to the beam energy squared and magnetic field strength squared, X-rays from the HER are the main concern.

\subsection{Beam injection}
Since the beam lifetime, which is limited mainly by Touschek losses, is of order \SIrange{10}{60}{\min}, SuperKEKB is operated using a so-called top-up injection scheme where the bunches are continuously refilled to keep the current stable. Due to optics mismatches between the injection chain and the main ring, injected particle bunches do not enter the aperture of the machine perfectly, resulting in higher particle losses from injected bunches until they stabilize ($\sim\SI{10}{ms}$).

\subsection{Sudden beam losses}
During otherwise stable machine operation, unexplained beam instabilities and beam losses may occasionally occur in one of the rings. Such losses can occur in one shot at a specific location around the ring due to machine element failure or due to the beam interacting with dust particles in the beam pipe. In extreme cases, high-intensity EM showers can lead to detector or collimator damage or superconducting magnet quenches. %Usually, only a few such catastrophic beam loss events happen per year. 
For example, eight such LER beam losses led to QCS quenching in the February -- June 2022 run period.

\subsection{Head-tail instabilities}
In linear accelerators (linacs), when a dense bunch of charged particles travels through the accelerator structure, the front part of the bunch (head) induces an EM wakefield which deflects the rear part of the bunch (tail). After some time, the tail oscillates around the head trajectory, which remains unperturbed. These oscillations may cause particle losses on the aperture of the machine. This process is referred to as Beam Break-Up (BBU) instability. In a synchrotron machine like SuperKEKB, the RF field induces a longitudinal focusing of the bunch, creating a collective transverse motion with head-tail modes of betatron and synchrotron oscillation coupling. Below a certain bunch current, this oscillation remains stable while at the threshold value, the growth rate of the tail is faster than the synchrotron period so that the head-tail modes are destroyed, and the beam becomes unstable. This process is referred to as the Transverse Mode Coupling Instability (TMCI)~\cite{Gareyte2001}. Since the TMCI threshold depends on the most narrow and steep aperture in the ring, the bunch current threshold is usually calculated considering only collimators. It can be defined as follows~\cite{Gareyte2001,Chao1999}:
\begin{equation}
I_{\rm th} = \frac{C_{\rm 1}f_{\rm s}E/e}{\sum_{\rm j}\beta_{\rm j}k_{\rm j}},
    \label{eq:eq1}
\end{equation}
where $I_{\rm th}$ is the upper limit on the bunch current, $f_{\rm s} = \SI{2.13}{kHz}$ or $f_{\rm s} = \SI{2.80}{kHz}$ is the synchrotron frequency for the LER or HER, respectively, $E$ is the beam energy, $e$ is the unit charge, $\beta_{\rm j}$ and $k_{\rm j} \sim d_\mathrm{j}^{-\frac{3}{2}}$ are the beta function and kick factor, as a function of the aperture $d_\mathrm{j}$~\cite{Yue2016physdesign}, for the $j$th collimator, respectively. Usually, the constant $C_{\rm 1}$ is equal to 8 or $2\pi$~\cite{Chao1999}; however, for SuperKEKB, $C_{\rm 1} \approx 4\pi$ is found to be a more realistic value based on machine studies~\cite{OhmiIshibashi2022} conducted in 2021 -- 2022. Additional machine elements with a variable aperture, such as the IR beam pipe, can be included in Eq.~\ref{eq:eq1} by extending the sum over $\beta_{\rm j}k_{\rm j}$ to include terms both for collimators and these additional elements.

\subsection{Countermeasures}
 In the last decade, we have developed a comprehensive set of countermeasures against each of these background sources. We use collimators at specific locations to suppress single-beam and injection backgrounds by stopping stray particles before they reach the IR. Vacuum scrubbing helps us reduce each ring's vacuum pressure, thus lowering the beam-gas scattering rate~\cite{Shibata:eeFACT2018-WEOAB01}. A heavy metal shield was installed to protect the inner detectors against EM showers from beam losses in the IR. A polyethylene+lead shield inside the ECL~\cite{Beaulieu2019} protects the ECL crystals and photodiodes from neutrons. Similarly, the inner layers of the ARICH were replaced with a boron-doped polyethylene shield against neutrons.  The incoming beam pipes of the IR collimate most of the SR photons and suppress their reflections thanks to the design geometry and a ridge structure on the inner beam pipe surface~\cite{Ohnishi2013}. To protect the vertex detector against residual SR, the inner surface of the IP beryllium beam pipe is coated with a \SI{10}{\micro m} thick gold layer.  The newly constructed Damping Ring (DR) reduces the emittance of injected positrons, thereby suppressing injection beam losses. We apply an injection veto on the Belle~II Level-1 trigger,  to avoid recording high beam losses when the injection bunch crosses the IR, for about \SI{10}{ms} after each injection. In addition, we continuously perform injection chain tuning to improve injection efficiency and reduce injection background. To monitor the beam losses in the IR, a system based on single-crystal synthetic diamond sensors, so-called Diamonds, is used since the start of the SuperKEKB operation. The system records the radiation dose-rates in positions close to the inner detectors of the Belle~II experiment, and protects both the detector and accelerator components against destructive beam losses, by participating in the beam-abort system~\cite{Ikeda2014}. It also provides complementary information for the dedicated studies of beam-related backgrounds~\cite{BACHER2021165157}. To protect Belle~II against abnormal beam losses, the beam abort system has been complemented by two new systems based on sCintillation Light And Waveform Sensors (CLAWS) detectors~\cite{Gabriel2021} near the IR and CsI-crystals around each ring. The CLAWS system provides an additional fast beam abort signal, while the CsI-based sensors help to locate the initial beam loss locations along the ring.

 Furthermore, as discussed in Ref.~\cite{Natochii2022}, during LS1, the machine and detector protection system is planned to be upgraded. This includes additional neutron and EM shielding around QCS magnets and Belle~II, heavy-metal shielding around IR bellows, new collimators, and a new IP beam pipe. So far, all mentioned upgrades have not been implemented in the official MC sample production code, leading to presumably overestimated detector BG rates at higher luminosity.
 
\section{Background status}
\label{sec:BackgroundStatus}

\begin{figure}[htb]
\centering
\includegraphics[width=\textwidth]{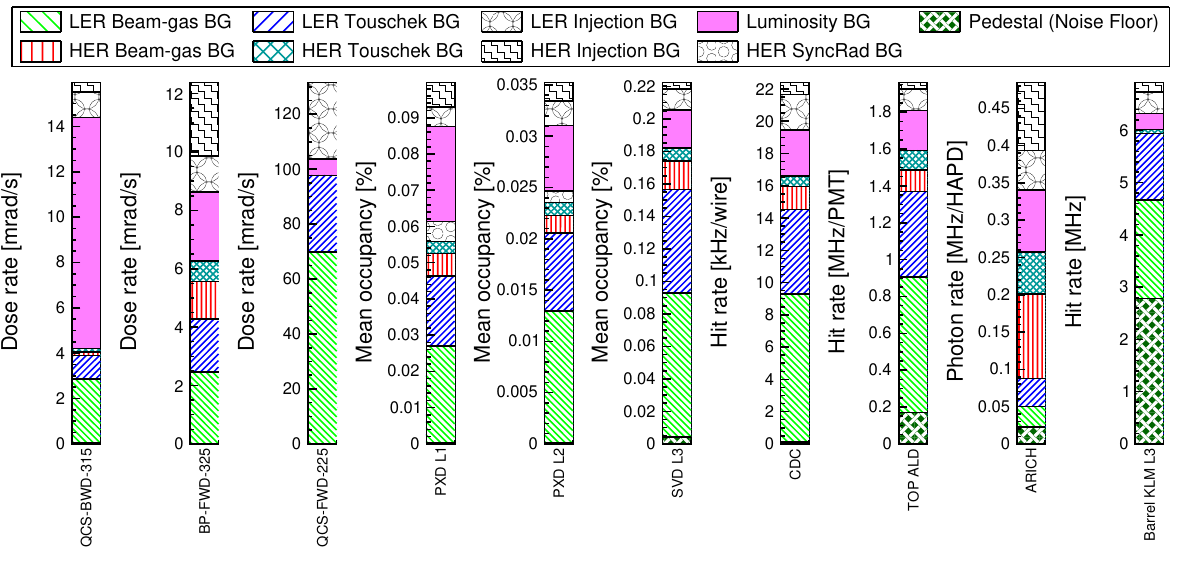}
\caption{Extrapolated Belle~II backgrounds from measurements in June~2021. Each column is a stacked histogram. From~\cite{NATOCHII2023168550}.}
\label{fig:BackgroundComposition}
\end{figure}

Since the start of the SuperKEKB commissioning with a fully assembled Belle~II detector in 2019, we have performed several dedicated beam background studies. Usually, the measurement starts from a single-beam configuration where only one beam is filled with a beam. Changing the number of bunches allows us to disentangle beam-gas and Touschek backgrounds during beam decay measurements when the injection is stopped. Repeating this measurement for the LER and HER, we can extract background sensitivities for each sub-detector separately. Then, by extrapolating the single-beam backgrounds to the collision study when both beams are filled, we subtract the single-beam contribution from the observable and estimate the luminosity background sensitivity. A detailed procedure of Belle~II background measurements is discussed elsewhere~\cite{NATOCHII2023168550}. For example, Belle~II background rates measured in June 2021 are shown in Fig.~\ref{fig:BackgroundComposition}. The data was recorded with a betatron function at the IP of $\beta^{\rm *}_{\rm y}=\SI{1}{mm}$, beam currents of $I_{\rm LER} = \SI{732.6}{mA}$ and $I_{\rm HER} = \SI{647.2}{mA}$, with ring-averaged beam pipe pressures of $\overline P_{\rm LER} = \SI{88}{nPa}$ and $\overline P_{\rm HER} = \SI{24}{nPa}$, $n_{\rm b} = 1174$ bunches in each ring, and a luminosity of $\mathcal{L} = \SI{2.6e34}{cm^{-2}.s^{-1}}$. Current background rates in Belle~II are acceptable and, in most cases, well below the limits listed in Table~\ref{tab:DetectorLimits}. The TOP limit before LS1 is related to the replacement of TOP conventional PMTs planned for LS1. At the same time, the limit after LS1 is associated with the replacement of ALD PMTs in LS2 and the longevity of life-extended ALD PMTs. The TOP luminosity background is assumed to be \SI{0.925}{MHz/PMT} per \SI{1e35}{cm^{-2}.s^{-1}}. The upper background rate limit quoted for the Diamond read-out electronics can be increased by selecting a lower signal amplification. The KLM detector limit corresponds to the muon reconstruction efficiency drop of about 10\%. Rates in the KLM appear high, but this is an analysis artifact: the barrel RPCs have a high pedestal rate associated with the noise floor. The KLM is a rather robust detector, with gradual, temporary efficiency reduction but no permanent damage at high rates. A dedicated analysis that suppresses the pedestal in KLM-RPC layers is in progress.

%%%%%%%%%%%%%%%%%%%%%%%%%%%%%%%%%%%%%%%%%%%%%%%%%%%%%%%%%%%%%%%%%%%%%%%%%
\begin{table}[htb]
\centering
    \caption{\label{tab:DetectorLimits}Background rate limits for different Belle~II detector sub-systems. The third column shows the total measured background rate in June~2021, excluding the pedestal rate.}
    \begin{tabular}{lccc}
    \hline\hline
    Detector & \multicolumn{2}{c}{BG rate limit} & Measured BG\\
    \hline
    Diamonds & \multicolumn{2}{c}{\SIrange{1}{2}{rad/s}} & $<\SI{132}{mrad/s}$\\
    PXD & \multicolumn{2}{c}{\SI{3}{\%}} & \SI{0.1}{\%}\\
    SVD L3, L4, L5, L6 & \multicolumn{2}{c}{\SI{4.7}{\%}, \SI{2.4}{\%}, \SI{1.8}{\%}, \SI{1.2}{\%}} & $<\SI{0.22}{\%}$\\
    CDC & \multicolumn{2}{c}{\SI{200}{kHz/wire}} & \SI{22.3}{kHz/wire}\\
    ARICH & \multicolumn{2}{c}{\SI{10}{MHz/HAPD}} & \SI{0.5}{MHz/HAPD}\\
    Barrel KLM L3 & \multicolumn{2}{c}{\SI{50}{MHz}} & \SI{4}{MHz}\\
    & \multicolumn{2}{c}{non-luminosity BG} &\\\cline{2-3}
    & Before LS1 & After LS1 &\\\cline{2-3}
    TOP ALD & \SI{3}{MHz/PMT} & \SI{5}{MHz/PMT} & \SI{1.8}{MHz/PMT}\\
    & \multicolumn{2}{c}{+ luminosity BG} &\\
    \hline\hline
    \end{tabular}
\end{table}
%%%%%%%%%%%%%%%%%%%%%%%%%%%%%%%%%%%%%%%%%%%%%%%%%%%%%%%%%%%%%%%%%%%%%%%%%

Figure~\ref{fig:BackgroundComposition} shows that dominant backgrounds for all sub-systems are the LER single-beam (beam-gas and Touschek) and luminosity backgrounds. The HER single-beam background is at the level of \SI{10}{\%} for most of the detectors except for the ARICH, which is placed on the FWD side of the Belle~II and hence sees more forward-directed particle losses in the IR from the electron beam. QCS-BWD-315, BP-FWD-325, and QCS-FWD-225 indicate backward QCS, beam pipe, and forward QCS Diamond detectors, respectively, with the higher dose rate. Barrel KLM L3 corresponds to the innermost RPC layer in the barrel region of the KLM detector. TOP ALD shows the averaged background over ALD-type MCP-PMTs, slots from 3 to 9.

The current level of the SR is of no concern in terms of occupancy for the innermost layers of the vertex detector. However, in the case of a large increase, SR may cause inhomogeneities in PXD module irradiation, which would make it more difficult to compensate by adjusting the operation voltages of the affected modules.

%%%%%%%%%%%%%%%%%%%%%%%%%%%%%%%%%%%%%%%%%%%%%%%%%%%%%%%%%%%%%%%%%%%%%%%%%%%
%%%%%%%%%%%%%%%%%%%%%%%%%%%%%%%%%%%%%%%%%%%%%%%%%%%%%%%%%%%%%%%%%%%%%%%%%%%

The high injection background observed in 2022b led to a visible degradation of CDC performance shortly after injection, caused by a voltage drop in the HV resistor chain and space charge effects.

Although currently at an acceptable level, some detectors started to see single-event upsets (SEU) of FPGAs electronics boards in 2021 and 2022. These SEUs are presumably from neutrons created in EM showers. Therefore, during LS1, we plan to introduce an additional Belle~II shielding against neutrons and EM showers originating from the accelerator tunnel and final focusing magnets~\cite{Natochii2022}.

During the same running period, studies revealed single-beam and single-bunch instabilities that cause vertical emittance blow-up in the LER. This occurred at lower bunch currents ($\sim\SI{1}{mA/bunch}$) than predicted by a standard TMCI theory ($\sim\SI{2}{mA/bunch}$), limiting the bunch current at the time. Machine studies showed that the instability threshold relates to the collimator settings and the vertical beam size inside the QCS beam pipe structure. This observation implies that the beam instability observed in the LER is not simply a TMCI phenomenon and might force us to eventually open collimators at higher beam currents and accept higher detector backgrounds. Moreover, dedicated studies held in 2022 show that the bunch current threshold before the instabilities occur also depends on the interplay between the machine impedance, which is mainly dictated by collimator settings and the IR beam pipe aperture, and the bunch-by-bunch feedback system~\cite{Tobiyama2016} adjusted to dump the frequency of the 0~mode (vertical betatron tune: $\nu_\mathrm{y}$), while the vertical beam size blow-up is attributed to oscillations caused by the --1~mode ($\nu_\mathrm{y} - \nu_\mathrm{s}$, where $\nu_\mathrm{s}$ is a synchrotron tune)~\cite{Terui:2022fli}.

In order to reach the luminosity of \SI{4.7e34}{cm^{-2}.s^{-1}} in 2022, we gradually increased beam currents above \SI{1}{A}. This consequently induced more frequent sudden beam losses (SBLs) in both rings. The SBLs usually appeared at the bunch current above \SI{0.7}{mA/bunch}, limiting the maximum beam currents and hence the collision luminosity. Some of the larger beam loss events caused severe damage to collimators, which made it challenging to effectively control beam backgrounds in Belle~II. Moreover, some events also induced high radiation doses in the IR, damaging \SIrange{4}{5}{\%} of switcher channels in the PXD detector and causing QCS quenches.

\begin{figure}[htb]
\centering
\includegraphics[width=\textwidth]{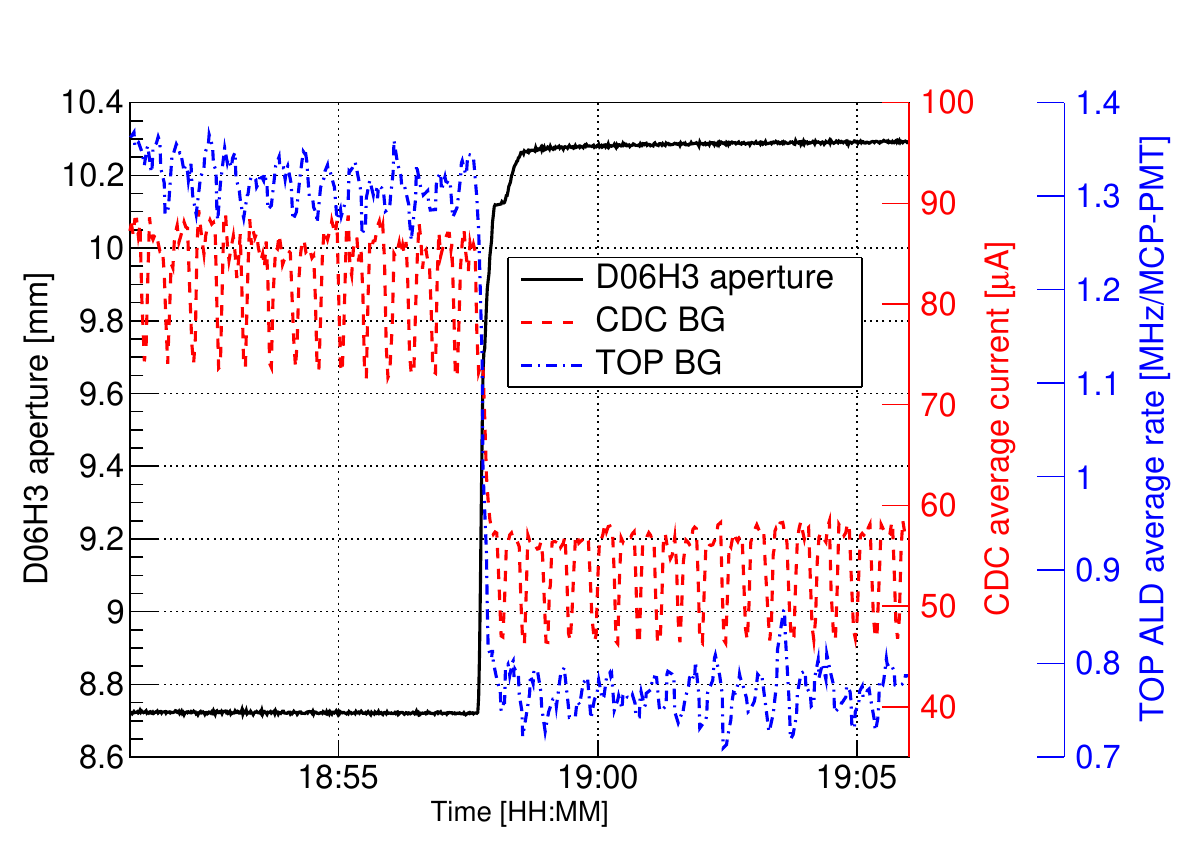}
\caption{The measured on December~2, 2021 (JST) Belle~II background as a function of the LER horizontal collimator aperture. The data were taken during LER single-beam operation at the beam current of \SI{750}{mA} with 1174~bunches. The periodic structure of the background observables is due to top-up injection.}
\label{fig:DamagedCollimatorImpactOnBG}
\end{figure}

Figure~\ref{fig:DamagedCollimatorImpactOnBG} illustrates the impact of the damaged horizontal collimator in the LER (D06H3) on Belle~II backgrounds. The collimator surfaces closest to the beam were damaged due to the injection kicker failure steering some part of the main beam horizontally toward the first horizontal collimator D06H3, which is located about \SI{1.2}{km} from the IP. By increasing the collimator aperture (black, solid line), which is defined as the distance between collimator jaws, the measured background for the most vulnerable Belle~II sub-systems, CDC (red, dashed line), and TOP (blue, dot-dashed line), dropped by about 40\%. 

Although the D06H3 collimator was not damaged in SBL events, Fig.~\ref{fig:DamagedCollimatorImpactOnBG} clearly demonstrates that opening the damaged collimator helps reduce Belle~II backgrounds. Therefore, in 2022 after more than 50~SBLs, several damaged collimators were opened, which led to a significant Belle~II background increase. 

%%%%%%%%%%%%%%%%%%%%%%%%%%%%%%%%%%%%%%%%%%%%%%%%%%%%%%%%%%%%%%%%%%%%%%%%%%%
%%%%%%%%%%%%%%%%%%%%%%%%%%%%%%%%%%%%%%%%%%%%%%%%%%%%%%%%%%%%%%%%%%%%%%%%%%%

\section{Background simulation procedure}
\label{sec:BackgroundSimulationProcedure}

We use several dedicated Monte-Carlo (MC) simulation tools to predict beam backgrounds at different machine settings and investigate possible mitigation measures. In addition, the prepared MC samples, stored in so-called \textit{BG overlay} files, are used for software training and detector performance studies.

For the simulation of beam-gas and Touschek scattering, we first use the Strategic Accelerator Design (SAD)~\cite{SAD2022} software developed at KEK. Bunches of scattered particles ($e^{\rm +}$ or $e^{\rm -}$) are generated at different locations around the ring and tracked for 1000 machine turns. The code collects the coordinates of particles, which reach the aperture of the machine, i.e., the inner beam pipe wall or collimators. These particles are then considered as lost. We recently improved the simulation of how beam particles interact with collimators. We have implemented a more realistic collimator profile and now include particle scattering off collimator materials. The measured vacuum pressure distribution around the ring is now also used to improve the beam-gas background simulation. Here, the average ring residual gas pressure at the center of the beam pipe is defined as: 
\begin{equation}
    \bar{P} = \bar{P}_\mathrm{0} + 3 \times \mathrm{d}\bar{P}/\mathrm{d}I \times I,
    \label{eq:eq2}
\end{equation}
where the \textit{base} pressure, $\bar{P}_\mathrm{0}$, is the average ring pressure when there is no beam ($I = \SI{0}{A}$), and the \textit{dynamic} pressure, $\mathrm{d}\bar{P}/\mathrm{d}I \times I$, is the average pressure increase per unit current, physically caused by gas molecules being released from the inner beam pipe walls. Both the base and dynamic pressures are estimated from vacuum pressure measurements with cold cathode gauges (CCG) installed around each ring. The factor 3 in the ring pressure formula is used to take into account the vacuum conductance between the beam pipe and CCGs, while the measured base pressure is assumed to be the same as seen by the beam. Further details about the multi-turn particle tracking in SAD for SuperKEKB can be found in Ref.~\cite{NATOCHII2023168550,PhysRevAccelBeams.24.081001}. 

At the next step, the lost particles in SAD are passed to Geant4~\cite{AGOSTINELLI2003250,1610988,ALLISON2016186} for further simulation. We only keep particles lost near the interaction region, out to $\pm\SI{30}{m}$ from the IP. The Geant4 software is embedded into the Belle~II Analysis Software Framework (basf2)~\cite{Kuhr:2018lps,basf2-zenodo}. A realistic description of the geometry of the detector and accelerator tunnel is used to simulate the resulting electromagnetic showers and secondary particles produced inside and outside the detector. Energy deposits in detectors are digitized and can then be compared against dedicated single-beam and luminosity background measurements.

To produce the MC samples for the luminosity background due to radiative Bhabha scattering and the two-photon process, we use the same Geant4-based software and geometry, but SAD is not required.

\begin{figure}[htb]
\centering
\includegraphics[width=0.48\textwidth]{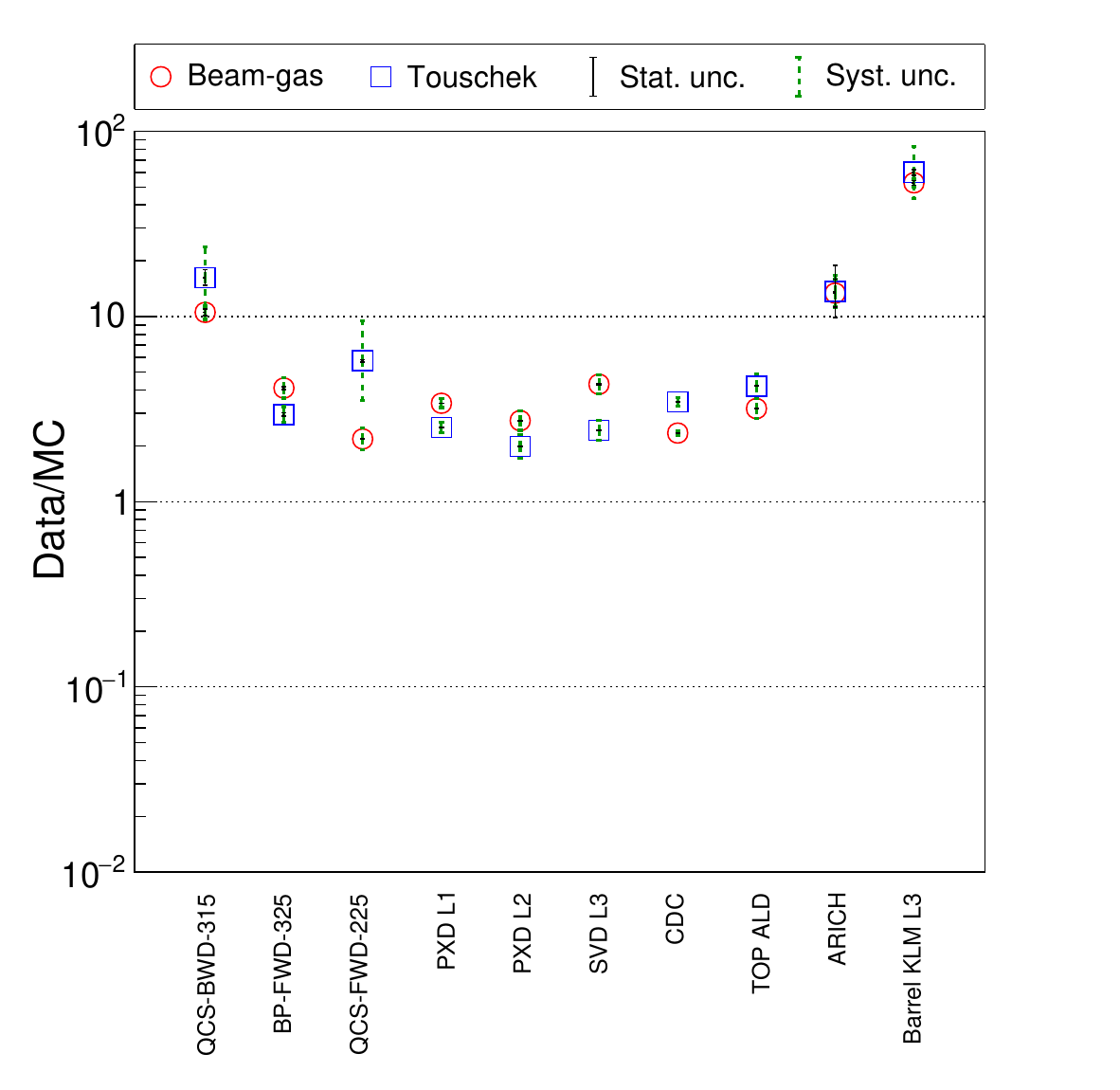}
\includegraphics[width=0.48\textwidth]{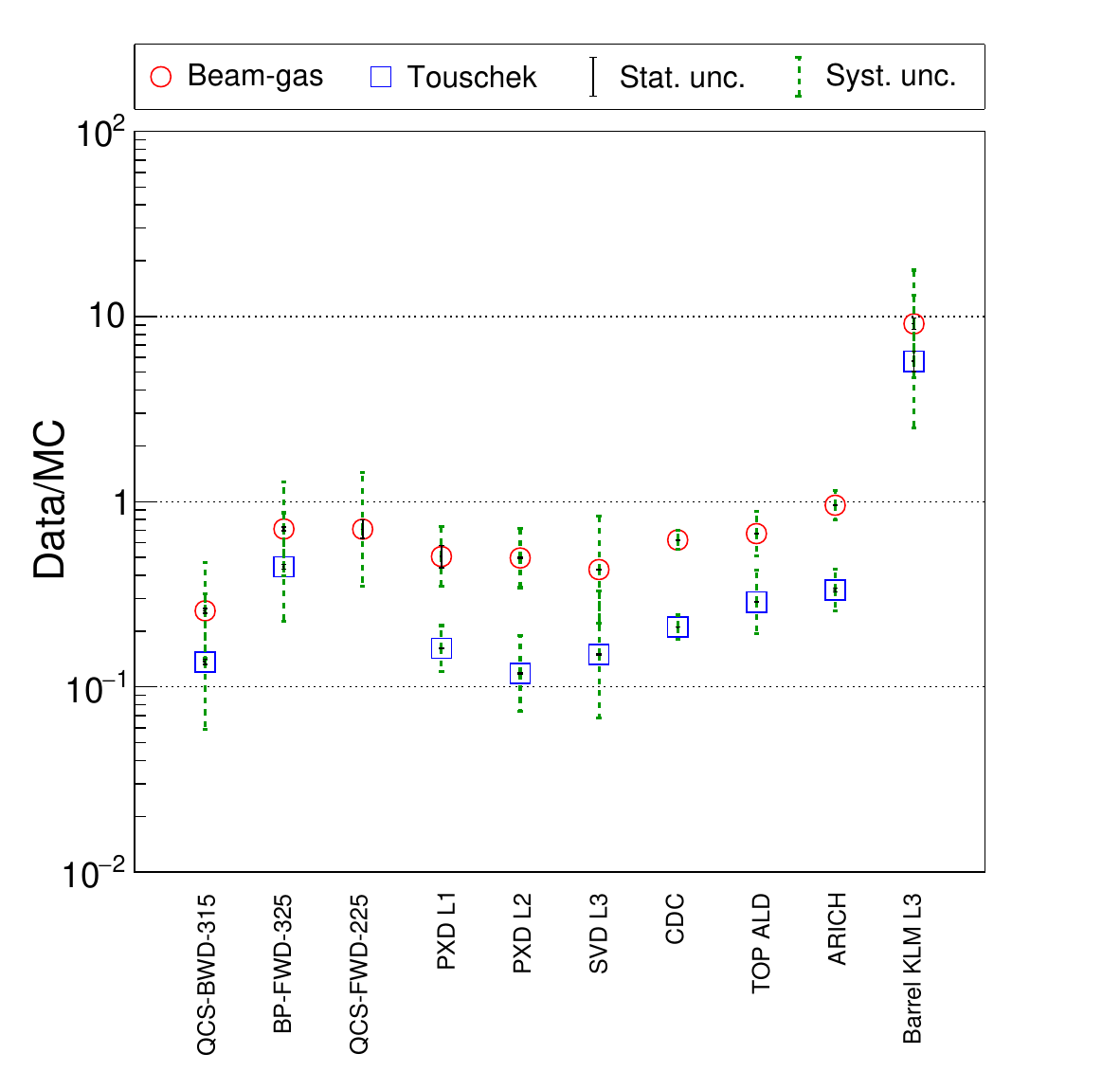}\\
\includegraphics[width=0.48\textwidth]{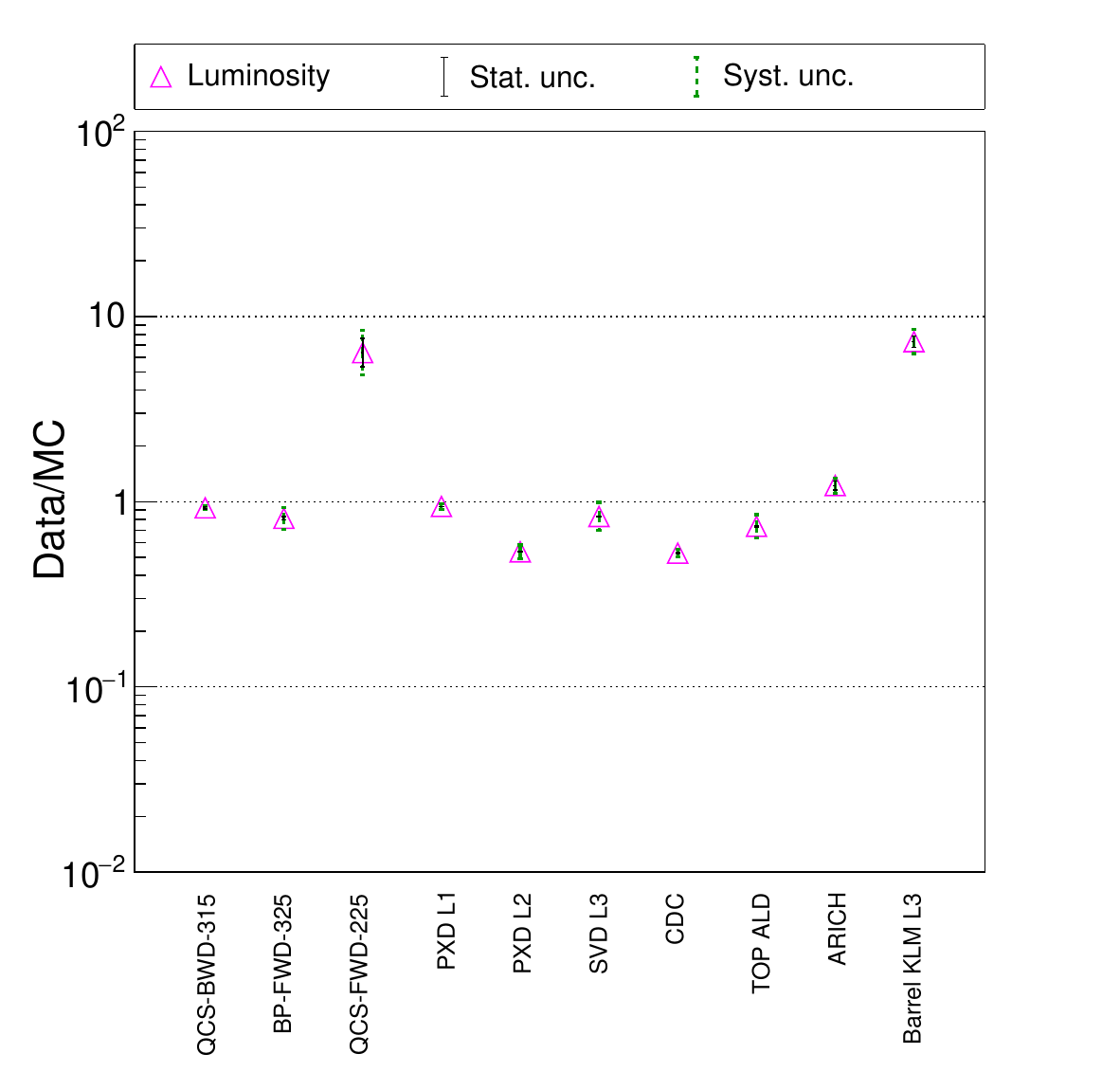}
\caption{Combined Data/MC ratios for LER~(top, left), HER~(top, right) single-beam and Luminosity~(bottom) backgrounds.}
\label{fig:DataOverMCplots}
\end{figure}

The current accuracy of our beam background simulation is illustrated by Fig.~\ref{fig:DataOverMCplots}, which shows ratios of measured to simulated background rates, so-called Data/MC ratios. The ratios are shown separately for different background components. These ratios are generally within one order of magnitude from unity, except for the shown innermost Barrel KLM RPC layer, where the simulated BG is still vastly underestimated. This is a substantial improvement compared to the early SuperKEKB commissioning phases in 2016~\cite{LEWIS201969} and 2018~\cite{LIPTAK2022167168}.

Since the neutron shielding around Belle~II is not ideal, neutrons can penetrate the detector and cause performance degradation. Our MC simulation predicts fast neutrons traveling from the accelerator tunnel towards Belle~II due to i) single-beam losses at the collimators closest to the IR ($\sim\SI{15}{m}$ from the IP) and ii) radiative Bhabha scattering producing photons parallel to the two beams at the IP. These photons follow the beams out of Belle~II, resulting in localized hot spots in the beam pipes just outside of the detector ($\sim\SI{10}{m}$ from the IP)~\cite{SCHUELER2022167291}. However, as discussed in Ref.~\cite{Natochii2022}, during LS1, we plan to install additional concrete and polyethylene shielding around Belle~II, which according to preliminary simulation, should reduce the neutron flux from the accelerator tunnel by up to a factor of 2.

%%%%%%%%%%%%%%%%%%%%%%%%%%%%%%%%%%%%%%%%%%%%%%%%%%%%%%%%%%%%%%%%%%%%%%%%%%%
%%%%%%%%%%%%%%%%%%%%%%%%%%%%%%%%%%%%%%%%%%%%%%%%%%%%%%%%%%%%%%%%%%%%%%%%%%%

\section{Background extrapolation}
\label{sec:BackgroundExtrapolation}

In this section, we discuss the Monte-Carlo simulation setup to predict beam backgrounds in Belle~II at the luminosity above \SI{3e35}{cm^{-2}.s^{-1}}, so-called \textit{after LS2} operation. The outcome of this exercise is the predicted Belle~II background rate in different sub-detectors at higher luminosity with expected detector safety factors\footnote{The safety factor is defined as a ratio between the detector limit and predicted background rate. It shows how much the background level can increase before reaching the detector limit.}.

%%%%%%%%%%%%%%%%%%%%%%%%%%%%%%%%%%%%%%%%%%%%%%%%%%%%%%%%%%%%%%%%%%%%%%%%%
\begin{table}[htbp]
\centering
    \caption{\label{tab:PredictedBeamParameters}SuperKEKB parameters used for the background extrapolation study. $\beta^\mathrm{*}$, $\mathcal{L}$, $I$, $\bar{P}_\mathrm{eff.}$, $n_\mathrm{b}$, $\varepsilon$, $\sigma_\mathrm{z}$, and $CW$ stand for the betatron function at the IP, luminosity, beam current, average beam pipe pressure, number of bunches, equilibrium beam emittance, bunch length, and Crab-Waist sextupoles, respectively. The numbers in columns Before LS2 and Target are from Ref.~\cite{NATOCHII2023168550}, while Design column numbers are adopted from Ref.~\cite{Ohnishi2013}.}
    \begin{tabular}{lcccccc}
    \hline\hline
    Setup & Before LS2 & Target & Design\\
    \hline
    $\beta^\mathrm{*}_\mathrm{y}$(LER/HER)~[mm]& 0.6/0.6 & 0.27/0.3 & 0.27/0.3\\
    $\beta^\mathrm{*}_\mathrm{x}$(LER/HER)~[mm]& 60/60 & 32/25 & 32/25\\
    $\mathcal{L}$~[$\times 10^{35}\rm~cm^{-2}s^{-1}$]& 2.8 & 6.0 & 8.0\\
    $I$(LER/HER)~[A] & 2.52/1.82 & 2.80/2.00 & 3.6/2.6\\
    $\bar{P}_\mathrm{eff.}$(LER/HER)~[nPa] & 48/17 & 52/18 & 133/133\\
    $n_\mathrm{b}$~[bunches] & 1576 & 1761 & 2500\\
    $\varepsilon_\mathrm{x}$(LER/HER)~[nm] & 4.6/4.5 & 3.2/4.6 & 3.2/4.6\\
    $\varepsilon_\mathrm{y}/\varepsilon_\mathrm{x}$(LER/HER)~[\%] & 1/1 & 0.27/0.28 & 0.27/0.28\\
    $\sigma_\mathrm{z}$(LER/HER)~[mm] & 8.27/7.60 & 8.25/7.58 & 6.0/6.0 \\
    $CW$ & ON & OFF & OFF \\
    \hline\hline
    \end{tabular}
\end{table}
%%%%%%%%%%%%%%%%%%%%%%%%%%%%%%%%%%%%%%%%%%%%%%%%%%%%%%%%%%%%%%%%%%%%%%%%%

Currently, the target instantaneous luminosity is \SI{6.0e35}{cm^{-2}.s^{-1}}. It should allow us to collect \SI{50}{ab^{-1}} of data within the next decade. However, at the time we are preparing this CDR, we do not have a so-called \textit{working machine lattice} to reach the target at the design optics with $\beta^{*}_\mathrm{y} = \SI{0.3}{mm}$, see Table~\ref{tab:PredictedBeamParameters} (Target column). Moreover, the final design of the IR magnet system and beam pipes is not completed, and it is still under discussion; see Section~\ref{sec:IRdesign}. According to the dedicated studies discussed in Refs.~\cite{Natochii2022,NATOCHII2023168550}, the highest possible luminosity that could be achieved at the current SuperKEKB is \SI{2.8e35}{cm^{-2}.s^{-1}} at $\beta^{*}_\mathrm{y} = \SI{0.6}{mm}$, see Table~\ref{tab:PredictedBeamParameters} (Before LS2 column). 

Despite the marginal discrepancy between the machine parameters before LS2, as outlined in Table~\ref{tab:PredictedBeamParameters} proposed long ago for the detector background study, and those defined by the SuperKEKB team in Table~\ref{tab:machineParams} (second stage) based on the most recent machine operation experience, we maintain the perspective that these disparities are relatively minor. We are of the opinion that these variations do not significantly impact the outcomes of our study or alter our final conclusions. However, the use of a more aggressive scenario with $\mathcal{L} = \SI{2.8e35}{cm^{-2}.s^{-1}}$ (Table~\ref{tab:PredictedBeamParameters}) instead of $\mathcal{L} = \SI{2.4e35}{cm^{-2}.s^{-1}}$ (Table~\ref{tab:machineParams}) might be considered as a conservative approach for the detector backgrounds simulation in the CDR.

\begin{figure}[htb]
\centering
\includegraphics[width=\textwidth]{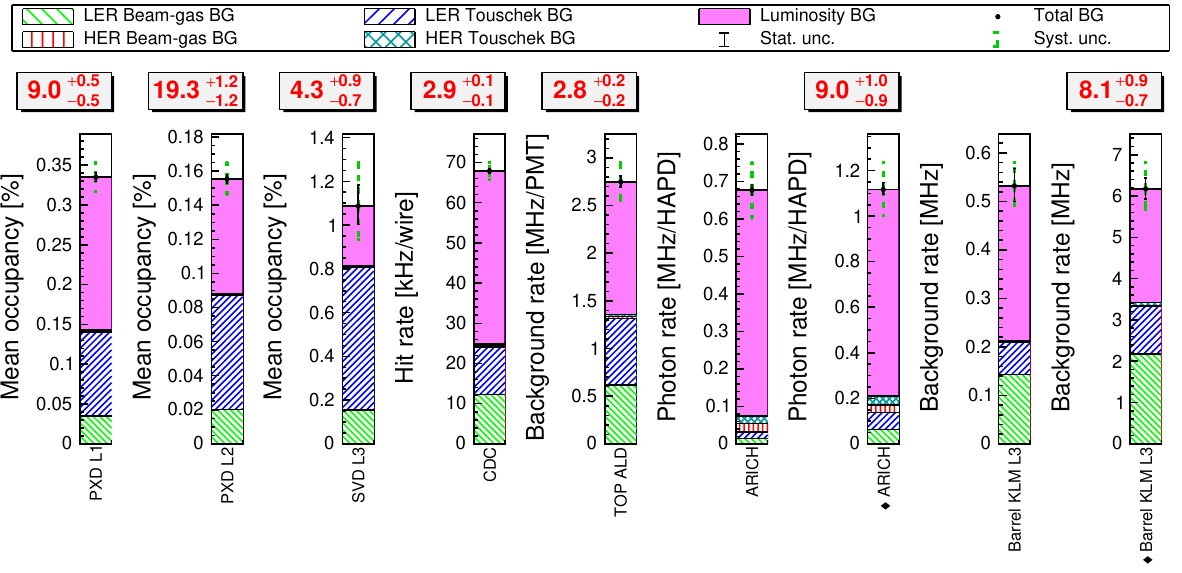}
\caption{Estimated Belle~II background composition for predicted beam parameters before LS2. Each column is a stacked histogram of BG rates from dedicated MC samples produced with FTFP\_BERT\_HP Geant4 Physics List and scaled with average Data/MC ratios listed in Table~\ref{tab:ScalingFactors}, except for the column marker with a $\blacklozenge$ symbol, where individual Data/MC ratios from Fig.~\ref{fig:DataOverMCplots} are used instead. The red numbers in rectangles are detector safety factors, showing that Belle~II should be able to operate safely until a luminosity of \SI{2.8e35}{cm^{-2}.s^{-1}}, with some important caveats, discussed in the text.}
\label{fig:BackgroundComposition_betay06_nlc}
\end{figure}

The prediction for the BG rates before LS2, with known machine configuration, is made with the full BG simulation described before, re-scaling each BG component with the  measured average Data/MC ratios, shown in first column of Table~\ref{tab:ScalingFactors}. Belle~II background rates expected with these conditions, shown in Fig.~\ref{fig:BackgroundComposition_betay06_nlc}, are high but acceptably below detector limits listed in Table~\ref{tab:DetectorLimits}.

On the contrary, given all the mentioned uncertainties in the machine configuration after LS2 (i.e., machine-detector protection upgrade during LS1, final focusing magnet redesign during LS2, collimator relocation, and the injection chain tuning), an alternative solution to roughly estimate background rates in the detector was used for this extrapolation. The background predicted for conditions before LS2 are used as a starting point and scaling factors are applied to the various components to extrapolate them to after LS2 in the following way. In addition to Data/MC ratios, the following three scaling factors are considered for single-beam background components: i) \textbf{x2} -- \textit{optimistic} \textbf{Scenario-1}, comes from a simple scaling of all single-beam background components from \textit{before LS2} to \textit{target} beam parameters (see Table~\ref{tab:PredictedBeamParameters}) following heuristic fit formulas discussed in Ref.~\cite{NATOCHII2023168550}, ii) \textbf{x5} -- \textit{intermediate} \textbf{Scenario-2}, and iii) \textbf{x10} -- \textit{conservative} \textbf{Scenario-3}, an arbitrary factor assuming that all single-beam backgrounds will be elevated by order of magnitude after LS2. Luminosity BG terms are instead correctly simulated for the target luminosity after LS2 over the luminosity before LS2, applying the measured Data/MC ratio.

%%%%%%%%%%%%%%%%%%%%%%%%%%%%%%%%%%%%%%%%%%%%%%%%%%%%%%%%%%%%%%%%%%%%%%%%%%%
%%%%%%%%%%%%%%%%%%%%%%%%%%%%%%%%%%%%%%%%%%%%%%%%%%%%%%%%%%%%%%%%%%%%%%%%%%%

\begin{figure}[htb]
\centering
\includegraphics[width=\textwidth]{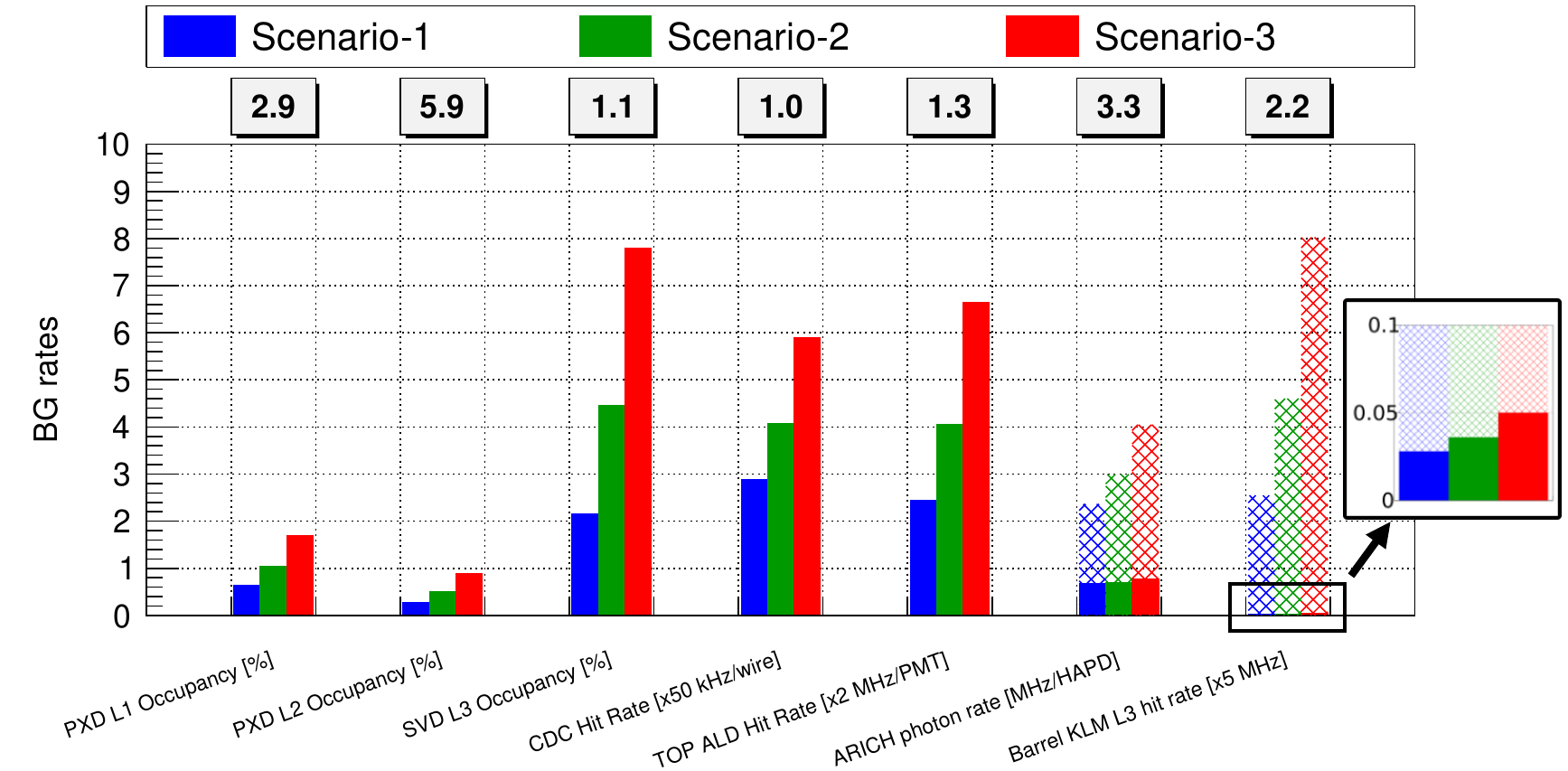}
\caption{Estimated beam background rates in Belle~II for after LS2 operation at luminosity of \SI{6.0e35}{cm^{-2}.s^{-1}}. Solid bars represent BG overlay results with average Data/MC ratio and default Belle~II Physics List, while hatched bars for ARICH and KLM stand for the BG levels with accurate neutron physics and individual Data/MC ratios. The histograms are not stacked. The numbers in rectangles are detector safety factors for Scenario-2: PXD, SVD, CDC, and TOP factors are calculated using solid bar numbers, while ARICH and KLM factors are shown as hatched bars. The units in brackets contain scaling factors used for better visualization, fitting all bars within the same vertical axis range; for example, the estimated intermediate (green, Scenario-2) CDC Hit Rate should be read as $4 \times 50 = \SI{200}{kHz/wire}$.}
\label{fig:BgLevels}
\end{figure}

Figure~\ref{fig:BgLevels} shows background rates in the crucial sub-detectors estimated with basf2 (prerelease-07-00-00d) for after LS2 conditions at the luminosity of \SI{6.0e35}{cm^{-2}.s^{-1}}, for the three different Scenarios described previously. Table~\ref{tab:ScalingFactors} summarizes the various scaling factors applied to produce BG overlay files for after LS2 extrapolations starting from the before LS2 setup.

%%%%%%%%%%%%%%%%%%%%%%%%%%%%%%%%%%%%%%%%%%%%%%%%%%%%%%%%%%%%%%%%%%%%%%%%%%%
%%%%%%%%%%%%%%%%%%%%%%%%%%%%%%%%%%%%%%%%%%%%%%%%%%%%%%%%%%%%%%%%%%%%%%%%%%%

%%%%%%%%%%%%%%%%%%%%%%%%%%%%%%%%%%%%%%%%%%%%%%%%%%%%%%%%%%%%%%%%%%%%%%%%%
\begin{table}[htbp]
\centering
    \caption{\label{tab:ScalingFactors}Scaling factors used for the BG overlay MC sample production after LS2 at the luminosity of \SI{6.0e35}{cm^{-2}.s^{-1}} for three Scenarios (Sc.).}
    \begin{tabular}{lccccccc}
    \hline\hline
    Background & Average & \multicolumn{3}{c}{Single-beam scaling} & \multicolumn{3}{c}{Total scaling}\\\cline{3-8}
    component & Data/MC & Sc.-1 & Sc.-2 & Sc.-3 & Sc.-1 & Sc.-2 & Sc.-3\\
    \hline
    Beam-gas LER & 3.46 & 2 & 5 & 10 & 6.92 & 17.30 & 34.60\\
    Beam-gas HER & 0.63 & 2 & 5 & 10 & 1.26 & 3.15 & 6.30\\
    Touschek LER & 3.44 & 2 & 5 & 10 & 6.88 & 17.20 & 34.40\\
    Touschek HER & 0.18 & 2 & 5 & 10 & 0.36 & 0.90 & 1.80\\
    Luminosity & 0.81 & 1 & 1 & 1 & 0.81 & 0.81 & 0.81\\
    \hline\hline
    \end{tabular}
\end{table}
%%%%%%%%%%%%%%%%%%%%%%%%%%%%%%%%%%%%%%%%%%%%%%%%%%%%%%%%%%%%%%%%%%%%%%%%%

We note that there are a number of issues related to the scaling factors listed in Table~\ref{tab:ScalingFactors} and BG rates shown in Fig.~\ref{fig:BgLevels} that must be taken into account:

\begin{itemize}
    
    \item \textbf{Data/MC ratios}\\
    After producing BG overlay MC samples with the scaling factors listed in Table~\ref{tab:ScalingFactors}, we found a bug in the basf2 code used for the PXD occupancy calculation. This bug affected how particles interact with the detector, and it led to the overestimation of simulated BG rates used for the Data/MC analysis. In addition, there was a bug related to the heuristic fit formula used for the estimation of the measured BG components in the Belle~II sub-systems. Instead of using estimated base pressure in the ring (\SIrange{7}{15}{nPa}~\cite{NATOCHII2023168550}), the code used a default value (\SI{10}{nPa}). Thus, fixing the mentioned bugs results in a set of new, slightly different Data/MC ratios shown in Table~\ref{tab:DataOverMcRatios}, which should be used for future campaigns of BG overlay files production. In Table~\ref{tab:DataOverMcRatios}, we assume the luminosity Data/MC ratio equals one based on our observation for the PXD detector since, applying all the mentioned bugfixes here, the luminosity ratio tends to be closer to unity. Moreover, the PXD BG is dominated by the luminosity component, while in outer sub-detectors, the relative luminosity contribution is less significant. Therefore, applying the average sub-detector ratio for the luminosity BG, which is 0.81 (before bugfixes, see Table~\ref{tab:ScalingFactors}) and 0.82 (after bugfixes, see Ref.~\cite{NATOCHII2023168550}), we could largely underestimate the PXD BG. While the conservative ratio equal to unity only slightly affects outer detector BGs.
    
    \item \textbf{Geant4 Physics List}\\
    There is a difference in calculated BG rates using MC samples produced with \textit{Default Belle~II} Physics List~\cite{Belle2PhysicsList}, which is used for the BG overlay files production, and \textit{FTFP\_BERT\_HP} Physics List which is recommended by Geant4 for collider physics applications~\cite{G4PhysicsListGuide} and used by the beam background group for the Data/MC analysis~\cite{Natochii2022,NATOCHII2023168550}. The origin of this difference is unknown and under investigation. This issue leads to at least an order of magnitude higher neutron rate in the KLM for FTFP\_BERT\_HP compared to Default Belle~II Physics List.

    \item \textbf{Beam-gas BG}\\
    An additional bug in the BG overlay files production scripts was found. The code that mixes base and dynamic components of the beam-gas BG in proportions of 1:3 (see Eq.~\ref{eq:eq2}) failed. This led to the underestimation of beam-gas BGs in Belle~II by up to a factor of 4, with the fraction of the beam-gas BG in the total BG of \SIrange{10}{20}{\%}.
    
\end{itemize}

The listed issues impact our estimation of beam-induced BGs in Belle~II at the level of 30\%, except for the ARICH and KLM, where the effect is much stronger. Due to i) the Geant4 Physics List issues mentioned above and ii) the usage of the average Data/MC ratios instead of individual ratios, the KLM BG rates are presumably underestimated by about two orders of magnitude in BG overlay MC files. In addition, the average Data/MC ratios underestimate the ARICH BG estimation by a factor of three. Therefore, for accurate safety factor estimation, in addition to BG overlay levels shown as solid bars in Fig.~\ref{fig:BackgroundComposition_betay06_nlc}, we added BG levels in hatched bars calculated with accurate neutron physics FTFP\_BERT\_HP and individual Data/MC ratios.

Nevertheless, the extrapolated BG rates for the three scenarios are still justifiable given the uncertainties associated with i) unknown machine and beam parameters at the target luminosity and ii) the set of assumptions regarding the usage of average Data/MC ratios (Table~\ref{tab:ScalingFactors}), instead of individual ratios (Fig.~\ref{fig:DataOverMCplots}), for all detectors in addition to empirical scaling factors for single-beam components of 2, 5, and 10. Therefore, we use the currently available BG overlay MC samples, which correspond to BG rates shown in Fig.~\ref{fig:BgLevels} at the luminosity of \SI{6.0e35}{cm^{-2}.s^{-1}}, for the detector performance study. Moreover, we note that previously mentioned machine and detector protection system upgrades planned for LS1 are not included in the current background simulation, which presumably leads to overestimated detector BG rates at higher luminosity.

%%%%%%%%%%%%%%%%%%%%%%%%%%%%%%%%%%%%%%%%%%%%%%%%%%%%%%%%%%%%%%%%%%%%%%%%%
\begin{table}[htbp]
\centering
    \caption{\label{tab:DataOverMcRatios}Updated average Belle~II Data/MC ratios after the basf2 code bug fixing.}
    \begin{tabular}{cccc}
    \hline\hline
    Background & LER & HER & Comment\\
    \hline
    Beam-Gas & ${3.94}_{-0.74}^{+0.92}$ & ${0.59}_{-0.18}^{+0.25}$ & From Ref.~\cite{NATOCHII2023168550}\\
    Touschek & ${3.67}_{-0.92}^{+1.22}$ & ${0.21}_{-0.07}^{+0.10}$ & From Ref.~\cite{NATOCHII2023168550}\\
    Luminosity & \multicolumn{2}{c}{${1.00}$} & Assumption\\
    \hline\hline
    \end{tabular}
\end{table}
%%%%%%%%%%%%%%%%%%%%%%%%%%%%%%%%%%%%%%%%%%%%%%%%%%%%%%%%%%%%%%%%%%%%%%%%%

According to the BG rates estimation for Scenario-2, which is a compromise between optimistic Scenario-1 and conservative Scenario-3, Belle~II backgrounds for the PXD, ARICH, and KLM are expected to stay well below detector limits (i.e., safety factors $> 1$), while the SVD, CDC, and TOP are expected to be closer to the detector limits listed in Table~\ref{tab:DetectorLimits} (i.e., safety factors $\sim 1$). Although the TOP background is close to its limit, it will not constrain the luminosity increase but shorten the lifetime of MCP-PMTs, which in principle can be replaced during a normal summer shutdown. In addition, even without considering possible performance degradation due to high injection background, the CDC background level is estimated to be at the detector limit, which was defined based on Belle/KEKB operation and potentially is not applicable for Belle~II/SuperKEKB operation. Therefore, we plan to study the latest MC samples to accurately update the limit for the CDC. Moreover, we still need to investigate the injection and SR BGs at higher luminosities by developing dedicated simulation models and conducting machine studies. Therefore, at the current stage of our understanding, we prefer to keep both injection and SR BG rates not exceeding the measured level in June 2021 (see Fig.~\ref{fig:BackgroundComposition}), which corresponds to \SIrange{10}{20}{\%} and \SIrange{5}{10}{\%} of the total BG, respectively.

Furthermore, for more accurate background estimation and Belle~II performance study at the target luminosity of \SI{6.0e35}{cm^{-2}.s^{-1}}, the working machine lattice at $\beta^{*}_\mathrm{y} = \SI{0.3}{mm}$ is essential. Here is a list of known shortcomings and missing ingredients ordered by importance of the future BG estimation at the target luminosity after LS2:

\begin{itemize}
    \item Lack of working machine lattice at $\beta^{*}_\mathrm{y}=\SI{0.3}{mm}$;
    \item Unsimulated detector shielding planned for LS1;
    \item Unsimulated collimation system after LS1 upgrades;
    \item Unclear vacuum pressure evolution at higher beam doses;
    \item Lack of accurate SR and injection background simulations.
\end{itemize}

We are working on improving Belle~II background measurements and predictions and closely collaborating with other accelerator laboratories around the globe on optimizing SuperKEKB in order to reach the target luminosity.
\section{Radiation monitoring and beam abort system}
\label{sec:RMBA}
\editor{L.Vitale}
%\subsection{Introduction}  useless to have a subsection here. 

In this section we discuss the potential upgrades of the radiation monitoring 
and beam abort system around the interaction region. 
We first describe the present system, 
based on 28 single crystal CVD diamond sensors and its custom read-out electronics 
and briefly recall its excellent performance.  
We underline the lessons learned from the first five years of operations 
and notable features to be preserved in future,  
in view of possible medium and long term upgrade. 
Then we critically review its main limitations and possible improvements. 
Finally we discuss the specifications and technical solutions for an upgraded electronics and for the sensors. 
The present beam abort system has been also complemented in 2021 by the CLAW scintillator system. This system cannot provide a steady monitoring and recording of the dose rate, but it can detect with a very good time resolution sudden beam losses. It was thus optimized using the diamond data to provide a faster beam abort signal. Its possible upgrade will also be mentioned in the sensors part.

\subsection{The present system and features to be preserved}

The present Belle II radiation monitoring and beam abort system is based on single-crystal diamond sensors, grown by chemical vapour deposition (sCVD sensors), and read-out by custom-designed electronics \cite{RMBA:RequirementsNote2015,RMBA:DiaPerformance}. 

Diamond is known as an extremely radiation-resistant material since a very long time. 
Among the synthetic diamonds now available, 
the electronic grade sCVD diamond is the one with the highest purity, 
presenting unique characteristics that make it 
an exceptional material for radiation detectors, such as its efficiency, stability, 
rise time, temperature-independent response \cite{RMBA:sussmann2009cvd,RMBA:berdermann2014diamond}.

Our detectors are made of electronic grade sCVD sensors with a size 
$4.5\times4.5\times0.5~\mathrm{mm^2}$ and two (Ti + Pt + Au), 
$(100+120+250)~\mathrm{nm}$ electrodes, mounted on a compact Rogers printed circuit board, providing both the mechanical support and the electrical screening, completed by a thin aluminium cover. 
The final package of the detectors has a size of $18.0\times12.0\times3.1~\mathrm{mm^2}$.

For Phase 3 a total 28 radiation detectors were mounted in several key points 
of the vertex detector and of the beam pipe (BP). 
In particular 8 detectors were mounted around the cooling blocks of the BP 
to be as close as possible to the interaction point (IP) 
and at about the same radius of the PXD, 12 detectors on the SVD support cones, 
and the remaining 8 detectors on the BP bellows, close to the QCS magnets. 

Two coaxial cables, one for the signal current and the other for the bias voltage, 
are routed for about $25~\mathrm{m}$ to the diamond electronics system. 
The electronics consists of seven purpose-designed modules, 
Diamond Control Unit (DCU).
Each DCU controls and reads out groups of four diamond detectors, moreover it provides the bias voltages individually to the four diamonds.

The diamond currents are amplified by trans-impedance amplifiers, digitized by a 16-bit ADC and processed by FPGAs in the DCUs. Three amplifier gain values can be selected by resistors in the feedback loop of the front-end operational amplifier, to provide three different current measurement ranges, indexed as 0, 1, and 2 in the following, as shown in Table~\ref{tab:DCU_ranges}.  The analogue bandwidth at the lowest gain (range 2) is about 10 MHz, matched to large and fast signals; it is reduced to the order of 10 kHz at the highest gain (range 0) used for monitoring smaller signals at 10 Hz. 
More than 10 orders of magnitude for signal currents can be covered 
in the three ranges, but the switching between ranges can be done 
by a slow control procedure that requires rather long times (order of few minutes). 

\begin{table}
\centering
\caption{
Selectable current-measurement ranges of diamond control units, with the corresponding rms noise values, measured in $100$~kHz data (third column) and in $10$~Hz data (fifth column). The typical corresponding ranges and rms noise values in dose-rate units can be obtained multiplying by the average calibration factor $k = 35$~(mrad/s)/nA. The effective dynamic ranges expressed in decibel at $100$~kHz and $10$~Hz are also shown in the fourth and sixth columns.
}
\label{tab:DCU_ranges} 
\begin{tabular}{crrrrr}
\hline\noalign{\smallskip}
Range & Current & Rms noise & dyn.range & Rms noise & dyn.range \\
index  & range  & @100 kHz & @100 kHz & @10 Hz & @10 Hz\\
\noalign{\smallskip}\hline\noalign{\smallskip}
0 & $36$~nA & $0.23$~nA & $44$~dB & $0.8$~pA & $93$~dB \\
1 & $9$~$\mu$A & $3$~nA & $70$~dB & $70$~pA & $102$~dB \\
2 & $4.5$~mA & $0.22$~$\mu$A & $86$~dB & $40$~nA & $101$~dB \\
\hline\noalign{\smallskip}
\noalign{\smallskip}
\end{tabular}
\end{table}

The electronics is located in the Electronic Hut, outside of the radiation controlled 
area of the Belle II detector, and accessible even during the collisions, in case 
of maintenance, replacements or even in case of a full upgrade of the system itself. 

This system can provide a steady monitoring of the radiation dose 
during initial injection, collisions and continuous injections 
with a running efficiency close to 100\%. 
In case of sudden beam losses the system can trigger a fast beam abort 
signal on a time scale shorter than the revolution time ($10~\mathrm{\mu s}$).
Moreover the system records the dose rate with a $2.5~\mathrm{\mu s}$ granularity 
before the abort, for post-abort analysis. 

In Phase 3 run 1 (2019-2022), 
four of the eight detectors mounted on the beam pipe were dedicated 
for the abort function, with range 2 selected in the corresponding DCU; 
the remaining 24 diamond detectors were dedicated to monitoring, 
and the corresponding DCUs preset on the most sensitive current 
measurement range 0. Occasionally four of the eight QCS bellows diamonds 
were used for injection and beam abort studies 
and their corresponding DCU set to range 1 or 2 respectively. 

In this way this system provides a substantial protection not only for 
the vertex detector, but for the full Belle II detector 
and also for some accelerator components. 
The experience of the first years of running for SKB and Belle II 
showed that sudden beam losses can for instance 
damage vertex PXD detector parts and collimator jaws. 
They can also quench the super-conducting magnets of the QCS.
An early and efficient beam abort helps to prevent or at least 
reduce impact and the risk for such potentially catastrophic occurrences \cite{RMBA:DiaPerformance}. 

The compact size of this detectors and their radiation hardness 
at least up to integrated doses of the order of $0.1-0.2~\mathrm{MGy}$
are well-suited for radiation monitoring in the interaction region both 
for the present Belle II detector and for its future upgrades; 
moreover their temperature-independent response allows operation 
without additional temperature monitoring and related corrections, 
as for example needed for silicon diodes. 

In summary the notable features of the present system to be preserved also for future upgrades are: 
\begin{itemize}
    \item Fast beam abort cycle of the order of $2.5~\mathrm{\mu s}$ or better. 
    \item Recording of pre-abort dose rates with same time resolution. 
    \item Large dynamic range of dose rate measurement over at least 10 orders of magnitude (now limited by the slow switch among the three available ranges). 
    \item Flexibility, several configurable options, separate thresholds for beam abort and masking for each channel. 
    \item Performance of the sensors in terms of stability, fast response and linearity over a wide range of stationary and transient radiation. 
    \item Radiation hardness of the diamond sensors to the future integrated doses with large margins. 
    \item Overall reliability of the full system.    
\end{itemize}

\subsection{Limitations and possible improvements}

As underlined in the previous section, the present system performed very well, 
and for many aspects even better than in the original specifications 
\cite{RMBA:RequirementsNote2015}.
However some limitations have also emerged, mainly in the read-out system. 

%Limitations as underlined by EOI.
\renewcommand{\labelenumi}{\alph{enumi})}
\begin{enumerate}
    \item {Dynamic range:}

The DCU dynamic range (Table~\ref{tab:DCU_ranges}) is not sufficient to accommodate all needs. In particular: (a) accelerator tuning needs measurements of dose-rates down to the order of $1$~mrad/s, with a $10\%$ precision of about $0.1$~mrad/s, corresponding on average to about $3$~pA; (b) beam losses correlated with aborts can reach or exceed $10- 100$~krad/s, corresponding to currents of the order of several milli-ampere; (c) in continuous injection mode, the beam-losses peak during short time intervals at values exceeding the average losses by more than two orders of magnitude. The separation of the functions of $24$ detectors ($6$ DCUs) dedicated to monitoring smaller currents in range 0, and four detectors (one DCU) dedicated to generate beam aborts on large signals in range 2, is a compromise solution, working only to some extent; in particular, injection spikes still saturate range 0. A wider dynamic range would be desirable.

\item {Common mode noise spikes:}

The noise quoted in Table~\ref{tab:DCU_ranges} does not take into account some high-frequency common-mode noise, not completely filtered out by averaging the $50$~MHz ADC data at $10$~Hz. The origin of this effect has been studied; its mitigation 
has been discussed here \cite{RMBA:EoI}.

\item {Memory read:}

The reading of the DCU internal buffer memory, containing the $400$~kHz data, is designed as a memory dump after the arrival of an abort signal from SuperKEKB. The abort ``freezes'' the memory by stopping the increase of the memory pointer, and allows the subsequent read-out of the appropriate memory section by EPICS. This mechanism is too rigid and has some drawbacks. During the readout time interval the abort function (computation of moving sums and comparison with abort thresholds) is suspended: memory read and abort are incompatible. This is not a problem after simultaneous aborts of both beams, as those generated now by the diamond-based system. However it makes it impossible, for VXD detector safety reasons, to read the memory after a single-beam abort or during normal operations with circulating beams, while periodically obtaining the information of dose-rates from diamond detectors with full $2.5$~$\mu$s time resolution might be relevant (for instance to study the injection losses with such time resolution, or during background studies). 

\item {Synchronization:}

The synchronization with SuperKEKB is limited to the exchange of abort signals: abort requests from DCUs to SuperKEKB and the abort confirmation timing signals from SuperKEKB. A rough alignment of $400$~kHz data recorded by the DCUs with SuperKEKB data and events is only possible through the SuperKEKB abort timing signal.
A tighter synchronization using an accelerator clock and some sort of time stamps from SuperKEKB had been considered as possible part of the initial specifications, but was abandoned due to lack of feedback from SuperKEKB and for the sake of simplicity. For the future, at least an additional timing signal from SuperKEKB injection should be considered. In this case, a precision of the order of 100~ns should be sufficient. 

\item {Injection-related dose rates:}

As mentioned previously, injection-related dose-rate peaks are clearly visible in the $400$~kHz data, read out after aborts or for some special studies of range 0 saturation effects. The time resolution of $2.5$~$\mu$s is largely sufficient to observe the beam-loss increase and oscillation patterns following injection, studied by CLAWS in Phase 1~\cite{RMBA:Lewis:2018ayu}. Upgraded electronics should be able both to avoid saturation and to integrate the radiation dose due to injection separately, using SuperKEKB injection timing signals as time reference, for proper monitoring of injection quality.

\item {HV control:}

The high voltage (HV) modules included in the DCUs are quite stable and reliable. However an internal reading of the output HV value is not available at present; a manual measurement of the voltage at the output connector is needed to cross-check the voltage setting. In one occasion, an EPICS software bug resulted in a wrong HV setting for a substantial period of time, during which diamond detector signals were reduced by about a factor two. Recorded data could be corrected for this effect afterwards, but clearly a better protection against such accidents is needed. 

\item {External logics:}

The type conversion and distribution of SuperKEKB abort signals to DCUs and the fan-out of DCU abort requests to SuperKEKB, CLAWS and PXD are performed by NIM modules hosted in a NIM crate. This solution is simple, flexible and reliable, but the availability of spares for the NIM crate and modules might be questioned in the long term. 
%The DCUs also have an output connector with eight logic signals (two abort thresholds for each of four diamond detectors), which is not used at the moment. We developed an external logic including signal level adapters and a programmable CAEN FPGA for added flexibility in defining abort conditions and correlations among different DCUs. It is available at INFN Trieste for future installation.
\end{enumerate}

%The most relevant limitations of the current diamond system are:
%\begin{itemize}
%    \item Limited dynamic range of dose rate measurement. For each sensor it must be determined in advance whether it is to be used for accurately measuring the dose rate (high sensitivity) or for generating beam abort signals (low sensitivity). 
%    \item Saturation of the most sensitive range during continuum injection.
%    \item Monitor and abort functionalities have to be deactivated during post-mortem readout of buffer memories.
%\end{itemize}
%Other limitations are 
%\begin{itemize}
%    \item Separate abort thresholds during collisions and top off injections: 
%    %under evaluation and probably implementable during LS1. 
%    \item Separate accounting for dose during collisions and top off injections. 
%    \item Common-mode noise not averaged out in some conditions.
%\end{itemize}

\subsection{Specifications for the new electronics}
\label{subsec:specifications}

To avoid confusion deriving from different calibration factors for the diamond detectors, we will use current units everywhere, keeping in mind the average conversion factor $35$~(mrad/s)/nA to dose rate units, and the relative variation of individual factors of up to $\pm 50\%$ with respect to the average.

The new electronics for the readout and control of diamond detectors will have the same functions as the DCUs, in particular: (1) individual HV control for the diamond detectors; (2) amplification and digitization of current (dose-rate) signals; (3) integration and recording of charge (dose) samples in a ring buffer memory at $400$~kHz;
(4) computation of moving sums of programmable numbers for the samples recorded in memory; (5) comparison with abort thresholds to generate logical signals; (6) generation of abort requests, using programmable logical combinations of these logical signals, and programmable masks; (7) read-out of ring buffer memory; (8) computation of $10$~Hz monitoring data from sums of samples recorded in memory, and their read-out; (9) initialization of parameters such as abort thresholds etc.

The DCU limitations listed in the previous sections will be overcome by complying with the requirements listed here; the corresponding numbers are summarized in Table~\ref{tab:requirements}.

\begin{table}
\centering
\caption{
Summary of dynamic range parameters, assuming an average current to dose-rate conversion factor $k = 35$~(mrad/s)/nA. Explanations are given in the text.
}
\label{tab:requirements} 
\begin{tabular}{lrr}
\hline\noalign{\smallskip}
Parameter & value & sampling \\
\noalign{\smallskip}\hline\noalign{\smallskip}
Dose rate monitoring & & \\
\hline\noalign{\smallskip}
Current (dose-rate) rms noise & $3$~pA ($0.1$~mrad/s) & $10$~Hz \\
Max. current (dose-rate) & $5$~mA ($175$~krad/s) & $10$~Hz \\
Total dynamic range & $184$~dB & $10$~Hz \\
Max. injection peaks & $20$~$\mu$A ($0.7$~krad/s) & $10$~Hz \\
Continuous injection: dynamic range & $136$~dB & $10$~Hz \\
\hline\noalign{\smallskip}
Beam abort generation & & \\
\hline\noalign{\smallskip}
Current (dose-rate) rms noise & $0.2$~$\mu$A ($7$~rad/s) & $400$~kHz \\
Max. current (dose-rate) & $5$~mA ($175$~krad/s) & $400$~kHz \\
Dynamic range & $88$~dB & $400$~kHz \\
\hline\noalign{\smallskip}
\noalign{\smallskip}
\end{tabular}
\end{table}

The crucial reliability of this system requires the introduction of self-test features in the electronics, generating a SuperKEKB interlock (beam abort request) if faults are detected.

The present method of generating two beam abort request signals for HER and LER (computation of moving sums from $400$~kHz data, comparison with abort thresholds, channel masking and combinatorial logics) has proven to be adequate and does not need to be changed or upgraded. The operating experience shows that the useful time intervals for the computation of the moving sums do not exceed about $1$~ms, corresponding to $400$ buffer memory locations at $400$~kHz. The main possible modification concerns the limited dynamic range, which does not allow the use of each individual diamond detector for both monitoring (smaller signals) and beam abort (higher signals), as explained below.

\begin{enumerate}
    \item {Dynamic range:}

The total dynamic range required to accept the input current signals exceeds about $180$~dB, that is more than nine decades: from a minimum of about $3$~pA, set by rms noise and requested precision on small signals, to about $5$~mA, which should be measurable in the $10$~Hz data. Present measurements of beam-loss spikes corresponding to injection, extrapolated to future injection conditions, indicate that a lower dynamic range of about $136$~dB would be needed to cover currents from the minimum set by noise up to the about $20$~$\mu$A ($0.7$~krad/s) of projected injection peaks. To keep the present flexibility for abort thresholds settings, a dynamic range of at least $88$~dB is needed in the $400$~kHz data: from about $0.2$~$\mu$A up to $5$~mA. These numbers are summarized in Table~\ref{tab:requirements}.

\item {Bandwidth and noise rejection:}

The accepted signal bandwidth should be of the order of $1$~MHz to match the $400$~kHz data sampling in the ring buffer memory and the radiation dose integration requirements. A reduction in common-mode noise pick-up might be obtained by an optimization of cabling and of the grounding scheme. Moreover the introduction of bi-polar ranges in the sampling of the analogue currents should be considered since in the present system, the choice of uni-polar ranges increased the noise, due to the lack cancellation symmetric noise in the averaging of the over-sampled currents. 

\item {Memory read:}

The internal ring buffer memory should store data at $400$~kHz, and have a depth corresponding to at least $600$~ms (the present one has a depth corresponding to 85~s). It must be readable at any time, without interruption of the computation of the moving sums and the generation of abort signals. This may be obtained by writing the $400$~kHz in two separate memories, one of which must be reserved for the beam abort function. The other memory can then be read out at any time, in particular: (a) after a beam abort the reading is triggered by the SuperKEKB abort timing confirmation; (b) for injection background studies, the reading must be triggered by the SuperKEKB injection timing signal (to be added as an input); (c) for other purposes, an operator should be allowed to initiate the reading of a memory section any time. 

\item {Synchronization:}

At present only a rough synchronization with SuperKEKB is obtained by the SuperKEKB abort timing confirmation signal. At least another synchronization signal must be available as input, the SuperKEKB injection timing. This signal initiates an adjustable veto in the Belle II trigger system; the information on the veto width must also be available as a hardware signal or via programming of slow controls. It should be clarified if a deeper integration with the SuperKEKB control system is needed and possible, in particular if the accelerator clock should be used and if some time-stamp mechanism could be used to identify local events in the SuperKEKB context.

\item {Injection monitoring:}

The numerical integration of digitized charge (dose) data by the FPGA firmware should take advantage of the improved synchronization including injection timing. In particular, the doses in the injection veto time intervals and in the complementary time intervals should be separately available.

\item {HV control:}

At present, all diamond detectors are biased at the same HV value: $100$~V. In principle, the requirements of separate HV adjustment for each detector might be relaxed. However, an important feature must be added: the possibility of reading back the bias voltage value without manually measuring it at the output HV connector.

\item {External logics}

The new system design should minimize the amount of external logics needed to interface the system with SuperKEKB and should avoid the use of NIM crates and modules, which may pose problems for long-term availability.
\end{enumerate}

\subsection{Technical solutions}
\label{subsec:technical_solutions}

The basic building blocks of an upgraded system will be similar to those of the existing electronics, described in \cite{RMBA:DiaPerformance,RMBA:EoI} 
(a) a digital core based on a FPGA complemented by external memory; (b) an analog front-end with amplifiers and analog-to-digital conversion; (c) HV modules; (d) communication interface for commands exchange, parameters initialization, and for readout of $10$~Hz and $400$~kHz data.

A preliminary technical choice concerns the modularity and form factor of the system. 

An approach based on some standard crate with power supply, hosting modules that exchange signals via a back-plane, has some advantages. In particular, it may allow the use of existing modules restricting the new design to some custom part. External cabling is minimized in this approach. It is also well suited to large systems with many channels, where the management of repairs and spare modules may be simplified. However, the overhead of crate infrastructure for test stations may significantly increase the cost for systems with a limited number of channels.

The approach of the present system may be advantageous for a relatively small number of readout channels: the self-contained DCUs, each dealing with four detectors, include all the components (power supply, FPGA board, analog front-end, HV modules, communications board). It is easy to set up a stand-alone bench for tests and maintenance; the drawback in this case is the need of some external cabling and interfacing to exchange signals among DCUs and between SuperKEKB and DCUs.

The major challenge in the upgrade project is undoubtedly the re-design of the analog front-end with a wider dynamic range, so R\&D is primarily required to investigate a viable solution. 
Some preliminary R\&D work showed that the required dynamic range can be achieved by the simultaneous introduction of a fourth range and a controlled system of fast switching 
devices that commute from one range to the other in about $1~\mathrm{\mu s}$ 
taking into account also noise introduced by the commutation itself. 
What seems still difficult to achieve is an overall fast commutation in the lower current ranges due to their slow rise-time for a sudden increase of current 
(intrinsic bandwidth limitation for the required accuracy of the lower ranges). 
A workaround must be found to overcome this limitation. 
Also the digital part would need some preliminary R\&D, in particular the management of two memories by the FPGA, to fulfil the requirement of compatibility between memory-read and beam-abort. 

A continuation of the collaboration with Elettra, initiated with the development and construction of DCUs, has been established in 2022 mainly because this could speed up the R\&D and design phase, since some parts of the present design could be re-used.

\subsection{Radiation sensors}

%[ Sensors for the upgrade. Diamonds. SiC studies, CLAWS. Radiation hardness. 
%Possibility for re-use present diamond detectors. Phase 2 detectors used for the QCS bellows in Phase 3. In LS1 we produced 10 new diamond detectors and installed 8 in the new beam pipe for PXD2. ]

Radiation detectors made with electron grade single crystal CVD diamond sensors have proven to be very good solution for the current systems both for the radiation monitoring and for the beam abort functionalities. 
This solution seems to be adequate for the upgrade. 
Moreover the present detectors can be used for a short and medium term upgrade or even be re-used for a longer term one depending on the actual integrated dose up to the upgrade time and the expected dose for the new usage. The present irradiation studies show no damage up to $0.1~\mathrm{MGy}$, at least with electrons and photons. 

However more studies could be useful such as: 
\begin{itemize}
    \item radiation hardness studies above $0.1~\mathrm{MGy}$ with different radiation sources, since large beam losses can occur any time on some spots. 
    \item calibration with steady radiation over several order of magnitudes 
    \item short and intense pulses to study the response of sensor in such transients where large saturation effects may occur. 
\end{itemize}

Their main drawback are the costs, the few available providers for the sensors and the metallization, and the long calibration procedure. 

If more detector would be needed for instrument other Belle II sub-detectors, the QCS or accelerator other options should be considered such as SiC or other radiation hard materials. 

For the beam abort only functionality, the CLAW scintillator system has proven to be very effective.

\chapter{Physics Performance}
\label{sec:Physics}
\editor{A.~Gaz}
%% Physics performance
\section{Physics performance motivations}
\label{sec:pp-motivations}

The physics case of \belletwo\ is extensively presented in Refs.~\cite{Belle-II:2018jsg}
and \cite{Belle-II:2022cgf}. It covers a large number of precision measurements on
observables of $B$- and $D$-meson and $\tau$-lepton decays that are related to
fundamental parameters of the standard model and could be sensitive to the existence
of particles and couplings not envisioned by the theory. Another part of the
physics program consists of searches for exotic multiquark states that challenge the
traditional QCD models and of fundamentally new particles that are connected to a dark
sector extension of the standard model.

In order to fully complete this ambitious project and to probe new areas of research
that might become interesting in the near future, \belletwo\ must ensure that the
detector performance matches or exceeds the expectations drawn at the start of the
experiment.

Unfortunately, the machine related background levels proved to be significantly
more severe than expected in the first years of data taking, so the focus of this
section will be mostly on the ways to ensure that the main physics goals of the
experiment will be achieved despite the harsher than anticipated data taking conditions.

From the point of view of the tracking of charged particles (see e.g.
Ref.~\cite{BelleIITrackingGroup:2020hpx} for an overview of the current implementation and
performance), the CDC is the sub-detector that has exhibited the highest sensitivity to
machine related background during the first years of data taking. The most evident effect
has been a loss of reconstruction efficiency for the tracks that primarily rely on the CDC
(most notably the pions originating from the \KS\ that decay beyond the fourth layer of the
VXD) due to the large increase in chamber current during and just after injection, resulting 
in a voltage drop of the HV resistor and space charge effects. Other detrimental issues have
been a degradation of the $dE/dx$ measurement accuracy and loss of trigger efficiency.

We do not plan any modification of the CDC hardware in the near future (while technically
feasible, the replacement of the innermost layers of the chamber is very risky and will
not be further considered), so the interventions to keep the tracking efficiency high
must be concentrated on:
\begin{enumerate}
\item{} upgrades in the readout electronics (see Sec.~\ref{sec:CDC});
\item{} developments for the tracking software. As of now, we are considering some ideas,
  such as \emph{inverting} the tracking chain (so that we start from the silicon tracker
  and extrapolate outward to the CDC), \emph{filtering} the CDC hits with small electronic
  signal (and possibly add them back to the reconstructed track, to improve the $dE/dx$
  measurement), and activating a \emph{local CDC finder}, to recover the reconstruction
  efficiency for very displaced tracks. These ideas are not mutually exclusive and will
  be tested in the near future.
\end{enumerate}  
It will be critical in any case to monitor the background conditions and make sure
that thresholds motivated by physics performance are not exceeded for long periods
of time.

\begin{figure}[htbp]
  \centering
    \includegraphics[width=0.9\linewidth]{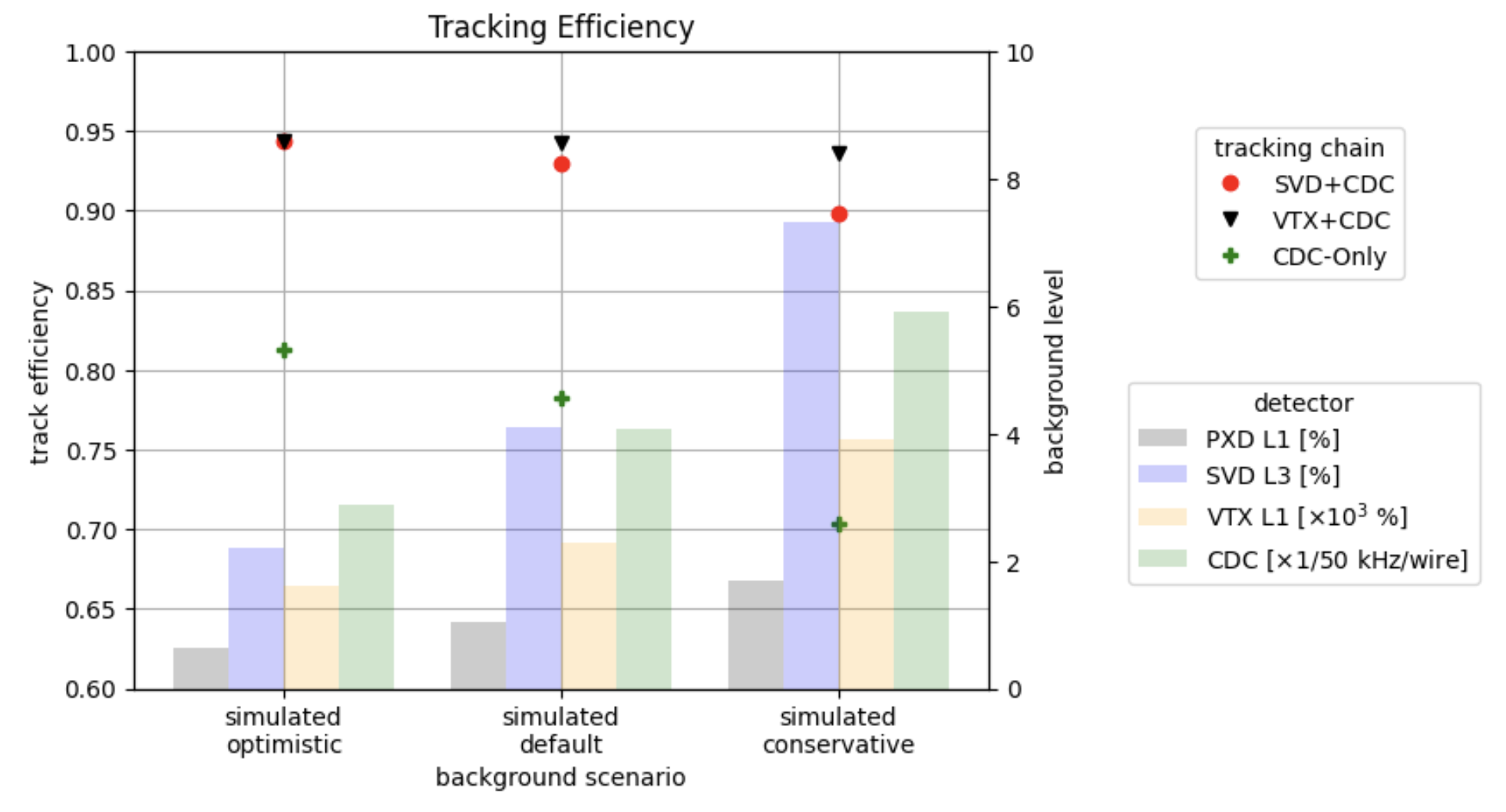}
    \caption{
      Occupancies (colored bars) of the current and planned tracking devices as
      a function of the level of beam background and corresponding tracking
      efficiencies (colored markers) for the SVD+CDC, VTX+CDC and CDC-only detector
      configuration. It is possible to see that the VTX can almost completely
      compensate the dramatic loss of efficiency of the CDC-only case (which the
      current SVD can only partially mitigate).
      \label{fig:pp-tracking_eff}}
\end{figure}

Tracking efficiency, particularly for low momentum tracks, can be improved with a
new silicon vertex detector with finer granularity both in space and in time.
This new concept would ensure higher efficiency at low momentum, better impact parameter
resolution, and more importantly more robustness against machine backgrounds. Given the
limitations set by the CDC, the interplay between the gas chamber and the silicon tracking
device will be of paramount importance to maintain the tracking efficiency (at least for
the charged particles that do not originate far from the interaction region),
as shown in Fig.~\ref{fig:pp-tracking_eff}. Advanced stage designs for such a sub-detector
already exist, see e.g. Sec.~\ref{sec:vxd-structure}, and the replacement of the currently
installed VXD detector will be one of the priorities of the next stage of the 
\belletwo\ upgrade, especially if restrictions in the detector envelope will be
required by the upgrades in the accelerator. Improvements in the vertexing
capabilities do not only mean better precision on time dependent analyses of $B$ and
$D$ mesons, but provide also handles for discriminating among different decay
topologies.  Moreover, the parameters at the point of closest approach of the tracks in
the rest of the event enter the most refined flavor tagging algorithms, giving
significant contributions to their performance, see e.g. Ref.~\cite{Belle-II:2021zvj}.

For the Particle IDentification (PID), no major hardware improvements are foreseen for the
ARICH and TOP, if we exclude regular maintenance and possible replacement of
photo-detectors. Also in this area the CDC is the most vulnerable sub-detector
as in high background conditions the measurement of the specific energy loss ($dE/dx$)
is affected by a bias that has a dependence on the time that elapsed since the last
$e^+/e^-$ bunch injection into the main ring. This effect can be partially corrected with 
a dedicated procedure (currently under test) aimed at determining calibration constants 
that depend on the time since the last injection. Further improvements in the PID 
performance will be achieved thanks to a more pervasive use of machine learning techniques 
applied to the reconstruction software.

For the reconstruction of the neutral particles, we do not plan any significant hardware
modifications of the ECL, while we are considering an upgrade of the KLM sub-detector,
replacing the resistive plate chambers in the barrel section with scintillators, adding
also \textit{time of flight} capabilities in the readout, so that a (10 -- 20\%)
measurement of the momentum of the neutral particles detected in the KLM
can be achieved. While there are no serious concerns about the muon reconstruction
efficiency and the performance stability of the current detector, the upgraded
version of the KLM could achieve better \KL\ reconstruction efficiency. This
would be very relevant not only to increase the signal yield of modes with
\KL's in the final state (most notably the eigenstates studied in time
dependent $CP$ violation analyses, such as $J/\psi \KL$, $\eta^{\prime} \KL$, etc.),
but also to improve the \KL\ veto capabilities for analyses with neutrinos or other
particles escaping detection in the final state (an undetected \KL\ would behave like a
neutrino, and thus increasing the \KL\ veto efficiencies would translate in a reduction of
the physics background). Examples of modes that would benefit from this include the
\belletwo\ flagship analyses $B \to K^{(*)} \nu \bar{\nu}$, $B \to D^{(\star)} \ell \nu$,
$B \to \tau \nu$, and inclusive $|V_{ub}|$ measurements.

In Sec.~\ref{sec:pp-parameters} we will discuss the expected improvements
to key physics performance parameters that we foresee from some of the upgrade
scenarios that are currently under consideration.

One of the areas in which \belletwo\ needs to quickly find strategies to improve
performance is that of trigger. With the current background conditions and luminosity
scenarios, the experiment might find itself in the difficult position of having to sacrifice
bandwidth for important physics channels, first of all those characterized by
low-multiplicity final states (very few tracks/photons with possibly missing energy/momentum
and/or displaced vertices). Under extreme background conditions even some low
charged-multiplicity \BB\ events (e.g. those in which one \Bz\ decays to $\piz\piz$) will
suffer from reduced trigger efficiency. Short-term measures will allow us to survive for a
while longer, but for the long term we must radically change our approach. In
Sec.~\ref{sec:pp-trigger} we present examples of radically new approaches, in which machine
learning is expected to play a decisive role.

In Sec.~\ref{sec:pp-benchmarks} we will conclude this chapter presenting the
improvements in the performance that we expect in some key areas of our Physics program.

\section{Key detector performance parameters for different options}
\label{sec:pp-parameters}

For the VTX upgrade, we study the performance of the upgrade geometries, and compare them
to that of the current \belletwo\ detector, utilizing a full Geant4 simulation of
the detector in the study of specific decay modes of interest. This approach, coupled
with the state-of-the-art predictions of machine backgrounds presented in
Chapter~\ref{sec:BeamBackgroundsAndCountermeasures}, ensures that all the effects and
correlations are properly taken into account (with the exception of possible correlated 
losses due to the unsimulated injection background).

We investigate two different VTX geometries, which differ only in the location
of the third layer. The \emph{nominal} geometry has the third layer placed at a distance
of $3.9$ cm from the interaction point (IP), and is expected to maximize the track impact
parameter resolution, while in the \emph{alternative} geometry, the third layer is
displaced at $6.9$ cm from the IP, in order to improve the \KS\ reconstruction efficiency.
The results are presented in Sec.~\ref{subsec:pp-pisoft_eff}-\ref{subsec:pp-KS_eff}.

For the KLM upgrade scenario, we do not have yet a full Geant4 geometry implemented in
the \belletwo\ software framework, thus we will limit ourselves to some qualitative
statements in Sec.~\ref{subsec:pp-KL_eff}.

\subsection{Low momentum tracking efficiency and impact parameter resolution}
\label{subsec:pp-pisoft_eff}

One of the areas in which we see a striking improvement of the VTX upgrade scenarios
over the nominal \belletwo\ detector is the tracking of low-momentum particles, most
notably the \emph{soft pions} from the decays of $D^{* \pm}$ mesons.

We study the low momentum tracking efficiency by reconstructing the decay chain
$\Bz \to D^{*-} \ell^+ \nu$, with $D^{*-} \to \overline{D}^0 \pi^-_{\rm soft}$ and
$\overline{D}^0 \to K^+ \pi^-$ or $K^+ \pi^- \pi^+ \pi^-$. We report here the
results only for the $\overline{D}^0 \to K^+ \pi^-$ decay mode, the performance
with $\overline{D}^0 \to K^+ \pi^- \pi^+ \pi^-$ is perfectly consistent.

\begin{table}[htbp]
  \caption{\label{tab:pp-PiSoftRecoEff} Summary of the reconstruction
    efficiency and purity for the $\Bz \to D^{*-} \ell^+ \nu$ decay chain,
    with $D^{*-} \to \overline{D}^0 \pi^-_{\rm soft}$ and $\overline{D}^0 \to K^+ \pi^-$,
    for the nominal \belletwo\ geometry with the intermediate v2 background scenario
    and the nominal VTX configuration in all three background scenarios.
  }
  \begin{center}
    \begin{tabular}{l|c|c|c|c}
      \hline\hline
      & \belletwo\ (v2) & VTX (v1) & VTX (v2) & VTX (v3) \\
      \hline\hline
      Generated events & 32533 & 32559 & 32559 & 30255 \\
      \hline
      Correctly reconstructed signal & 10059 & 16913 & 16848 & 15583 \\
      \hline
      Self-crossfeed & 28495 & 51375 & 51826 & 47527 \\
      \hline\hline
      Efficiency & 30.9\% & 51.9\% & 51.7\% & 51.5\% \\
      \hline
      Purity & 26.1\% & 24.8\% & 24.5\% & 24.7\% \\
      \hline\hline
    \end{tabular}
  \end{center}
\end{table}

Table~\ref{tab:pp-PiSoftRecoEff} shows the yields for correctly reconstructed signal and
self-crossfeed (i.e. signal events in which some of the particles assigned to the signal
candidate are actually decay products of the other $B$ meson in the event) after the
application of minimal analysis cuts on the quality of the reconstructed tracks, masses of the
resonances involved, and overall vertex probability of the reconstructed chain. The nominal
\belletwo\ geometry with intermediate v2 beam backgrounds scenarios is compared to the nominal
VTX configuration (the one with Layer3 closest to the beam line) in all three background
hypotheses.

We see a remarkable improvement (of almost a factor 1.7) in the reconstruction efficiency
for the VTX geometry over nominal \belletwo, with comparable purity (defined as the ratio
between the number of correctly reconstructed signal events and the total number of
candidates). It is also apparent that the performance is very stable against
the increasing background conditions.

\begin{figure}[htbp]
  \begin{subfigure}[t]{0.48\textwidth}
    \includegraphics[width=1.0\linewidth]{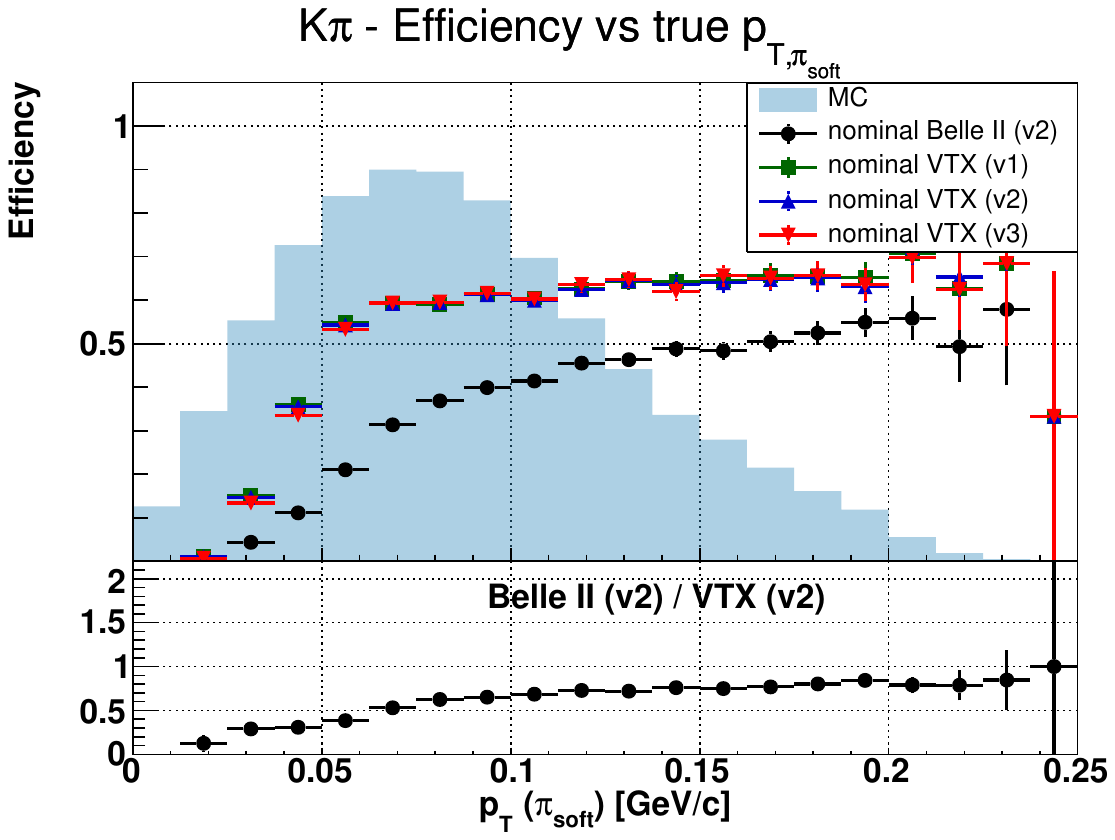}
  \end{subfigure}
  \hfill
  \begin{subfigure}[t]{0.48\textwidth}
    \includegraphics[width=1.0\linewidth]{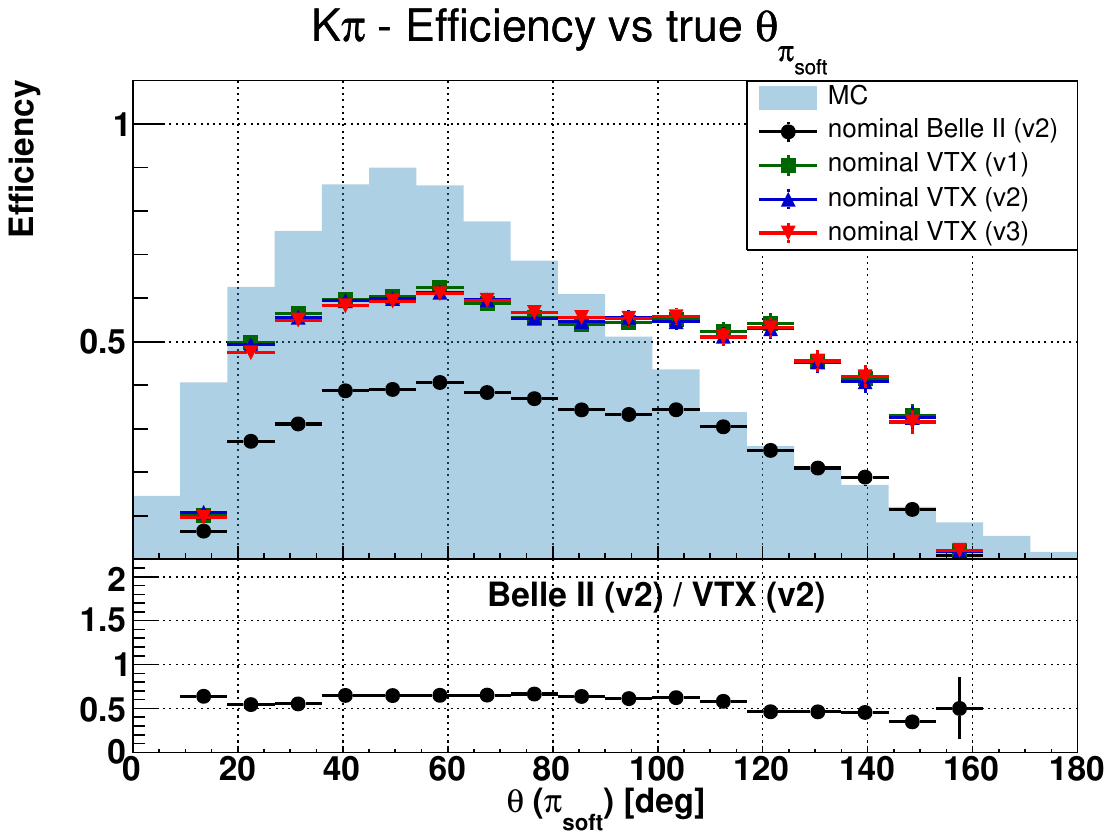}
  \end{subfigure}
  \caption{Reconstruction efficiency of $\Bz \to D^{*-} \ell^+ \nu$ as a function
    of the transverse momentum of the $\pi^-_{\rm soft}$ from the $D^{*-} \to \overline{D}^0 \pi^-_{\rm soft}$
    decay (left plot) and cosine of the polar angle of the $\pi^-_{\rm soft}$ track (right).
    We compare the performance of the nominal \belletwo\ detector (black dots) with
    intermediate scenario-2 background and the nominal VTX configuration in the optimistic
    background scenario-1 (v1, green squares), intermediate scenario-2 (v2, blue upward pointing triangles), and conservative scenario-3  (v3, red downward pointing triangles). 
    The shaded blue histograms 
    display the momentum spectrum of the $\pi^-_{\rm soft}$.
    The bottom plots show the ratio between nominal Belle II and nominal VTX in the
    intermediate scenario-2 background.
    \label{fig:pp-PiSoftEfficiency}
  }
\end{figure}

The much better reconstruction efficiency is mostly due to the enhanced tracking
efficiency for the $\pi^-_{\rm soft}$ candidates, as can be seen in Fig.~\ref{fig:pp-PiSoftEfficiency},
where we see a dramatic improvement for transverse momenta below $0.05$ \gevc\ and
only for $p_T \gtrsim 0.2$ \gevc\ we see the reconstruction efficiency of the nominal
\belletwo\ detector approach that of the VTX upgrade scenario.

\begin{figure}[htbp]
  \begin{subfigure}[t]{0.48\textwidth}
    \includegraphics[width=1.0\linewidth]{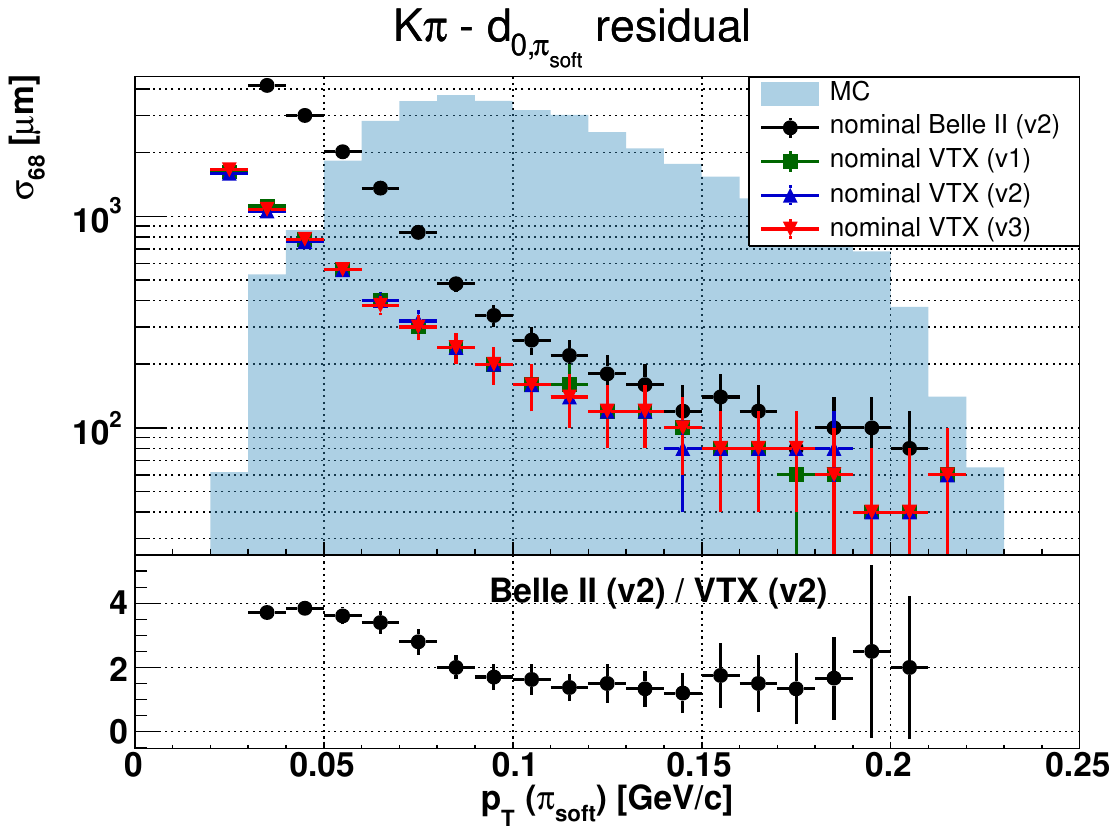}
  \end{subfigure}
  \hfill
  \begin{subfigure}[t]{0.48\textwidth}
    \includegraphics[width=1.0\linewidth]{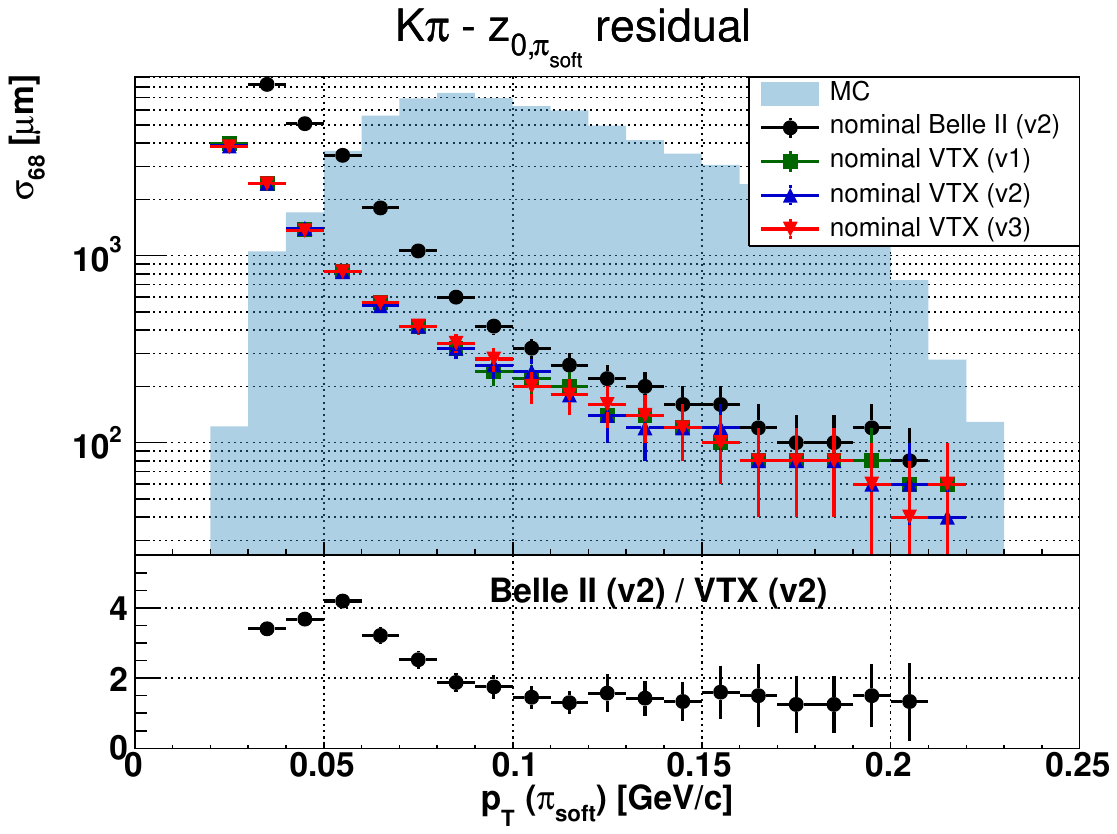}
  \end{subfigure}
  \caption{Width ($\sigma_{68}$) of the distribution of the residuals of $d_0$ (left plot)
    and $z_0$ (right) as a function of the $\pi^-_{\rm soft}$ momentum.
    We compare the performance of the nominal \belletwo\ detector (black dots) with
    intermediate scenario-2 background and the nominal VTX configuration in the optimistic
    background scenario-1 (v1, green squares), intermediate scenario-2 (v2, blue upward pointing triangles), and conservative scenario-3  (v3, red downward pointing triangles). 
    The shaded blue histograms 
    display the momentum spectrum of the $\pi^-_{\rm soft}$.
    The bottom plots show the ratio between nominal Belle II and nominal VTX in the
    intermediate scenario-2 background.
    \label{fig:pp-PiSoftResiduals}
  }
\end{figure}

The improvement in the tracking quality of the soft pions can also be seen from
Fig.\ref{fig:pp-PiSoftResiduals}, where we compare the width of the distributions
of the residuals of $d_0$, the signed distance from the $z$ axis of the Point Of Closest
Approach (POCA), and $z_0$, the $z$-coordinate of the POCA. The VTX geometry improves
by a factor of up to $\sim 4$ over the nominal \belletwo\ and does not show any
significant sensitivity to the background conditions.

\subsection{Momentum resolution}
\label{subsec:pp-momentum_res}

For tracks with transverse momentum greater than $\sim 0.3$ \gevc, the momentum
resolution is dominated by the CDC, so it is not surprising that no appreciable
improvement in the track momentum resolution is observed for the daughter
leptons of the $J/\psi$ in the above mentioned sample.

A moderate improvement (a few \%) is seen for the momentum resolution of the
\KS\ candidates, for the VTX upgrade scenarios, mostly due to the improved
vertexing capabilities.

\subsection{Vertexing resolution}
\label{subsec:pp-vertexing_res}

We study the vertexing performance of the upgrade scenarios by analyzing samples of
one million $B^0 \to J/\psi \KS$ events generated and reconstructed with all the
possible combinations of detector geometries and background scenarios.

\begin{figure}[htbp]
  \begin{subfigure}[t]{0.48\textwidth}
    \includegraphics[width=1.0\linewidth]{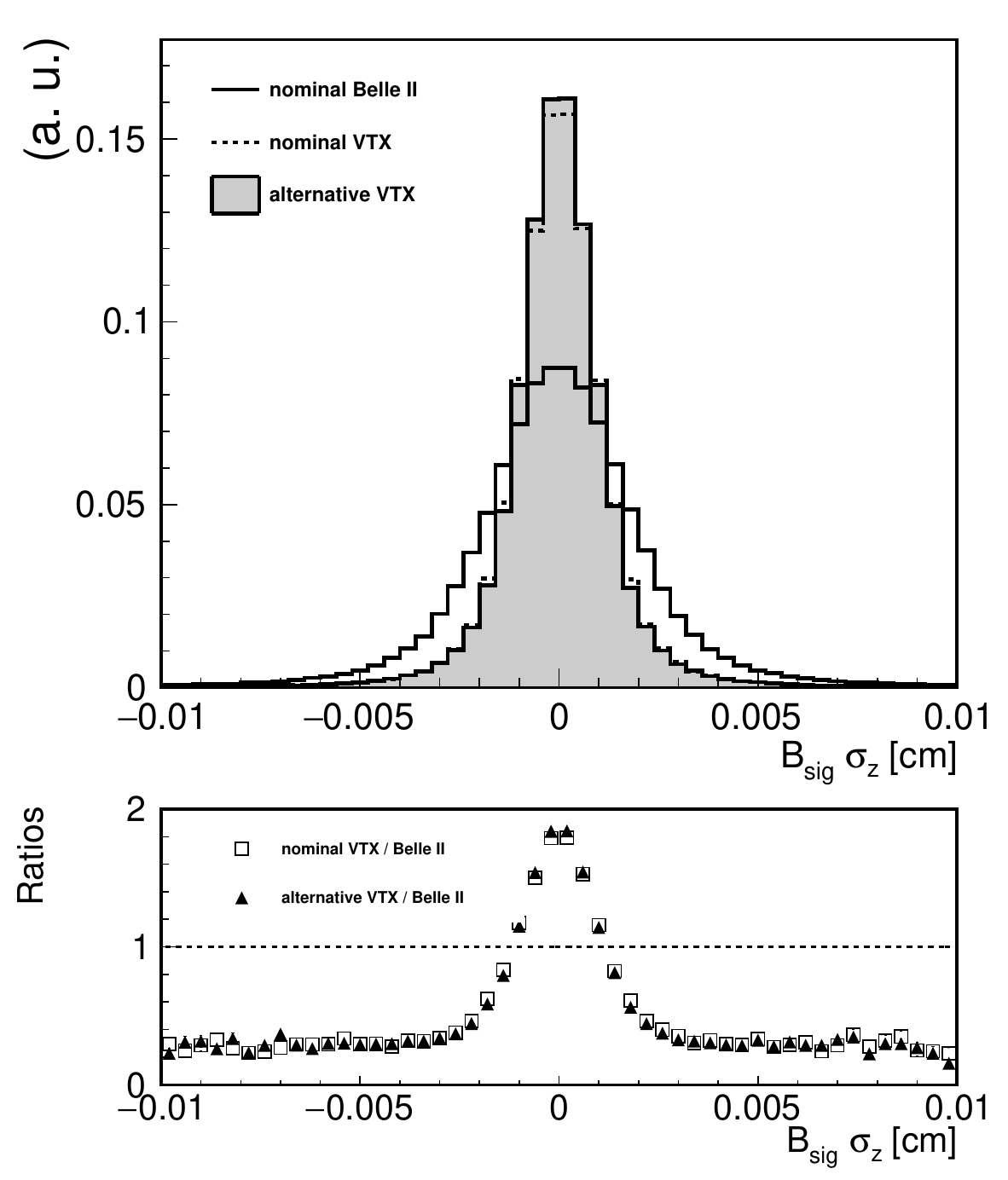}
  \end{subfigure}
  \hfill
  \begin{subfigure}[t]{0.48\textwidth}
    \includegraphics[width=1.0\linewidth]{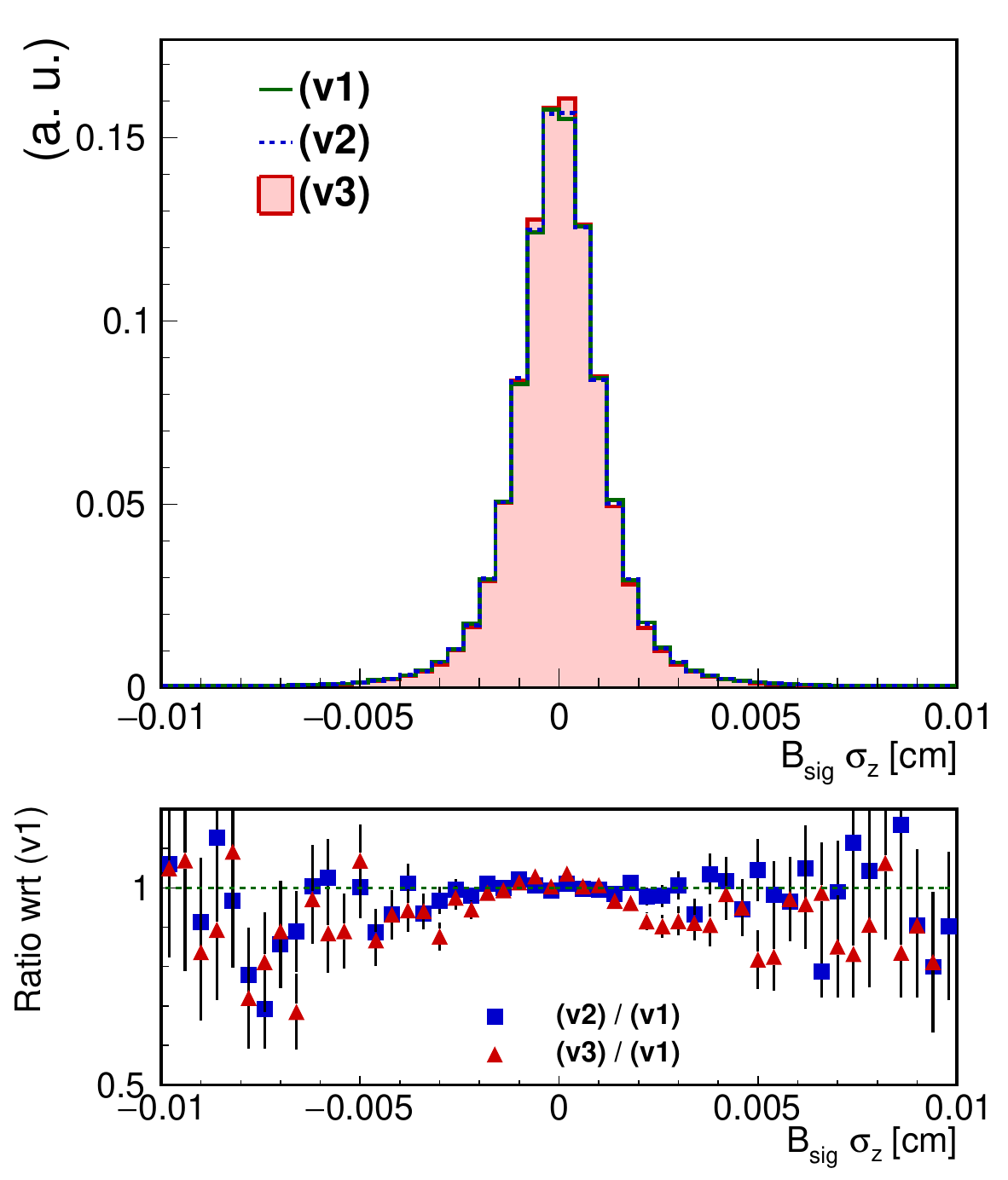}
  \end{subfigure}
  \caption{Left figure: comparison of the $B$ decay vertex resolution along the $z$ axis
    in $B^0 \to J/\psi \KS$ events for the nominal \belletwo\ detector (solid line),
    nominal VTX geometry (dotted line) and alternative VTX geometry (filled grey histogram).
    The bottom plot shows the ratio between the VTX geometries (empty squares for the
    nominal and filled triangles for the alternative) and nominal \belletwo.
    Right figure: $B$ decay vertex resolution along the $z$ axis for the nominal VTX
    geometry in the three background scenarios: optimistic scenario-1 (v1, green solid
    line), intermediate scenario-2 (v2, blue dotted line), and conservative scenario-3
    (v3, red filled histogram). The bottom plot shows the ratio between the two higher
    background scenarios and the optimistic one.
    \label{fig:pp-BsigVtxResZ}
  }
\end{figure}

Figure~\ref{fig:pp-BsigVtxResZ} shows the distribution of the decay vertex resolution
$\sigma_z = \sigma(z_{\rm meas} - z_{\rm true})$ (the width of the distribution obtained by
taking the difference between the measured and the true positions) along the $z$ axis of
the signal $B$ decay for the three detector configurations in the intermediate background
scenario and for the nominal VTX geometry in the three background scenarios.
The results are summarized also in Table~\ref{tab:pp-BsigVtxResZ} and show
that the upgraded geometries achieve a $\sim 35\%$ better resolution on the $B$
decay vertex and, contrary to the nominal \belletwo\ detector, they do not show
any significant degradation as a function of the background conditions.

\begin{table}[htbp]
  \caption{\label{tab:pp-BsigVtxResZ} Signal $B$ vertex resolution along the $z$ axis
    (in $\mu$m) for the three detector geometries and three background scenarios
    investigated.}
  \begin{center}
    \begin{tabular}{l|c|c|c}
      \hline\hline
      $B_{sig}$ $z$ vertex resolution ($\mu$m) & Bkg (v1) & Bkg (v2) & Bkg (v3) \\
      \hline\hline
      \belletwo & 21.9 & 23.0 & 24.9 \\
      \hline
      Nominal VTX & 14.5 & 14.4 & 14.1 \\
      \hline
      Alternative VTX & 14.4 & 14.3 & 14.0 \\
      \hline\hline
    \end{tabular}
  \end{center}
\end{table}

Similar results (not shown) are obtained for the vertexing resolution along the $x$ and
$y$ directions: the VTX upgrade guarantees significantly better performance and is
almost insensitive to the levels of machine background.

\begin{figure}[htbp]
  \begin{subfigure}[t]{0.48\textwidth}
    \includegraphics[width=1.0\linewidth]{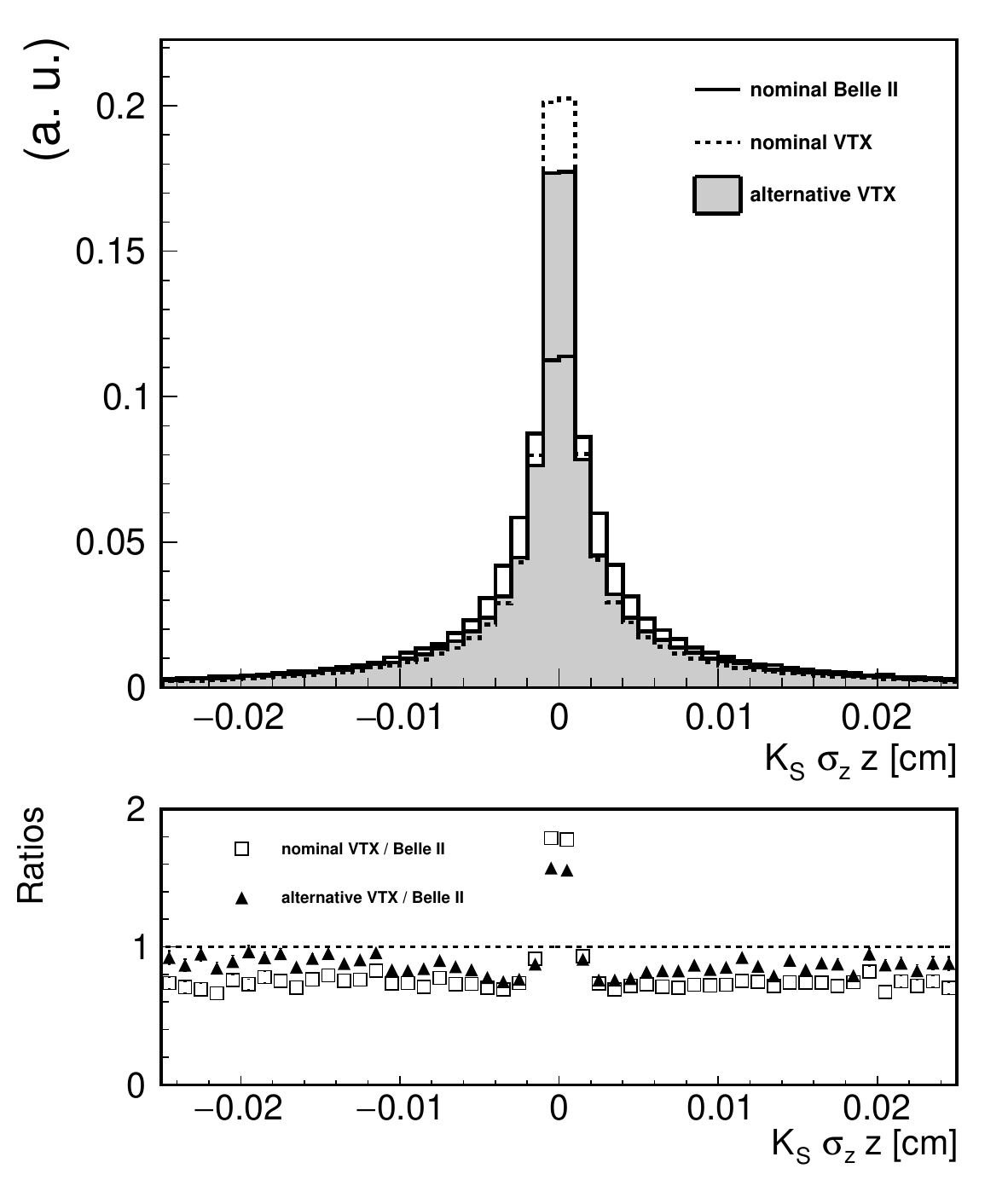}
  \end{subfigure}
  \hfill
  \begin{subfigure}[t]{0.48\textwidth}
    \includegraphics[width=1.0\linewidth]{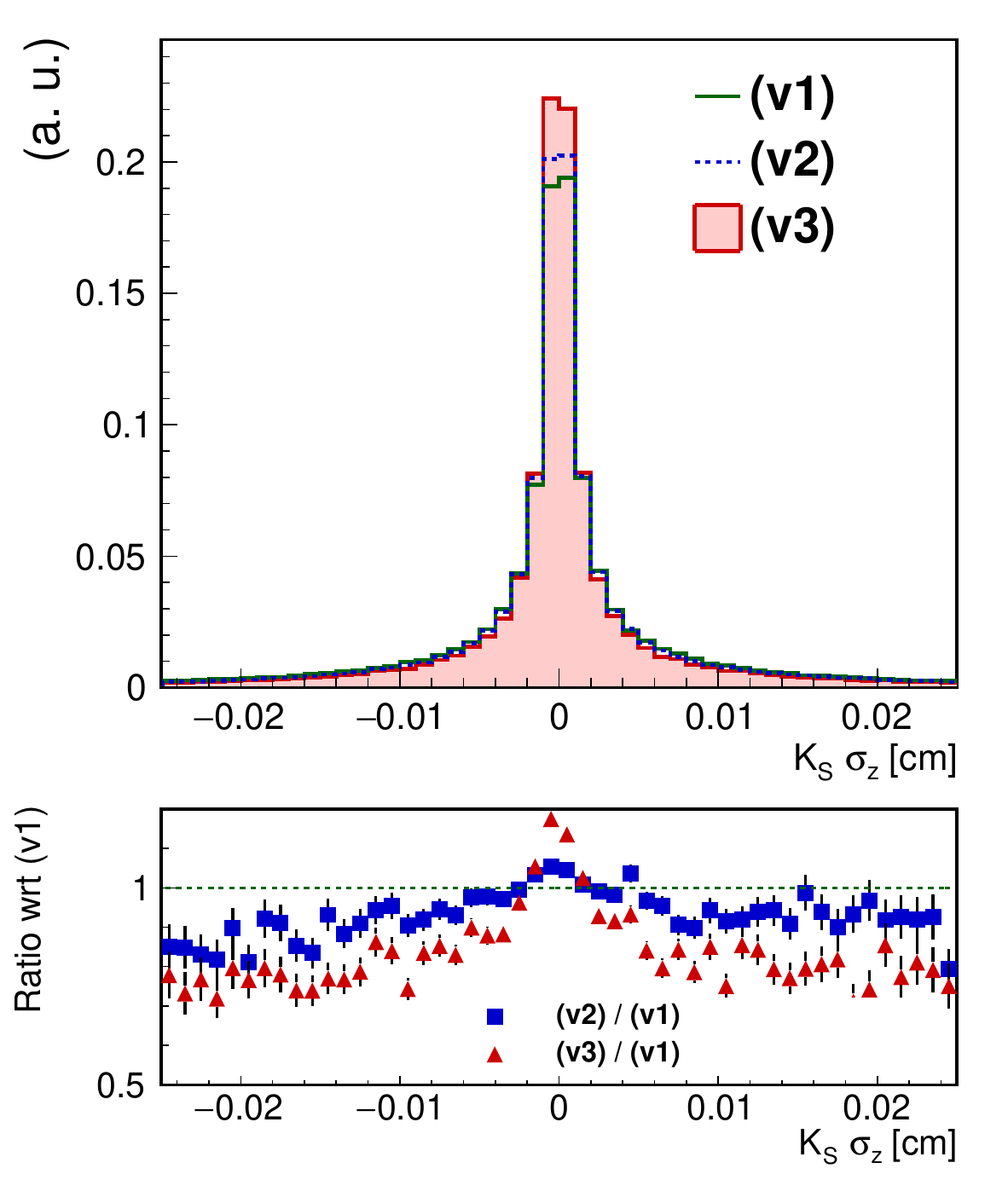}
  \end{subfigure}
  \caption{Left figure: comparison of the \KS\ decay vertex resolution along the $z$ axis
    in $B^0 \to J/\psi \KS$ events for the nominal \belletwo\ detector (solid line),
    nominal VTX geometry (dotted line) and alternative VTX geometry (filled grey histogram).
    The bottom plot shows the ratio between the VTX geometries (empty squares for the
    nominal and filled triangles for the alternative) and nominal \belletwo.
    Right figure: \KS\ decay vertex resolution along the $z$ axis for the nominal VTX
    geometry in the three background scenarios: optimistic scenario-1 (v1, green solid
    line), intermediate scenario-2 (v2, blue dotted line), and conservative scenario-3
    (v3, red filled histogram). The bottom plot shows the ratio between the two higher
    background scenarios and the optimistic one.
    \label{fig:pp-KsVtxResZ}
  }
\end{figure}

Analogous studies for the \KS\ decay vertex resolution are displayed in
Fig.~\ref{fig:pp-KsVtxResZ}.
Also in this case, the upgraded geometries show a better vertexing resolution compared
to the nominal \belletwo\ detector and do not appear to suffer any degradation in
performance due to the increasing machine background. The apparent improvement
of the vertexing resolution with the harsher background scenarios that is visible
in the right plot of Fig.~\ref{fig:pp-KsVtxResZ} is a spurious effect that is due to
the loss of reconstruction efficiency for candidates with large flight distance (that
are thus affected by poorer vertex resolution) in the higher background scenarios.

\subsection{\KS\ reconstruction efficiency}
\label{subsec:pp-KS_eff}

\begin{figure}[htbp]
  \centering
    \includegraphics[width=0.8\linewidth]{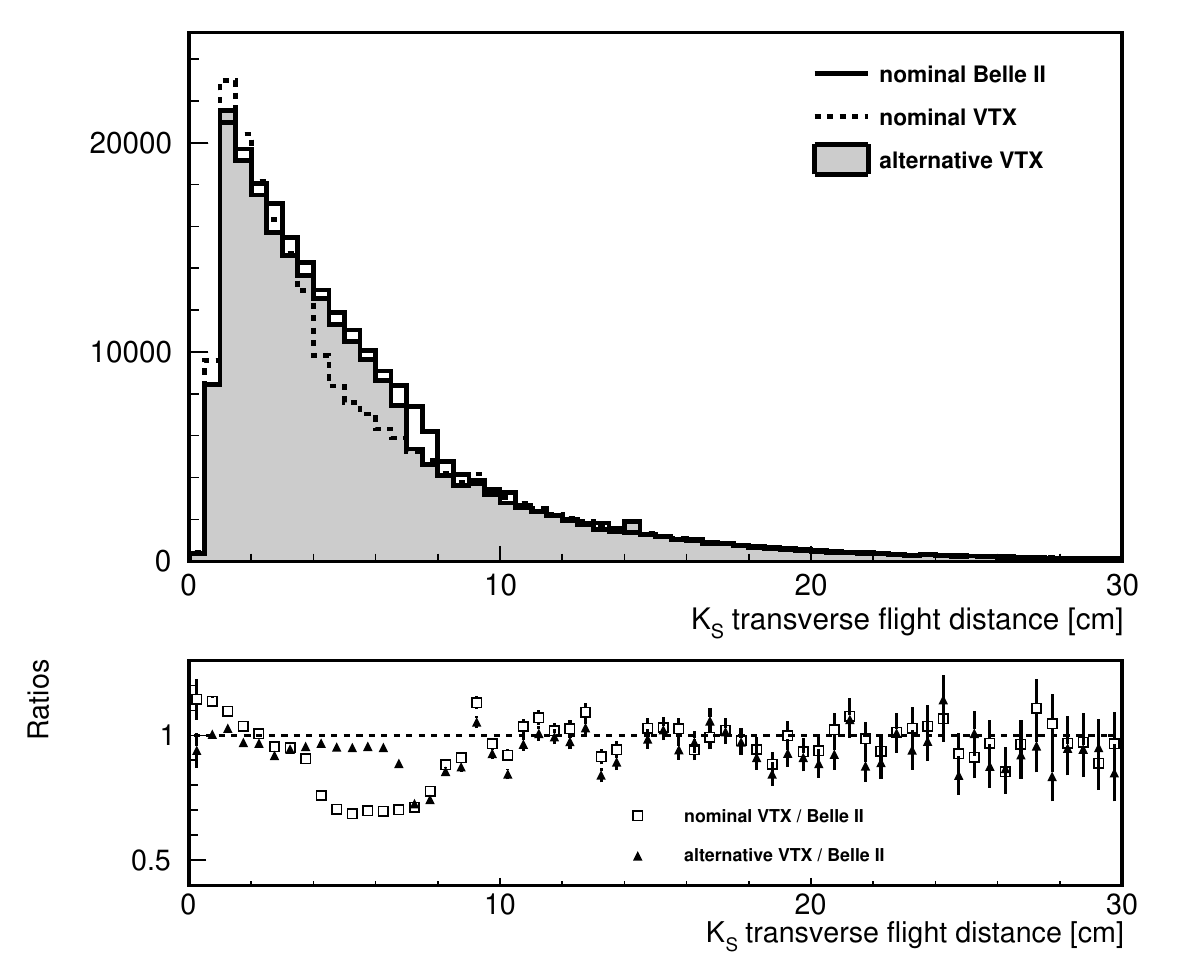}
    \caption{Distributions of the flight length, projected on the $(x, y)$ plane, of \KS\
      candidates from $B^0 \to J/\psi \KS$ events reconstructed with the nominal \belletwo\ detector
      (solid line), nominal VTX geometry (dotted line) and alternative VTX geometry (filled
      grey histogram). The bottom panel shows the ratio between the nominal (empty squares)
      or alternative (black triangles) VTX geometry and the nominal \belletwo\
      configuration. The intermediate (v2) background hypothesis has been utilized for
      these plots.
      \label{fig:pp-KsFlDistXY}}
\end{figure}

We study the \KS\ reconstruction efficiency using the $B^0 \to J/\psi \KS$ samples
defined above. Figure~\ref{fig:pp-KsFlDistXY} shows the distributions of the
measured flight length, projected on the $(x,y)$ plane, of the \KS\ candidates
reconstructed with the three different detector configurations under study.
We only consider candidates that are matched with the generated
\KS\ particles.

Standalone VXD/VTX tracks require at least 2.5 3-dimensional hits (a hit on each
side of a double-sided sensor counts as 0.5) in the detector: due to this
requirement the efficiency rapidly falls as a function of the transverse flight
distance. The nominal VTX geometry has a more pronounced falloff
(compared to the others) in correspondence of the position of the
third layer. For the alternative geometry, a similar feature appears
a few cm further from the $z$ axis of the detector (and thus the overall
impact on the efficiency is smaller). Beyond $\sim 10$ cm, the \KS\
reconstruction is dominated by the CDC, so no significant differences
between the geometries can be seen.

Table~\ref{tab:pp-KsRecoEff} summarizes the reconstructed yields in all the investigated
configurations. Besides some common selection criteria on the properties of the $B$ and
$J/\psi$ candidates, we require that the three-dimensional flight length of the \KS\
candites be greater than 1 cm: this is applied to reduce the contribution from
random combinations of tracks giving an invariant mass similar to that of the \KS.
Furthermore, in order to make the comparison more meaningful, we require that the
projection of the flight length on the $(x, y)$ plane be less than 14 cm, so that
the decay vertex of the \KS\ candidate is within the volume of the silicon vertex
detector.

The nominal \belletwo\ configuration shows the best performance in terms of overall
reconstruction efficiency and also in terms of resilience to the increasing background 
conditions. The alternative VTX geometry, shows only slightly worse resilience against beam 
background and exhibits a $\sim 5\%$ lower reconstruction efficiency. Significantly worse 
results are obtained with the nominal VTX geometry, especially in the higher background
scenarios.

It should be stressed that the main reason for the VTX geometries to have significantly 
worse performance is that they have one less silicon layer compared to the nominal \belletwo\
configuration. Alternative 6-layers configurations are currently being considered for the
next phases of the upgrade project.

\begin{table}[htbp]
  \caption{\label{tab:pp-KsRecoEff} Number of reconstructed \KS\ candidates
    for the three detector geometries and background scenarios that we are considering.
    For each of the nine configurations, we report the yield obtained by requiring that
    the three-dimensional flight length be greater than 1 cm (this is typically done to
    reduce the combinatorics) and by requiring in addition that the flight length in
    the transverse plane be less than 14 cm (to restrict to the candidates that decay
    within the volume of the silicon vertex detector).
    The percentages in parentheses under the yields in the last two columns indicate
    the loss of efficiency of the intermediate (v2) and conservative (v3) background
    hypotheses compared to the optimistic (v1) scenario.    
  }
  \begin{center}
    \begin{tabular}{l|c|c|c}
      \hline\hline
      & BG (v1) & BG (v2) & BG (v3) \\
      \hline\hline
      \multicolumn{4}{c}{Nominal Belle II} \\
      \hline
      flLen $ > 1 $ cm & 272\,845 & 243\,396 & 181\,395 \\
      & - & (-10.8\%) & (-33.5\%) \\
      flLen$_{xy} < $ 14 cm & 237\,021 & 217\,761 & 169\,687 \\
      & - & (-8.1\%) & (-28.4\%) \\
      \hline\hline
      \multicolumn{4}{c}{Nominal VTX} \\
      \hline
      flLen $ > 1 $ cm & 254\,357 & 222\,243 & 154\,258 \\
      & - & (-12.6\%) & (-39.4\%) \\
      flLen$_{xy} < $ 14 cm & 219\,103 & 196\,974 & 143\,109 \\
      & - & (-10.0\%) & (-34.7\%) \\
      \hline\hline
      \multicolumn{4}{c}{Alternative VTX} \\
      \hline
      flLen $ > 1 $ cm & 259\,563 & 230\,514 & 171\,998 \\
      & - & (-11.2\%) & (-33.7\%) \\
      flLen$_{xy} < $ 14 cm & 225\,559 & 206\,002 & 160\,890 \\
      & - & (-8.7\%) & (-28.7\%) \\
      \hline\hline
    \end{tabular}
  \end{center}
\end{table}

\subsection{\KL\ reconstruction efficiency}
\label{subsec:pp-KL_eff}

Neutral $K$ particles enter many channels of the \belletwo\ Physics program.
Most channels containing one $K^0$ in the final state can be studied by reconstructing
only $K^0 \to \KS \to \pi\pi$ candidates and then assuming that the same results hold true
for $K^0 \to \KL$.

Notable exceptions are the channels in which we have multiple $K^0$'s in the final state
(where we cannot assume that e.g. $\KS\KL X$ behaves as $\KS\KS X$) and channels utilized in
analyses of time-dependent $CP$ violation, for example $B^0 \to J/\psi K^0$, which are still
limited by statistical power and thus the inclusion of $\KL$'s candidates, which typically
account for $\sim 30\%$ of the total yield, improves the precision on the $CP$ violation
parameters. The intended KLM upgrade is expected to improve the reconstruction efficiency and
will also allow for a measurement of the \KL\ momentum with a resolution of $\sim 13\%$ for the
momentum range relevant for \CP\ violation measurements. Traditional $B$-factory analyses
of final states containing \KL's apply a kinematic constraint (for example imposing that the
beam-constrained mass of the candidate coincides with the mass of the $B$ meson, or that the
difference between the measured and expected energy of the $B$ candidate $\Delta E$ is null)
to supply for the missing information on the \KL\ momentum. This kinematic constraint allows
to obtain a narrow signal peak over a much broader background component. The additional
constraint deriving from the direct measurement of the \KL\ momentum will allow to better
characterize the signal component and to significantly reduce the combinatorial backgrounds,
but the final impact on the sensitivity to the $CP$ violation parameter will likely be small.

More promising results could be obtained in analyses in which \KL's are produced as part of a
background process that must be vetoed. These analyses typically target final states with one
or more neutrinos in the final state, and are thus among the \emph{golden channels} of the
\belletwo\ physics program. Events with unreconstructed \KL's (that would mimic the behavior
of a neutrino) constitute one of the major sources of background in these studies.

Example analyses include the inclusive measurement of $|V_{ub}|$, in which decays
originating from $b \to c \to \KL$ transitions must be kept under control. In Belle
and BaBar analyses (see e.g.~\cite{Belle:2003vfz} and ~\cite{BaBar:2005acy}) uncertainties related to the
\KL\ reconstruction efficiency are among the leading sources of systematic errors.
Another class of analyses that might benefit from increased \KL\ veto capabilities
is that of (semi)tauonic $B$ decays, such as $B \to \tau \nu$ or $B \to D^{(*)} \tau \nu$.
The Belle analysis \cite{Belle:2012egh} explicitly applies a \KL\ veto,
which results in a $7.3\%$ uncertainty out of $14.7\%$ total systematics. 

Also in the $B \to K^{(*)} \nu \nu$ and $D \to \pi \nu \nu$ analyses, the ability to veto
\KL\ candidates is critical to control peaking background. In the recent
$B^+ \to K^+ \nu \nu$ analysis~\cite{Belle-II:2023esi} from \belletwo\
a sizable fraction of the backgrounds, for both the inclusive-tag and the hadronic-tag
approaches, contains \KL's in the final state. Moreover, the $B^+ \to K^+ \KL \KL$ process,
is very difficult to distinguish from the expected standard model signal in case both
\KL's are unreconstructed.
For all the analyses mentioned above, having an increased \KL\ detection efficiency
would not only reduce the background contamination, but it would allow also to
gather higher-statistics and more reliable data control samples, further reducing
our systematic errors.

A higher KLM efficiency will also translate into more effective photon/neutron veto,
which is currently a limiting factor for some dark sector searches, most notably
the search for dark photons in the single photon final state, which is severely
limited by the inadequate $\gamma$-veto capabilities of the current detector.
Searches for specific dark photon final states with long-lived particles that can
be detected in the KLM and whose momentum can be measured with time-of-flight
capabilities would also benefit from an upgraded detector.

At the moment we do not have a full simulation of the \belletwo\ detector incorporating
the upgraded KLM, thus we cannot provide quantitative statements on the expected
improvement on the physics reach in the channels mentioned above.

\section{Trigger}
\label{sec:pp-trigger}

\subsection{Machine learning}
\label{sec:TRG_ML}
\editor{T.Ferber}

The existing Belle II trigger system mostly relies on conventional reconstruction algorithms
that utilize local information and rather simple algorithms to fit the hardware constraints
of currently used FPGAs. One notable exception is the NeuroZ track trigger at Belle II that
uses conventionally constructed track segments as inputs for a simple fully-connected neural
network to infer transverse momentum and $z$-impact parameter. With the advent of significantly
more powerful FPGAs and more tools to simplify the translation of Python-level ML solutions to
FPGAs, more complicated network architectures and higher dimensional inputs are of interest for
real-time applications. 

\begin{figure}[htbp]
  \centering
    \includegraphics[width=0.95\linewidth]{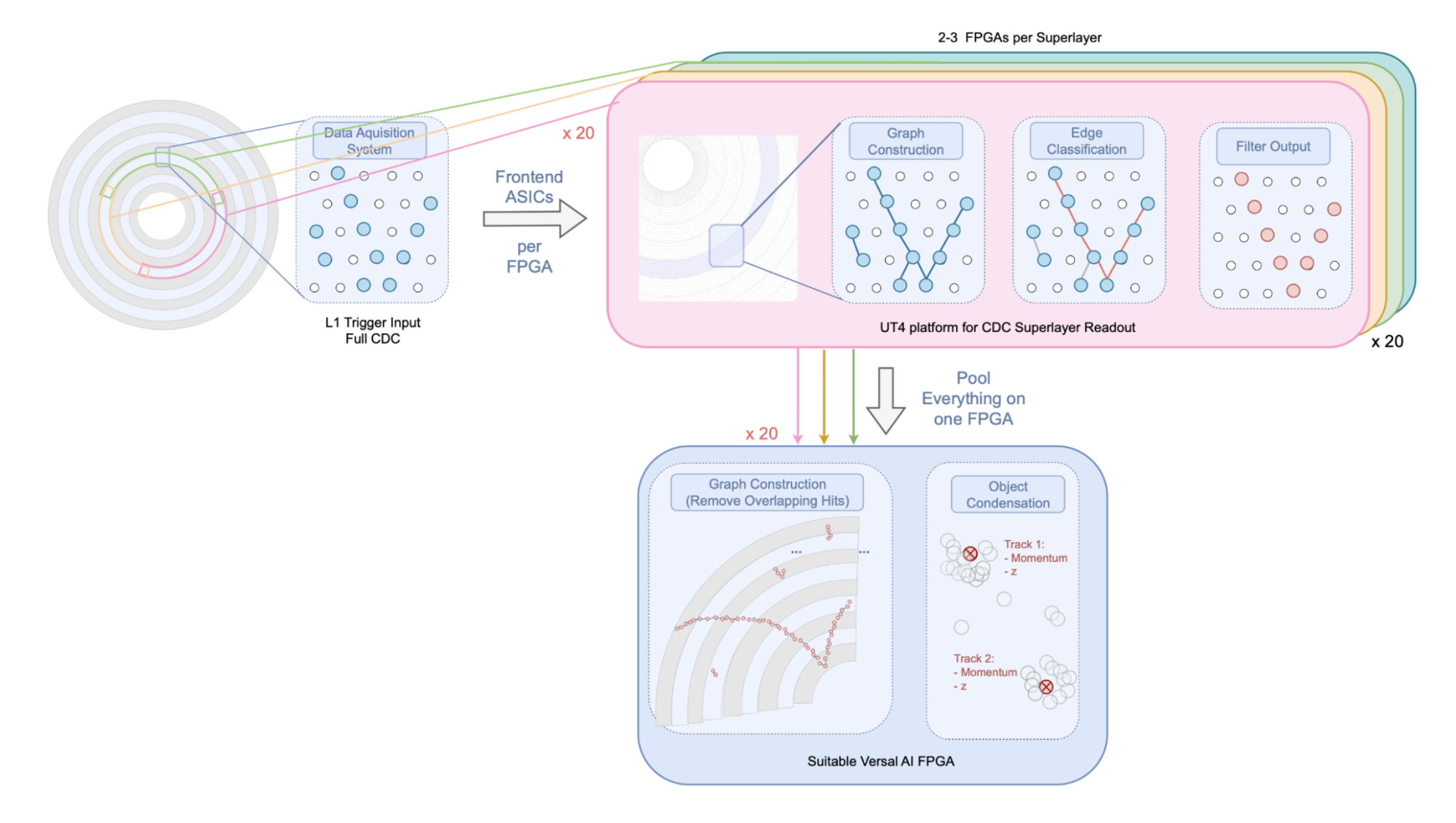}
    \caption{Schematic representation of the Graph Neural Network based track finding and
      fitting algorithm.
      \label{fig:pp-trigger1}}
\end{figure}

Track reconstruction is a crucial part of the L1 trigger in the Belle II experiment. Track-finding
algorithms are designed to balance the requirements of high efficiency in finding charged particles
with good track parameter resolution, a low rate of spurious tracks, and reasonable demand of
resources. We started development of a complete Graph Neural Network track finding and fitting
algorithm. A graph representation of CDC detector hits, including the 2D position, timing
information and pulse height, is used to find tracks and infer their track parameters. In order to
identify the varying number of tracks per event, we use GNN-based object condensation for track
finding. While in principle the inference of the algorithm aims at low latency single-step
inference using for example GravNet layers, the large number of CDC-hits per event prevents the
building of a single event-wide graph. Instead, we plan to first run a cleaning step using smaller
GNNs that utilize local graphs in 2-3 sectors per superlayer on about 20 individual UT4 FPGAs in
parallel (see Fig.~\ref{fig:pp-trigger1}). For simulated beam backgrounds for final luminosity
conditions, we find that 99\% of all generic \BB\ events contain less than 300 hits per sector.
These GNNs will classify hits into signal- and background-like and reject background-like hits
based on locally built graphs using edge classifiers. This clean-up step will also be responsible
to provide data that is rather robust against changes of beam background conditions for the final
tracking step. The cleaned output is passed to a single large FPGA on which a global graph is built,
which is then passed to a GNN with object condensation loss \cite{Kieseler:2020wcq} for track
finding and parameter inference.  We have demonstrated that real-time graph building on FPGAs is
possible within the existing throughput and latency requirements of the Belle II experiment’s
hardware-based trigger system for the CDC \cite{Neu:2023sfh}. First studies show that the large
field-of-view of a global GNN provides superior performance for signatures with shallow track angles,
tracks with low transverse momentum, overlapping tracks, and tracks originating from displaced
vertices \cite{chep23reuter}. A full implementation requires a co-design of graph building, cleanup
GNN, and final track finding GNN to optimize efficiency and resolution metrics. The current trigger
system does not provide pulse height information which proved powerful for background rejection.
Conceptually, it is also possible to include a vertexing step by changing the object condensation
loss targets.

ECL clustering is another area where GNNs can be used to exploit information from large fields of
view. The main interests for possible improvements are better spatial resolution, separation of
overlapping clusters, and low energy cluster identification. The current ECL trigger algorithm is
based on so-called trigger cells (TCs) that contain about 4x4 crystals, depending on their location.
The energy of all TC crystals is combined into an analog sum and read out if it exceeds a threshold
of 100 MeV.  The actual clustering algorithm combines TCs into clusters but is not able to separate
neighbouring TCs into separate clusters or provide actual spatial reconstruction. For simulated beam
backgrounds for final luminosity conditions, we find that 99\% of all generic \BB\ events contain less
than 89 TCs. This number is small enough to directly feed it into a global GNN: a graph representation
of ECL TCs, including the 2D position, timing information and energy, is used to find clusters and
infer their energy and cluster centroid position. In order to identify the varying number of clusters
per event, we use GNN-based object condensation. First studies show comparable efficiency and fake rate
for photons with $E > 100$ MeV, but a better energy and position resolution \cite{chep23haide}. The
latter can be used to improve Bhabha-vetoes that rely on angular deviations from back-to-back topologies
of ECL clusters. Using the full information of nearby energy depositions, we have shown that local GNNs
are able to separate overlapping energy depositions \cite{BelleII:2023egc} that are a signature of
axionlike particles (ALPs) or high energy neutral pions even in very high background environments. In
particular the ALPs $a$ produced in $ee\to \gamma a(\to \gamma\gamma)$ are reconstructed as
$ee\to\gamma\gamma$ events for low mass ALPs and are prescaled on trigger level if the decay photons are
not isolated. In addition to replacing the current clustering algorithm with a GNN approach, two
hardware upgrades are promising to provide a robust calorimeter trigger at high beam background levels:
first, the readout threshold is limited by electronics noise that could be improved by dedicated redesign
of the Shaper-DSP boards.
%The number of TCs increases to about XXX for a reduced energy threshold of
%80 MeV which is still suitable for a single FPGA.
The GNN performance generally improves for lower readout thresholds and shows an improved energy and
position resolution. Second, the large size of 4-by-4 crystals in each TC limits the position resolution
and the capability to resolve overlapping clusters. We study a Shaper-DSP upgrade with reduced TC size
of about 2-by-2 crystals with 25 MeV readout threshold. This increases the number of input nodes for the
GNN but provides a handle to run neutral particle identification on trigger level exploiting shower
shapes as additional object condensation loss targets.

The GNNs that show the best robustness against beam backgrounds for both tracking and clustering relied
on learning a latent space representation, followed by a $k$-nearest neighbour algorithm to select graph
nodes of interest. kNN implementations on FPGAs are currently not able to cope with the large number of
input nodes and additional studies to either replace kNNs or to implement approximate kNNs on FPGAs are
ongoing.

\section{Physics benchmarks}
\label{sec:pp-benchmarks}

In Section~\ref{sec:pp-parameters}, we showed that the area in which we expect
the biggest impact in physics performance is the improvement in reconstruction
efficiency and quality for the $\pi_{\rm soft}$ produced in the decay of
charged $D^*$ mesons.

This improvement will be very beneficial in at least two major areas of the
\belletwo\ physics program. In the first place, the $D^{*+} \to D^0 \pi^+_{\rm soft}$
decay is the main process through which we reconstruct and determine the
flavor ($D^0$ or $\overline{D}^0$) of the neutral $D$ mesons that are utilized
in measurements of the $D\overline{D}$ mixing and in searches for mixing-induced
$CP$ violation in the charm sector, which are still very much limited by the 
statistical uncertainties and for which \belletwo\ will have unique sensitivity
in decay modes containing one or more $\pi^0$'s. Secondly, the reconstruction
of $D^*$ candidates is of the utmost importance for $B$-physics channels exploiting
the Full Event Interpretation (FEI) \cite{Keck:2018lcd}. In these analyses, one of
the two $B$ mesons in the event is reconstructed through one of many hadronic or
semileptonic decay chains. Charged $D^*$ mesons feature in about $\sim 50\%$ of
the reconstructed $B$ mesons in the tag side, thus an improvement in the
$\pi_{\rm soft}$ reconstruction efficiency would immediately translate in an increased
sensitivity for the channels that need FEI (most of which are still severely
limited by statistical uncertainties). These analyses typically target
final states containing one or more neutrinos and are in many cases among the
\emph{golden channels} of Belle II.

Significant improvements over \belletwo\ will be seen also on the vertexing
precision. These will translate into better sensitivity to time dependent analyses
in the $B$ and charm sectors. As an example, we report the estimated sensitivity
to the time dependent $CP$ asymmetry in the $B \to \KS \pi^+ \pi^- \gamma$
channel, which is potentially sensitive to right-handed contributions from physics
beyond the standard model.

\begin{figure}[htbp]
  \begin{subfigure}[t]{0.50\textwidth}
    \includegraphics[width=1.0\linewidth]{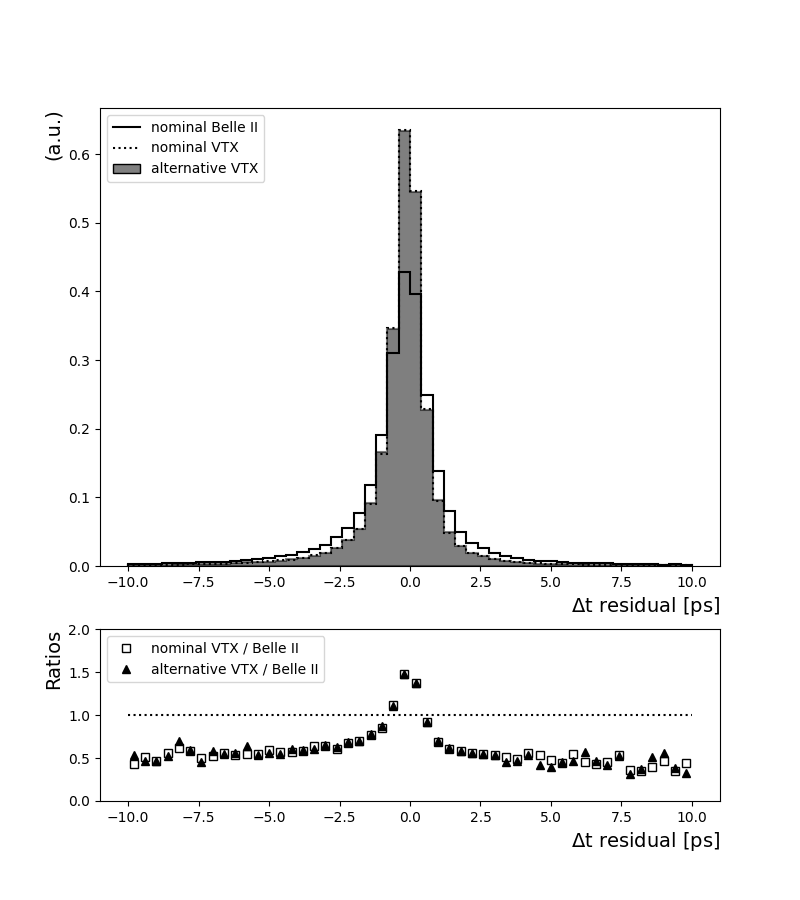}
  \end{subfigure}
  \hfill
  \begin{subfigure}[t]{0.50\textwidth}
    \includegraphics[width=1.0\linewidth]{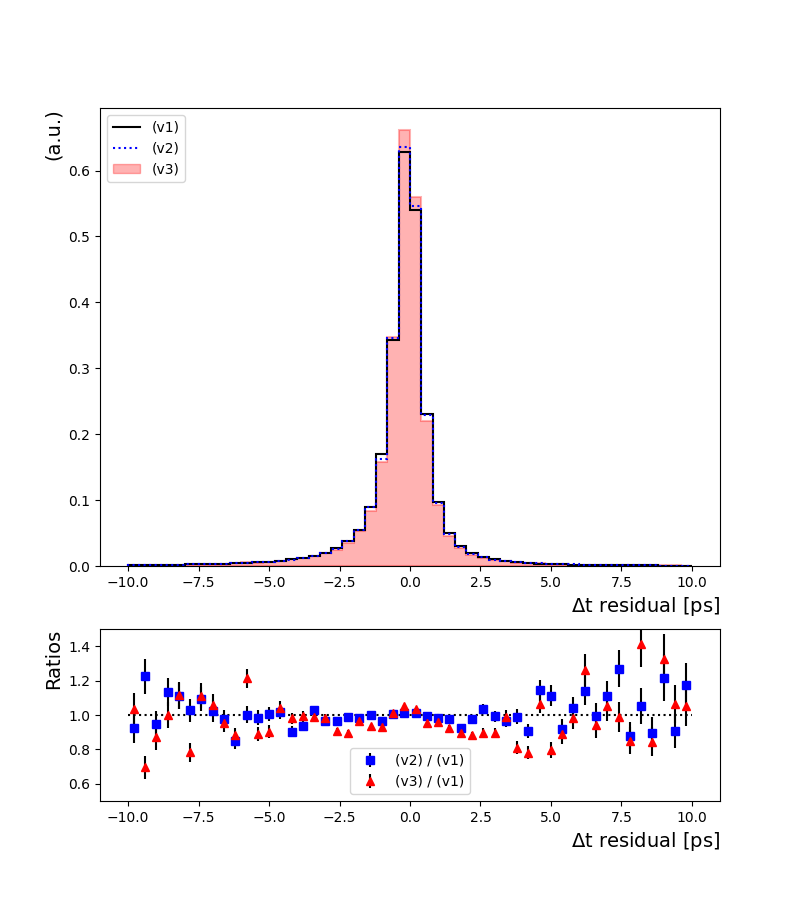}
  \end{subfigure}
  \caption{Left figure: $\Delta t$ resolution for $B^0 \to \KS \pi^+ \pi^- \gamma$ events
    for the nominal \belletwo\ detector (solid line), nominal VTX geometry (dotted line)
    and alternative VTX geometry (filled grey histogram). The bottom plot shows the ratio
    between the VTX geometries (empty squares for the nominal and filled triangles for
    the alternative) and nominal \belletwo. Right figure: $\Delta t$ resolution for the
    nominal VTX geometry in the three background scenarios: optimistic scenario-1
    (v1, green solid line), intermediate scenario-2 (v2, blue dotted line), and conservative
    scenario-3 (v3, red filled histogram). The bottom plot shows the ratio between the two
    higher background scenarios and the optimistic one.
    \label{fig:pp-KspipiDtReso}
  }
\end{figure}

We estimate the sensitivity of the analysis by generating 1000 toy experiments
for each combination of detector geometry and background scenario, assuming
an integrated luminosity of 50 ab$^{-1}$. The critical observable for this class
of analyses is the proper decay time difference between the two $B$ mesons in the
event: $\Delta t = t(B_{\rm sig}) - t(B_{\rm tag})$, where $t(B_{\rm sig})$ is the decay
time of the signal $B$ candidate (the one decaying to $\KS \pi^+ \pi^- \gamma$ in
this case) and $t(B_{\rm tag})$ is the decay time of the other $B$ meson produced
in the event. The decay time difference $\Delta t$ is determined from the
spatial displacement along the boost direction of the two decay vertices.
Fig.~\ref{fig:pp-KspipiDtReso} shows the $\Delta t$ resolution obtained for
the different vertex detector geometries and same intermediate (v2) background
scenario (left plot) and the resolution for the nominal VTX geometry and different
background levels (right plot). It is apparent that the performance of the
VTX is significantly superior to that of the current detector and shows little
sensitivity to the background conditions.

The events in the toy experiments are generated with a smearing on the generated
value of $\Delta t$ determined by a fit to the distributions of the residuals for
each configuration; the fitting function we utilize is the sum of three gaussians.
The much better $\Delta t$ resolution of the VTX configurations more than
compensates the small loss of efficiency in the \KS\ reconstruction that we
discussed in Sec.~\ref{subsec:pp-KS_eff}. The final expected uncertainty on the
time-dependent $CP$ violation parameter $S$ is 0.028 for the current \belletwo\
detector, 0.023 for the nominal VTX geometry, and 0.025 for the alternative VTX.
The differences between those results might seem unimpressive, but it is worth
noting that the current \belletwo\ detector would need a $\sim 40\%$ increase
in the available integrated luminosity in order to match the result of the nominal
VTX upgrade, i.e. the upgrade is estimated to be worth the equivalent of 20\invab\
of data on the design luminosity.

\chapter{Inner Detector}
\label{sec:inner}
\editor{C.~Marinas}
 
\section{Vertex Detector}
\label{sec:VXD}

\subsection{VXD upgrade baseline} 
\label{sec:vxd-baseline}

While a number of scenarios are being considered as a part of the machine upgrade for achieving the intended final peak luminosity, several of these options result in modifying the spatial envelope of the interaction region. This step would require a complete overhaul of the Belle II innermost subdetectors. Preparing for a vertex detector upgrade also in the scenario in which the interaction region is maintained has also strong strategic value, for example to replace the current VXD in case of a severe accident or to exploit the advantages of having a better performing inner detector system in certain physics channels.

Among the various possibilities considered by the Belle II upgrade group, the development of a CMOS-based successor to the PXD and SVD detectors has emerged as the most promising solution. The monolithic active pixel sensor CMOS technology has the potential for providing higher granularity, thinner, intelligent detectors at a lower overall cost. The monolithic CMOS option is supported as the baseline of the VXD upgrade proposal and is implemented through the development of the OBELIX sensor discussed in section~\ref{sec:vxd-obelix}.\\
This detector concept would improve the tracking robustness against large machine backgrounds and allow the detector envelope to be adapted to possible changes to the QCS systems. In addition, track reconstruction for low momentum particles and vertex position resolution are expected to improve. Other performance for physics should stay at the level of the present detector. Quantitative estimates obtained with baseline design described in this chapter are presented and discussed in section~\ref{sec:pp-parameters}.

The studies presented here assume the new vertex detector would occupy the same volume of the current VXD (between \SI{14}{\milli\meter} and \SI{135}{\milli\meter} in radius) and, although the final geometrical configuration is still being optimized, the system is being designed flexibly to adapt also to any possible modifications of the machine-detector boundaries. A good particle hit spatial resolution below \SI{15}{\micro\meter} is required, along with a low material budget around $0.2\% X_0$ per layer for the inner radii and $0.8\% X_0$ per layer for the outer radii.
To ensure good long term performance the reference hit rate and radiation levels  used in the design of the system are the following values for the innermost layer:
\begin{itemize}
    \item Hit rate capability: \SI{120}{\mega\hertz/\cm\squared}
    \item Total ionizing dose: \SI{100}{\mega rad}
    \item NIEL fluence: \SI{5 e 14}{n_{eq}/cm^2}
\end{itemize}

Although there is an engineering effort ongoing towards a 6-layered configuration, the current VTX baseline design consists of five detection layers, all equipped with the same monolithic sensor. Nevertheless, the design and integration of the layers depends on the radius and two separate concepts have emerged. Layers below \SI{30}{\milli\meter} radius are grouped into the so-called iVTX design, while layers at outer radii follow the oVTX configuration.\\
Section~\ref{sec:vxd-obelix} describes the main sensor option, which relies on the CMOS technology benefiting from a large expertise and support in the R\&D community. The VTX structure, including ladder design and service concepts, is detailed in section~\ref{sec:vxd-structure}. Finally, schedule and cost estimate for the baseline option are discussed respectively in \ref{sec:vxd-schedule} and sections~\ref{sec:vxd-cost}.\\

%%%%%%%%%%%%%%%%%
\subsection{OBELIX sensor}
\label{sec:vxd-obelix}
\label{sec:obelix}

OBELIX stands for Optimised BELle II monolithic pIXel sensor and is the name of the monolithic active pixel CMOS sensor (or MAPS) proposed as the main option to equip all VTX layers. The introductory section \ref{sec:VXD} reviews the main requirements on the detector layer in terms of detector performance and the practical constraints related to the integration of the sensor into detection modules. These requirements and constraints lead to the sensor specifications, as explained in section \ref{sec:vxd-obelix-spec}. The evaluation of the forerunner sensor, TJ-Monopix2, comes next in section \ref{sec:vxd-monopix2}. Then the sensor implementation is presented in section \ref{sec:vxd-obelix-implementation} along with the development schedule.

%%%%%%%%%%
\subsubsection{Sensor technical specifications}
\label{sec:vxd-obelix-spec}

The sensor technical specifications are reviewed in the following order: the pixel matrix, the trigger logic and track-triggering, the single event effect tolerance, the power dissipation, the sensor interface with the outside world and finally the sensor dimensions. All specifications are summarized in table~\ref{table:vxd-obelix-spec}.

The design of the {\bf pixel matrix} of monolithic sensors drives several key performances. The position resolution, detection efficiency and radiation tolerance depend on the depletion and depth of the sensitive volume, the pixel size, the analogue pre-amplification and detection threshold settings. A spatial resolution below $15\mum$ typically requires a pixel pitch below $40\mum$. In order to match the material budget between 0.2\%$X_0$ and 0.8\%$X_0$ per layer and taking into account support structures and cables, the sensor thickness in the fiducial volume has to be below $100\mum$. Consequently the depleted sensitive thickness should be lower than $50\mum$, well in line with available thickness of  MAPS. Such thickness will lead to MIP signals on pixels as low as few thousand electrons and calls for equivalent detection thresholds in the range 100-200 electrons to ensure detection efficiency close to 100\%.

The overall {\bf hit rate capability} of the sensor results from the combination of two factors. First comes the rate of fired pixel addresses and associated timestamps that the read-out architecture in the pixel matrix can actually reconstruct efficiently. The second factor is the performance of the logic in the matrix periphery that correlates in time these fired pixel with the triggers available to the sensor.\\
The hit rate value of \SI{120}{\mega\hertz/\cm\squared} quoted in section \ref{sec:vxd-baseline} corresponds to an average rate exceeding by a safety factor of 4 the expectation for the innermost layer presented as the conservative scenario-3 in section \ref{sec:BackgroundExtrapolation}. This average value sets the requirements for the overall hit rate, encompassing the performance of the pixel matrix and the periphery treatment. Hence, the corresponding hit rate requirement is referred to as {\it Hit rate capability at the sensor output} in table~\ref{table:vxd-obelix-spec}.  \\
During SuperKEKB injections, short spikes of very high rates are expected. A maximum rate is presently set to be \SI{600}{\mega\hertz/\cm\squared} for a period of $0.5 \mus$ every $10 \mus$, the beam revolution period in the collider. Only the pixel matrix should sustain this maximum value during periods where no triggers are expected, leading to a requirement labeled as {\it Hit rate capability in the matrix} in table~\ref{table:vxd-obelix-spec}.\\
The {\bf trigger logic} storing the hit information coming from the matrix in order to compare their timestamp to the arrival times of triggers, should offer a memory large enough to contain the hit information up to a maximal trigger delay of 10~\mus (extending the current limit of 5~\mus) with a hit rate of \SI{120}{\mega\hertz/\cm\squared}. In addition, the logic should handle trigger rates up to \SI{30}{\kilo\hertz}, which is the unchanged maximal L1 trigger rate set in the \belletwo\ technical design report \cite{Abe:2010gxa}.\\

A key requirement for the new sensor consists in achieving a high granularity simultaneously in space and time. While the pixel pitch limit fixed previously addresses the former, the latter leads to a maximum integration time of 100~ns from the trigger arrival time. This limit on the overall {\bf integration time} actually applies to the combination of the timestamp precision with any fluctuation over the matrix of the timestamp reference.\\
Providing better position information on tracks than the main tracker (central drift chamber, CDC) within a time useful for the trigger system would improve the Belle~II track-trigger capability. Setting precise specifications for the space-time granularity of such information is not possible without a detailed study of the trigger algorithm, which is not yet available. As an starting point, the sensor will be divided azimuthally into 10 segments. The information about which of these regions are hit by particles within about 10 nanoseconds should then be continuously transmitted out of the sensor.\\

Tolerance to Total Ionizing Dose (TID) and Non Ionizing Energy Loss (NIEL) fluence have been discussed already. But Single Event Effects (SEE) generated by radiation might affect the sensor. Since no Single Event Latchup (SEL) were observed in the past operation years of \belletwo, no specification is given. However Single Event Upsets (SEU) were observed and will be a concern to some extent, but are not yet quantified. Consequently, the first specification is to ensure that the control system allows resetting the sensor registers to operational values at least every few minutes. The actual frequency will be chosen after the measurement of the SEU cross section with the OBELIX sensor and the comparison to the occurrence distribution of large linear energy transfer in the experiment.\\

The material budget per layer does not only depend on the sensor thickness but mainly on the services and support structures required to operate the sensor. The chip should then feature a {\bf pad ring} facilitating the connection with a single light cable. The sensor {\bf power dissipation} is also desired to be as low as possible and constrained within 200~$\rm{mW}~{cm}^{-2}$, a value which is anticipated to allow air-cooling for the small area of the two inner layers and single phase liquid coolant with small radius polyamide pipes for the outer layers.\\
For the sake of the simplicity of the {\bf system integration}, it is also specified that the sensor is powered through on-chip voltage regulators and features a single rather low data bandwidth ($<$200~$\rm{Mbps}$).\\

\begin{figure}
\centering
\includegraphics[width=0.5\textwidth]{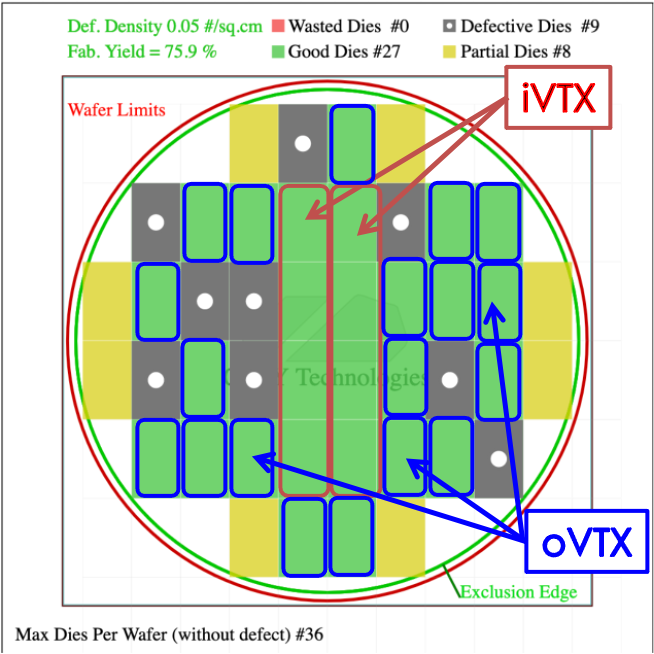}
\caption{Sketch of the OBELIX reticule occupancy over an 8 inch diameter wafer. Red rectangles indicate possible dicing area including 4 consecutive sensors for iVTX ladders, while blue rectangles stand for individual chip dicing for oVTX stave population.}
\label{fig:vxd-obelix-wafer}
\end{figure}

The sensor overall dimensions follow from various constraints. First, the chosen CMOS technology offers a maximal reticule size of $30 \times 25 \mma$. Considering the need to cover a large area (about $1 \ma$ for the  outer layers) with a minimal number of chips, matching this size for OBELIX seems benefical. However, the iVTX concepts uses 4 consecutive chips simultaneously diced out of the wafer to make a ladder. Taking into account an expected yield of 60\% for individual die, it is critical to maximise the number of such potential 4 consecutive dies on the wafer in order to minimise the number of wafers needed to obtain the 20 iVTX ladders. The result of the optimisation is displayed in figure \ref{fig:vxd-obelix-wafer} and leads to a chip dimension of $30 \times 19 \mma$.\\
Assuming a $3 \mm$ width for the digital periphery, as an educated guess from previous sensors in the same technology, the sensitive area or pixel matrix size reaches $30 \times 16 \mma$.

%\begin{landscape}
\begin{sidewaystable}
\centering
\caption{OBELIX sensor specifications, compared to the relevant specification of the TJ-Monopix2 sensor.}
\label{table:vxd-obelix-spec}
\begin{tabular}{lc|c}
\hline
\hline
 & Specification & TJ-Monopix2\\
\hline
Pixel pitch & $< 40$\mum & $< 33 \mum$ \\
Sensitive layer thickness & $< 50 \mum$ & $30 \mum$ and $ 100\mum$ \\
Sensor thickness & $< 100$\mum & - \\
\hline
Hit rate capability in the matrix & $> 600$~MHz~cm$^{-2}$ & $> 600$~MHz~cm$^{-2}$ \\
Hit rate capability at the sensor output & $>120$~MHz~cm$^{-2}$ & $\gg 100$~MHz~cm$^{-2}$ \\
Trigger delay & $>10 \mus$ & - \\
Trigger rate & $30$~kHz & - \\
%(optional) Track-trigger capability & \rem{to fill} & - \\
Overall integration time & $<100$~ns & - \\
(optional) Time precision & $<50$~ns & - \\
\hline
Total ionizing dose tolerance & 100~Mrad & - \\
NIEL fluence tolerance & \SI{5 e 14}{n_{eq}/cm^2}  & \SI{1.5 e 15}{n_{eq}/cm^2} \\
SEU tolerance & frequently (~min$^{-1}$) flash configuration & - \\
\hline
Matrix dimensions & around $30 \times 16 \mma$ & $19 \times 19 \mma$ \\
Overall sensor dimensions & around $30 \times 19 \mma$ & $20 \times 19 \mma$\\
Powering & through voltage regulators & - \\
Outputs & one at $<200 \rm{ MHz}$ & one at $160$~MHz \\
\hline
\hline
\end{tabular}
%\caption{bandwidth with average hit rate and max hit rate}
\end{sidewaystable}
%\end{landscape}

The specifications regarding the pixel matrix established for the Belle II sensor overlap significantly with the ones used for the development of monolithic pixel sensors targeting the upgrade of the inner tracker of the ATLAS experiment. Within this program, the TJ-Monopix family sensors, depleted MAPS designed in the TowerJazz~180nm process, match particularly well the Belle~II requirements as summarized in table \ref{table:vxd-obelix-spec}. The first sensor, TJ-Monopix1 has already shown promising detection performance. The second version TJ-Monopix2, featuring a larger matrix and coming with a number of improvements, was fabricated in 2021.

It was then decided to use the pixel matrix of TJ-Monopix2 as a starting point for the OBELIX sensor in order to inherit from the results obtained during the development of the TJ-Monopix series. An additional benefit from this choice is that TJ-Monopix2 offers all the functionalities needed for an application-ready sensor: fast output, triggering, slow-control interface. Consequently, TJ-Monopix2 can be considered as the first large size prototype for the VTX detector concept. As a consequence, the test system of TJ-Monopix2 will be re-used for the later OBELIX sensor.

%%%%%%%%%%
\subsubsection{TJ-Monopix2 as prototype}
\label{sec:vxd-monopix2}
%\rem{low threshold testing, detection efficiency, radiation tolerance, in-time efficiency}

\paragraph*{Sensor description}

\begin{figure}
\begin{subfigure}[t]{0.48\textwidth}
\centering
\includegraphics[width=0.99\textwidth]{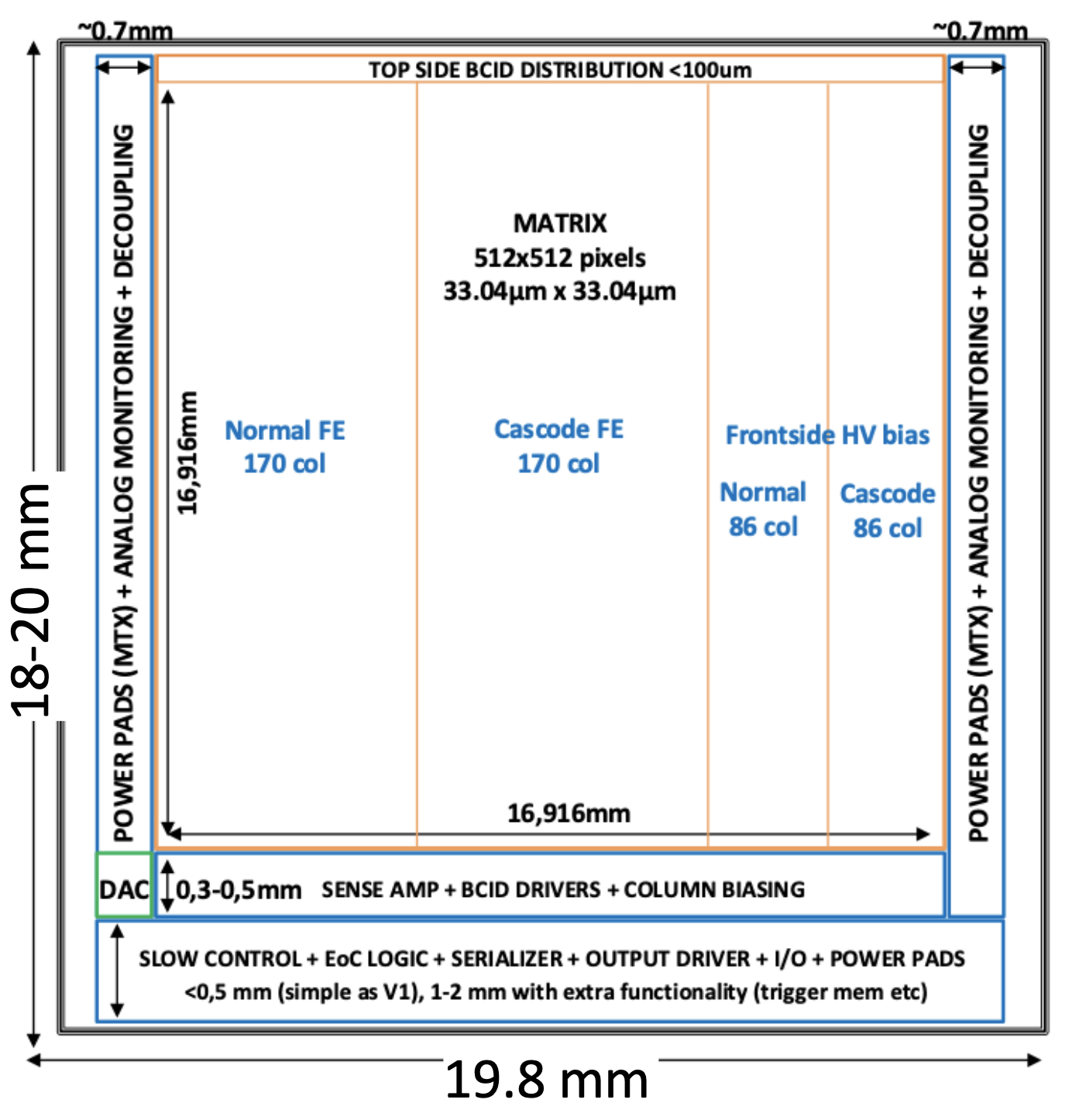}
\caption{General layout of the sensor functional blocks.}
\end{subfigure}
\hfill
\begin{subfigure}[t]{0.48\textwidth}
\centering
\includegraphics[width=0.99\textwidth]{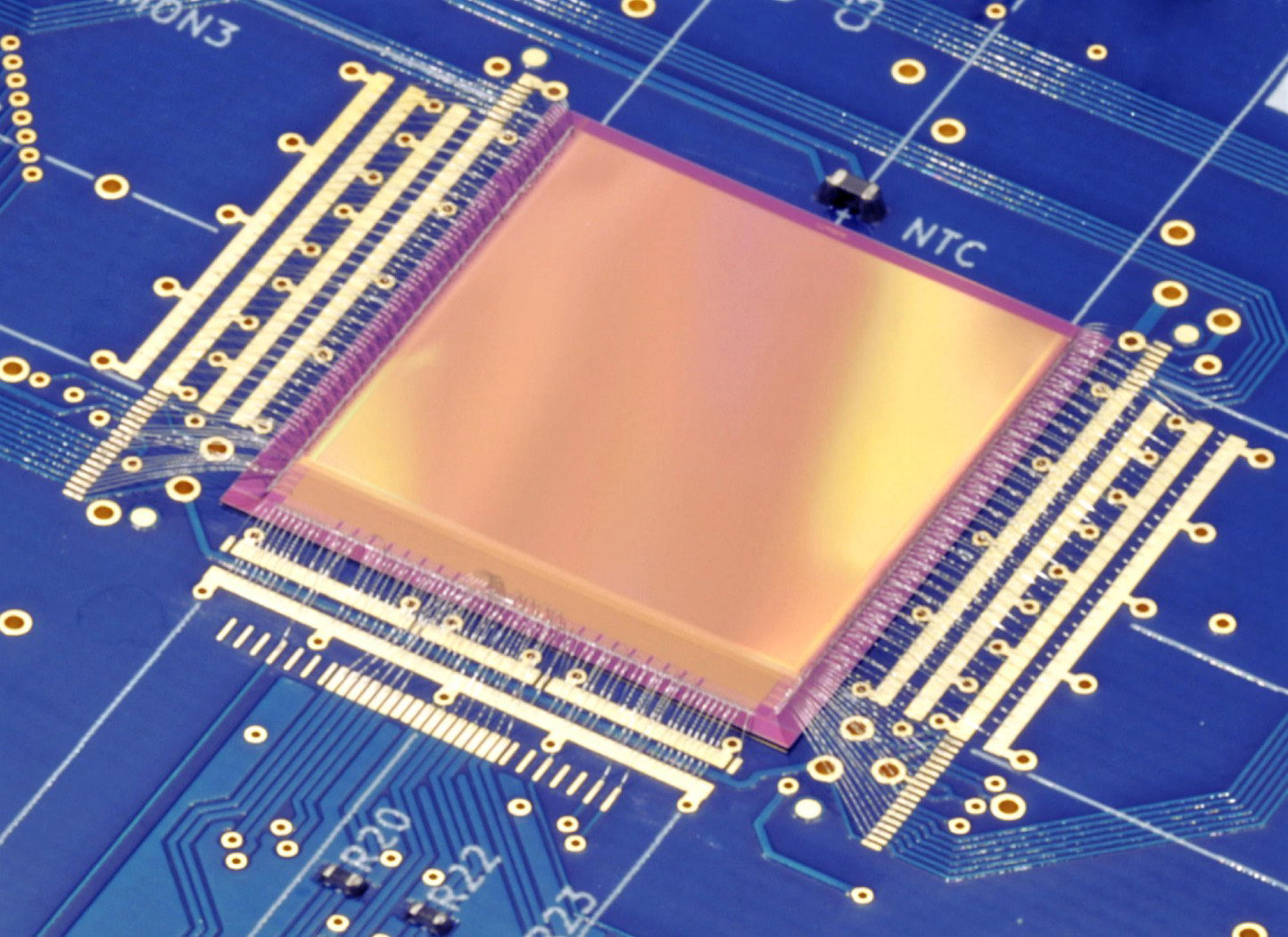}
\caption{Photograph of the sensor bonded on a test board.}
\end{subfigure}
\caption{Two different views of the TJ-Monopix2 sensor.}
\label{fig:vxd-monopix2}
\end{figure}

The TJ-Monopix2 general view is depicted in figure~\ref{fig:vxd-monopix2}.
The in-pixel collection node and front-end of the TJ-Monopix pixel matrices inherit from the ALPIDE sensor concept \cite{aglieri_rinella_alpide_2017}, featuring a small diode and low-threshold discriminator. The fabrication process from Tower Semiconductor, CMOS Image Sensor 180~nm, has been further modified \cite{snoeys_process_2017} in order to enhance the radiation tolerance beyond $10^{13} \rm{ n}_{\rm{eq}}\rm{cm}^{-2}$. The discrimination threshold can be adjusted for each pixel through a 3-bit DAC and any individual pixel can be masked from the read-out.\\
Four front-end flavours are implemented in four different regions of the pixel matrix \cite{bespin_dmaps_2020}, all aiming for enhanced radiation tolerance compared to the original ALPIDE design. The standard or normal version follows the ALPIDE implementation with only transistor size modification. A version targeting higher gain uses a cascode scheme amplification. The two last versions correspond to the introduction of capacitive coupling between the charge-collection electrode and the pre-amplification, normal or cascode. The capacitor allows biasing the collection node at a few tens of volts, way beyond the 1.8~V supply of the technology. These so-called HV-normal and HV-cascode flavours are expected to increase the depletion of the sensitive volume compared to the non-HV flavours. Without the AC coupling, the depletion growth depends on the voltages applied on the p-type wells surrounding the collection diode and on the substrate, which have limited excursions.\\

The read-out architecture implemented within the matrix uses the established synchronous column-drain mechanism developed for the ATLAS FE-I3 readout chip \cite{peric_fei3_2006}. For each fired pixels, the leading and trailing edge times are stored in two 7-bit counters clocked at 40~MHz. The end-of-column logic uses these counters to quantify the charge collected by the pixel with the time-over-threshold and to time-stamp the signal arrival with a precision of 25~ns.\\

The digital logic located at the periphery of the matrix contains an integrated read-out controller, LVDS drivers and receivers as well as the command decoder compatible with the RD53B readout chip \cite{garcia-sciveres_rd53b_2019}, which allows configuring all the chip registers via only four differential lines.\\

\paragraph*{Evaluation}

The detailed evaluation of the TJ-Monopix2 sensor is currently being performed \cite{bespin_performance_2023}. 
Operation of selected parts of the pixel matrix have been demonstrated with detection threshold from 100~$e^-$ a temporal noise around 6~$e^-$ equivalent noise charge and a fixed pattern noise (taken as the standard deviation of the threshold distribution) corresponding to about 3~$e^-$ equivalent noise charge after tuning the pixels. Both results match with the design values.\\

\begin{figure}
\begin{subfigure}[t]{0.48\textwidth}
\centering
\includegraphics[width=0.99\textwidth]{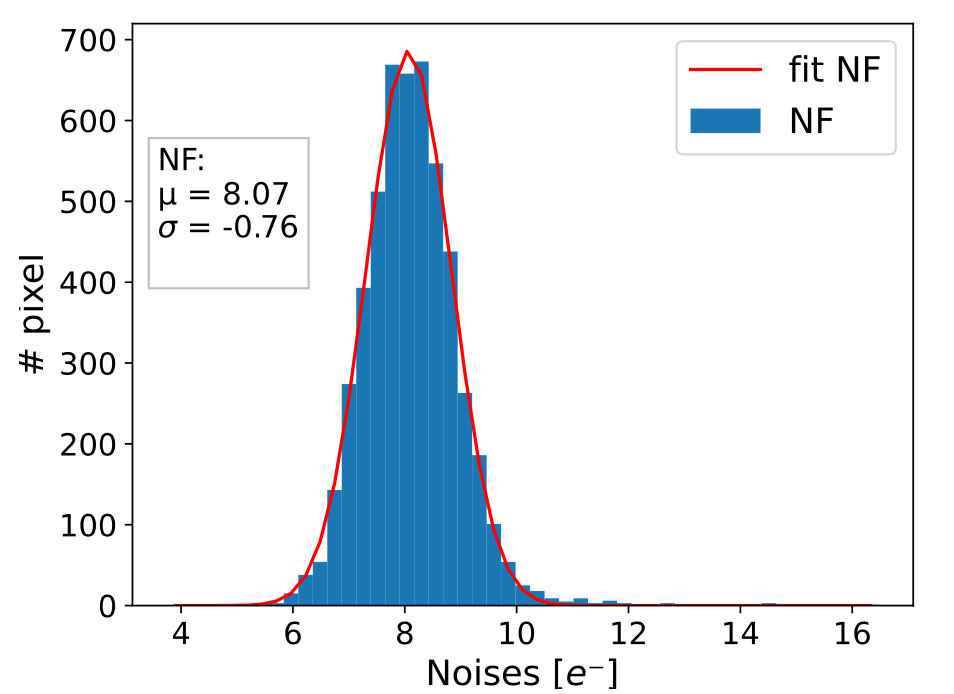}
\caption{Noise distribution.}
\end{subfigure}
\hfill
\begin{subfigure}[t]{0.48\textwidth}
\centering
\includegraphics[width=0.99\textwidth]{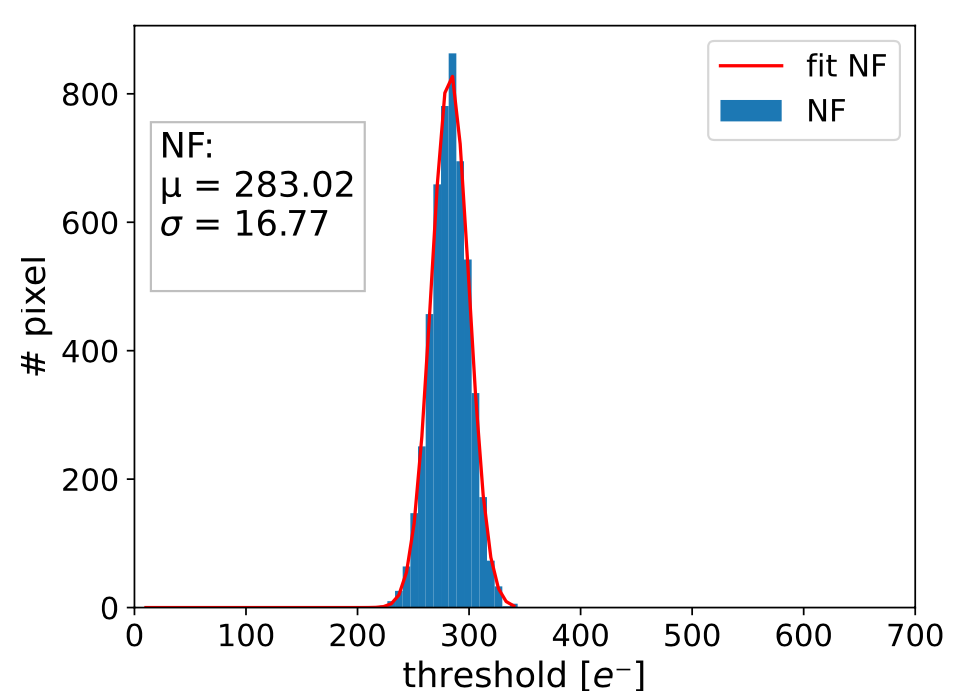}
\caption{Threshold distribution.}
\end{subfigure}
\caption{Threshold and noise distribution obtained on the normal front-end of the TJ-Monopix2 sensor, from \cite{schwickardi_upgrade_2023}.}
\label{fig:vxd-monopix2-lab-perf}
\end{figure}

The VTX collaboration has also conducted optimization of the TJ-Monopix2 chip \cite{Massaccesi_upgrade_2023, babeluk_cmos_2023, schwickardi_upgrade_2023} to acquire experience with operating this sensor technology and find the most performance sensitive parameters.\\
Control and data acquisition system is based on the BDAQ53 setup. Since the OBELIX sensor will embed the same control interface, see \ref{sec:vxd-obelix-implementation}, this system will require minimal adaptation for its use for OBELIX characterization. The firmware will essentially need an update to take into account the larger number of pixels and potentially slightly different format for the pixel data.\\
Typical results after tuning the matrix are displayed in figure \ref{fig:vxd-monopix2-lab-perf} where the distributions of the thresholds and temporal noise for the normal front-end are presented. Considering all pixels of each sub-matrix and for thresholds between 200 to 300~$e^-$, the average temporal noise varies from 7 to 8~$e^-$ equivalent noise charge depending on the front-end flavor and the temporal noise varies from 10 to 17~$e^-$.

\begin{figure}
\centering
\includegraphics[width=0.5\textwidth]{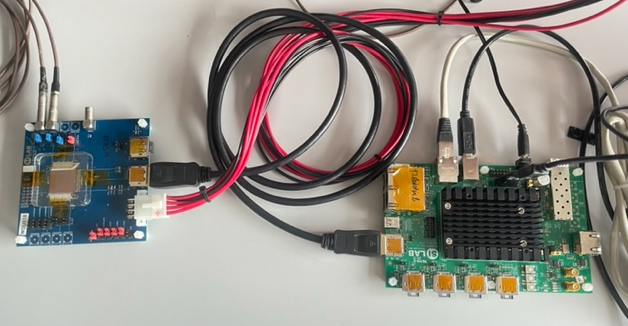}
\caption{Test setup for TJ-Monopix2 using the BDAQ53 system.}
\label{fig:vxd-monopix2-bdaq53}
\end{figure}

\begin{figure}
\begin{subfigure}[t]{0.48\textwidth}
\centering
\includegraphics[width=0.99\textwidth]{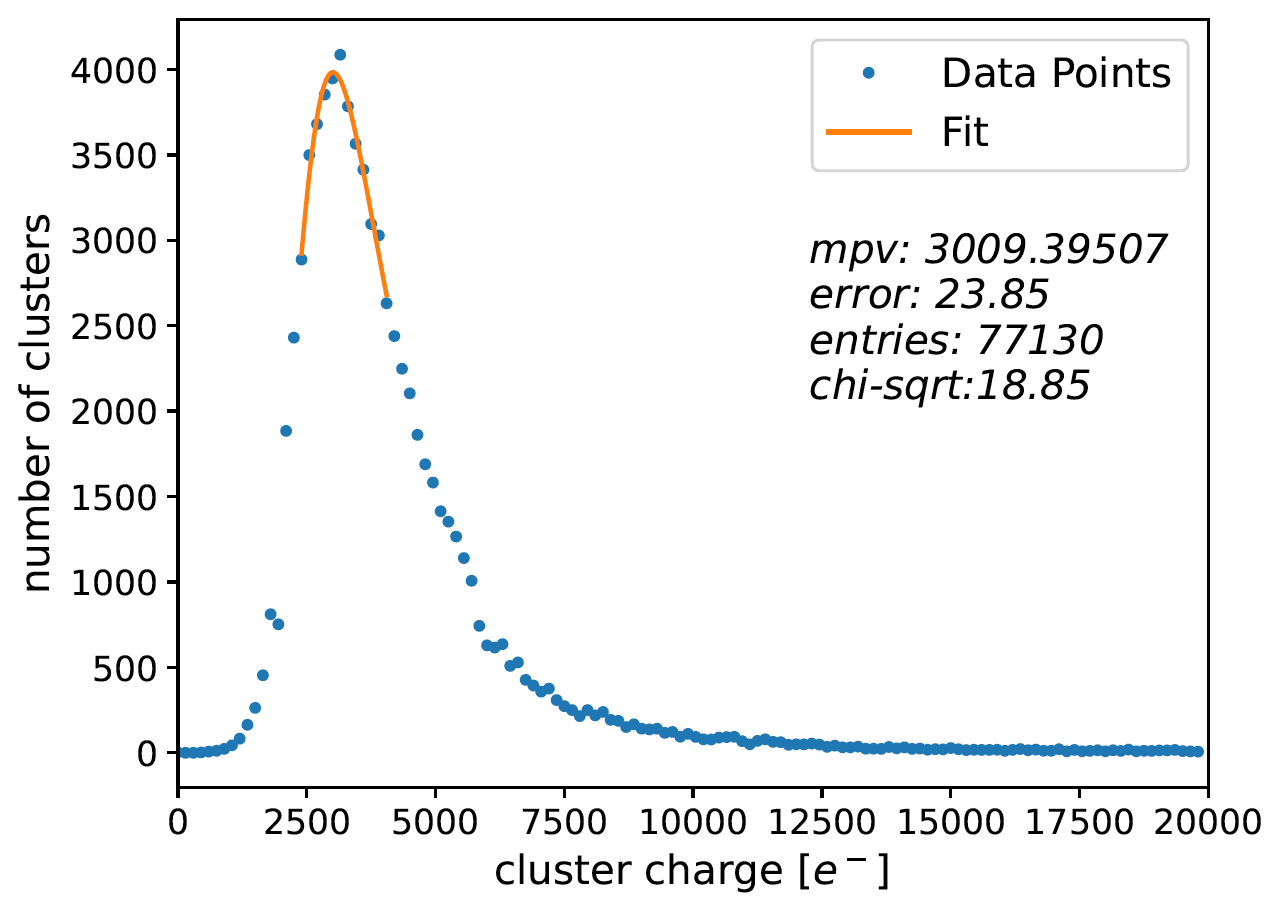}
\caption{Cluster charge.}
\end{subfigure}
\hfill
\begin{subfigure}[t]{0.48\textwidth}
\centering
\includegraphics[width=0.99\textwidth]{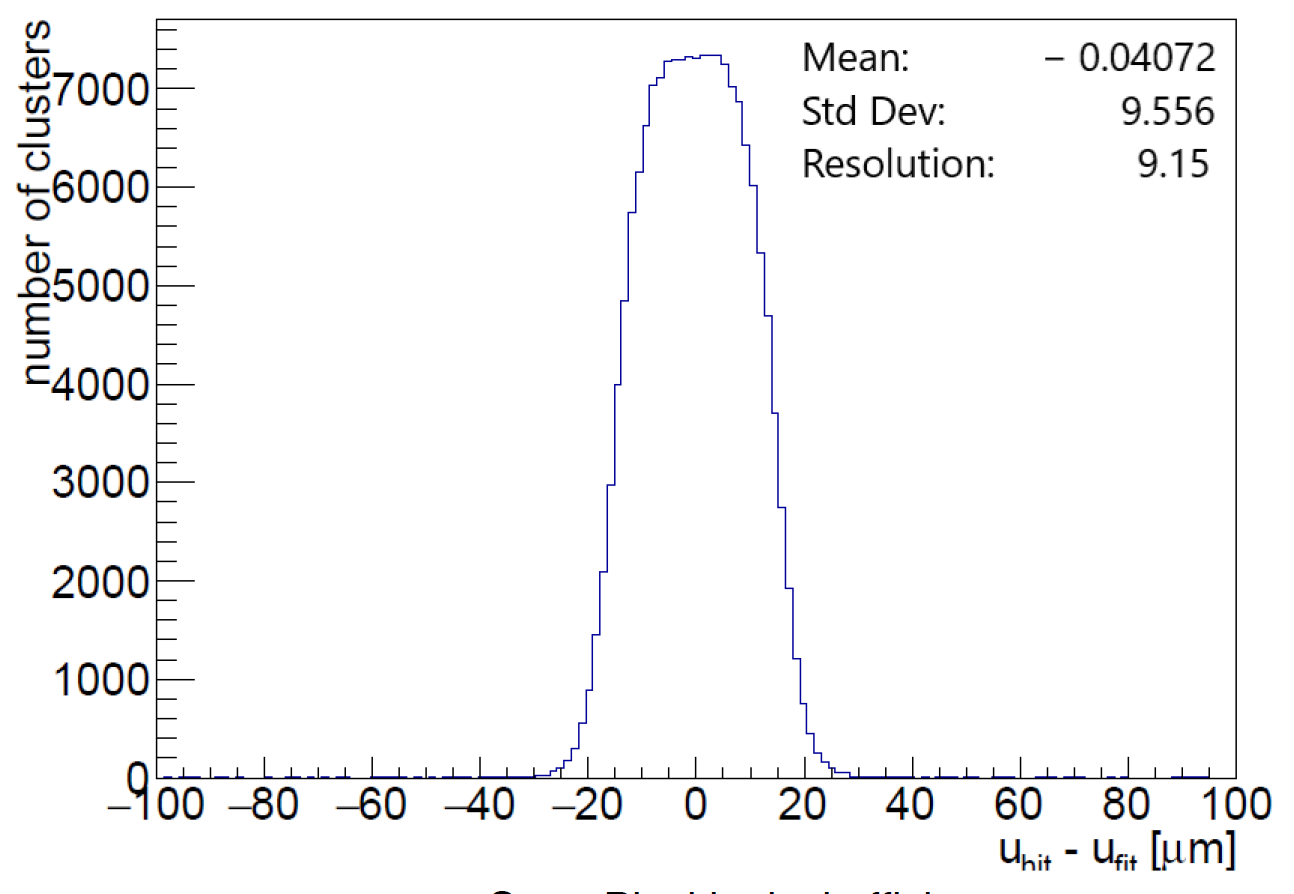}
\caption{Residual between the track extrapolation and the measured position by TJ-Monopix2.}
\end{subfigure}
\caption{Distributions obtained with the TJ-Monopix2 sensors recording data from a 4~GeV~e- beam at normal incidence, with a threshold of 500~$e^-$, from \cite{schwickardi_upgrade_2023}.}
\label{fig:vxd-monopix2-beam-perf}
\end{figure}

Two test campaigns at the DESY 3-5~GeV electron beam were conducted. Soon after the initial commissioning of TJ-Monopix2 in the laboratory, first detection performance with minimum ionising particles were explored in June 2022 and using an EUDET-type telescope \cite{jansen_performance_2016}. Due to the early stage of the sensor tuning at that time, only high-threshold operation was established. Figure \ref{fig:vxd-monopix2-beam-perf} illustrates the typical cluster charge distribution and residuals. Still, with a threshold set at 500~$e^-$ equivalent charge and an unirradiated sensor featuring a high resistivity Czochralski layer, it was possible to obtain above 99.9\% detection efficiency.\\
The spatial resolution was estimated to $9.15\mum$, slightly better than the digital resolution expected from the $33\mum$ pixel pitch and consistent with the small cluster size induced by the relatively high detection threshold. 
We determined the sensitive depth of the sensor to be $30\mum$ (systematic uncertainty ~$2\mum$) via an angular study. This observation justifies the base choice for the starting material of OBELIX-1, namely an epitaxial layer of $30\mum$ thickness.\\
These results were used to implement a realistic sensor response into the Belle~II software (basf2). The existing model reproducing the digitized signals on each pixel fired by a particle has been tuned to reproduce these beam test results \cite{fillinger_detector_2021}.\\

In July 2023, the second test period with beam occurred after the tuning of the TJ-Monopix2 sensor was refined. This campaign focused on sensors featuring a $30\mum$ epitaxial layer and included a chip irradiated to a fluence\footnote{Irradiation with 24~MeV proton beam} of $5\times10^{14}\rm{ n}_{\rm{eq}}\rm{cm}^{-2}$. The detection thresholds were tuned in a range from 200 to 300~$e^-$.\\
The data analysis is on-going while this text is being finalised. Only early results are briefly quoted below.
The detection efficiency was measured to exceed 99.9\% for all un-irradiated front-end flavours. For the irradiated sample, 99.5\% detection efficiency can be reached for the AC-coupled front-end when the high voltage is raised to 30~V and the detection threshold lowered to about 200~$e^-$.\\

This test beam campaign is the last one before the submission of the \BTWO\ dedicated sensor OBELIX-1 based on the TJ-Monopix2 matrix. The early results so far confirms the role of TJ-Monopix2 as a forerunner.\\

The careful evaluation of the various front-end performances from the last beam test forms the basis for limiting the number of front-end variants (possibly to only 2 versions) in OBELIX-1.

%%%%%%%%%%
\subsubsection{Sensor technical implementation}
\label{sec:vxd-obelix-implementation}

This section describes the technical implementation of the OBELIX sensor in order to match all specifications listed in table \ref{table:vxd-obelix-spec}.\\

\begin{figure}
\begin{subfigure}[t]{0.48\textwidth}
\centering
\includegraphics[width=0.99\textwidth]{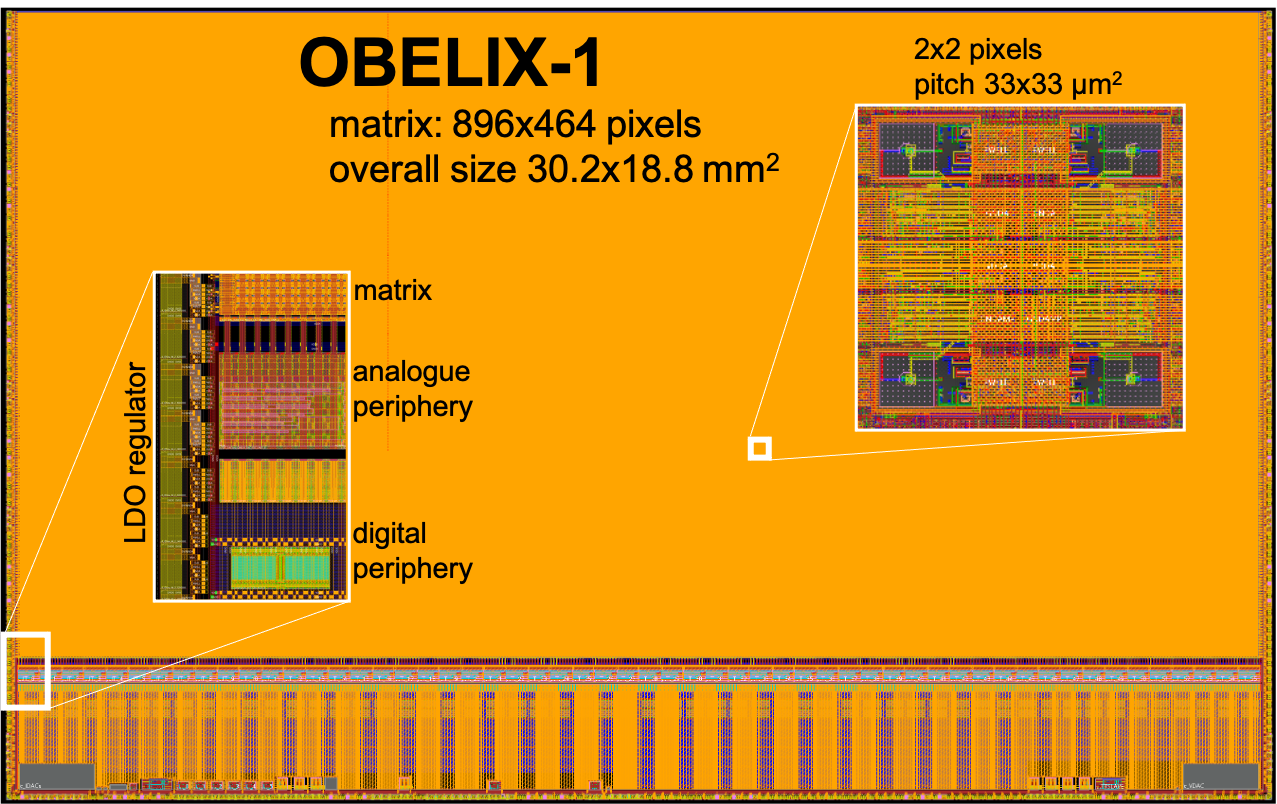}
\end{subfigure}
\hfill
\begin{subfigure}[t]{0.48\textwidth}
\centering
\includegraphics[width=0.99\textwidth]{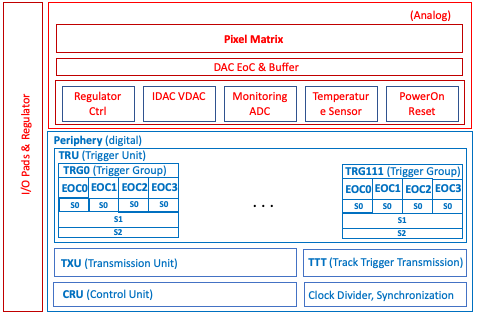}
\end{subfigure}
\caption{Layout of the OBELIX sensors, right: functional layout, left: physical layout with the various blocks described in the text.}
\label{fig:vxd-obelix-layout}
\end{figure}

\paragraph*{Pixel matrix}

As explained previously, the pixel matrix in OBELIX reuses the architecture developed for the TJ-Monopix2 sensor described in detail in reference \cite{moustakas_design_2021}. Only the main features are underlined here as well as modifications when appropriate.\\

The pixel conserves exactly the same pitch, $33\times 33\muma$, with the same layout for the analogue and digital parts. Based on current characterization results, we decided to choose the Cascode front end flavor and for the next iteration we will keep DC and AC coupling versions on equal area for final verification on the full size prototype. Pixels are still powered from horizontal rails in order to prevent performance gradients throughout the matrix, inducing the need for power pads on the short side of the sensor.\\

The signal digitization over 7~bits, through the Time-Over-Threshold method, and the column-drain read-out architecture implemented for pairs of columns stay unchanged. Also, the 3-bit register to trim the threshold value at the individual pixel level is still included, but the range of the possible correction is enlarged. However, each column-pair is equipped with its individual End-of-Column logic, which actually transmits ToT and timestamp information from each fired pixel to the memory stage, which is discussed in the trigger logic paragraph just below. Following the TJ-Monopix2 nomenclature adapted to LHC, the timestamp is also named BCID for bunch crossing identification. The clock frequency driving the precision of the ToT and the BCID changes to a lower value in order to reflect the targeted integration time, typically from $40$ down to $20$~MHz. The current baseline for OBELIX, is then to timestamp fired pixels with a precision of $50$~ns.\\

The matrix read-out logic has been demonstrated \cite{moustakas_design_2021} to sustain a hit-rate beyond $600$~MHz~cm$^{-2}$, covering all the currently expected use-cases in \belletwo\ operation\footnote{Additional simulations and test with the fabricated OBELIX-1 sensor will be conducted to estimate the hit-rate limit beyond which the hit detection efficiency for subsequent triggers might be affected and at which level.}. The sensor peripheral logic is designed to transmit data to the outside world up to a hit-rate of 120~MHz~cm$^{-2}$, as explained in the TRU module description.\\ 
% Following sentence removed since the mechanism is finally not implemented, The excess rate is absorbed by the TRU
% When the hit rate is too large, for instance during beam injection, it is possible to briefly decouple the matrix from the next stage in order to avoid any bottleneck once the data transmission is allowed again.\\

\paragraph*{Trigger logic (TRU, TXU, TTT)}

Two entirely new modules are related to the \belletwo\ trigger. First the trigger logic unit or TRU has the key task to select among the fired pixel information coming from the matrix, those related in-time with the triggers sent by the \belletwo\ system. A second unit, track trigger transmitter (TTT), is in charge to transmit to the trigger system useful information to build a trigger decision based on tracks. \\

\begin{figure}
\centering
\includegraphics[width=0.85\textwidth]{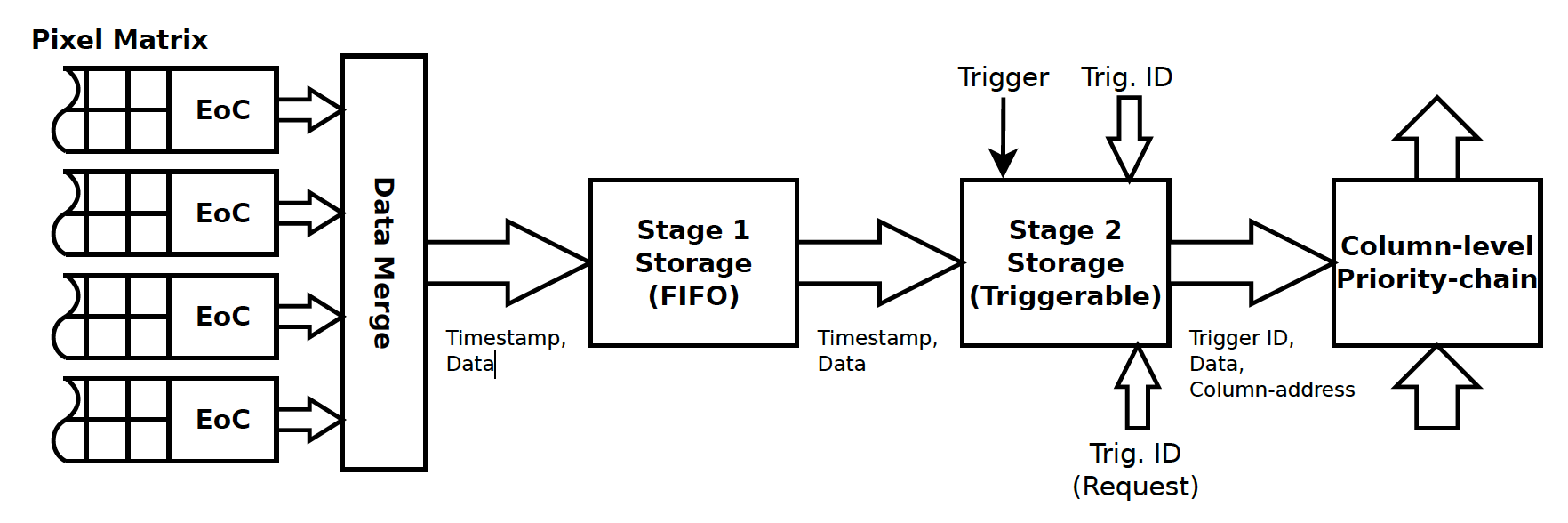}
\caption{Schematic illustrating the principle of operation of the trigger logic unit (TRU).}
\label{fig:vxd-obelix-tru}
\end{figure}

The TRU module handles the incoming data from the pixel matrix with a two stages of memory as depicted in figure \ref{fig:vxd-obelix-tru}. The first stage memory simply buffers the pixel information, during the trigger latency. In the second stage memory, the fired pixel BCIDs are compared to each trigger time present in a dedicated global memory keeping track of received triggers. When matched, the pixel information is passed to the transmission unit (TXU). The size of both TRU memories have been optimized in order to ensure full efficiency at the maximum hit-rate of 120~MHz~cm$^{-2}$, taking into account the $30$~kHz trigger rate and $10\mus$~latency and to minimize the power dissipated. It was verified that the memory size can also absorb the spikes of excess hit rate at 600~MHz~cm$^{-2}$.\\
The matching criteria associates to a single trigger fired pixels with a BCID overlapping with the trigger time and also with the next BCID. Since the BCID precision is $50$~ns, the duration of the integration window of OBELIX is then $100$~ns. This strategy accounts for the possibility that a physics hit associated to a trigger is timestamped with a later BCID due for instance to timewalk effect.\\

The TTT module profits from the presence of a fast signal for each column of the matrix, corresponding to the OR of all pixel outputs along the column. Trading the spatial granularity for speed, the TTT divides the matrix into 8 (still to be confirmed) regions and produces a one byte word indicating which of these regions is fired. The light load represented by a single byte (or even two if the granularity needs to be extended) allows to transmit this information within typically $100$~ns on a line parallel to the main data output of the sensor. The ensemble of such lines can then be input to the trigger system for elaborating a decision based on the vertex information, of course with coarse position precision.\\
This functionality is rather exploratory in its current state and may only be used on the oVTX layers due to power budget constraints. It still requires further simulations in correlation with the trigger and full VTX system behaviours, which might lead to further optimisation.\\

\paragraph*{Control (CRU, synchro)}

The control of the sensor is implemented in the control unit (CRU) exploiting the protocol developed for the RD53 ASICs \cite{garcia-sciveres_rd53b_2019}, as in the case of TJ-Monopix2. It allows receiving the configuration instructions (i.e. list of masked pixels, threshold settings, etc.) and the in-coming trigger information as well as sending out the data prepared by the TXU module.\\

The anticipated data throughput per sensor is estimated to not exceed $100$~Mbps in the innermost layer, allowing to use a single physical output line clocked at about $160$~MHz. \\

The OBELIX sensor is configured by setting a number of registers, which define the sensor operation and necessary biases. The protocol is fast enough to allow for reconfiguration of the sensor on a regular basis mitigating potential mis-configurations from single radiation events.\\

Complementing the control unit, the synchronisation unit uses the SuperKEKB radio-frequency clock ($509$~MHz), which is provided to one of the sensor input pads, to derive all other needed clocks. The most important ones are the $21.2$~MHz clock for the BCID (meaning that the timestamp precision is actually $47$~ns) and the $169.7$~MHz clock for the data output (allowing a 340 Mbps throughput).\\

%\underline{\bf Monitoring}
%\rem{Quote rough spec and goals for the ADC, present the architecture being implemented by Jose}

\paragraph*{Powering and dissipation estimate}

To simplify the power distribution along a single ladder, which may contain up to 24 sensors (see table~\ref{table:vxd-parameters}), a low dropout (LDO) voltage regulator is included in each sensor. As depicted in the left part of figure~\ref{fig:vxd-obelix-layout}, one pad-ring and one associated LDO-regulator dedicated to powering the chip is located on each side of the pixel matrix. That arrangement aims to insure uniform bias supplies throughout the matrix.\\

\begin{figure}
\centering
\includegraphics[width=0.9\textwidth]{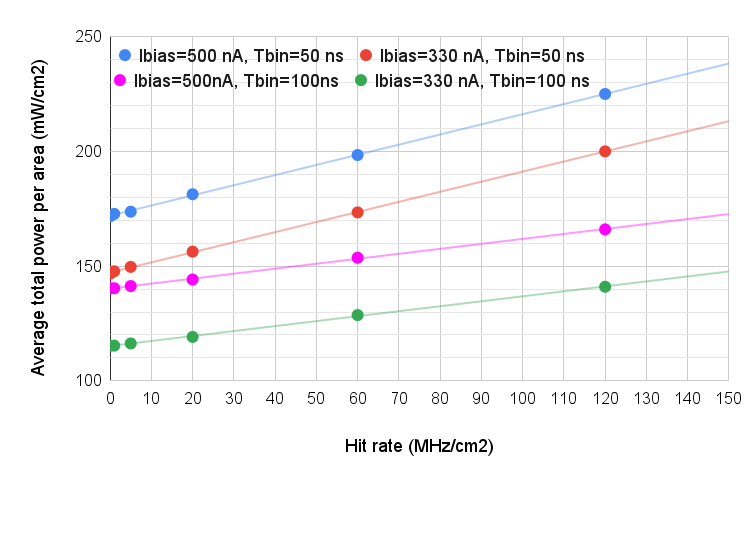}
\caption{Estimated power of the OBELIX sensor depending on the main front end current (Ibias), timestamping precision (tbin) and hit-rate.}
\label{fig:vxd-obelix-power}
\end{figure}

The existing description of the various units composing the sensor allows to estimate the power dissipation with the tools provided by the electronic design automation software (EDA). Three main parameters drive the chip current consumption; the current flowing into the front-end, the BCID clock frequency, which drives the timestamping precision and the hit rate. The estimated dependence is showed by figure \ref{fig:vxd-obelix-power}. This evaluation comes with a 20\% uncertainty, which reflects the current status of the sensor design.\\ 

It appears that the power budget of $200$~mW~cm$^{-2}$ is exceeded by 10~\% at the maximum average hit-rate of $120$~MHz~cm$^{-2}$, corresponding to the most pessimistic scenario, and for the best timestamping precision of $50 \ns$. The power can be recovered within budget by compromising the timing precision either by worsening the time binning to $100 \ns$ or with lower biasing current to the in-pixel amplifier, which degrades the time walk figure.\\

\paragraph*{ Monitoring}

The temperature and supplied voltages within the sensor are monitored through dedicated thermal probes and an analogue to digital converter featuring 12 bits (10 bits resolution targeted) and a sampling rate of about 1.7~kHz.\\

\paragraph*{OBELIX schedule}

The first version of the sensor, OBELIX-1, is currently being designed by a team composed of six full time designers (about ten people spread over CPPM-Marseille, Dortmund University, HEPHY-Vienna, IFIC-Valencia, IPHC-Strasbourg and SiLab-Bonn) under the coordination of the IPHC-C4Pi\footnote{c4Pi is a facility with 22 engineers devoted to the development of CMOS pixel sensors.}, which takes in charge the ASIC overall integration and submission. The OBELIX-1 chip will be submitted for fabrication in the third quarter of 2024. It is expected for such a complex ASIC that a single submission is not enough to reach the sensor quality required for building an experiment. Consequently a second improved version, OBELIX-2, will become the production version.\\

OBELIX-1 is expected to return from fabrication in early 2025 and the evaluation of its main performances will require about 6 months. OBELIX-1 featuring the already final size of the VTX sensor, the chip will also be used to prototype final detection modules and layers.\\ 

While the design of OBELIX-2 will start in 2025, its submission can only occur after the characterisation of OBELIX-1, hence at the earliest after Q3-2025. 

%%%%%%%%%%%%%%%%%%%%%%%%%%%%%%%%%%%
\subsection{VTX structure}
\label{sec:vxd-structure}
The VTX proposal consists of five barrel-shaped layers, equipped with the same sensor. The design of the ladders  differs due to their active length, varying from 12 to 70 cm to cover the required 
acceptance ($17^{\circ} <\theta<150^{\circ}$). 
The two innermost layers (iVTX) follow an all-silicon ladder concept, the outermost (oVTX) layers 
require carbon fiber support structures and flex print cables.
The VTX targets an overall material budget of about 2.0\%~$X_0$ and will be the first light and compact MAPS-based vertex-tracker operating at a high-rate $e^+e^-$ collider.\\  
Table~\ref{table:vxd-parameters} details the main structural parameters of the VTX detector. As discussed in the physics performance section~\ref{subsec:pp-KS_eff}, the two options currently under consideration for the radius of the middle layer appear explicitly.

One additional option is being considered, although not explicitly described in this document, to replace the entire VXD with CMOS sensors keeping current machine-detector boundaries as backup option should fatal accidents happen.

\begin{figure}[htb]
\begin{subfigure}[t]{0.59\textwidth}
\centering
\includegraphics[width=10cm]{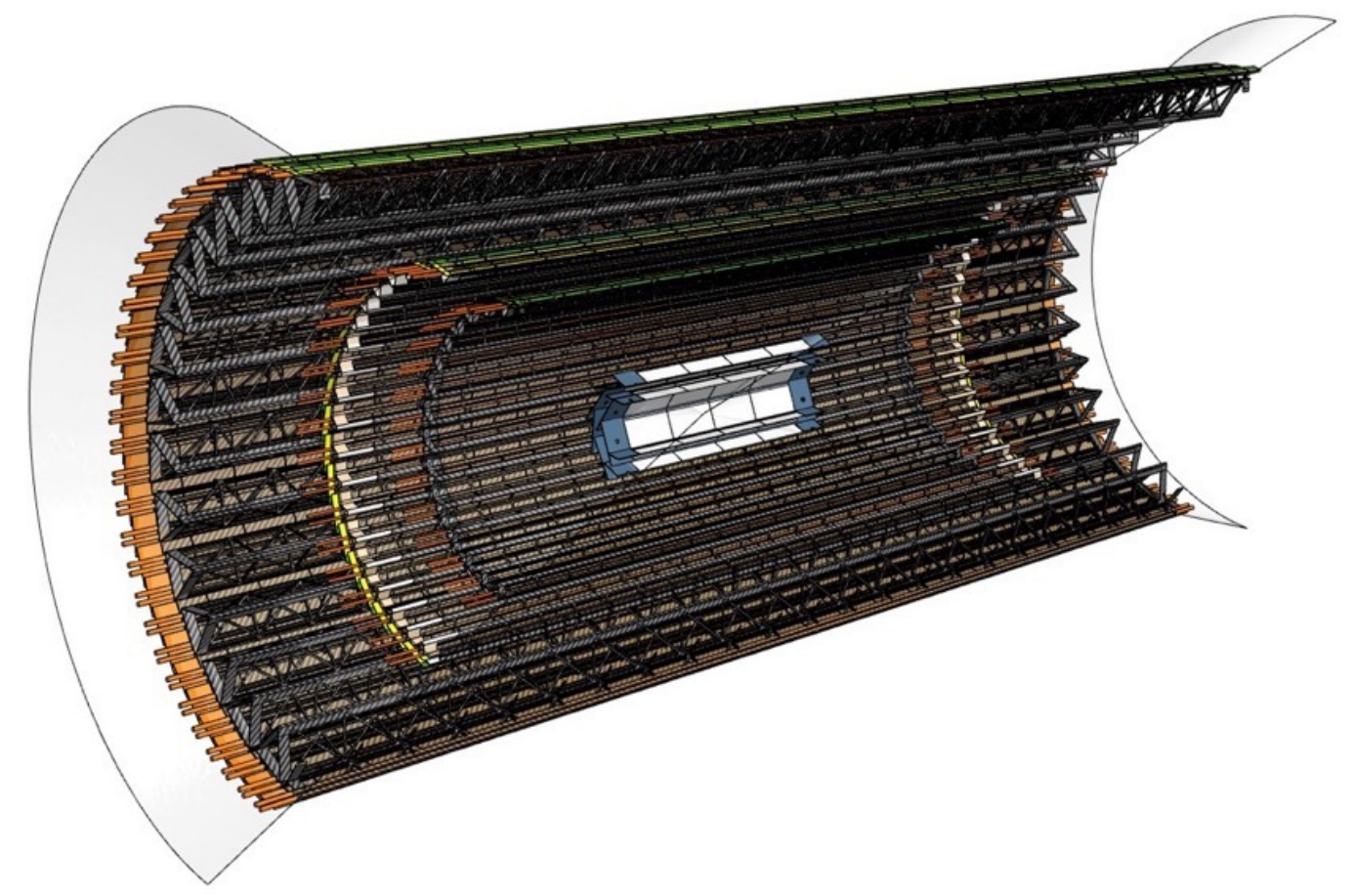}
\caption{Cut 3D view.}
\end{subfigure}
\hfill
\begin{subfigure}[t]{0.39\textwidth}
\centering
\includegraphics[width=0.99\textwidth]{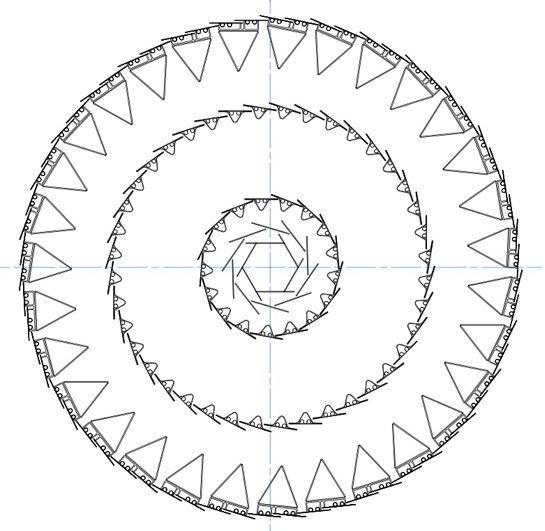}
\caption{End view.}
\end{subfigure}
\caption{Two different views of the VTX detector layout for version 3a. The five layers are represented with their support structure, their radii are: \SI{14}, \SI{22}, \SI{39}, \SI{89}, \SI{140}{\milli\meter}.}
\label{figure:VTX_cut_view}
\end{figure}

\begin{table}
\centering
\begin{tabular}{l|c|c|c|c|c||rc|c||r}
\hline
Layer &     1   &   2   &   3a   &   4   &   5   &   \bf{Total} & \hspace*{5mm} &     3b &   \bf{Total}\\
\hline  
\hline
Radius (mm) & $14.1$ & $22.1$ & $39.1$ & $89.5$ & $140.0$ &  & & \bf{$69.1$} &  \\   
\hline
\# Ladders    & $6$ & $10$ & $17$ & $40$ & $31$ & \bf{104} & & $30$ & \bf{117}  \\
\hline
\# Sensors/ladder & $4$ & $4$ & $7$ & $16$ & $2\times 24$ & \bf{2311}      & & $12$ & \bf{2552} \\ 
\hline
Mat. budget (\% $X_0$) & $0.2$ & $0.2$ & $0.3$ & $0.5$ & $0.8$ & $2.0$      & & $0.4$ & $\mathbf{2.1} $\\   
\hline
\end{tabular}
\caption{The VTX detector main parameters with the two geometrical options (3a and 3b) for the radius of the middle layer.}
\label{table:vxd-parameters}
\end{table}

%%%%%%%%%%%%%%%

\subsubsection{Ladder concepts}
\begin{figure}[htb]
\centering
\includegraphics[width=14cm]{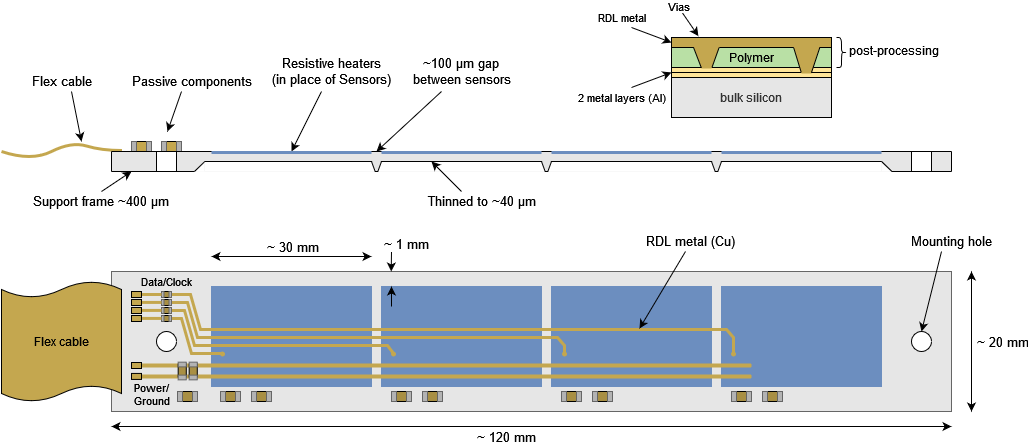}
\caption{Schematic view of the iVTX ladder design.}
\label{figure:iVTX}
\end{figure}

%4) From chip to ladder/stave: Large area objects, power distribution, system integration.
\label{sec:vxd-ladders}
The two innermost layers of iVTX are placed at radii 14 and 22 mm respectively. They target an individual material budget of about $0.2$\%~$X_0$ per layer. Such light layer is possible due to the moderate overall surface of these two layers, below 400~cm$^2$, the low sensor power dissipation and the few connections needed for the sensor operations. Such conditions make air cooling a viable solution. The monolithic structure is essentially made of four contiguous OBELIX sensor blocks diced out of the production wafers and thinned to 50~$\mu$m except in some border areas, that are a few hundred $\mu$m thick to insure mechanical stability. \\
In more detail, a post-processing step etches additional metal strips on a so called redistribution layer (RDL) to interconnect sensors along the ladder and to provide a unique connector at the ladder backward end. 
The RDL will be produced in a thin-film process with photo lithographic structuring of polymer dielectric and metal layers. Vias connect the sensor bond pads to the metal traces, which route power and data via impedance-controlled transmission lines to the end of the ladder, where a flex cable is connected. Exposed pads are created on top of the RDL to mount surface mounted components (SMD) like bypass capacitors.
After the RDL has been processed, the backside of the ladder will be thinned selectively. A thickness of 40-100 $\mu$m is foreseen underneath the active sensor area, the outer perimeter of the ladder remains unthinned for mechanical rigidity. Mounting holes will be added via laser-cutting.
The first ladder demonstrator is currently under production (see figure~\ref{figure:iVTX}), to explore the feasibility of the two techniques, a large size signal RDL on top of the sensors and the selective backside thinning of the ladder.
The demonstrator is based on a blank wafer with dummy heater structures in place of the sensors, allowing to characterize the electrical, mechanical and thermal performance of the ladder design. A working plan, increasing complexity in several steps, is being developed together with IZM (Berlin).

\begin{figure}[htb]
\centering
\includegraphics[width=14cm]{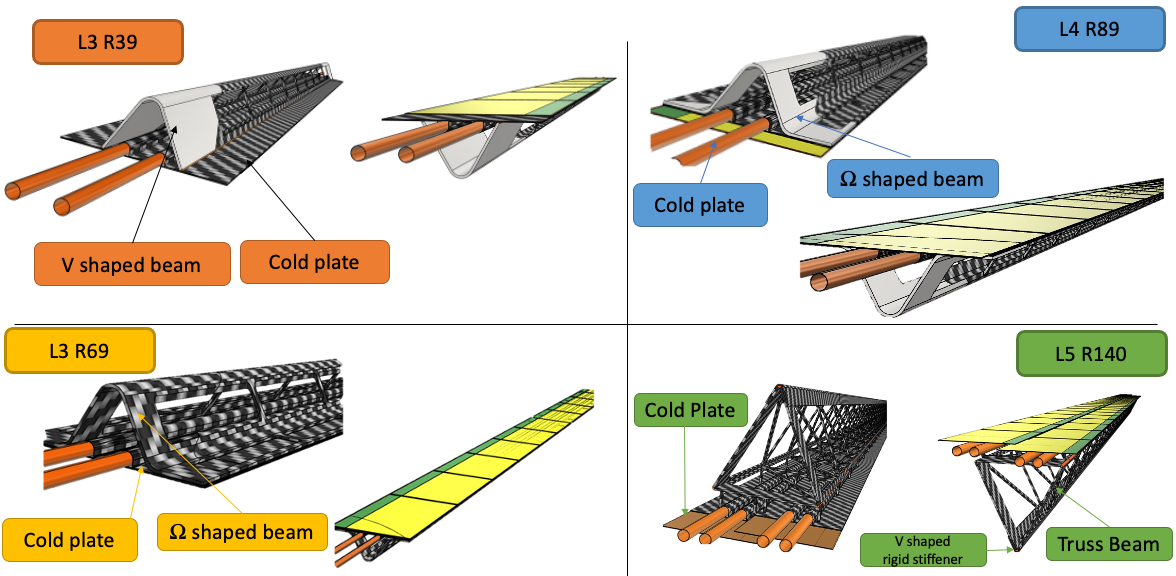}
\caption{Schematic view of the oVTX ladders: for Layer 3 there are the two solutions at radius 39 mm and 69 mm. 
Layer 4 and Layer 5 are placed at radii 89 and 140 mm respectively. The green region on the chip corresponds to the 
periphery, i.e. not sensitive, while the yellow region indicates the sensitive area.}
\label{figure:oVTX}
\end{figure}

A more traditional approach drives the oVTX ladder concept for the outer layers, strongly inspired by the design successfully developed for the ALICE ITS2~\cite{fantoni_upgrade_2020}. Each ladder (see figure \ref{figure:oVTX}) is made of a light carbon fiber support structure, a cold plate including tubes for liquid coolant circulation by negative pressure, a row of sensors glued on the cold plate and finally two flex print cables (see figure~\ref{figure:L5_flex}) connecting each half  ladder to a connector.  For Layer 3 a single flex exiting on the backward side of the detector can be foreseen in order to allow more room for machine components on the forward. Depending on their radius, the material budget of individual oVTX ladders ranges from 0.3 to 0.8\%~$X_0$, matching the requirements set in section~\ref{sec:vxd-baseline}.
%matching the values reported in the table~\ref{VTX:thickness_simulation}.  (I am assuming that in the performance studies there is a table reporting the thickness in \% $X_0$ of the 5 layers. But I couldn't find that.)
% Jerome: indeed there is none, so I changed it to a ref to the initial VXD section 
\\

\begin{figure}[htb]
\centering
\includegraphics[width=14cm]{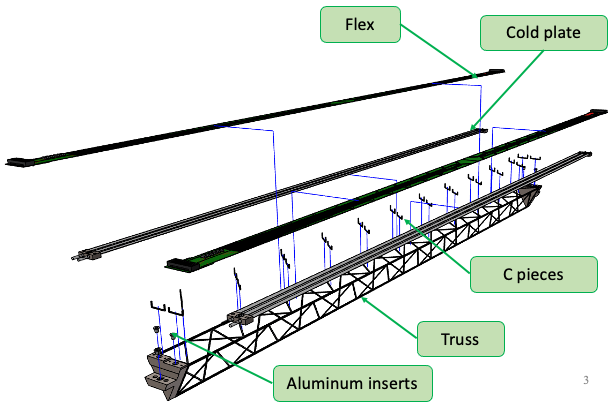}
\caption{Exploded view of the Layer 5 ladder.}
\label{figure:oVTX_L5}
\end{figure}

A drawing of the parts composing the Layer 5 ladder is reported in figure~\ref{figure:oVTX_L5}. \\
The $r-\phi$ overlap between adjacent ladders is ensured to be at least 10 pixel pitches for all the ladders, 
with a clearance between the ladders in radial direction of the order of 1 mm. \\

%%%%%%%%%%%%%%%
\subsubsection{Thermomechanics, services and data acquisition }
%5) Thermomechanics. Services (power, data, cooling). 
%6) Data acquisition. Interface to Belle II DAQ.
\label{sec:vxd-services}
A specific power density baseline of 200~mW/cm$^2$ is assumed in the thermo-mechanical studies. For iVTX the power could be anyway reduced by operating the front-end at a lower current and/or lower frequency, in case the extreme hit rate of 120~MHz/cm$^2$ is reached in the innermost layer, while power dissipation could be even lower for oVTX, due to the significant lower hit rate in the external layers.
 
The proposed VTX detector is expected to be operated at room temperature. The sensor temperature will be monitored both internally, see section~\ref{sec:vxd-obelix-implementation}, and with additional thermal sensors as implemented in the current VXD~\cite{adamczyk_design_2022}. The usage of the same sensor all over the detector allows a unique control and power system.  An important reduction of services with respect to the current VXD  is expected, due to smaller cross sections of the data cables with the extensive use of optical fibers, and a less complex cooling (i.e. single phase) system.  Consequently, simpler overall mechanical support, cabling and acquisition system are required. In particular, the standard PCIe40~\cite{daq:pcie40} acquisition boards used in Belle~II match well the data throughput requirement.

The iVTX full-silicon concept is currently being assessed with industrial partners, first using dummy silicon wafers and in future with real sensors. 
Thermal studies, considering power is uniformly dissipated on the sensor surface, 
demonstrated that air at ${15^{\circ}\rm C}$ flowing at a speed of $10\; {\rm m/s}$ is effective 
to cool a single inner module. The maximum temperature reaches 
about ${20^{\circ}\rm C}$. \\
At system level, the bottleneck is the space to pass the tubes to bring the air to the IP region: a rough estimate shows that an equivalent cross section of 6 tubes of 10 mm in diameter is needed to evacuate the ~65 W of heat source of the inner layers. 
An experimental verification of such studies on a dedicated thermal test bench is foreseen in order to
produce a mechanical design compatible with the new interaction region.  

Prototypes of key elements of the oVTX ladder, the longest carbon fiber support structure and flex cable have been designed and fabricated.  For Layer 4 and Layer 5 services (electrical connections and cooling) 
can be provided both on forward and backward sides, naturally dividing the ladders in two electrical units. To limit the space required on the forward side, giving more room for the accelerator components, for the Layer 3 ladder also the solution with the only access on backward has been investigated. Traditional flex circuits can be used to distribute power and for data output for the oVTX.
A multiline power bus has been designed and realized first on four-layer process with copper in order to have each OBELIX chip along the ladder powered by a dedicated positive voltage (VDD) and ground (GND) line pair. Trace widths are trimmed to fulfill the same maximum voltage drop requirement (200 mV) for all the chips. The electrical tests on signal integrity performed on the prototypes demonstrated that the bandwidth is large enough to allow more than the needed 160 Mbps baseline data throughput.\\

\begin{figure}[htb]
\centering
\includegraphics[width=8cm]{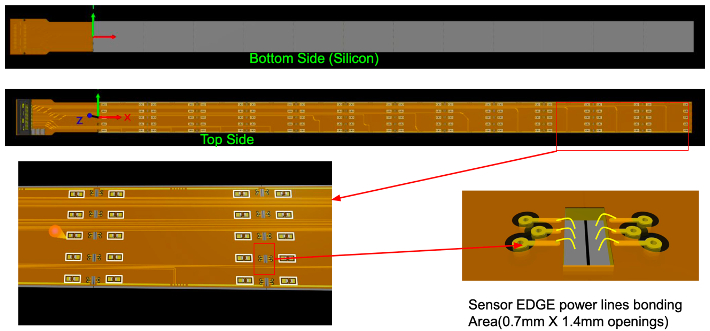}
\caption{Drawing of the prototype Al flex for the layer 5 half-ladders.}
\label{figure:L5_flex}
\end{figure}

For the flexible printed circuit in aluminum, the design (see figure~\ref{figure:L5_flex}) is following the floorplan of the 
chip, to allow, after the characterization, also a final test by eventually connecting the final chip. 

A full scale Layer 5 prototype, with kapton heaters mimicking the power dissipation from the chips, has been assembled and characterized on the thermofluidynamic bench.  
We set the coolant (demineralized water) temperature at $10^{\circ}C$ at the inlet, the enviromnent at $20^{\circ}C$ and a negative pressure of $0.2 \;{\rm bar}$. Three flow configurations (monodirectional, bi-directional and with an U-turn at one end) have been tested and the results have shown an average temperature of $24^{\circ}C$ with a maximum gradient $\Delta T\approx 4 ^{\circ}C$ along the full length of the ladder, matching the specifications required to safely operate the chip.\\   

Results of the FEA studies on the longest Layer 5 ladder have been confirmed by mechanical measurements on the sagitta distorsion. 
Also vibrational FEM analysis matched the experimental tests on the truss, showing the first resonance frequency of the structure at $200 \;{\rm Hz}$, safely far from the typical earthquake frequency.
All these tests allowed a validation of design of the longest (i.e. most challenging) Layer 5 ladder. 
%as a result of the R\&D conducted under general assumptions.

%%%%%%%%%%%%%%%%%%%%%%%%%%%%%%%%%%%
\subsection{Roadmap towards realization}
\label{sec:vxd-schedule}

The previous sections \ref{sec:vxd-obelix} and \ref{sec:vxd-structure} described the intensive R\&D program to establish the technical solutions for the sensor and the detection element structures (electric, mechanical and thermal). Both are expected to converge at the end of 2024. Additional developments for the systems to control and acquire the data will start and advance in 2024 in order to identify the main solutions, but not to complete the entire development which may be finalised during the detector production phase .\\

It is hence expected that the VTX critical R\&D phase will close at the end of 2024, allowing to enter a production and construction phase should the project be accepted.\\

The production of all the VTX parts is foreseen to last for slightly more than two years. This estimation stems from the following assumptions for the needed part quantities. All parts are produced in quantities including roughly 20\% spares. A sixth layers might be added which would typically correspond to retain both options 3a and 3b from table~\ref{table:vxd-parameters}. the VTX geometry requires then the assembly of 20~iVTX ladders and 140~oVTX ladders, which correspond to about 3000 sensors. The ASIC fabrication yield is taken as 50~\% and 30 dies are expected per wafer, hence the total number wafers to produce is  195.\\
It is also considered that among the 9 Belle~II groups currently involved in the development of the VTX design (Austria: Vienna; France: Marseille, Orsay, Strasbourg; Germany: Bonn, Göttingen; Italy: Pisa; Japan: KEK, Spain: Valencia) and a few additional groups that may join, about 8 laboratories will join production activities.
Finally we elaborate the schedule assuming 4 work days per week and 3 work weeks per month.
\begin{itemize}
    \item 6 month for the wafer production, followed by 6 months for the probe testing (the integration of sensors onto modules can obviously start before the full probing step is finished). The probing occur a minimum of 2 sites, with 1 wafer tested per day.
    \item 10 months for the iVTX ladder production, assuming 2 ladders per month at a single production site (IZM, Berlin).
    \item 24 months for the oVTX ladder production, assuming 2 ladders per month at 3 sites. It has to be noted that a potential intermediate step will most probably needed, i.e. a ladder being composed of modules. In this case, 2 additional sites will be needed for the module production.
    \item 12 month for the mounting of the ladders onto partial detector structures, like semi-cylindrical shells representing half-layers, which are then assembled around the beam-pipe outside the \belletwo\ detector. This operation might be done entirely in Japan at KEK or partially in one laboratory in Europe.
\end{itemize}
iVTX and oVTX production can be conducted in parallel and start mid-way (after 3 months) of the wafer testing phase. The shell production can begin after 6 months from the oVTX production start. This very notional plan leads to a total production length of: 
6~(wafer prod) +~3~(delay oVTX prod) +~6~(delay full assembly) +~12~(full assembly) = 27~months.\\

From the point where the VTX is attached to the beam pipe, an additional 6 to 12 month period is needed for installing and commissioning the detector in the \BTWO\ experiment.\\

Under this preliminary schedule, the VTX can be ready to take data between 2.5 to 3 years after the final sensor (OBELIX-2) is submitted to fabrication and without contingency.

%%%%%%%%%%%%%%%%%%%%%%%%%%%%%%%%%%%
\subsection{VTX cost estimate}
\label{sec:vxd-cost}

The VTX cost is estimated from various sources, first the experience of the previous ALICE-ITS2 detector cost, then the part prices actually found during the R\&D phase (typically for the cost of the sensors, the ladder structure and flex cables). Cross-checking the estimates obtained from these two sources, the cost breakdown is provided in 
table~\ref{tab:vxd-vtxcost}.\\

\begin{table}[!ht]
\centering
    \caption{Breakdown of the current cost estimate of the VTX in kEUR without any contingency.}
    \label{tab:vxd-vtxcost}
    \begin{tabular}{l|cc|c}
    \hline
    \textbf{Component}  & \textbf{Development} & \textbf{Production} & \textbf{Total (kEUR)} \\
    \hline
    \hline
    \textbf{Sensors} & 380 & 1130 & \textbf{1510} \\
    \hline
     \textbf{Ladders}  & 120 & 1400 & \textbf{1520} \\
    \hline
     \textbf{Assembly}  & 130 & 630 & \textbf{760} \\
    \hline
     \textbf{DAQ \& services}  & 280 & 1060 & \textbf{1340} \\
    \hline
     \textbf{Installation}  & - & 100 & \textbf{100} \\
    \hline
    \hline
     \textbf{Total} &
     \textbf{910} & \textbf{4320} & \textbf{5230} \\
    \hline
    \end{tabular}
\end{table}

Considering the development stage of the project and the possibility of large price fluctuations from the semiconductor industry, it is reasonable to apply a rough 50~\% contingency on the production cost and estimate that an envelope of 7800~kEUR will cover the project finalisation.

\subsection{R\&D and technology options}
\subsubsection{DuTiP sensor}
\label{sec:SOI}

One of the promising technology options is a monolithic sensor using Silicon-On-Insulator (SOI) technology. The sensor concept optimized to Belle II upgrade, called DuTip (Dual Timer Pixel), was invented~\cite{Ishikawa:2020gpo} based on PIXOR striplet detector~\cite{Ono:2013rka}. 
The DuTiP prototypes are fabricated using the SOI technology.

\paragraph*{General concept of DuTiP}
Under large track density environments, a global shutter readout mode based on level~1 trigger would be adopted to realize a fast readout with low occupancy allowing for small data transfer rates. 
Hit information is to be stored in the pixel detector during the trigger latency. Even if a second hit arrives at the same pixel during this period, this information would be also kept. To satisfy these requirements, DuTiP concept was invented. 

\begin{figure}[htbp]
\begin{center}
\includegraphics[width=1.0\textwidth]{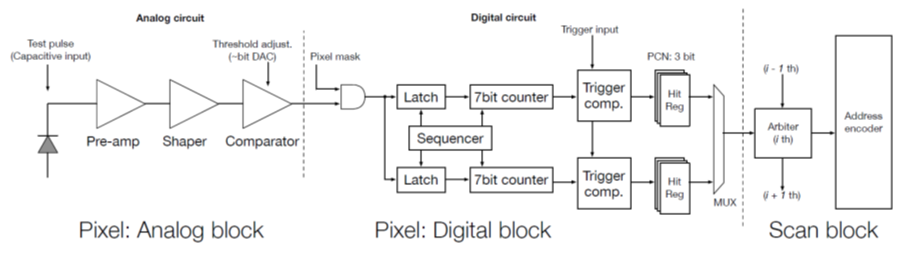}
\end{center}
\caption{Analog, Digital and Scan blocks for DuTiP detector. The analog and digital circuits are on a pixel while the Scan block is on a periphery.}
\label{fig:DuTiP}
\end{figure}

Fig.~\ref{fig:DuTiP} shows the block diagram of the DuTiP. 
The analog part consists of a pre-amplifier, shaper, and comparator which are the usual configuration for a binary detector. 
The digital circuit is the most important part of the DuTiP concept. 
When the binary hit signal is sent to the digital part, one of the timers starts counting down. 
The starting time is set as trigger latency plus one clock cycle. 
If the trigger signal is received when the time is 1~(2/0), the signal is readout as current (next/previous) timing (Previous/Current/Next timings are denoted as PCN timings). 
If trigger signal is not received at the PCN timings in the pixel, the timer is reset. 
To take into account the second hit in a pixel during trigger latency, a sequencer and two timers are equipped. 

The concept does not specify the technology however the complicated digital circuit should be fabricated on each pixel, thus an SOI technology is chosen as a baseline for the development of DuTiP sensor. 

\begin{figure}[htbp]
\begin{center}
\includegraphics[width=0.46\linewidth]{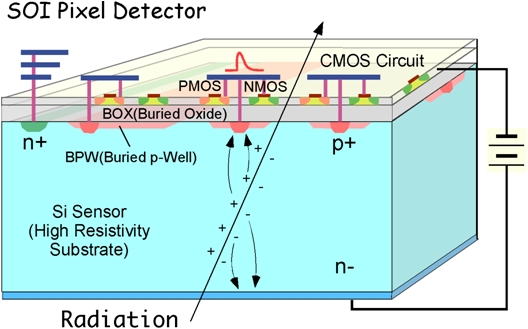}
\includegraphics[width=0.46\linewidth]{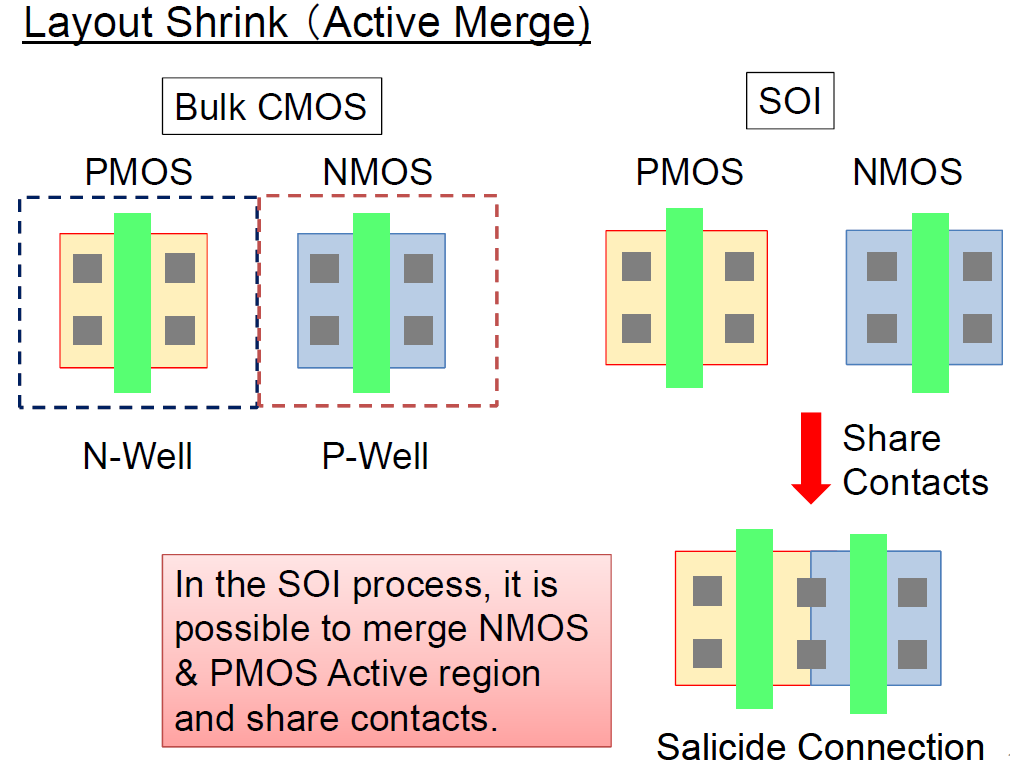}
\end{center}
\caption{(Left) Schematic view of SOI. The circuit silicon layer~(light yellow) is on the insulator SiO$^2$ layer~(grey). The sensor silicon layer~(sky blue layer) is electrically connected with the circuit layer with through-silicon vias. (Right) Active merge technique to merge PMOS and NMOS.}
\label{fig:SOI}
\end{figure}

\paragraph*{SOI Technology}
SOI technology has the potential to build thin and low power monolithic sensors for its use in pixel vertex detectors in particle physics experiments.
Figure~\ref{fig:SOI} (Left) shows a schematic view of the SOI detector. 
We choose the Lapis semiconductor 0.2~$\mu$m FD-SOI CMOS technology for the development since many pixel detectors are already developed, for example, SOFIST for the ILC~\cite{Ono:2019yrh}, and pixel detectors for CEPC~\cite{Wu:2019nqx} and CLIC~\cite{Bugiel:2018ckn, Bugiel:2021nag}. 

To fabricate a complicated circuit, reduction of the circuit area is important even for the SOI. 
Since active regions of NMOS and PMOS can be merged and the contacts can be shared with the SOI~\cite{Arai}, the circuit area can be smaller than bulk CMOS with the same process rule (Fig.~\ref{fig:SOI} (Right)).
While the SOI implementation is tolerant against neutrons and single event effects, the most important concern is related to surface damage. Charged particles create holes in the SiO$_2$ insulator layer which causes the backgate effect. 
Nevertheless, this had been solved incorporating a double SOI structure and test structures have shown radiation tolerance up to 1~MGy~\cite{Hara:2019dfo}.

\begin{figure}[htbp]
\begin{center}
\includegraphics[width=0.48\linewidth]{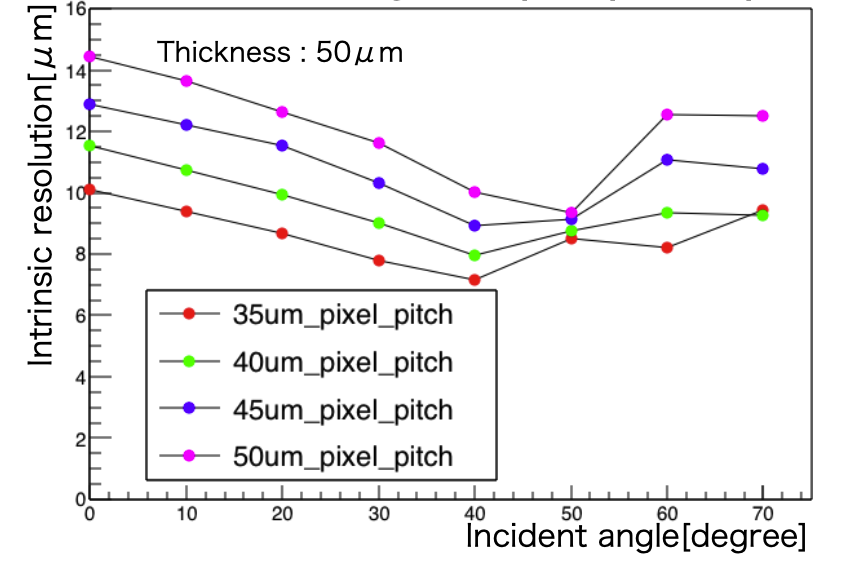}
\includegraphics[width=0.48\linewidth]{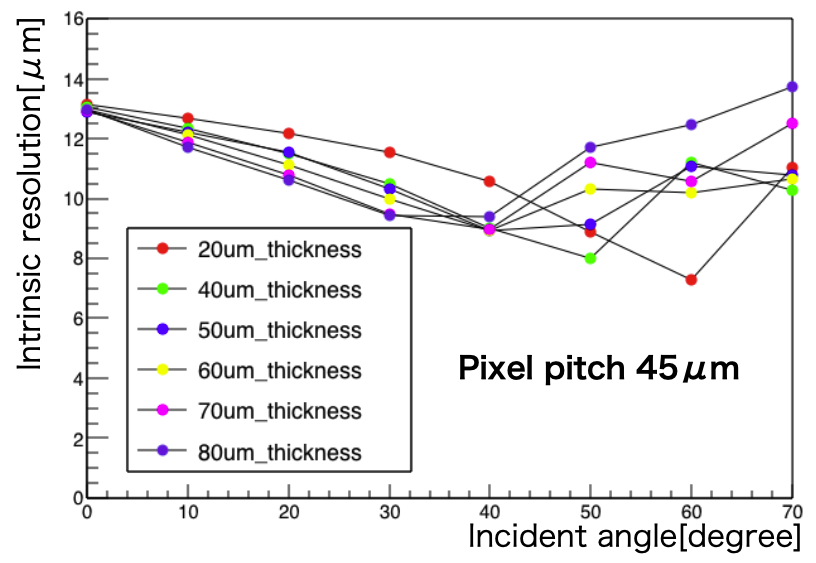}
\end{center}
\caption{Intrinsic resolutions as a function of incident angle with different (left) pitch and (right) thickness using GEANT MC.}
\label{fig:IR}
\end{figure}

\paragraph*{DuTiP sensor technical specifications}
Table~\ref{table:DuTiPSpec} shows DuTiP sensor specification. To achieve an intrinsic resolution better than 15~$\mum$, pixel pitch is chosen to be 45~$\mum$~(Figure~\ref{fig:IR}). 
The sensor active thickness define the amount of generated charge carriers in the bulk.
We choose 50~$\mum$ as a baseline however up to 80~$\mum$ meets the requirement for intrinsic resolution.
The clock speed is 62.9~ns which is determined by SuperKEKB clock 509~MHz divided by 32. 
For analog circuits, ALPIDE analog circuit with some modifications for optimizing to the SOI is adopted~\cite{Suljic:2016bmm}. To achieve faster shaping time, the supply current is increased to 200~nA which gives about 500~ns pulse width for the output signal. The equivalent noise charge~(ENC) is expected to be less than 100 electrons. 

The DuTiP sensor is equipped with a 300~MHz LVDS interface which has a data transfer rate of 600~Mbps. 
Since DuTiP sensor transfers only hit information synchronized with trigger timing, the data rate is expected to be low even for the innermost layer.
The sensor size is determined by the reticle size for the SOI process.
The Lapis Semiconductor SOI process has a successfully demonstrated stitching technique. 
Three sensors for X-ray imaging were combined into single chip with the stitching technique.
DuTiP can also adopt the same technique to produce larger chips which should be more mechanically stable when making ladders.

\begin{table}
\centering
\caption{DuTiP sensor specifications.}
\begin{tabular}{lc}
\hline
\hline
 & DuTiP L Sensor (S Sensor)\\
\hline
Pixel pitch & $45\mum$ \\
Sensitive layer thickness & $50 \sim 80 \mum$ \\
Sensor thickness & $< 100$\mum \\
\hline
Clock speed & $62.9$~ns\\
Trigger latency & $8 \mus$ \\
Trigger rate & $<30$~kHz  \\
Time precision & $<16$~ns  \\
\hline
data transfer rate & $600$~Mbps \\
Pixel array size & $21.6 \times 28.8 \mma$ ($14.4 \times 28.8 \mma$) \\
Sensor size      & $24.4 \times 29.6 \mma$ ($17.2 \times 29.6 \mma$)\\
\hline
\hline
\end{tabular}
\label{table:DuTiPSpec}
%\caption{bandwidth with average hit rate and max hit rate}
\end{table}

\begin{figure}[htbp]
\begin{center}
\includegraphics[width=0.40\linewidth]{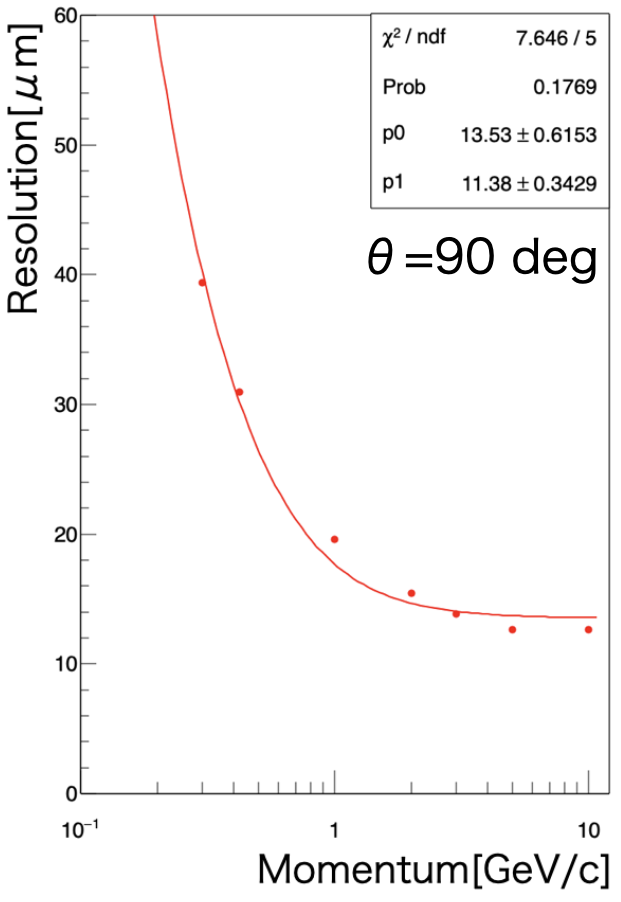}
\includegraphics[width=0.40\linewidth]{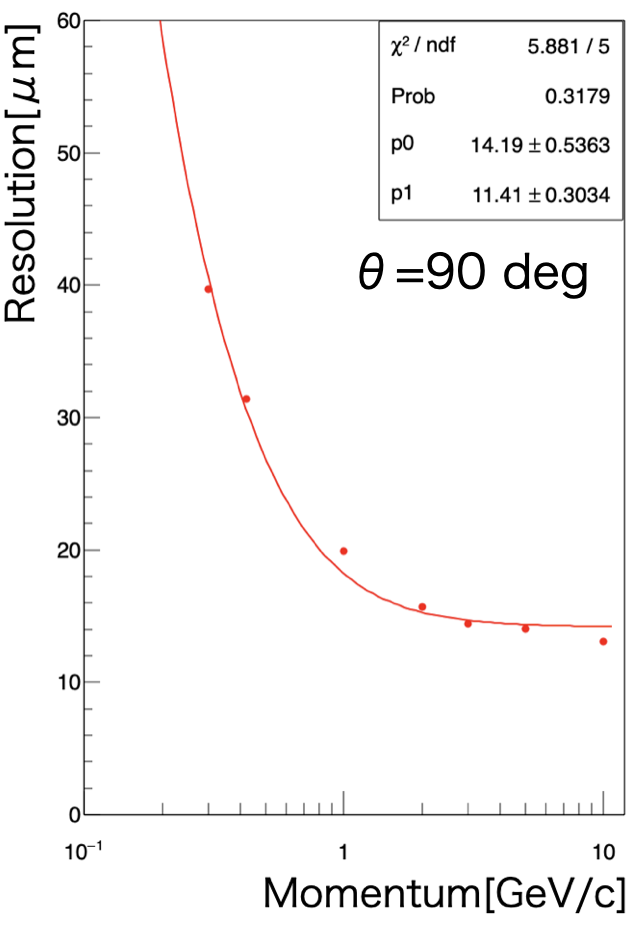}
\end{center}
\caption{Impact parameter resolutions for (left) $r-\phi$ and (right) Z directions.}
\label{fig:IPres}
\end{figure}

\paragraph*{DuTiP performance with GEANT MC}
The basic performance of the DuTiP sensor is evaluated with simple GEANT MC.
The PXD is replaced by two layers of DuTiP sensors, while keeping the beam pipe and SVD untouched.
The impact parameter resolution is studied using pions with different momentum~(Figure~\ref{fig:IPres}). 
The fit results for incident angle of 90 degree are $\sigma_{r-\phi}(\mu m) = 13.5 \oplus 11.4/p(\rm{GeV})\sin^{3/2}\theta$ and $\sigma_{Z}(\mu m) = 14.2 \oplus 11.4/p(\rm{GeV})\sin^{5/2}\theta$ which are enough good to measure the $B$ meson decay vertices.

\begin{figure}[htbp]
\begin{center}
\includegraphics[width=0.24\linewidth]{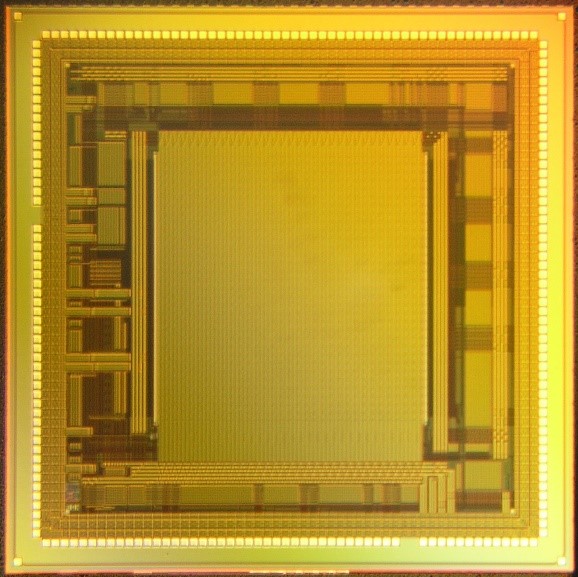}
\includegraphics[width=0.72\linewidth]{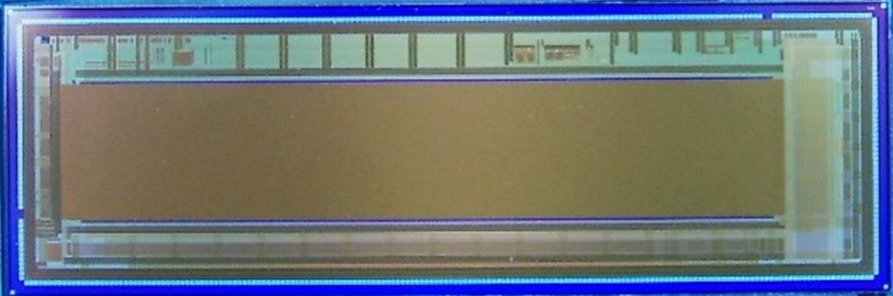}
\end{center}
\caption{(Left)~Picture of the DuTiP1~(6$\times$6mm${}^2$). (Right)~Picture of the DuTiP2~(6$\times$17.2mm${}^2$).}
\label{fig:DuTiPPicture}
\end{figure}

\begin{figure}[htbp]
\begin{center}
\includegraphics[width=0.50\linewidth]{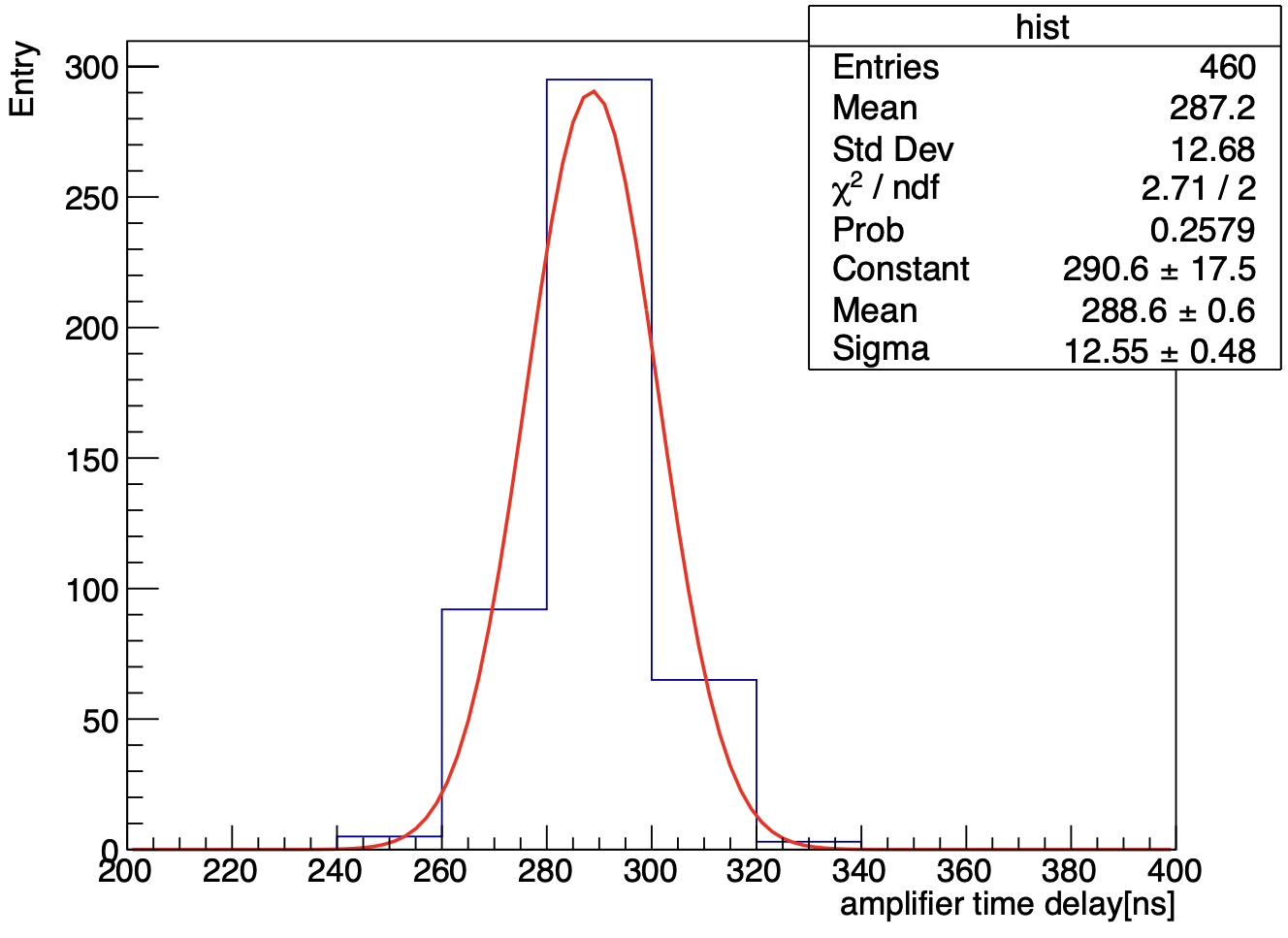}
\end{center}
\caption{Time resolution of DuTiP. The fit result of 12.6~ns includes a finite binning effect.}
\label{fig:DuTiPTimeRes}
\end{figure}

\paragraph*{DuTiP Prototype Sensors}
The first prototype DuTiP1 were delivered in 2021~(Fig.~\ref{fig:DuTiPPicture} left). 
This prototype was developed for demonstration of DuTiP concept thus all functionalities in pixel were fabricated while scan block and fast readout system were not implemented. 
The chip size is 6$\times$6mm${}^2$ with a pixel array of 64$\times$64.
The active thickness is 300~$\mum$ which is larger than the baseline.
The digital circuit worked using a test pulse with 25~MHz clock.
The time resolution of DuTiP is checked with test pulse. About 5000 electrons are injected to the analog circuit and check the time resolution with faster 50~MHz clock~(Figure~\ref{fig:DuTiPTimeRes}). 
The time resolution is evaluated as 11~ns which is enough smaller than the time bucket of 62.9~ns.
The efficiency is checked using a $^{90}$Sr radioactive source with a scintillation counter as the reference. 
The efficiency is measured as $98 \pm 2$\%. 
The large uncertainty is associated with systematic effect of the configuration for the scintillation counter.
The efficiency and position resolution will be measured with a electron beam at KEK with reference telescope.

The second prototype DuTiP2 had been delivered in 2022~(Fig.~\ref{fig:DuTiPPicture} right). 
This prototype has all functionalities except for fast hit data collection to periphery, and is full size in column direction, chip size of 17.2$\times$6mm${}^2$ with a pixel array of 32$\times$320, thus we can perform the full characterization needed for operation (extending the chip size to row direction is trivial). 
The evaluation is now on-going.

\begin{figure}[htbp]
\begin{center}
\includegraphics[width=0.24\linewidth]{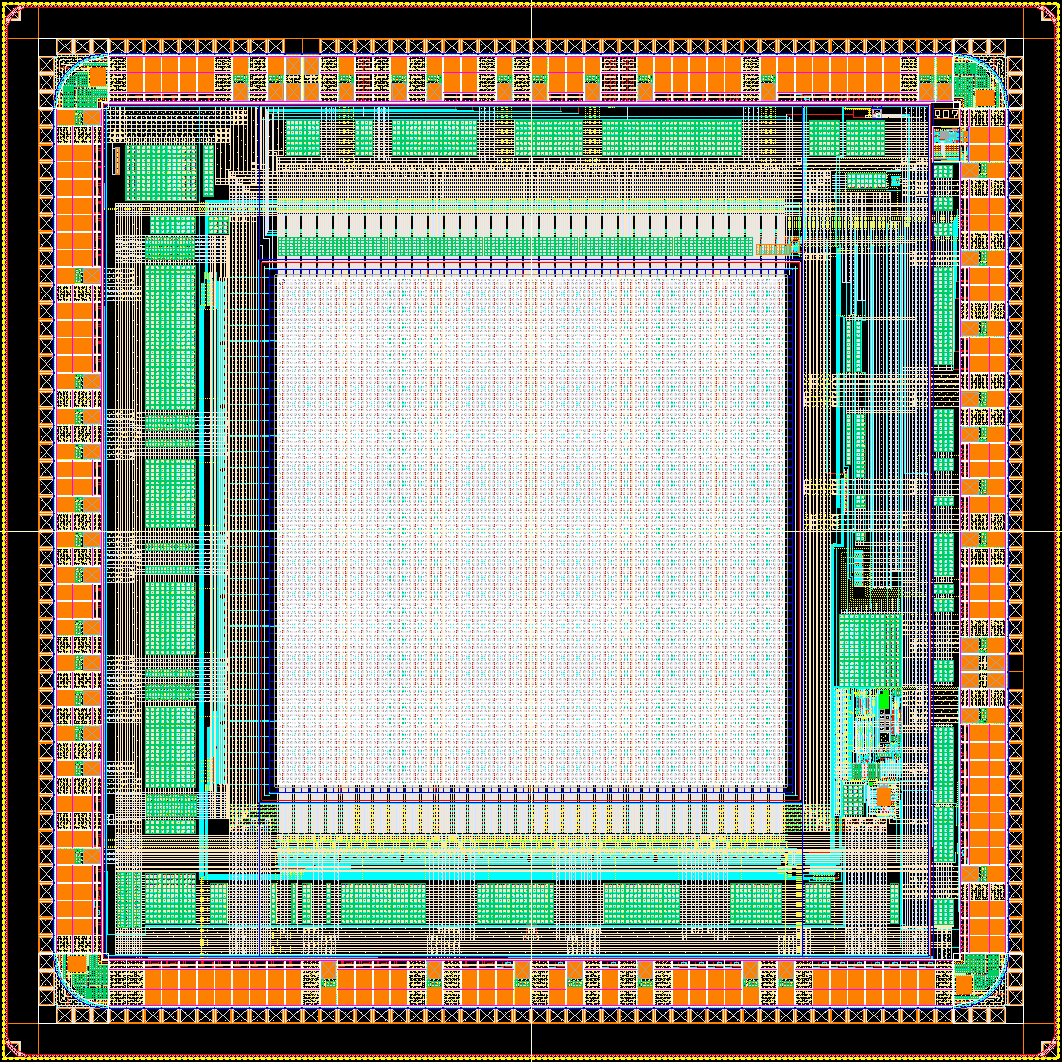}
\end{center}
\caption{Layout of the DuTiP3~(6$\times$6mm${}^2$).}
\label{fig:DuTiP3}
\end{figure}

The third prototype DuTiP3 had been submitted in 2023 and will be delivered in 2024~(Figure~\ref{fig:DuTiP3}). 
This prototype is based on DuTiP1 and equips the optimized amplifier and LVDS and PLL circuits.

We will submit the fourth prototype having all functionalities needed for real DuTiP in 2024.

\section{CDC}
\label{sec:CDC}
%\editor{Y. Nakazawa, N.Taniguchi}

% 1 Upgrade of Readout electronics
% 1.1 Introduction
% 1.2 New ASD-FADC ASIC
% 1.3 Radiation hardness
% 1.4 Potentially interested community in Belle II
% 1.5 Order of magnitude cost estimate

% will add
% - photo of the prototype board
% - performance of ASIC
% - FPGA candidate: Kintex-7
% - QSFP radiation test
% - Nakao option for remorte JTAG
% - ASIC mass-production

% 1.1 Introduction 
% 1.2 ASIC (including perfromace
% 1.4 Prototype (including perfromance
% 1.3 Radiaton (QSFP)
% 1.5 Potentially increased community in Belle II
% 2.0 Order of magnitude cost estimate

Assuming that the degradation in CDC performance associated with the injection background observed in 2022 can be mitigated by the SuperKEKB improvements implemented in LS1, the CDC detector is expected to withstand the background conditions after the Belle II upgrade through Runs 2 and 3. However there are concerns regarding the radiation hardness of components of the CDC readout. While dedicated ageing studies are being prepared to understand possible degradation of the chamber during its lifetime, this section discusses the upgrade path for the electronic components. We will use early Run 2 data to make more reliable CDC lifetime extrapolations and understand possible performance degradation.

\subsection{Upgrade of readout electronics}
\def\II{\rm I\hspace{-.1em}I}

\begin{figure}[h]
  \centering
  \includegraphics[height=6.0cm]{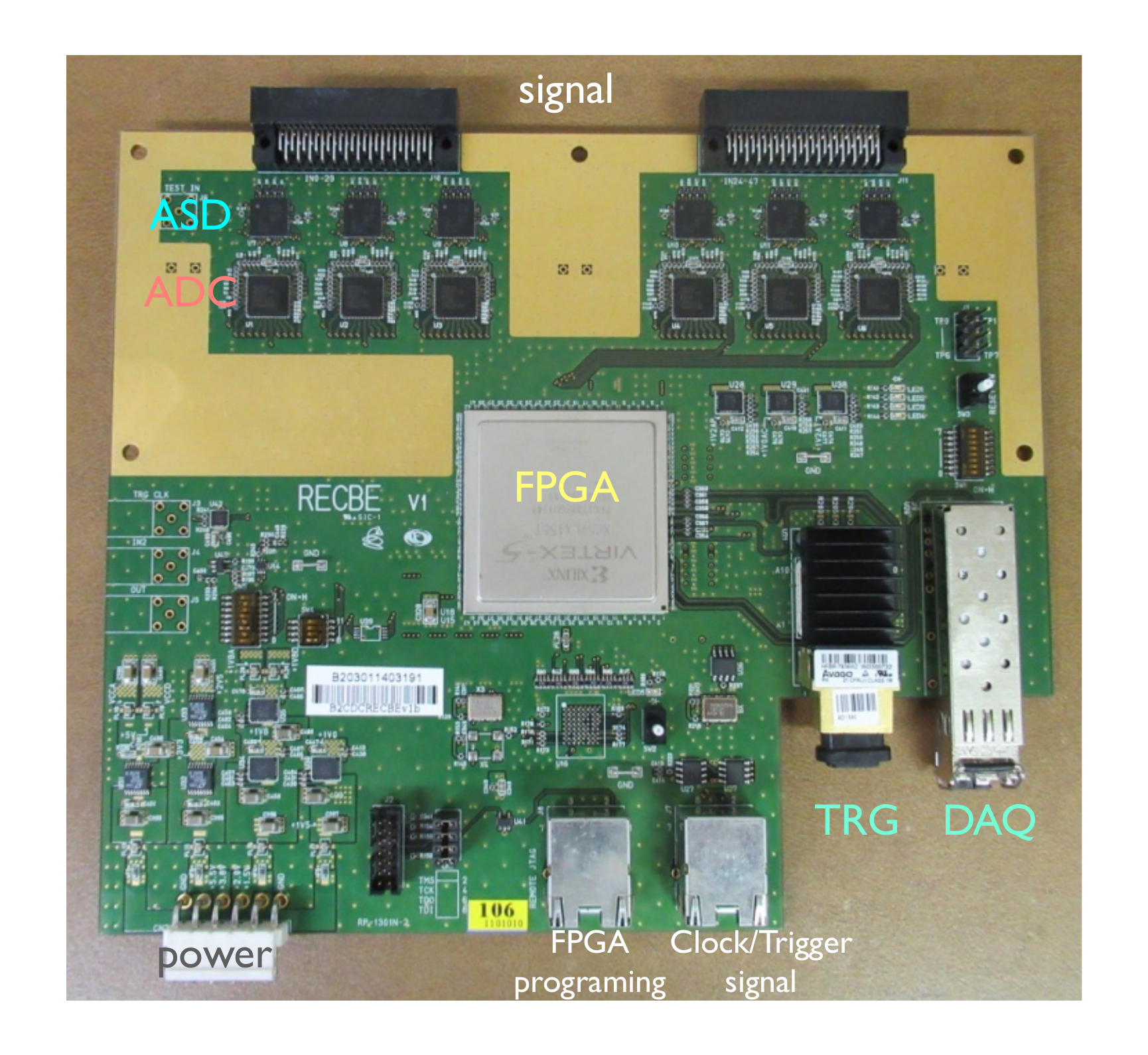}
  \caption{Photograph of the present CDC front-end module.} 
  \label{fig:05s6:cdc_ele_00}
\end{figure}

%\subsubsection{Introduction}
The front-end electronics located inside the detector read 14336 signal lines from the CDC.
Figure~\ref{fig:05s6:cdc_ele_00} shows a photograph of the current CDC front-end module.

With the design luminosity being an order of magnitude higher than the currently achieved, the operating environment is expected to become much harsher, especially in terms of radiation damage.
The expected final dose and fluence at the end of the physics run are, respectively, \SI{1}{kGy} of gamma rays and \SI{1.0 e 12}{n_{eq}/cm^2} of neutrons, being larger than initially expected at the time of the design of the current front-end modules.

In this respect, we plan to upgrade the front-end electronics to operate CDC readout system until the end of the the Belle $\II$ experiment.
The main critical item on the boards are the optical transceivers, that are very sensitive to the total ionizing dose. Therefore, these components will be replaced with a higher radiation tolerant optical transceivers.
The increase in beam currents and luminosity also demands the trigger system to effectively classify physics phenomena of interest under high background conditions.
To achieve this, the CDC readout boards are required to transmit more data about particle information while suppressing electrical noise such as on-board cross-talk.
A newly developed ASIC chip would enable to operate under such background levels.
The expected radiation tolerance of the new optical transceiver and ASIC are $\mathcal{O}(\SI{1}{kGy})$ and over \SI{1}{MGy}, respectively.
Thus, this upgrade plan will provide solutions to the above requirements for radiation hardness, cross-talk suppression, and faster data transfer.

The development of the ASIC has already been completed, and the first prototype board has been produced.
We are currently conducting performance tests on the prototype, including radiation tests.
In the following section, we give details about the new components and present the results of the performance tests of the prototypes.

\subsubsection{New ASD-FADC ASIC}
The new ASIC chip has been designed at KEK by the electronics system group using a 65-nm CMOS process.
This chip consists of eight sets of Amplifier-Shaper-Discriminator (ASD) and Flash Analog-to-Digital Converter (FADC) circuits.
In the ASD sector, a double-threshold technique is applied, where the lower threshold makes the time-walk effect small and the higher threshold cuts off electrical noise.
Charged particles that create large energy loss can induce smaller electrical signals, known as cross-talk, in the surrounding channels within an ASIC chip, compared to signals associated with tracks.
Cross-talk signals increase the fake rate of the track trigger as background levels increase.
The higher threshold suppresses cross-talk contamination in data, leading to improvements in both online and offline tracking reconstruction performance.
Data handling of FADC has about \SI{300}{\ns} delay from timing information in the present electronics.
FADC gives charge ($dE/dx$) information, but this is not used for the level-1 trigger decision due to the unacceptable long delay.
A trigger-performance study shows the use of charge information in the tracking trigger improves the trigger performance.
The FADC of the new ASIC reduces the delay due to data processing to about \SI{100}{\ns} and enables the use of charge information in the trigger system. % FIXME
The estimated power consumption of \SI{41.3}{mW/ch} is low compared to the total power consumption of the ASIC and ADC on the current board. % FIXME

% Note CM: Power consumption is still missing.

\subsubsection{Prototype of the upgraded module}
\begin{figure}[h]
  \centering
  \includegraphics[height=6.0cm]{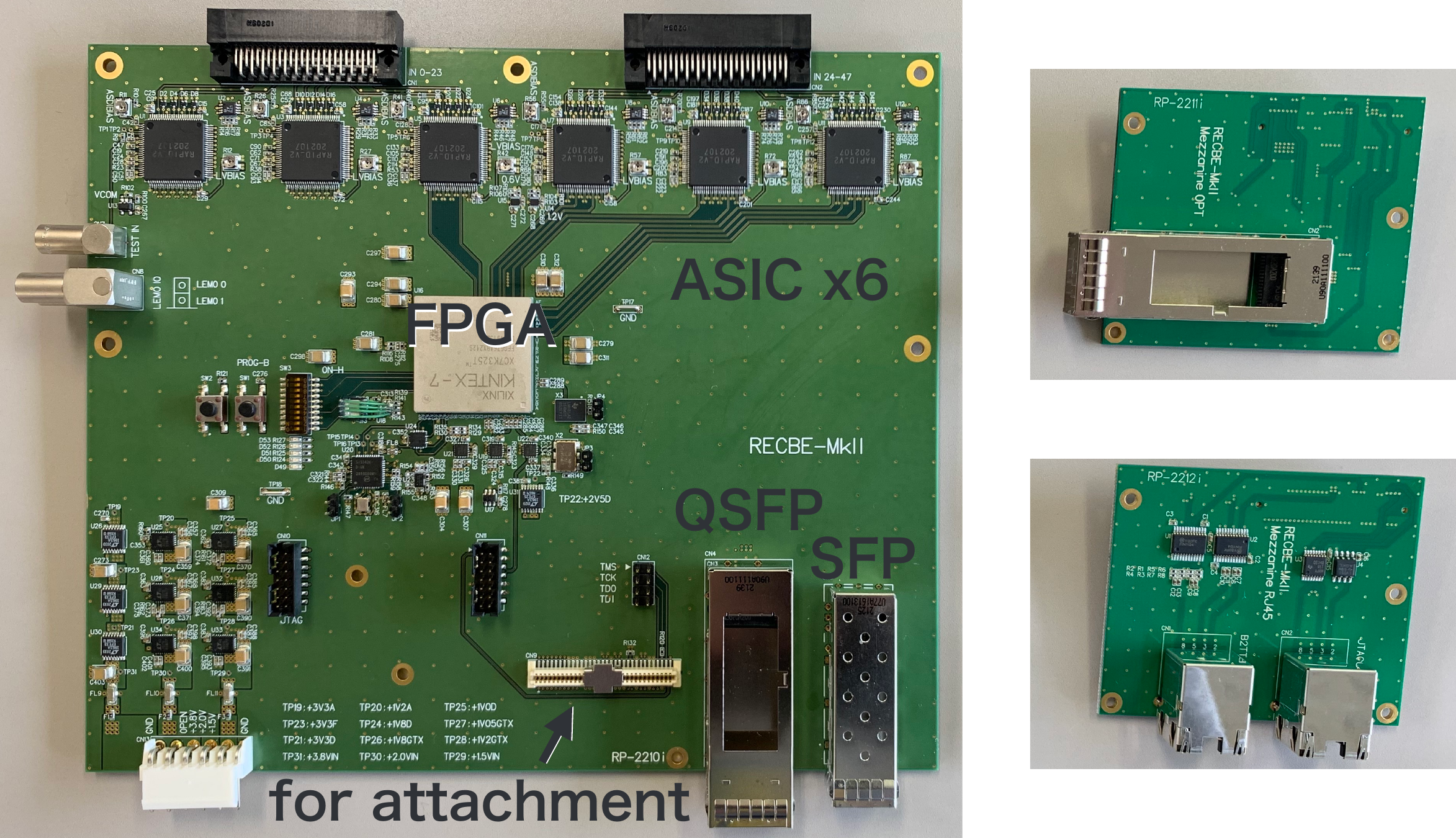}
  \caption{Prototype of the new CDC front-end module (Left) and the two attachment boards (Right).} 
  \label{fig:05s6:recbe_mkii}
\end{figure}

We have developed a prototype of the upgraded CDC front-end module that has the same functionality as the current module.
To test both, the current and upgraded functionalities, two attachment boards have been produced.
Figure~\ref{fig:05s6:recbe_mkii} shows a picture of the prototype of the new CDC front-end module along with the two attachment boards.
The module incorporates the following parts:
\begin{itemize}
    \item Six ASIC chips to measure signal timing and digitize waveform information
    \item A Field Programmable Gate Array (FPGA) for online data processing for trigger and data acquisition systems
    \item A Quad Small Form-factor Pluggable (QSFP) for data transfer to the trigger and data acquisition systems
    \item SFP connector for standalone performance tests (to be removed in future iterations) 
    \item A connector for the attachment boards to operate B2TT and B2JTAG
\end{itemize}
One attachment board is designed for the traditional slow control and time/trigger distribution operations with two RJ45 sockets, while the other one is designed to operate over three lanes using a QSFP module.
Firmware development is ongoing for the event-data format and link, trigger and timing separately, and the former is already available.

\begin{figure}[h]
  \centering
  \includegraphics[height=5.0cm]{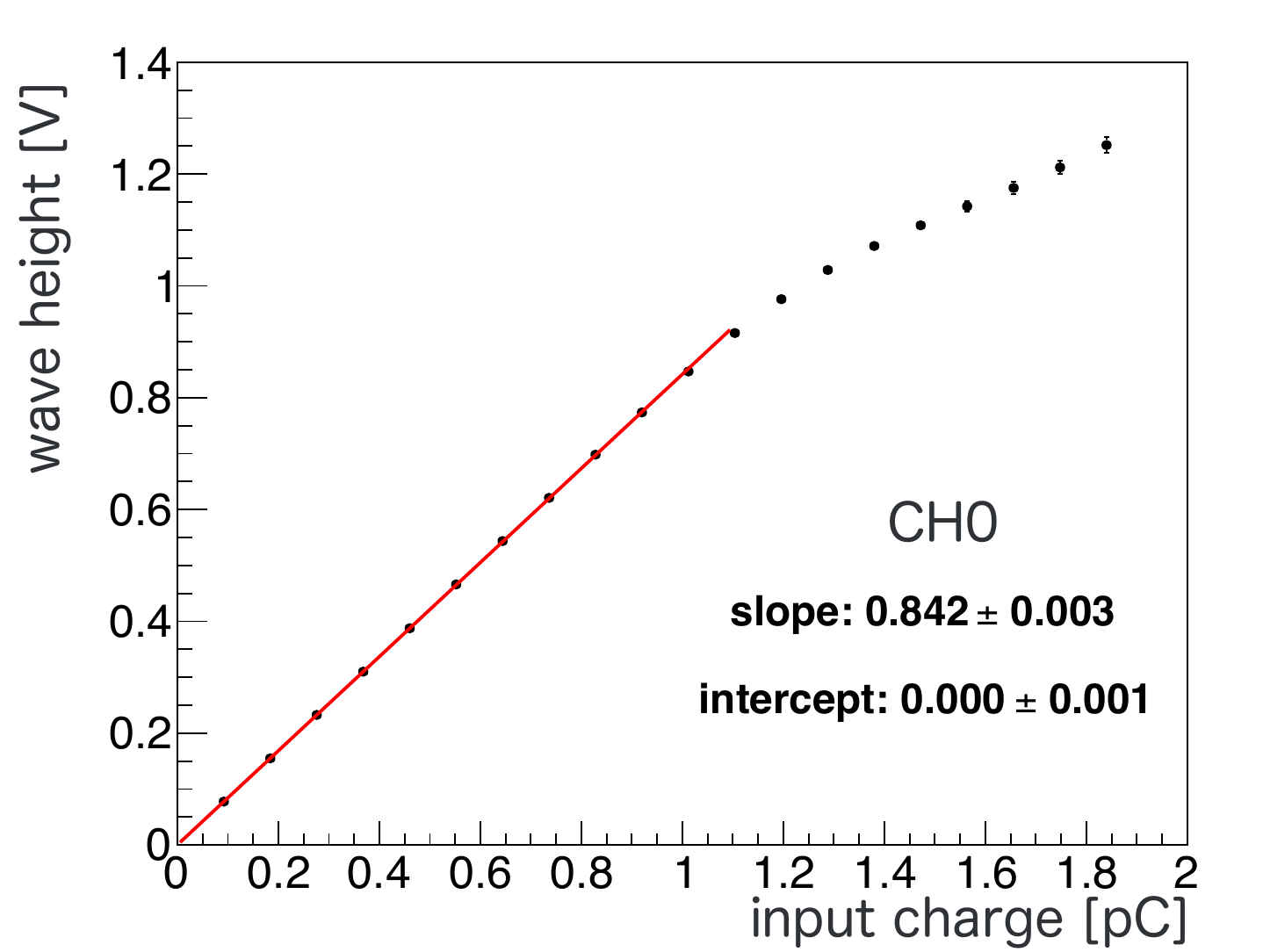}
  \includegraphics[height=5.0cm]{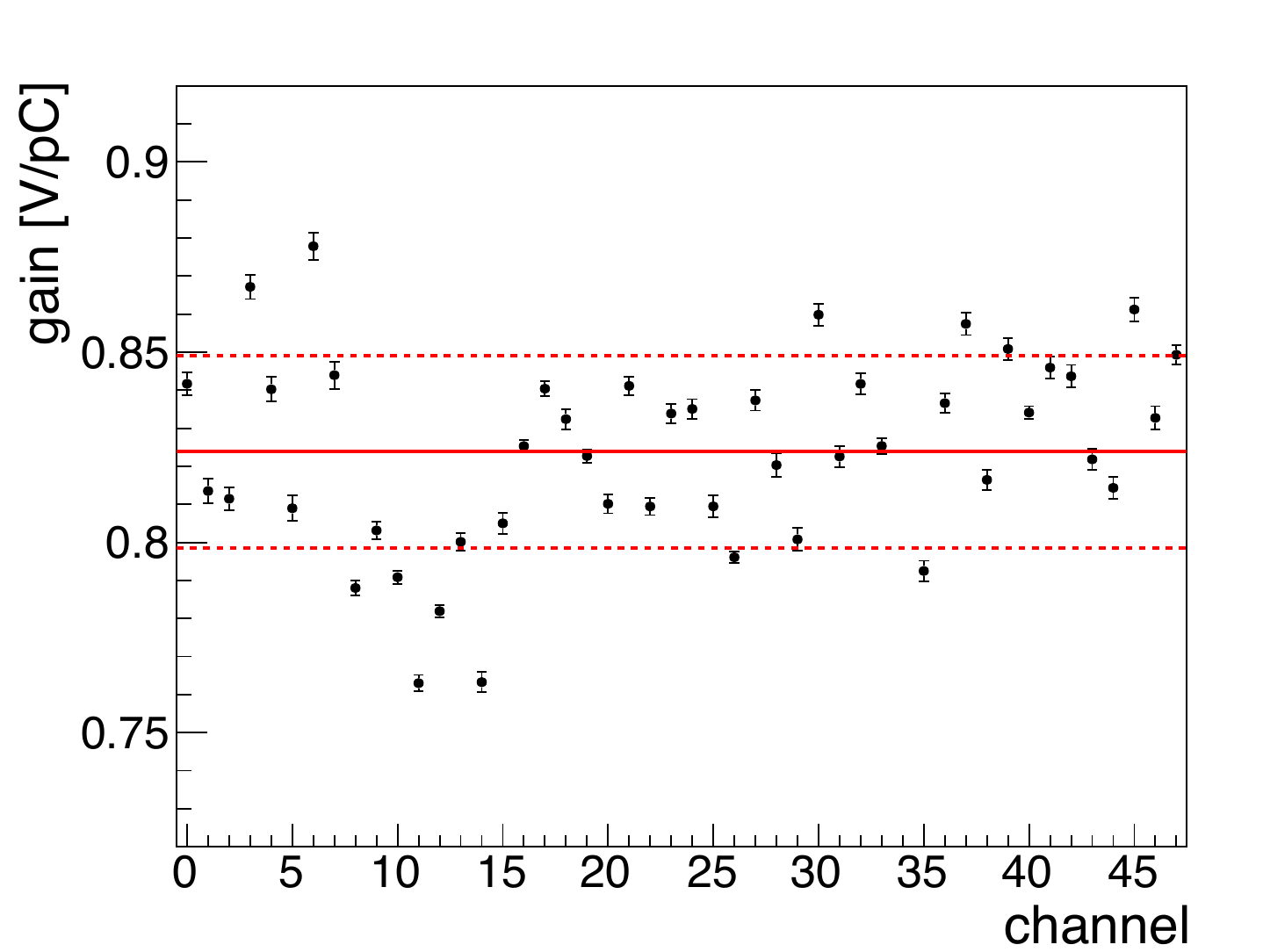}
  \caption{(Left) Relationship between wave height and input charge of channel 0. The red line represents the results of linear fitting. (Right) Gain values for each channel. The red solid and dashed lines represent the mean and mean$\pm$RMS values of the gain.} 
  \label{fig:05s6:dout_gain}
\end{figure}

The basic performance of the new module, specifically its linearity and gain, has been evaluated using a charge input board and a waveform generator.
The gain of the ASD component was measured by using a charge input board.
We measured the detection efficiency defined as follows:
\begin{equation}
  \mathrm{Detection\,\,Efficiency} 
  = \frac{\mathrm{\#\,\,of\,\,Events\,\,Exceeding\,\,Threshold}}{\mathrm{\#\,\,of\,\,Input\,\,Events}}\mathrm{.} 
\end{equation}
We varied the thresholds and calculated the detection efficiency, subsequently determining the pulse height by fitting it with an error function:
\begin{equation}
  \mathrm{erf}(x) = \frac{2}{\sqrt{\pi}} \int^x_0{\exp{\left( \frac{-(t-\mu)^2}{2\sigma^2} \right)}dt}
\end{equation}
where $\mu$ represents the wave height and $\sigma$ denotes the uncertainty in wave height.
The gain was determined by fitting relationship between wave height and input charge.
Figure~\ref{fig:05s6:dout_gain} illustrates the relationship between wave height and input charge and gain values for each channel.
The mean gain value is approximately \SI{0.83}{V/pC}, slightly lower than the designed value of \SI{0.9}{V/pC}.

\begin{figure}[h]
  \centering
  \includegraphics[height=6.0cm]{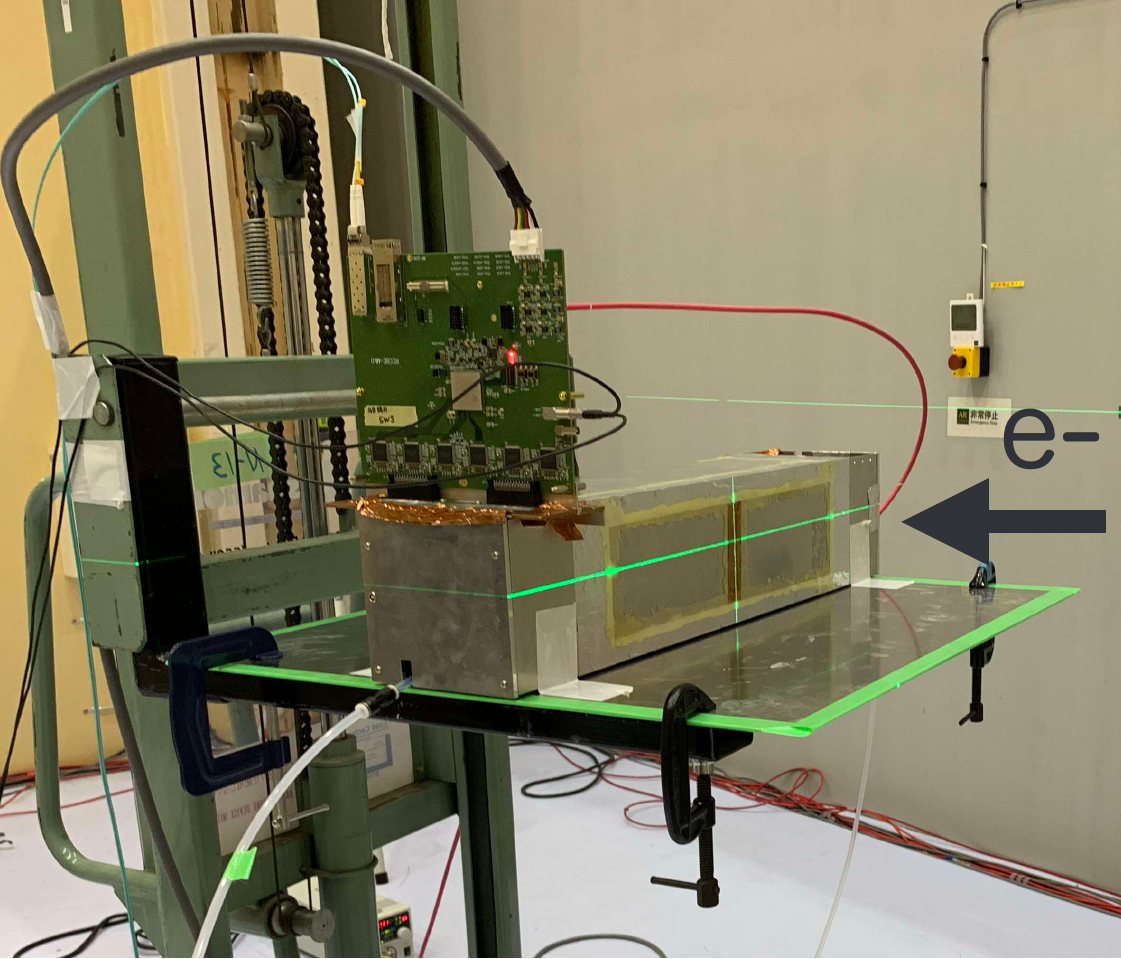}
  \caption{Photograph from the electron-beam test.} 
  \label{fig:05s6:test-setup}
\end{figure}

\begin{figure}[h]
  \centering
  \includegraphics[height=2.0cm]{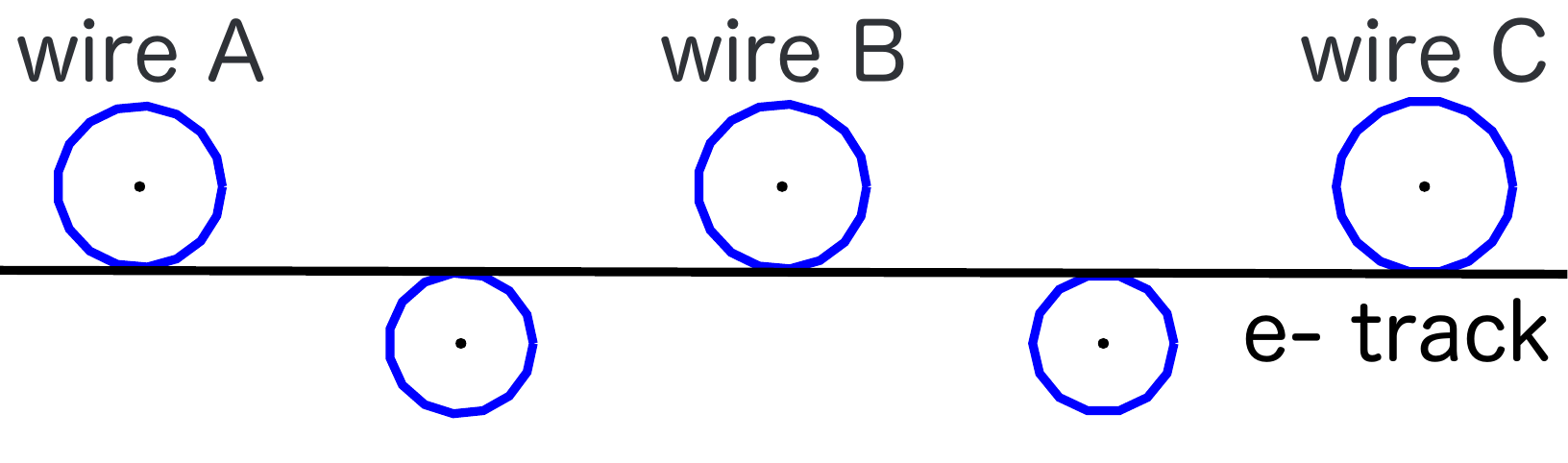}
  \caption{Wire configuration in the electron beam test. Black dots represent wires in the chamber and are surrounded by drift circles, with diameter corresponding to the drift time (blue circumferences). The black line indicates the reconstructed trajectory of an electron. The direction of the electron beam is perpendicular to the layer of the test chamber and parallel the wire plane so all the drift times should be equal.} 
  \label{fig:05s6:residual-cal}
\end{figure}

\begin{figure}[h]
  \centering
  \includegraphics[height=6.0cm]{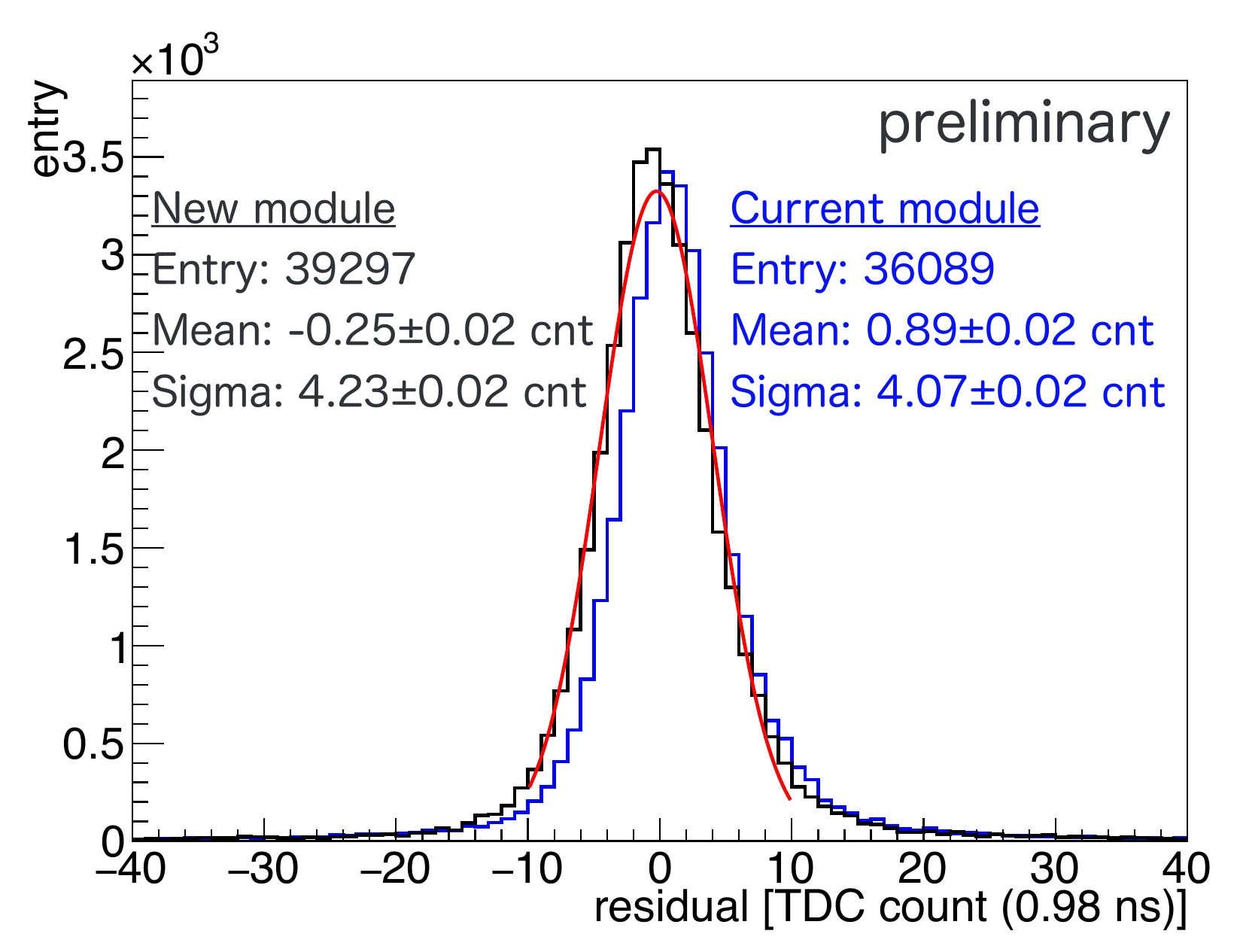}
  \caption{Residual distributions obtained at beam test. The red line corresponds to the Gaussian fit result of the black histogram representing the new module TDC counts.} 
  \label{fig:05s6:residual-dis}
\end{figure}

We assessed the time resolution using a CDC test chamber during a dedicated test-beam campaign.
The campaign was conducted at the AR Test Beam Line, KEK, in November 2022 with a $\sim$3 GeV electron beam.
Figure~\ref{fig:05s6:test-setup} provides an illustration of the beam test setup.
The chamber configuration is outlined as follows:
\begin{itemize}
  \item Cell: \SI{15}{mm} $\times$ \SI{15}{mm}
  \item Structure: \SI{8}{layers} and \SI{44}{wires}
  \item Gas mixture: \ce{He}:\ce{C2H6}=50:50
  \item High Voltage: \SI{2.37}{kV}
\end{itemize}
The direction of the electron beam is perpendicular to the layer of the test chamber.
Time resolution is determined through a fitting process applied to the residual distribution, which is calculated as:
\begin{equation}
  \mathrm{residual} = \left( t_{\mathrm{A}} + t_{\mathrm{C}} \right) \times \frac{1}{2} - t_{\mathrm{B}}
\end{equation}
where $t_{i}$ represents the timing of particle detection, and $i$ corresponds to the wire ID, as explained in Fig.~\ref{fig:05s6:residual-cal}.
For performance comparison between the new module and the existing one, data was also collected using the same setup with the current module.
The residual distributions are displayed in Fig.~\ref{fig:05s6:residual-dis}
The measured time resolution of the new module is \SI{4.23}{ns}.
This is slightly worse than the current module's performance, but since there are more charged particles at $\sim$\SI{1}{GeV} and multiple scattering dominates, the impact of the CDC performance is negligible.
Further detailed performance evaluations are still ongoing.

\subsubsection{Radiation hardness}
%To improve performance of track fitting in track trigger, we plan to transfer more information into trigger system by using feasible optical transceivers. 
The radiation tolerance of the front-end module and related optical modules should be examined through gamma-ray and neutron irradiation.
The required tolerance is a total dose of \SI{1}{kGy} and a neutron fluence of \SI{1.0 e 12}{n/cm^2} over the 10 years Belle $\II$ data acquisition period.
Some of the integrated circuits (ICs), including voltage regulators, are identical to those employed in the current module, and their radiation tolerance has already been verified.
The new components to be tested are the ASIC, QSFP, and FPGA.
The semiconductor technology integrated into the ASIC is acknowledged for its robust radiation resistance, adequately enduring the anticipated radiation levels of Belle $\II$.

% CM: Need to explicitly add the technology and citation if possible

% QSFP
Regarding the QSFP, typical optical module failures are generated by the total ionizing dose; therefore, we have to find QSFP candidates with sufficient radiation tolerance.
To this end, two gamma-ray irradiation tests were conducted at Tokyo Institute of Technology, in March 2022 and February 2023.
During these tests, radiation hardness was evaluated by monitoring errors during the high-speed communications including also variations between individual modules.
As a result, we found three QSFP candidates that meet the required radiation tolerance.
These candidates were also qualified against neutron damage, as described below.

% FPGA
As for the FPGA, neutrons can introduce soft errors, leading to unintended logic changes - a phenomenon termed single event upsets (SEUs).
The implemented firmware automatically corrects SEUs using a soft-error mitigation scheme provided from Xilinx Inc., but it cannot correct multi-bit upsets (MBUs).
Therefore, when MBUs are detected, data acquisition is temporarily stopped to reprogram the FPGA.
The duration of this MBU-induced reprogramming interval potentially constitutes a significant portion of data-acquisition downtime.
We conducted a neutron irradiation test at Kobe University using a TANDEM accelerator providing 2 MeV neutrons in March and September 2023.
The soft-error mitigation scheme successfully detected soft errors and corrected SEUs and no permanent damage was observed.
In the new module, the soft-error rate has reduced to 2/3 while the number of logic blocks is twice that of the current module. 
This is thought to be attributed to the transition from 65-nm technology to a 28-nm technology FPGA.
%An additional neutron irradiation test is scheduled for September 2023 to assess detailed performance outcomes.

\subsubsection{Potentially interested community in Belle II}
The design of the new ASIC and layout of the prototype board have been done by experts in KEK Esys group.
%The new ASIC of ASD and layout of a prototype of readout board is also don by experts in KEK Esys group.
% Mass production was done in Taiwan.
The mass production of the present readout modules has been conducted in Taiwan.
National Taiwan University (NTU) has experiences in performing comprehensive quality assessments across all of these modules.
It is anticipated that KEK and NTU will contribute on this upgrade.

\subsection{Order of magnitude cost estimate}
An estimation of the main cost is shown in Table.~\ref{cdc_tb1}. We include $10\%$ spare. 
Cost for mass production of electronics is based on the cost of FPGA.
Mass production of the ASIC chip will be completed by the end of JFY2023 (March 2024).

\begin{table}[h]
  \caption{Estimated cost for mass production of ASD ASIC and readout electronics board.}\label{cdc_tb1}
  \centering
  \begin{tabular}{|c|c|c|}
    \hline
    item & number & cost \\
    \hline\hline
    % mass production of ASIC & $\sim$ 2,000 chip  & 15 M JPY\\
    mass production of ASIC & $\sim$ 3,000 chips  & 25 M JPY\\
    \hline
    mass production of FE board & $\sim$330 FEs (including spare) & 100 M JPY\\
    \hline
  \end{tabular}
\end{table}

\subsection{R\&D and technology options}
\subsubsection{TPC-based tracking system}
%%%%%%%%%%%%%%%%%%%%

At sufficiently high luminosity drift chambers will suffer from high occupancy due to their two-dimensional nature. A highly segmented three-dimensional tracker could allow tracking at far higher luminosities. A natural solution would be a gas Time Projection Chamber (TPC) with micropattern gaseous detector readout, combining low material budget with fine three-dimensional segmentation. With existing technologies, the $\mathcal{O}(10^4)$ CDC cells could be replaced with $\mathcal{O}(10^{12})$ effective voxels. Accounting for the $\mathcal{O}(10^2)$-times longer drift time for primary electrons, such a TPC therefore promises an occupancy reduction of $\mathcal{O}(10^{6})$ relative to the CDC.

A preliminary conceptual design for a TPC-based tracking system for a longer-term Belle II upgrade was described in Ref.~\cite{tpc_whitepaper}. The envisioned TPC used an integrated grid (\textit{InGrid}) pixel readout with \SI{200}{\micro\meter} pixel pitch and continuous binary readout. In this concept, the VTX system is extended to to $r=\SI{44}{\centi\meter}$ and optional fast-silicon timing layers at $r=25$ or \SI{45}{\centi\meter} provide a trigger and low-momentum track PID via time-of-flight. The preliminary simulation-based studies suggest that such a system could provide comparable tracking performance to the current CDC even in the pessimistic background scenarios investigated, though considerable open questions remain. 

Ions generated during amplification drift $100$ times slower than the primary elections, so ion backflow (IBF) could lead to performance-degrading space-charge buildup. This is potentially most problematic with injection backgrounds due to the \SI{50}{\hertz} repetition rate of SuperKEKB's top-up scheme. The ion drift time of \SI{3}{\milli\second} is comparable to the stabalization time of the beam after injection, so the TPC integrates injection backgrounds over bunch revolutions but not over injections. It is not possible to add an active gating layer above the amplification without unacceptable dead-time. The preliminary results suggest that the ion density would be comparable to that of current or planned tracking TPCs due to the intrinsically low IBF of InGrid sensors. However, no injection background simulation was available.

Without a modification of the injection scheme, it would be prudent to minimize the IBF intrinsic to the amplification. One option is to add a second integrated amplification layer that also acts to screen backflowing ions. Such a \textit{TwinGrid} device~\cite{tpc_twingrid} could suppress IBF by a factor up to two orders of magnitude relative to InGrid sensors, to the level of about half an ion per primary electron~\cite{twingrid_ibf}. However, injection-background simulation followed by testbeam campaigns with prototypes would be needed to confirm acceptable space-charge densities. 

Regardless of the source, the overwhelming majority of hits are generated by low-energy electrons spiraling in the gas toward the readout plane. These \textit{microcurlers} arise mainly from Compton scattering of low-energy beam-induced background photons in the target gas. While these hits are not an occupancy concern, they could overwhelm the throughput of the system with useless data and obscure hits associated with tracks from collisions. However, microcurlers are very distinctive, and it is likely that they could be identified and filtered out at the front end using embedded machine learning without loss of tracking information. 

Although preliminary and not conclusive, these early findings suggest that a TPC-based tracking system combined with a radially extended vertex detector compared to the current one and the implementation of dedicated trigger and PID detector layers could be viable at luminosities beyond the limits of the CDC, and are promising enough to motivate a dedicated R\&D program to develop the concept further.

\chapter{Outer Detector}
\label{sec:outer}
%\subimport{.}{S1.TOP}
\section{TOP}
\label{sec:TOP}
\editor{E.Torassa}

\subsection{Introduction}

\begin{figure}[ht]
\begin{center}
\includegraphics[scale=0.55]{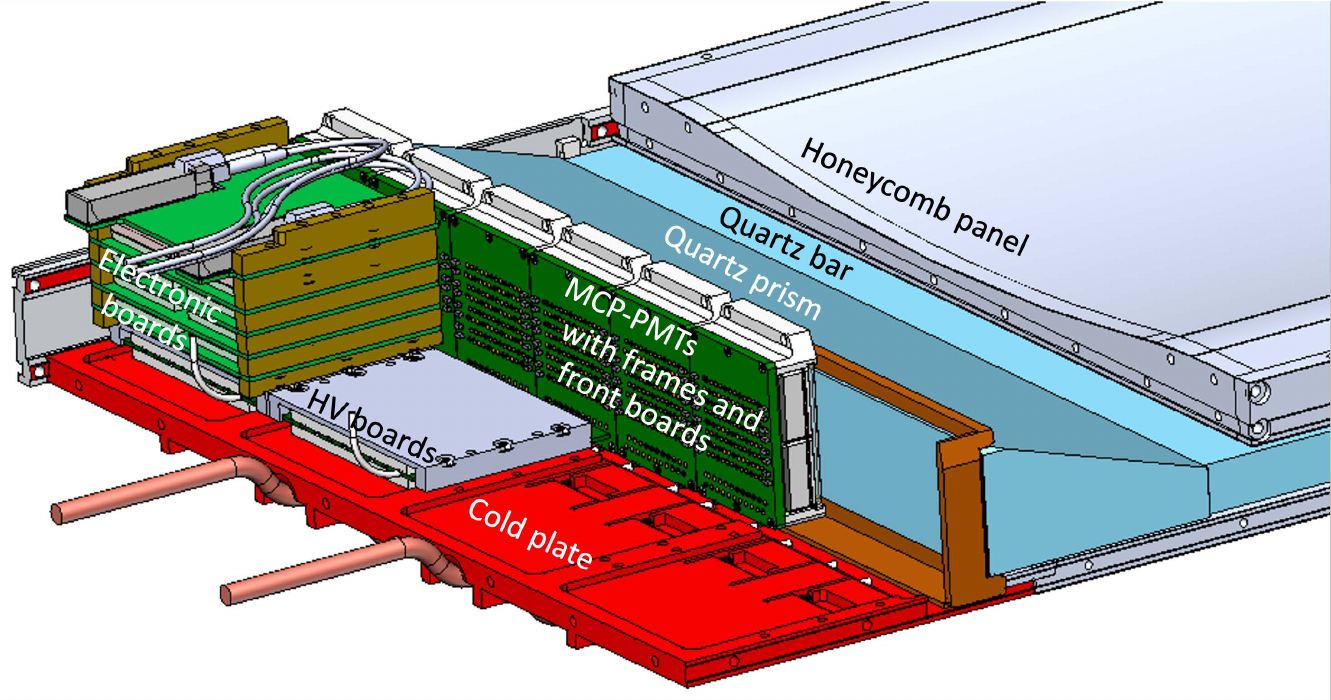}
\caption{\it Illustration of the back side of one TOP module where electronic boards and photodetectors are interfaced with the quartz prism and quartz bar.}
\label{fig:TOPmodule}
\end{center}
\end{figure}

Since the beginning of the SuperKEKB operation, the TOP detector (Fig.~\ref{fig:TOPmodule}) has been performing well without major issues. However, due to the limited lifetime of the Micro-Channel Plate Photomultiplier Tubes (MCP-PMTs), there are concerns about their ability to last through the entirety of the experiment. To address this, a replacement of the MCP-PMTs with lifetime-extended ALD type was partially conducted in 2023 during the Long Shutdown 1 (LS1), and will continue during the Long Shutdown 2 (LS2) to ensure sustained performance of the TOP detector for particle identification throughout the experiment.
%Fig. 1

\begin{figure}[ht]
\begin{center}
\includegraphics[scale=0.55]{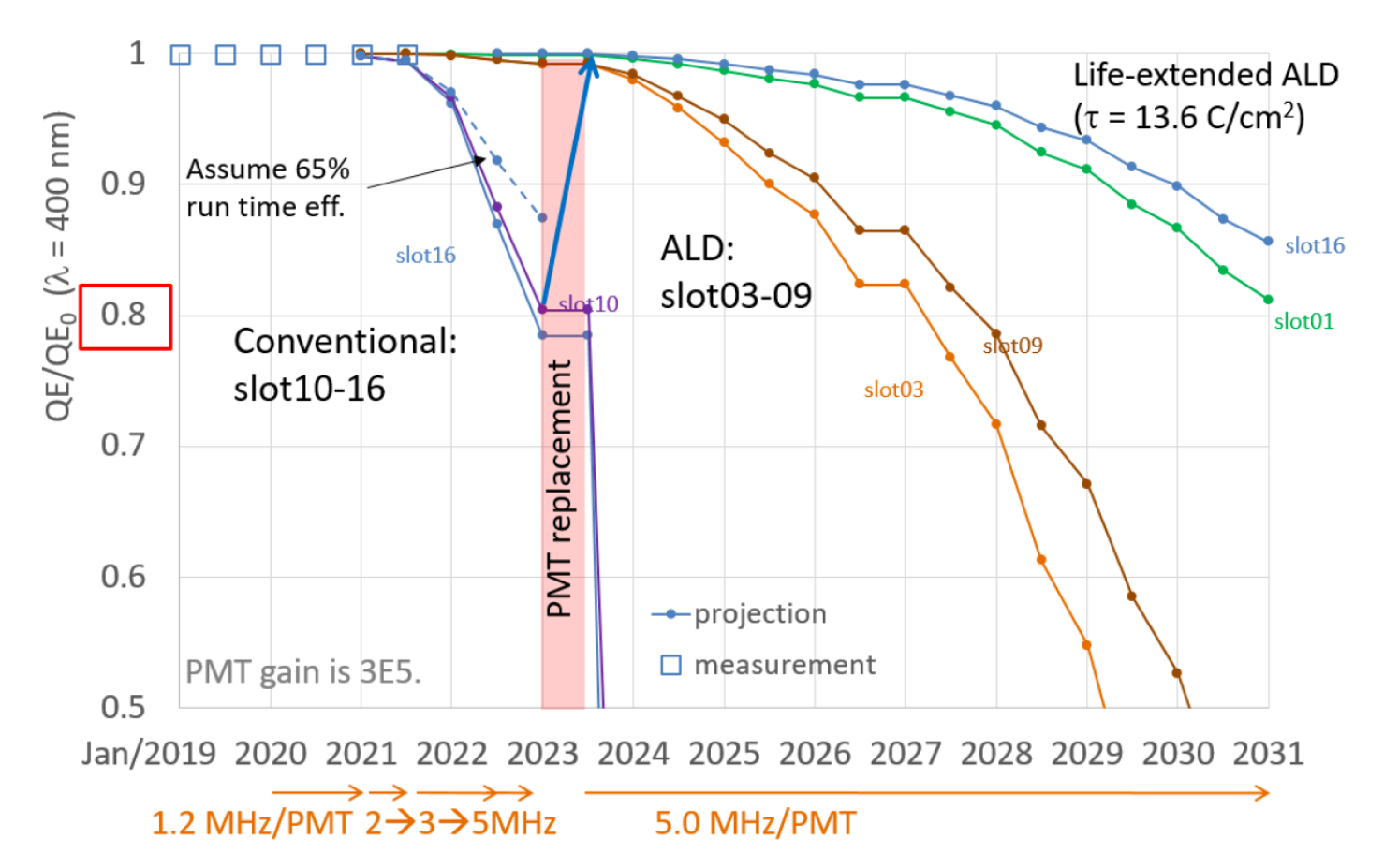}
\caption{Expected QE degradation for the three different generations of MCP-PMTs installed in the TOP detector: conventional, ALD and life-extended ALD.  Horizontal axis shows the year and vertical axis is the relative QE.}
\label{fig:QEprojection}
\end{center}
\end{figure}

After four years of operation, it has become evident that the beam background level at the Belle II detector is higher than anticipated. As a result, the hit rate of each PMT has been increased, accelerating their aging and causing degradation in quantum efficiency. Therefore, the lifetime of the MCP-PMTs has been carefully monitored, and lifetime projections have been re-evaluated, as illustrated in Figure~\ref{fig:QEprojection}.
%Fig. 2

In this figure, the background from the Monte Carlo simulation was determined for the components related to the luminosity (Bhabha scattering, two-photon processes) and a maximum limit of 5 MHz/PMT for the other components (beam-gas, Touschek). The expected gradual increase in peak luminosity described in~\cite{top:topupgrade} was taken into account. 

Overall, with the adoption of life-extended ALD-type MCP-PMTs, the TOP detector is expected to maintain its performance without significant degradation until the integrated luminosity target is achieved. Therefore, replacing the MCP-PMTs still not  lifetime-extended ALD type is currently considered as the default option for the TOP detector upgrade.

%In this case, the QE of normal ALD type PMTs will drop below 80~\% starting from 2027 and all PMTs will reach this level of QE degradation at around the current luminosity target.

Nevertheless, there is a possibility that the background level could also rise beyond current projections. In preparation for such a scenario, we also have initiated further R\&D to explore replacing the MCP-PMTs with Silicon Photomultipliers (SiPMs) as a backup option. For this option, the required cooling will reduce the space available for the readout electronics and limit its acceptable heat load. Thermal gradients and potential thermal and mechanical stresses need to be carefully evaluated.
The option of improving the current electronics was also considered.

%\subsection{QE degradation with high temperature}

%The timing for the replacement of the MCP-PMTs is related to the level of background during the next years. 
%Some MCP-PMTs showed a faster than expected reduction in QE considering their accumulated charge. One possible explanation is that they operated at a higher temperature for some time due to a temporary overheating of the electronics. 

%Quantum-efficiency measurements performed in the laboratory as a function of the accumulated charge show a faster degradation at higher temperatures (Fig.~\ref{fig:MCP-PMTheat}). New readout electronics with reduced power consumption can help maintain the MCP-PMT's lifetime.
% Fig. 3
%\begin{figure}[ht]
%\begin{center}
%\includegraphics[scale=0.62]{fig/TOP/MCP-PMTheat.pdf}
%\caption{Relative QE degradation as a function of the accumulated charge at $20\degrees C$ and $40\degrees C$ for two conventional MCP-PMTs.}
%\label{fig:MCP-PMTheat}
%\end{center}
%\end{figure}
%

%Planning has started for these upgrades, which could take place during a long shutdown to rebuild the interaction region on the timescale of roughly 5 years from now. 

\subsection{MCP-PMT production status}

For the photosensors, the Hamamatsu production line of life-extended ALD MCP-PMT is operating with an expected output of 10 MCP-PMTs/month. In total ~220 life-extended ALD MCP-PMT needs to be produced  to replace normal ALD type PMTs. The production has already restarted, 90 MCP-PMTs have been ordered, 30 of them have been delivered. A possible further increase of lifetime can be investigated for example by an optimization of the atomic layer thickness.

\subsection{Readout electronics upgrade}

\subsubsection{Motivation}
Some MCP-PMTs showed a faster than expected reduction in Quantum-efficiency (QE) given their accumulated charge. One possible explanation is that they operated at a higher temperature for some time due to temporary overheating of the electronics. 

\begin{figure}[ht]
\begin{center}
\includegraphics[scale=0.62]{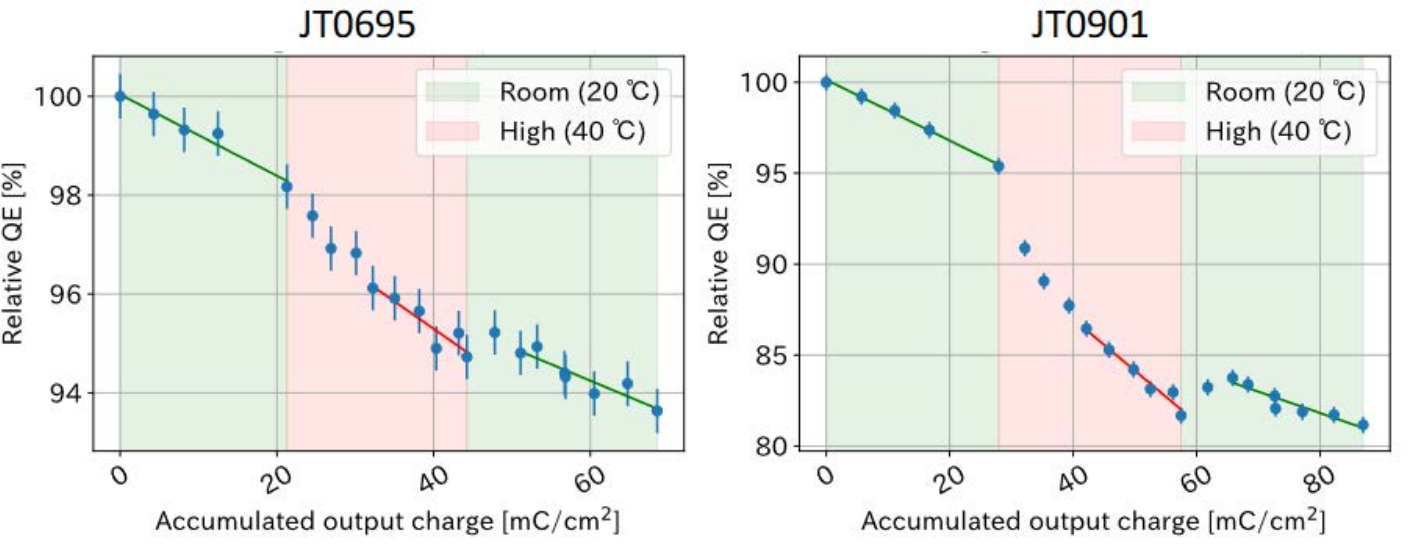}
\caption{Relative QE degradation as a function of the accumulated charge at $20\degrees C$ and $40\degrees C$ for two conventional MCP-PMTs.}
\label{fig:MCP-PMTheat}
\end{center}
\end{figure}

The QE measurements performed in the laboratory as a function of the accumulated charge show a faster degradation at higher temperatures (Fig.~\ref{fig:MCP-PMTheat}). By implementing new readout electronics with reduced power consumption, we can mitigate QE degradation and help extend the lifetime of the MCP-PMTs.
% Fig. 3

Moreover, the projected high rates of Single Event Upsets (SEUs) is also problematic, as some number of them are not recoverable and will lead to increasing data taking loss. This SEU issue also motivates the development of new electronics with a smaller FPGA on the readout board.

\subsubsection{Power consumption measurement}

\begin{figure}[ht]
\begin{center}
\includegraphics[scale=0.52]{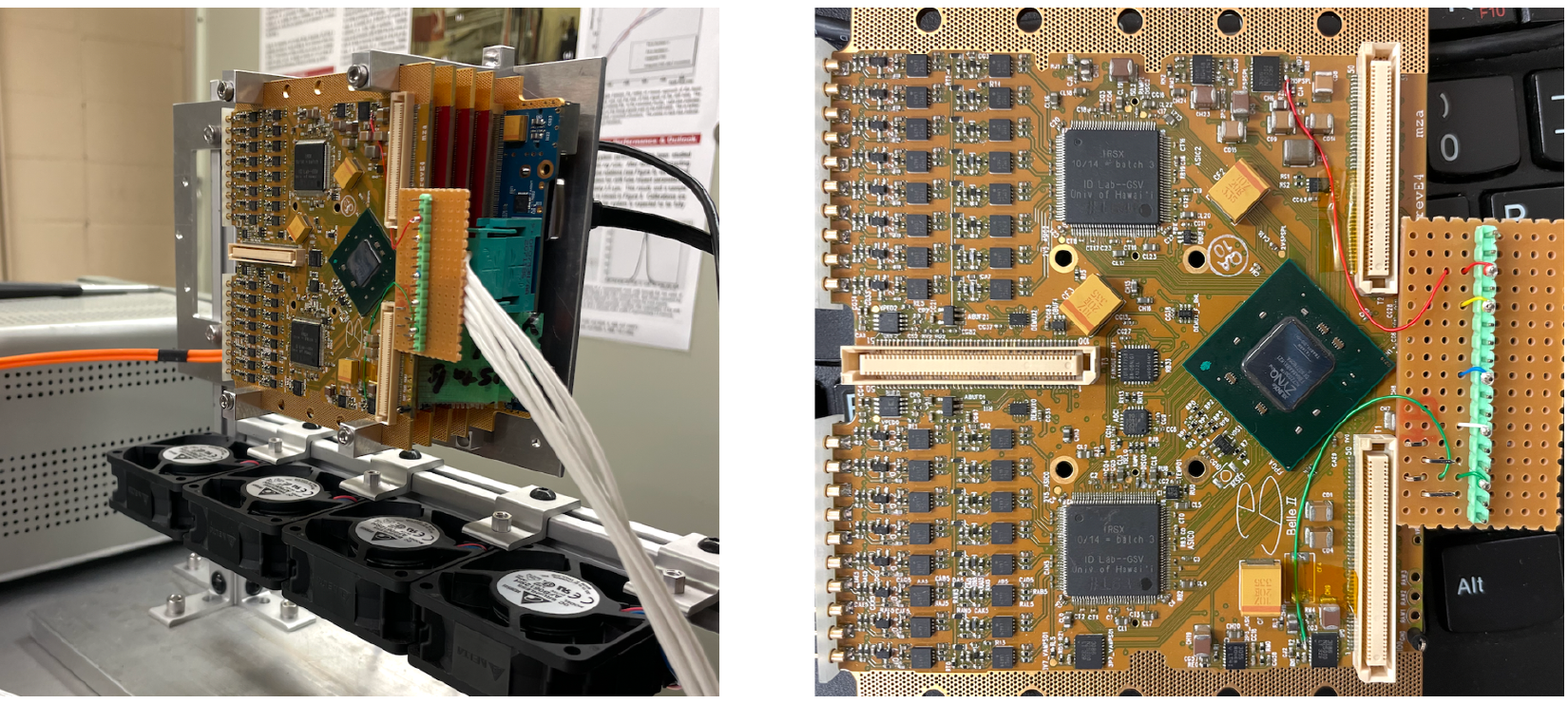}
\caption{Boardstack used for the measurement of the power consumption of our front-end electronics. To access some of the voltage regulators on the Carrier Board, soldering and wiring were required.}
\label{fig:pw_consumption}
\end{center}
\end{figure}

Before delving into a detailed study of electronics R\&D, the power consumption of the current front-end electronics, also known as the board stack, was measured. The setup used for this measurement is shown in Figure~\ref{fig:pw_consumption}. The board stack is composed of four Carrier Boards and one Standard Control Read-Out Data (SCROD) board. It was discovered that approximately 90\% of the power consumption originates from components on the Carrier Boards, such as IRSX (ASIC), IRSX regulators, FPGAs, and amplifiers, as indicated in Table~\ref{iTOP:pw_consumption}. These results underscore the potential benefits of upgrading the Carrier Boards, particularly the ASIC components, rather than the SCROD board, to achieve power reduction.

% Fig. 4

\begin{table}[htb]
\caption{Summary of power consumption on the current boardstack. Approximately 90\% of the consumption originates from the carrier boards.}
\centering
\begin{tabularx}{\textwidth}{ | X | X | X | X | }
\hline
\multicolumn{4}{|c|}{\bf Individual contribution to power consumption [mW] (Part 1)} \\ \hline
{\bf SCROD} & {\bf IRSX (ASIC)} & {\bf IRSX regulators} & {\bf Amplifiers} \\ \hline
4858 ~(6.7\%) & 10764 ~(14.8\%) & 17495 ~(24.0\%) & 16540  (22.7\%) \\ \hline
\end{tabularx}

\vspace{2em} % Add some vertical space between the tables

\begin{tabularx}{\textwidth}{ | X | X | X | X | }
\hline
\multicolumn{4}{|c|}{\bf Individual contribution to power consumption [mW] (Part 2)} \\ \hline
{\bf FPGAs on CBs} & {\bf Other regulators} & {\bf Unknown} & {\bf Total } \\ \hline
18520 ~(25.4\%) & 3507 ~(4.8\%) & 1138 ~(1.6\%) & 72822 ~(100\%) \\ \hline
\end{tabularx}
\label{iTOP:pw_consumption}
\end{table}

\subsubsection{R\&D status}

Regardless of the choice of photosensors, this necessitates an upgrade of our readout electronics to maintain the performance of the TOP detector. The primary goal of the upgrade is to reduce the power consumption required to operate the electronics. In the SiPM scenario, to ensure sufficient space for enhanced cooling, the readout board must be smaller to prevent an increase in dark-count rates.

%The goals of the baseline upgrade to the readout electronics are twofold: reduce significantly their size, as well as the power required. Both goals facilitate the implementation of the requisite cooling needed to operate the SiPMs in single-photon detection (though with significantly increased dark-count) mode.

In order to reduce cost and development time, much of the existing software, firmware and hardware infrastructure can be retained.  The boundary conditions considered for this costing exercise are:

\begin{itemize}
    \item retain the Standard Control Read-Out Data (SCROD) board, the existing cables, low-voltage power system and the existing infrastructure for the TOP-based online event time logic of the L1 trigger in TOP FEE;
    \item remove the HV divider chain/delivery board and associated cables;
    \item employ multiple Analog to digital converter System-on-Chips (ASoC) developed by Nalu Scientific.
%    per ASIC "Carrier" card, each with multiple Gbps connections to the SCROD FPGA
\end{itemize}

 Nalu Scientific, LLC (NSL) is a small business in Honolulu, HI and was originally a spin-off of the University of Hawaii’s High Energy Physics group back in 2016.  NSL works closely with US DOE laboratories and universities. It has developed commercially available variations of the IRSX and TARGETX chips that had been previously used for TOP and KLM readout.  Further development and refinement of the IRSX and TARGETX ASICs has led to a new generation of chips: the ASoC (Analog to digital converter System on a Chip)~\cite{top:ASoC} and the HDSoC (High Density digitizer System on a Chip)~\cite{top:HDSoC3}. The ASoC and HDSoC include on chip circuitry and functionality to perform sampling, digitization, triggering, region of interest extraction operations. Essentially, an analog channel enters the chip and packetized data exits the chip in high speed serial form. The chips operate with a high level of autonomy, which reduces the need for high-end FPGAs on the readout board. The ASoCv3 is a 4 channel digitizer with 16k samples of buffer per channel operating at 2.5-4 GSa/s channel. It is commercially available with a suite of test hardware, firmware, software and documentation. The version 3 has been tested and has improved power consumption compared to the IRSX. It can be used for bench testing and development purposes starting immediately.
A custom version of the ASoC (ASoCv4) is proposed for the TOP readout upgrade with 8 channels per chip, which is matched to the existing readout infrastructure and has improved speed and power consumption. This new ASIC would be developed for the TOP upgrade. 
The ASoC can communicate with less complex FPGAs (compared to IRSX) which will reduce size, cost, power consumption, and SEU issues while maintaining overall firmware backwards compatibility with the current IRSX based system. 
%Based upon this proposed modification, the upgraded readout system still consists of 64 "board stacks" and 4 ASIC "carrier" cards per "board stack".  The new ASIC "carrier" card will no longer have a FPGA to reduce radiation fault events and employ eight 4-channel readout AISCs per card with point-to-point Gbit serial links to the SCROD FPGA. 

\subsection{SiPM R\&D status}

SiPM R\&D is also in progress. It is important to find the best balance among different SiPM's characteristics:
single device size, single cell dimension, operating temperature.
Considering the expected neutron rate near the photodetector region at the target instantaneous luminosity of $6\times 10^{35}cm^{-2}s^{-1}$ is about \SI{2.0 e 10}{n/(cm^2 \cdot~year)}, we need to be able to distinguish single photons at least up to an irradiation of  \SI{1 e 11}{n/cm^2}, even more for higher luminosity. We need also to retain photon efficiency, a good time resolution and have acceptable thermal and mechanical stresses for the detector.
The single channel active area of the MCP-PMT with 16 channels and $1~in^2$ size is about $6 \times 6~mm^2$. Smaller SiPMs have been considered to limit the impact of the dark count rate to the single photoelectron spectrum. 
Sixteen Hamamtsu SiPMs  $1.3 \times 1.3~mm^2$ size and 50 $\mu m$ cells have been irradiated from \SI{1.03 e 09}{n/cm^2} to \SI{5.07 e 11}{n/cm^2}. 
%Fig. 4
\begin{figure}[h]
\begin{center}
\includegraphics[scale=0.60]{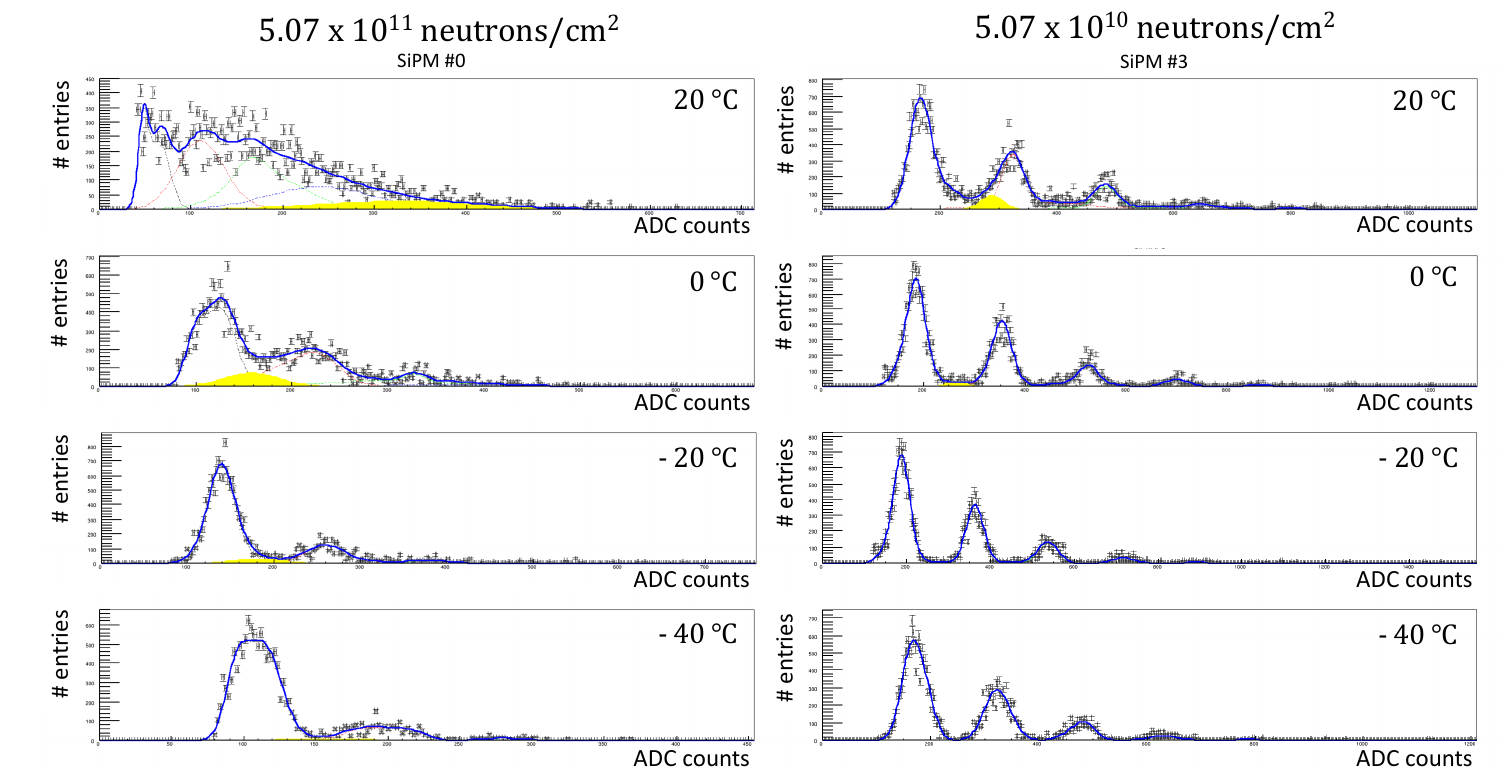}
\caption{Single photoelectron spectrum (number of entries vs signal amplitude in ADC counts) for an irradiation of \SI{5.07 e 11}{n/cm^2} (left) and  \SI{5.07 e 10}{n/cm^2} (right) at different temperatures. Signal peaks have been fitted with the convolution of a Poisson distribution with some double Gaussian functions (dashed lines), a Gaussian background (yellow) has been added to the fit.}
\label{fig:SinglePhoton}
\end{center}
\end{figure}
The single photon spectra are shown in Fig.~\ref{fig:SinglePhoton} for two different irradiations at different temperatures. In the most critical condition with \SI{5.07 e 11}{n/cm^2} and 20 \textdegree C the single photon peak cannot be distinguished from the dark count background.
From this preliminary test SiPMs operating at -20 \textdegree C is a possible solution also with a factor of 5 safety margin. However it should be noted that using  $1.3 \times 1.3~mm^2$ SiPMs will require  ~20 times more electronic channels 
or a multiplexing downstream of the SiPM amplification.

\subsection{TOP trigger upgrade}
The expected improvements in the SuperKEK instantaneous luminosity will require to tighten the readout timing windows in most detector subsystems. As the result, the TOP-based contribution to the online event time logic of the L1 trigger will become even more important. Significant backgrounds expected in the future call for novel machine learning and artificial intelligence algorithms to be implemented in the trigger firmware logic of the existing trigger boards UT4. We anticipate also to take full advantage of the new trigger board UT5 which is based on Versal FPGA architecture and is being developed at KEK. This work will also require to upgrade firmware of the trigger path in TOP FEE. Backend TOPTRG L1 work is further discussed in the TRG section of this CDR. 

\subsection{Design development and human resources}

The primary option for the photosensors is to replace the older ALD MCP-PMTs, when they reach their lifetime limit, with life-extended ALD MCP-PMTs. 

The design of the electronics for the readout of all the channels digitized by the new ASoCv4 is still to be defined. To properly estimate the power consumption reduction full details of the new boards are needed.

It is anticipated that  Nagoya U., INFN/U. Padova and INFN/U. Torino will work on the photosensor upgrade, while Indiana U, U. Hawaii, U. Pittsburgh and Nalu Scientific personnel will contribute to the electronics upgrade.

The TOP-based L1 trigger work in TOP FEE and in the TOP TRG system will be performed by U. Hawaii and U. Pittsburgh with possible contributions from other universities and laboratories.

\subsection{Cost and schedule}

It is anticipated that 1.5 FTE of engineering effort is required for the design of the carrier card board in years 1 and 2. In addition 1 postdoc and 1 graduate student will assist with assembly, test and documentation under the oversight of the U. Hawaii instrumentation frontier professor.
Board fabrication and assembly costs are based upon those realized during earlier prototyping. After the pre-production a yearly firmware development engineering support is costed during the 2 years of production, and at a reduced level during commissioning.
The estimated costs of the electronics upgrade are described in Table~\ref{iTOP:table1}. The estimated costs for the replacement of the 220 ALD MCP-PMTs are described in Table~\ref{iTOP:table2} for the life-extended ALD MCP-PMT option and in Table~\ref{iTOP:table3} for the SiPM option. 
The estimated cost of human resources for TOP-based TOP FEE and L1 trigger work is 0.5 FTE (from base funding). Estimates of the costs associated with the new trigger boards are discussed in the TRG section of this CDR. 
\begin{table}[htb]
\caption{TOP readout electronics upgrade cost estimate.}
\centering
\begin{tabularx}{\textwidth}{ | l | l | X | r | } \hline
\multicolumn{4}{|c|}{\bf ASoC ASIC readout Upgrade } \\ \hline
{\bf Phase}    & {\bf Resource}              & {\bf Basis of estimate} & {\bf Cost (k\$)} \\ \hline
Pre-production & Engineering \& test         & 1.5 FTE       & 610 \\ 
               & Parts procur. \& assembly   & cost summary  & 120 \\ \hline
Production     & Engineering \& test         & 1.5 FTE       & 610 \\ 
               & Parts procur. \& assembly   & cost summary  & 800 \\ \hline
System test    & Beam test         & past experience with IRSX   & 500 \\ \hline 
Commissioning  & Integration \& Installation & 1.0 FTE       & 445 \\ \hline
Subtotal       &     \multicolumn{2}{|r|}{}                  & 3085 \\ \hline
Contingency (20 \%) & \multicolumn{2}{|r|}{}                  & 617 \\ \hline
{\bf Total} & \multicolumn{2}{|r|}{}                  & {\bf 3702} \\ \hline
\end{tabularx}
\label{iTOP:table1}
\end{table}
\begin{table}[htb]
\caption{Life-extended ALD MCP-PMT upgrade cost estimate}
\centering
\begin{tabular}{ | l | l | l | r | } \hline
{\bf Phase} & {\bf Resource} & {\bf Basis of est.} & {\bf Cost (k\$)} \\ \hline
 Production & MCP-PMT Production & Engr. est. & 1500 \\  
           & MCP-PMT Test & Engr. est. & 15 \\
           & Assembly parts & Engr. est. & 50  \\  \hline 
 Commissioning & Integration & 0.5 FTE & 50 \\  \hline 
{\bf Total} & & & {\bf 1615} \\ \hline \hline
\end{tabular}
\label{iTOP:table2}
\end{table}
\begin{table}[htb]
\caption{SiPM upgrade cost estimate}
\centering
\begin{tabular}{ | l | l | l | r | } \hline
{\bf Phase} & {\bf Resource} & {\bf Basis of est.} & {\bf Cost (k\$)} \\ \hline
 Pre-production & Engineering & 0.5 FTE & 50\\ 
 & SiPM R\&D & Engr. est. &  10 \\ 
 & Cooling R\&D & Engr. est. & 50 \\ \hline
 Production & Engineering & 0.5 FTE & 50 \\  
 & SiPM + assembly parts & Engr. est. & 700 \\
 & Cooling system & Engr. est. & 100  \\  \hline 
 Commissioning & Integration & 0.5 FTE & 50 \\  \hline 
 Contingency (20 \%) &  &  & 202 \\  \hline 
 {\bf Total} & & & {\bf 1212} \\ \hline \hline
\end{tabular}
\label{iTOP:table3}
\end{table}

\subsection{TOP program summary}

For the photosensors the primary option is the replacement of normal ALD PMTs with life extended ALD MCP-PMTs once the installed PMTs approach the end of their lifetime.
With this option, we need ~220 new MCP-PMTs to replace normal ALD type PMTs, at an estimated costs of 1.6 M$\$$. The production of life-extended ALD MCP-PMT has already restarted, 90 MCP-PMTs have been ordered, 30 of them have been delivered. A further improvement in the ALD technology
and the MCP-PMTs replacement with cold SiPMs are the possible solutions if the increase of luminosity leads to a rapid loss of efficiency of the life extended ALD MCP-PMTs.

The rudimentary schedule and cost estimates for the electronic readout are based upon production and delivery of the 8k channels of TOP readout.  This endeavor leverages heavily two sizeable past and new investments: re-use of existing infrastructure and firmware, wherever possible and development of ASICs appropriate to the TOP upgrade.
In both cases the pre-production plus production timescale is approximately 2 years, with 6-9 months for installation and commissioning.  Based upon the preliminary prototyping being performed and costed separately from this, the total project cost is estimated to be 3.1 M$\$$ for upgrades to the TOP readout systems.

%%%%%%%%%%%%%%%%%%%%
\clearpage
\section{KLM}
\label{sec:KLM}
\editor{L.Piilonen, X.Wang, G.Varner, R. Peschke}

Two independent upgrades have been proposed for the Belle II 
KLM subdetector, which identifies $K_L$ mesons and muons.
The KLM is described briefly in Sec.~\ref{klm-geometry}.

The first upgrade adds the capability of energy measurement for the $K_L$
mesons via a time-of-flight measurement in new scintillator bars. This
upgrade would rebuild each detector panel and replace all existing resistive
plate chambers (RPCs) in the barrel with scintillators. The precision-timing
aspect of this upgrade is described in Sec.~\ref{klm-timing-upgrade}. At the
same time, the existing scintillator readout would be replaced by moving the
external analog and analog-to-digital readout elements into the detector panels,
simplifying immensely the external connections. This is described in
Sec.~\ref{klm-readout-upgrade}.

The second upgrade continues to use the existing RPCs, rather than replacing
them with scintillators, and operates them at somewhat lower voltage in
avalanche mode rather than in the extant streamer mode. The motivation here
is to improve the immunity to the ambient neutron background that would otherwise
lower the instantaneous efficiency of the RPCs. This upgrade is
described in Sec.~\ref{klm-avalanche-mode-upgrade}.

\subsection{KLM geometry and instrumentation}
\label{klm-geometry}

\begin{figure}[htbp]
\centering
\includegraphics[width=0.8\textwidth]{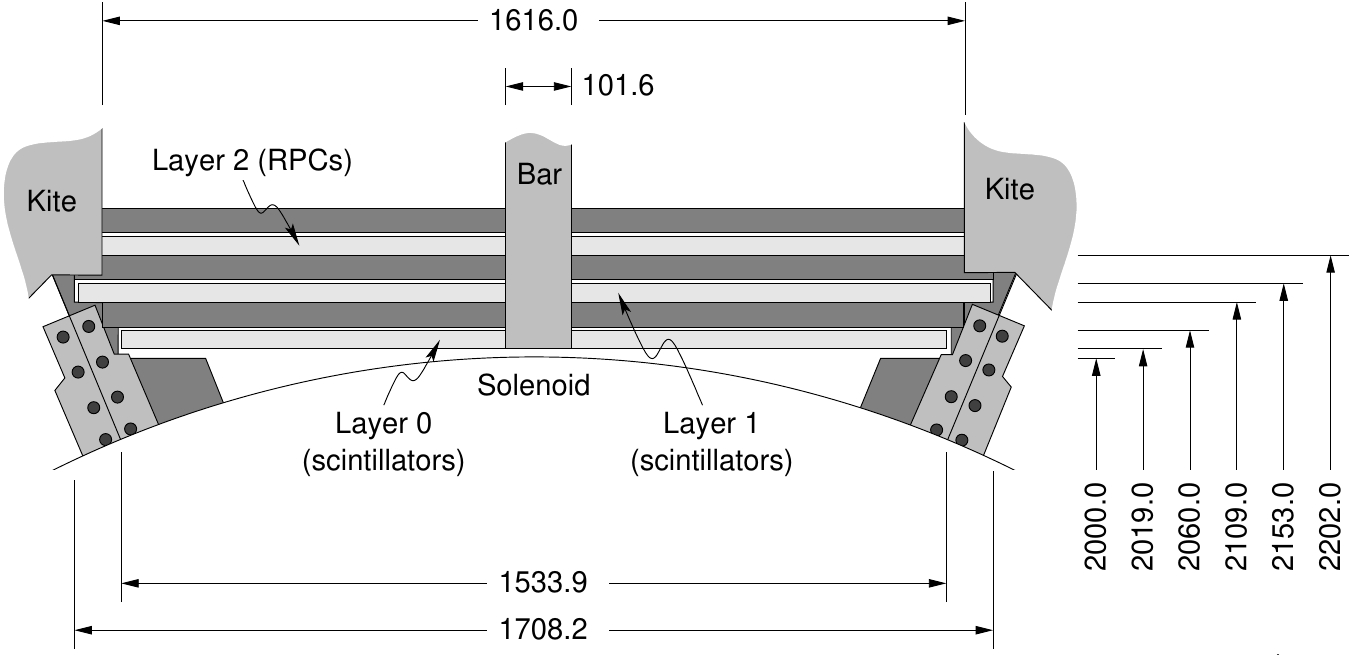}
\caption{End view of one octant of the KLM barrel, showing the thin detector
panels (light gray) sandwiched between the steel plates (dark gray) of the
solenoid's flux return. The
kites and bar are removable steel plates.}
\label{klm-barrel-endview}
\end{figure}

The KLM subdetector occupies the volume outside the superconducting 
solenoid, spanning radii of $200{}-{}340\cm$ in the octagonal 
barrel and $130{}-{}340\cm$ in the forward and backward 
endcaps. Large-surface-area detector panels of about 
3.1-cm thickness are sandwiched in the gaps between the 
steel plates of the solenoid's magnetic-field flux return (Fig.~\ref{klm-barrel-endview}). Each panel 
contains either RPCs (barrel layers 3--15) or scintillator strips (barrel 
layers 1--2, forward-endcap layers 1--14 and backward-endcap layers 
1--12). Each readout strip is about $4\cm$ wide, measured 
orthogonal to a line from the SuperKEKB $e^+ e^-$ collider's interaction 
point (IP); this matches the Moli\`ere radius of muons originating at the 
interaction point and traversing the electromagnetic calorimeter located 
just inside the solenoid. Each strip can be up to $2.7\m$ long.
Each detector panel provides a three-dimensional measurement of a
through-going particle's position via hits on two orthogonal readout strips
within the panel, constrained by the spatial location of the panel itself.
There are about 38,000 readout channels in the KLM, almost equally split
between RPCs and scintillators. 

Each RPC has two orthogonal planes of copper strips (on kapton sheets) that
image the streamer that forms within either---\textit{or both}---of the two
independently operated RPCs within the detector panel when a charged particle
passes through either RPC gas gap. Each strip acts as a transmission line that
delivers the imaged streamer-mode pulse (amplitude $\sim 200\,{\rm mV}$) to the panel's
readout end.

The cross section of each barrel scintillator strip is $4\cm \times 1\cm$, so
each panel in layers 1--2 contains 54 $z$-readout strips and between 38 and 45
$\phi$-readout strips, for a total of 92 to 99 strips. The cross section of
each endcap scintillator strip is $4\cm \times 0.7\cm$; each panel contains 75
$x$-readout strips and 75 $y$-readout strips for a total of 150 strips. Photons
from each scintillator strip are collected by an embedded wavelength-shifting
optical fiber (Kuraray Y11(200)MSJ) and detected by a Silicon PhotoMultiplier
(SiPM), also known as a Multi-Pixel Photon Counter (MPPC)---an avalanche photodiode
operated in Geiger mode. This solid-state device, the Hamamatsu S10362-13-050C,
is affixed directly to the scintillator strip at one end of the WLS fiber. The
fiber's other end is mirrored.

The RPC-strip and scintillator-MPPC signals (preamplified internally by a factor
of $\sim$10 for the scintillators) are routed via ribbon cables to external
readout electronics mounted on the magnet yoke.

\subsection{Upgrade path 1: precision timing for $K_L$ momentum}
\label{klm-timing-upgrade}

The motivations for pursuing the precision-timing upgrade path are twofold.
First, it improves our ability to use
timing information to reject out-of-time hits in the KLM arising from the ambient
neutron flux that would otherwise contribute to the background in any $B$-physics
or dark-matter related searches. Second, it adds the
capability to measure the four-momentum of $K_L$ candidates---not just the direction
as is presently done---so that $B$ or $D$ decay
modes with this particle in the final state can be added with a quality similar to
that attained in the modes with the $K_S$ meson. (In certain $CP$-sensitive studies,
the opposite $CP$ value for $K_L$ vs $K_S$ bolsters our confidence in the validity
of the measurements with the $K_S$ meson.)

There is no marked difference in the muon- or $K_L$-identification performance
measures for the existing RPCs and scintillators. Both are able to measure the
direction of the line joining the interaction point (IP) to a reconstructed hadronic
cluster to about $2\degrees$ in the polar and azimuthal angles (assuming that the
cluster was created by a parent $K_L$ meson that originated at the IP).

The precise hit-time measurement can be used to veto any hit that is
more than $\pm1\ns$ out of coincidence with the event's electron-positron collision
time, which would be
an improvement over the present $\pm 10\ns$ coincidence-window capability. This
would permit better rejection of out-of-time background hits that would otherwise
contaminate the prompt $K_L$-candidate population or any dark-matter-like event.

Additionally, better scintillators and readout electronics would provide
a precise time-of-flight measurement for the KLM clusters that arise from hadronic
interactions of a $K_L$ meson in either the yoke steel or in the ECL (with shower
leakage into the KLM). This time-of-flight measurement requires both the start time
of the electron-positron collision at the IP (known to better than $32\ps$) and the stop time at the point of the
$K_L$ hadronic interaction (or debris) in the KLM. This flight time and the
flight length from the IP can be converted to the
candidate $K_L$ meson's momentum magnitude, which is not measurable in the present
KLM. Combining this with the already measurable direction, the four-momentum of
this hadron can be determined. A stop-time resolution of $\delta t = 100\ps$ and a
flight length of $2\m$ from the IP provides a momentum resolution of $\delta p = 
0.19\gevc$ for a $K_L$ with a momentum of $1.5\gevc$, i.e., a fractional
resolution of $\delta p/p \approx 13\%$. The new ability to measure the four-momentum
improves the momentum resolution of the parent particle in physics modes with a
$K_L$ meson in the final state, such as $B^0 \to J/\psi K_L$, and reduces the
accidental background in such modes.

With several years of R\&D in hand, it is found that the combination of 
pure scintillators with long attenuation length \textit{and without an embedded 
wavelength-shifting fiber}, large-area photosensors, and custom preamplifiers 
with improved timing can achieve the aforementioned stop-time resolution. This R\&D
program, elaborated in the following subsections, encompasses the evaluation of
new large-area photosensors from Hamamatsu and NDL (Beijing) coupled to new
preamplifiers tailored for SiPM readout, with one or more SiPMs attached to a
single scintillator strip (up to $1\m$ in length); the Geant4 simulation of each
geometry variant (for different scintillator thicknesses and arrangements); the
assessment of various algorithms to extract the stop time
of the collection of hits within a cluster; and the resultant timing and momentum
resolution for the associated $K_L$ meson. 

\subsubsection{New SiPM and preamplifier}

\begin{figure}[htbp]
\centering
\includegraphics[width=0.8\textwidth]{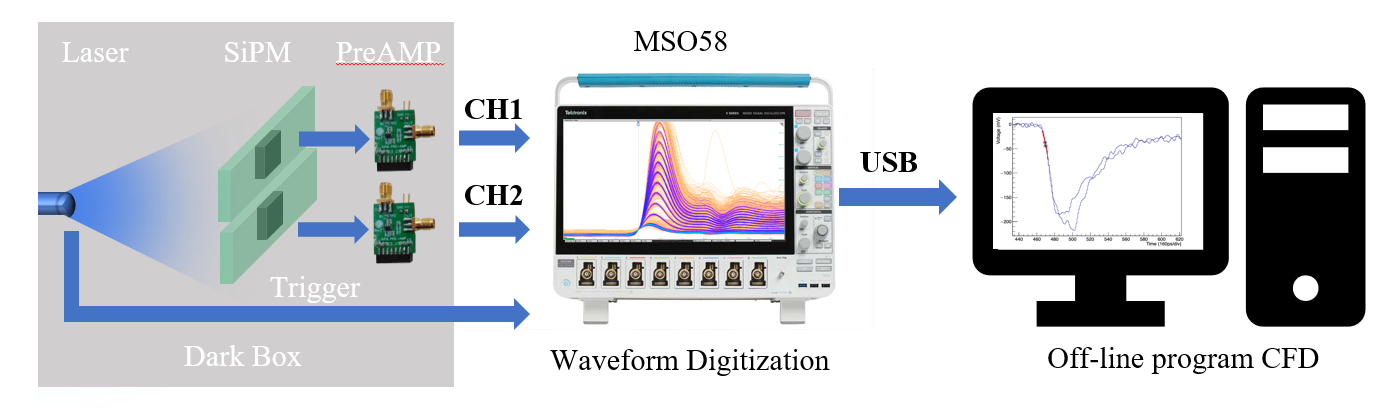}\\
\includegraphics[height=4.0cm]{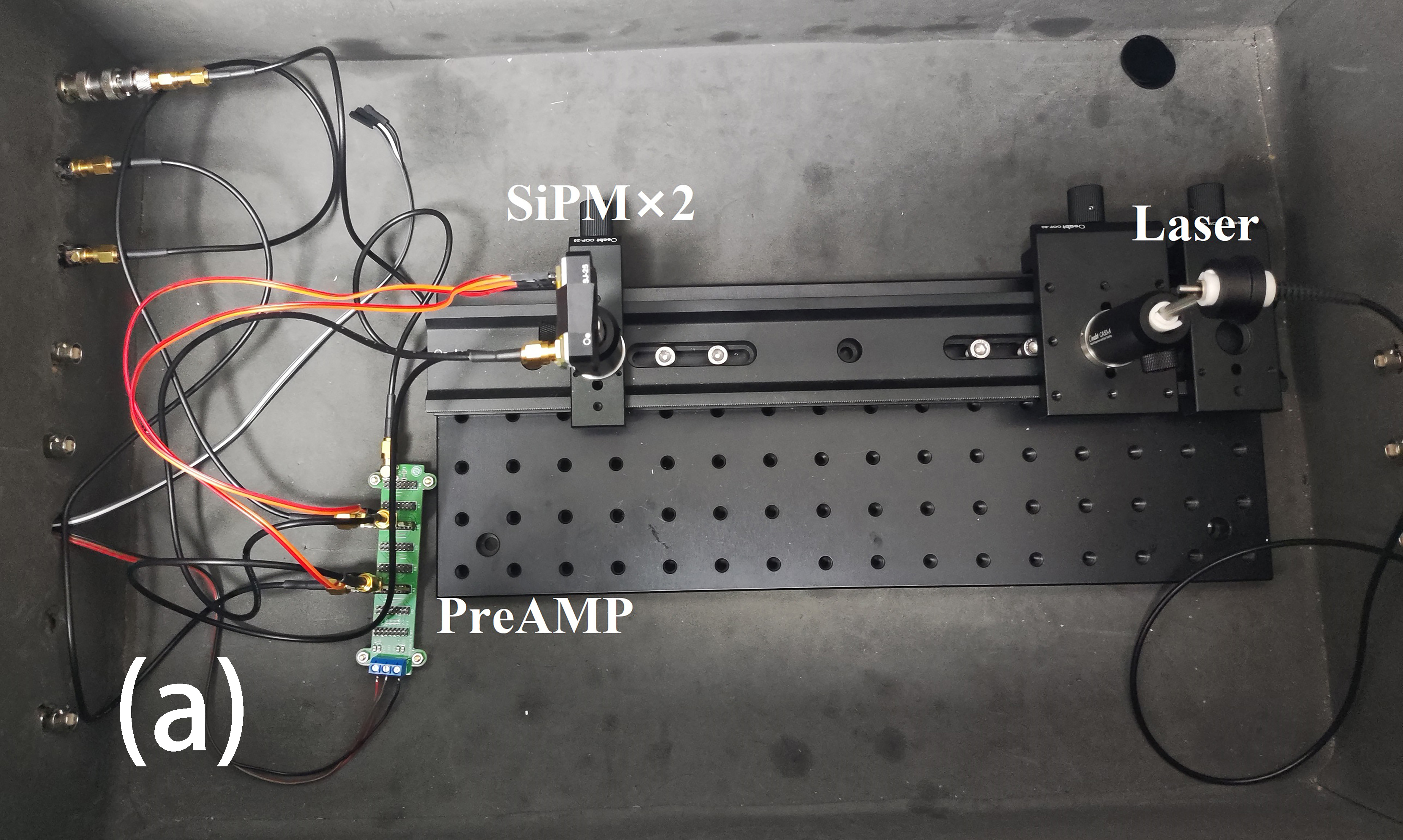}
\includegraphics[height=4.0cm]{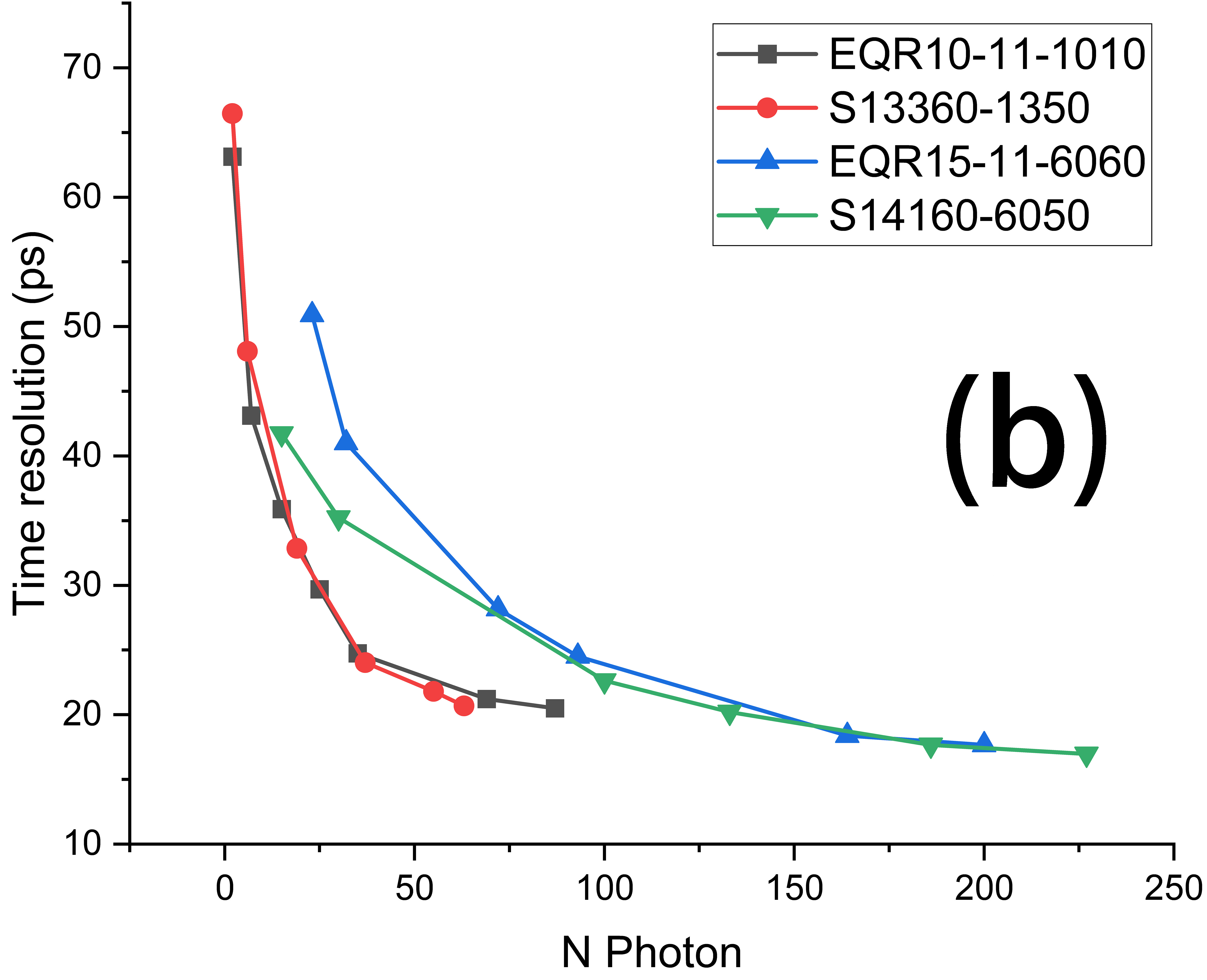}
\caption{Measurement of the intrinsic time resolution for SiPM-preamplifier
combinations. The top schematic shows the diagram of the apparatus while panel
(a) is a photograph of the interior of the dark box containing the laser, 
the SiPMs and the preamplifiers. Graph (b) shows the measured time resolution
versus the number of detected photons for several SiPMs.}
\label{reso_comb}
\end{figure}

The new $6\times 6\mma$ Hamamatsu and NDL SiPMs (S14610-6050HS and EQR15
11-6060D-S, respectively) are combined with a novel high-speed and low-noise
preamplifier~\cite{klm:preampSiPM}. This preamplifier has good gain stability, a
bandwidth of 426 MHz, a baseline noise of $\sim${}$0.6\,{\rm mV}$, and
a dynamic range for the input signal of up to $170\,{\rm mV}$. Figure~\ref{reso_comb}
shows the R\&D setup for studying the intrinsic time resolution for various SiPM
and preamplifier combinations illuminated directly by short laser pulses. A time
resolution of about $20\ps$ per cluster has been achieved for clusters with 50
or more pixels (i.e., detected photons). Pole-zero cancellation in the preamplifier
shortens the long tail of a typical SiPM pulse to about $5\ns$ so that no pile-up
is observed for laser-pulse frequencies up to 20 MHz.

\subsubsection{Time resolution of long scintillator bar}

The precision of the time-of-flight measurement for a $K_L$ candidate is improved
with more struck scintillators in its hadronic cluster. This requires multiple layers with
pure scintillators having very long attenuation length, rather than the extruded
scintillators with embedded wavelength-shifting fibers presently in use in the KLM.

\begin{figure}[htbp]
\centering
\includegraphics[width=1.0\textwidth]{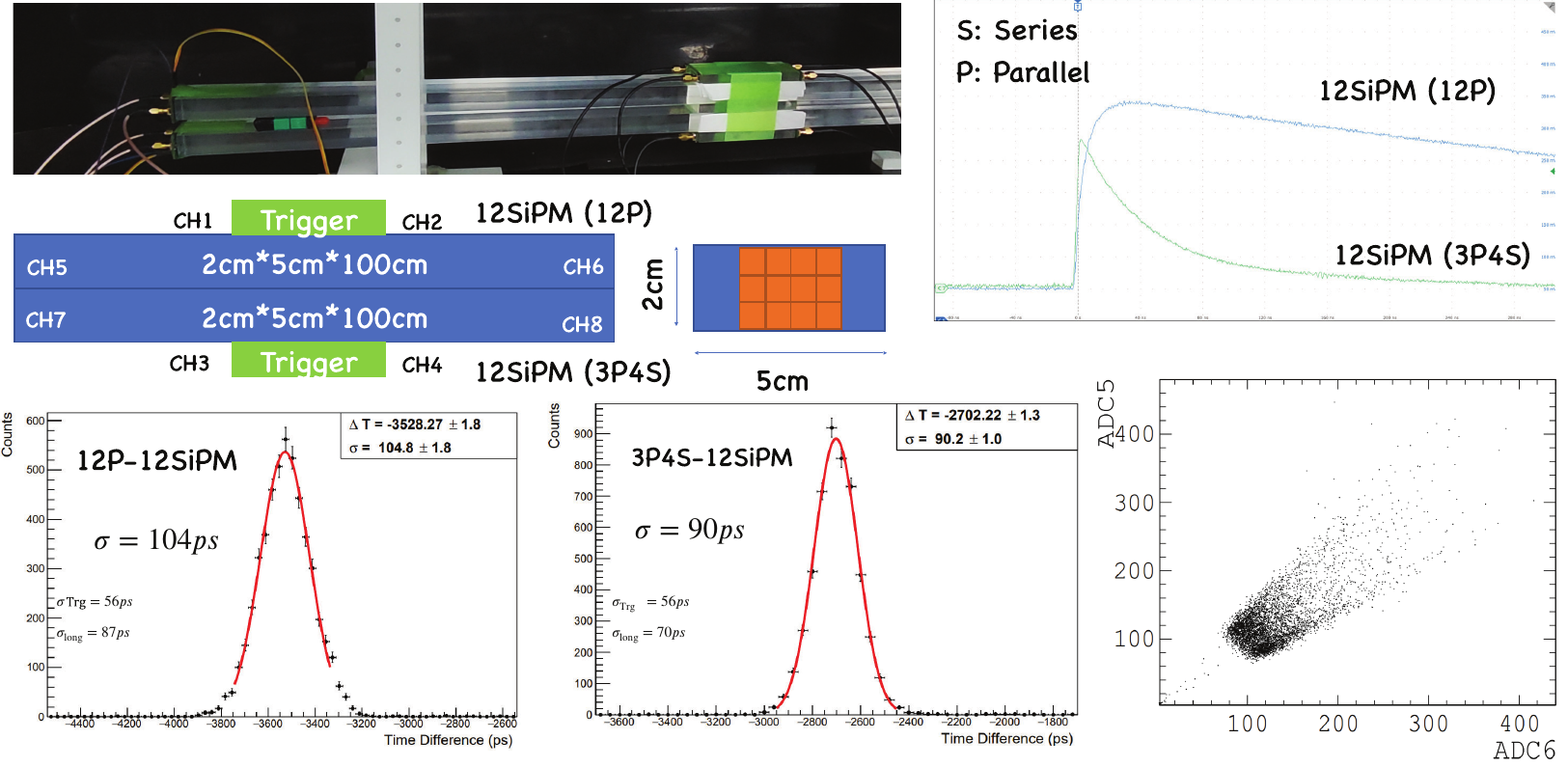}
\caption{Cosmic-ray test of two GaoNengKeDi $1\m$ long scintillators. Top left
shows the photograph and schematic of the apparatus, with a $3\times 4$ array
of SiPMs at each end of a scintillator. Top right shows typical signal pulses for
this (3P4S) and the parallel (12P) SiPM configurations. Bottom left graphs show the
time distributions and extracted resolutions for these two configurations. After
unfolding the trigger-scintillator resolution, the time resolutions are
$70\ps$ (3P4S) and $87\ps$ (12P). Bottom right scatterplot shows the range of
observed pulse heights at the two ends of a scintillator.} 
\label{fig:klmtof}
\end{figure}

Pure-scintillator samples from Bicron (BC420) and GaoNengKeDi are tested.
Figure~\ref{fig:klmtof} shows one test with two GaoNengKeDi strips ($2\cm
\times 5\cm \times 1\m$) and arrays of 12 SiPMs at the two ends of each strip.
The 12 SiPMs are arranged in parallel (``12P'') or parallel+series (``3P4S'') on a printed circuit board. Two
short scintillator strips ($10\cm$ length) are used to generate a
cosmic-ray trigger with a detection efficiency in the long strip of nearly 100\%.
SiPMs in series show a better time resolution of about $70\ps$ but with a smaller
pulse height compared to SiPMs in parallel. A hybrid arrangement of the SiPMs
shows a good balance between time resolution and pulse height.

To maintain the required timing precision over time, a robust calibration
protocol with frequent updates each day will be necessary for the TOF-like KLM.
This requires an integrated laser-calibration infrastructure. A cosmic ray test
with six strips ($\delta t \sim 130\ps$ per strip), calibrated with a laser split
among the six channels, yields a speed distribution of $(30.0 \pm 3.4)\cm/\ns$ or 11\%
for throughgoing cosmic rays (moving at almost the speed of light).

\subsubsection{Geant4 simulation for the $K_L$ momentum determination}

\begin{figure}[htbp]
\centering
\includegraphics[height=4.0cm]{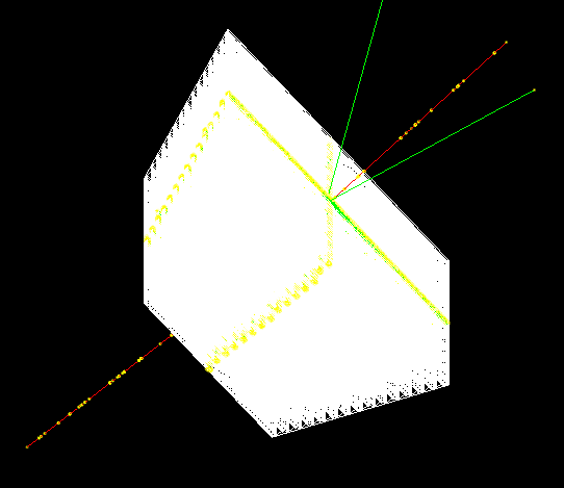}
\includegraphics[height=4.0cm]{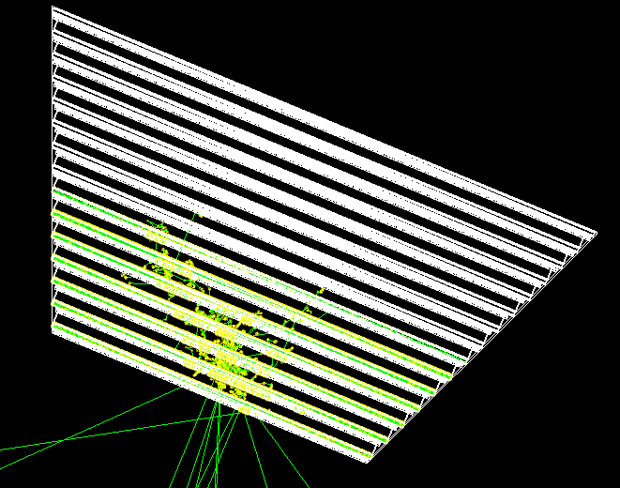}\\
\includegraphics[height=4.0cm]{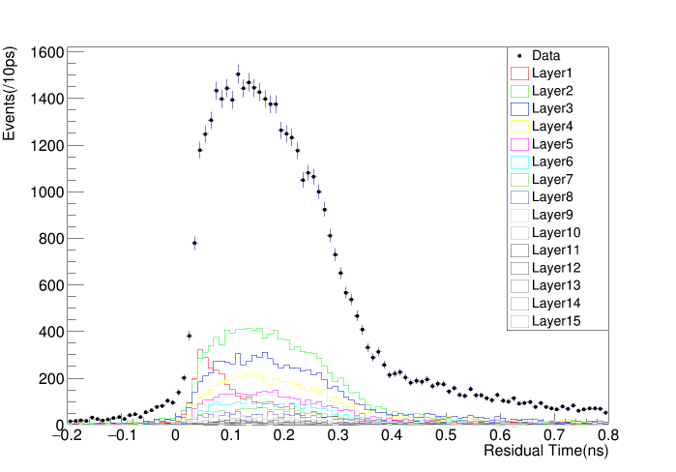}
\includegraphics[height=4.0cm]{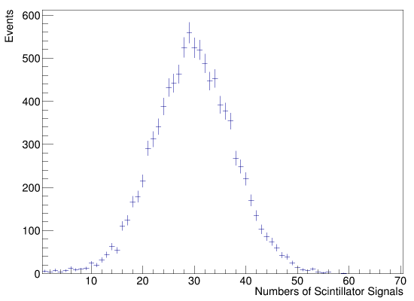}
\caption{Geant4 simulation results for one barrel octant of an upgraded KLM with
precision time-of-flight scintillators. Top left: response for one $1\gevc$
muon. top right: response for one $1\gevc$ $K_L$ meson. Bottom left:
time distribution of hits in each layer and cumulative for the $K_L$ cluster;
the resolution of $(95\pm 4)\ps$ is extracted from a fit to the leading-edge
(earliest) times in this graph. Bottom right: population distribution of
scintillator signals for the $K_L$ cluster.}
\label{klm_G4_simulations}
\end{figure}

One barrel octant of an upgraded KLM with 15 layers of precision-timing
scintillator panels is simulated with Geant4, as shown in
Fig.~\ref{klm_G4_simulations}. The active components are pure scintillators with
a long attenuation length and SiPM arrays at both ends of each scintillator strip.
The ECL is not included in this simulation. $K_L$ mesons and muons with a
momentum of $1\gevc$ are generated at the IP using the Geant4 particle gun. 
The figure shows the responses for a muon (clean track) and a $K_L$ meson (hadronic
cluster). A preliminary algorithm is developed to extract the  $K_L$ arrival time
from selected (earliest) hit times in the cluster and has a FWHM resolution of
about $240\ps$. A Gaussian fit to this leading-edge time distribution
yields $\sigma = (95 \pm 4)\ps$, consistent with the FWHM measure.

\subsubsection{Replacement of the RPC layers in the barrel}
\label{klm-rpc-replacement}
Concurrently with the upgrade of the innermost three barrel layers for the
precision-timing capability, it would be prudent to take this opportunity to
replace the RPCs in the outer twelve barrel layers with present-design
scintillator panels. The design of each panel would match
that of the existing scintillator panels in the two innermost barrel layers:
each $4\cm \times 1\cm$ scintillator strip is extruded with a central bore
for a wavelength-shifting optical fiber that captures the light generated
within the scintillator and transmits the wavelength-shifted light via total
internal reflection to one end that is instrumented with a single SiPM. A
protective and reflective TiO${}_2$-loaded polystyrene jacket is co-extruded
with the scintillator strip. 

\subsubsection{Improved scintillator front-end readout}
\label{klm-readout-upgrade}

Leveraging experience from the deployment and operation of the endcap and 
inner barrel KLM scintillators, there is a compelling reason to refactor 
the readout in a way that eliminates the many kilometers of twisted-pair 
ribbon cables. We propose a more compact readout placed inside each detector
panel that interfaces with the external DAQ system via a few optical fibers.
This readout upgrade would be implemented in conjunction with the TOF-like
KLM upgrade described earlier.

The new readout embeds the readout ASICs on circuit boards inside each new
detector panel, and incorporates a newly developed more compact SCROD
(Control and Readout) card that also resides inside the detector panel.
We propose to use the HDSoC ASIC, which has four times higher channel density
than the existing TARGETX. The HDSoC ASIC utilizes a high-speed serial
connection to the new SCROD FPGA. For the barrel KLM (15 layers), this new
readout system will consist of 480 HDSoC ASICs and 240 SCRODs. For the endcap
KLM (five new precise-timing layers at each end), this system will contain
120 HDSoC ASICs and 40 SCRODs.

\subsubsection{Materials cost estimate}

This cost estimate is based on a scenario in which the three innermost barrel
layers and five innermost layers in each endcap are replaced with precision-timing
scintillators while the RPCs in the outermost
twelve barrel layers are replaced with standard-design scintillators and SiPMs
(Sec.~\ref{klm-rpc-replacement}). All new detector panels would be
instrumented with the new readout described in Sec.~\ref{klm-readout-upgrade}.
Labor cost is not included here since some of the personnel might already be
on an institution's payroll.

Table~\ref{klm-cost-table} itemizes the estimated material costs for each element
of this upgrade. 
The precision-timing scintillators would cost \$0.3M for the three barrel
layers and \$0.5M for the ten endcap layers,
based on material costs in China. With an array of eight NDL SiPMs
at each end of a scintillator strip,
the photosensors would cost an additional \$0.8M for the barrel
and \$1.0M for the endcaps. Customized precision-timing
preamplifiers (one preamplifier at each end of a scintillator strip) would cost an
additional \$0.3M for the barrel and \$0.4M for the endcaps.
Contingency is not included here.

The replacement of the RPCs in the outermost twelve barrel layers with
current-design scintillator panels would cost about \$3.4M, based on the actual
costs for all of the components (strips, WLS fiber, photosensors, frame and
passive materials, HV, shipping) and escalation for the greater geometric size
and inflation at 3\% per year. This includes a 20\% contingency.

The associated front-end readout upgrade is estimated to cost between \$1.4M
(utilizing the HDSoC ASIC with internal amplification) and \$1.8M (if discrete
amplifiers are deployed instead). Development and prototyping of the new ASIC
and SCROD would add an estimated \$1.6M to the total readout project cost.

In summary, the rough estimate of the material cost of this upgrade is \$9.6M.

\begin{table}[hbt]
\caption{\sl KLM Upgrade material cost estimates. This assumes that the items
for the TOF upgrade would be procured in China while the items for the RPC-replacement
upgrade and the readout upgrade would be procured in the US.}
\centering
\begin{tabular}{|l|r|r|r|} \hline
 & Quantity & Unit cost & Cost (k\$)\\
\hline \hline
\textbf{Barrel TOF Upgrade (layers 1--3)} & & &\\
\quad Scintillator volume & 5.724\thinspace m${}^3$ & \$55000/m${}^3$ & 315 \\
\quad Photosensors (16/scintillator strip) & 75008 & \$10/item & 750 \\
\quad Preamplifiers (2/scintillator strip) & 9376 & \$30/item & 281 \\
\hline
\textbf{Endcap TOF Upgrade (layers 1--5)x2} & & & \\
\quad Scintillator volume & 9.302\thinspace m${}^3$ & \$55000/m${}^3$ & 315 \\
\quad Photosensors (16/scintillator strip) & 96000 & \$10/item & 960 \\
\quad Preamplifiers (2/scintillator strip) & 12000 & \$30/item & 360 \\
\hline
\textbf{Barrel RPC-to-scint (layers 4--15)} & & &\\
\quad Scintillator volume & 15.686\thinspace m${}^3$ & \$34000/m${}^3$ & 535 \\
\quad WLS fiber length & 2900\thinspace m & \$133/m & 386 \\
\quad Photosensors (1/scintillator strip) & 22856 & \$44/item & 1005 \\
\quad Al frames, HDPE sheets, etc & 192 panels & \$2700/item & 520 \\
\quad Shipping (US to KEK) & 28 crates & \$7000/item & 196 \\
\quad HV modules (CAEN A1510) & 16 units & \$10750/item & 172 \\
\quad HV mainframes (CAEN SY2527) & 2 units & \$20000/item & 40 \\
\quad Contingency (20\%)& &  & 571 \\
\hline
\textbf{Readout Upgrade} & & &\\
\quad Parts for above layers & & & 1400 \\
\quad HDSoC-related R\&D & & & 1600 \\
\hline\hline
\textbf{ESTIMATED MATERIAL COST} & & & \textbf{9603} \\
\hline
\end{tabular}
\label{klm-cost-table}
\end{table}

\subsubsection{Upgrade logistics}

This upgrade requires the roll-out of Belle II from the interaction region
and the opening of the forward and backward endcap doors to gain access to the
barrel and endcap KLM detector panels. In the roll-in position, the base of
the beamline-supporting concrete shield prevents access to the bottom three
sectors of the barrel KLM at the forward and backward ends. 

If all existing barrel detector panels (backward \textit{and} forward)
are removed from the forward end and the
new panels are inserted from this end, then all other-subsystem cables and
services that reside in the service channels at the forward face of the
barrel yoke would have to be peeled away
to permit these panel insertion/extraction operations, and then re-installed
at their conclusion. In this forward-access scenario, the other-subsystem
cables and services at the backward barrel yoke would not be disrupted.
The KLM cables and services at both ends of the barrel yoke can be
disconnected and reconnected without touching the other subsystems' cables
and services. The endcap KLM sectors can be accessed without disturbing
the other subsystems' cables and services.

We estimate that it would take one calendar year to accomplish these steps,
including the time needed during the new-panel installation for testing and,
if necessary, remediation.

\subsection{Upgrade path 2: RPC avalanche mode}
\label{klm-avalanche-mode-upgrade}

The second upgrade path does not replace any of the existing KLM detector panels
but instead addresses the inherent rate limitation of the barrel RPCs and their
related blinding by the ambient neutron flux in the KLM during high-luminosity
operation in the coming years. In this upgrade, the present streamer mode of RPC
operation is replaced with the avalanche mode. In this way, the blinding from the
localized discharge of the glass-plate electrodes by the streamer is avoided, since
the gas-gap electric field will not be reduced so dramatically by the avalanche
induced by an ambient neutron's interactions with the nearby material. Since the
imaged pulse height is reduced by a factor of about 10, the avalanche-mode
pulses will require new in-line preamplifiers at the detector-panel faces at the
forward and backward barrel yoke.

There are significant uncertainties in the expected particle flux at the design
luminosity and in the RPC efficiency as a function of particle flux. Operating the
RPCs in avalanche mode mitigates the dual risks of a higher-than-expected
background neutron flux and/or a worse-than-expected rolloff of the RPC efficiency
due to blindness from the high flux. 

%\begin{figure}[htbp]
%\centering
%\includegraphics[width=0.8\textwidth]{fig/KLM/RPC_rate_capability.png}
%\caption{TBD}
%\label{klm:RPC_rate_capability}
%\end{figure}

%\begin{figure}[htbp]
%\centering
%\includegraphics[width=0.8\textwidth]{fig/KLM/RPC_rate_capability2.png}
%\caption{TBD}
%\label{klm:RPC_rate_capability1}
%\end{figure}

%\begin{figure}[htbp]
%\centering
%\includegraphics[width=0.8\textwidth]{fig/KLM/RPC_Efficiency_vs_flux.png}
%\caption{TBD}
%\label{klm:RPC_Efficiency_vs_flux}
%\end{figure}

\subsubsection{RPC response in streamer and avalanche modes}
\label{klm-avalanche-response}

Here, we explore the shifts in detector response when operating the KLM RPCs
in avalanche mode. We also address the necessary modifications of the front-end
electronics, gas mixture, and operating voltage.

The RPC rate capability is inversely proportional to the average charge
$\langle Q\rangle$ produced in the gas gap by an incident particle 
\cite{klm:ImprovingTheRPCratecapability}:
$$\text{Rate Capability} \propto \frac{1}{\rho t\langle Q\rangle}$$
where $\rho$ is the bulk resistivity and $t$ is the total thickness of the 
electrodes. Given that the glass electrodes will not be replaced, this rate
capability can only be increased by reducing $\langle Q\rangle$. This approach
is constrained by the existing noise levels within the system and the sensitivity
of the readout electronics. We note that
the charge received by the front-end electronics is  significantly smaller than
the $\langle Q\rangle$ formed in the gas gap. In our RPCs, the far end of each
cathode strip is terminated to avoid pulse reflections: this implies a factor
of two reduction in the signal at the preamplifier.

\begin{figure}[htbp]
\centering
\includegraphics[width=0.8\textwidth]{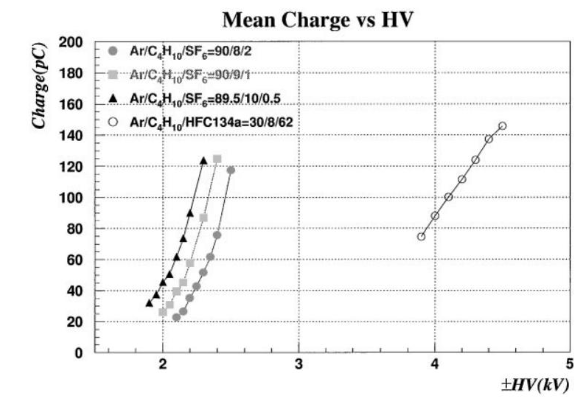}
\caption{Mean charge $\langle Q\rangle$ for the Belle-style RPCs operated in
streamer mode with various gas mixtures and over a range of high-voltage
settings.}
\label{klm:Charge VS HV}
\end{figure}

The KLM's glass-electrode RPCs, originally built for the first-generation Belle
experiment, have undergone thorough examination and analysis~\cite{klm:glassRPCSF6, 
klm:YAMAGA2000109, klm:2000Abe, klm:2000KLMdetectorsubsystem, klm:2002freonlessoperation}. The electrodes surround a $2\mm$ gas gap.
The $\langle Q\rangle$ values are depicted in Fig.~\ref{klm:Charge VS HV} for
streamer-mode operation. In these papers, it was established that the typical
streamer discharge has $\langle Q\rangle \sim \mathcal{O}(100\,{\rm pC})$.

This value is comparable to those in ATLAS for $2\mm$ gas gap RPCs with
Bakelite electrodes operating in streamer mode with various gas mixtures and
a range of high voltages:  
$\langle Q\rangle \sim 30\,{\rm pC}$~\cite{klm:ATLASStreamersuppression}, 
$70{}-{}120\,{\rm pC}$~\cite{klm:ATLASwithCO2} and 
$50{}-{}60\,{\rm pC}$~\cite{klm:ATLASwithEfriendlyGas}.

\begin{figure}[htbp]
\centering
\includegraphics[width=0.8\textwidth]{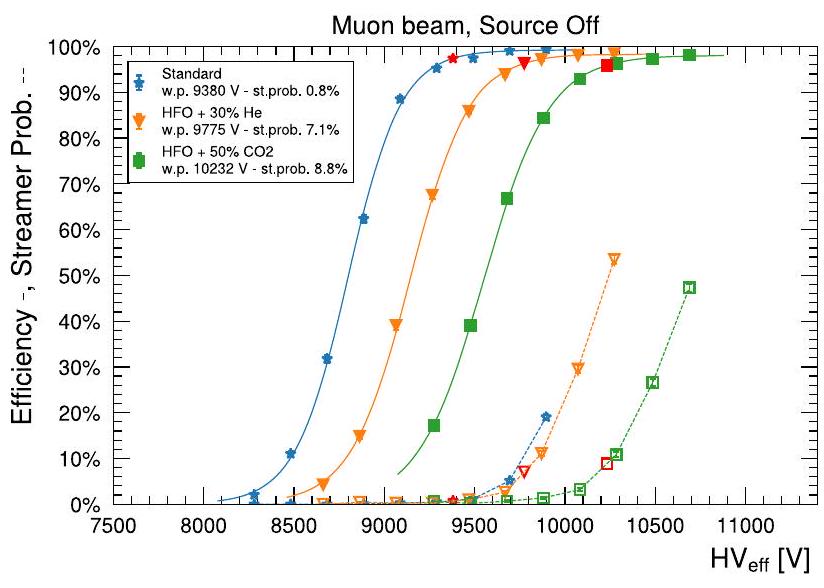}
\caption{Efficiency (solid) and streamer-probability (dashed) curves for the
standard gas mixture and two selected HFO-based gas mixtures in ATLAS RPCs
operated in avalanche mode with a muon beam and no background radiation.}
\label{KLM:Effieciency_streamer_prop_vs_HV}
\end{figure}

In the past two decades, it has become a prevailing practice to operate RPCs in
avalanche mode, wherein charge generation and amplification in the gas gap is
quenched during the avalanche phase, avoiding progression to the streamer phase.
This can be achieved through the use of an electronegative gas like SF${}_6$ as
a quenching agent and by reducing the operating voltage. As a result, avalanche
events are distinguished from streamer events by their notably smaller
$\langle Q\rangle$ and distinct timing characteristics. For the ATLAS
RPCs operated in avalanche mode~\cite{klm:ATLASStreamersuppression}, a range of
voltages with high efficiency \textit{and} low streamer probability (for each gas
mixture) can be seen in Fig.~\ref{KLM:Effieciency_streamer_prop_vs_HV}.

The mean charge in the ATLAS RPCs operated in proportional mode is
$\langle Q\rangle \sim 9\,{\rm pC}$ with these gas-mixture and HV modifications.
This has been reduced further to $\langle Q\rangle \sim 2\,{\rm pC}$ by the
deployment of a novel preamplifier, as detailed in Sec.~\ref{klm-avalanche-readout}.

\subsubsection{Gas mixture}
\label{klm-gas-mixture}

Currently, the RPCs use a gas mixture consisting of 30\% argon, 
8\% butane, and 62\% HFC-134a. Various studies have explored 
alternative gas mixtures with similar glass RPCs, revealing a noteworthy 
shift toward lower signal charge while maintaining high efficiency. In 
particular, the introduction of SF${}_6$, characterized by its high electron 
affinity of \qty{1.05}{\eV}, appears to play a crucial role in avoiding
streamer formation.

Ref.~\cite{klm:glassRPCSF6} primarily focuses on glass RPCs that closely resemble
ours. It conducts a comparative analysis of alternative gas mixtures, each of 
which contains SF${}_6$ and excludes HFC-134a. This study reveals that the overall
efficiency is only slightly lower than that of the current mixture, and, notably,
indicates a reduction in $\langle Q\rangle$ from $80\,{\rm pC}$ to
$25{}-{}40\,{\rm pC}$. Notably, these test RPCs had a single gas gap, in contrast
to the two independently-operated RPCs within one KLM detector panel. Thus, the
equivalent two-gap efficiency of the test RPCs would have been $\varepsilon_2
= 1 - (1 - \varepsilon_1)^2$. Recent advancements in front-end electronics might
enhance the overall performance significantly.

The R\&D task here is to find a suitable gas mixture and operating point for
the high voltage for our RPCs in avalanche mode to ensure high efficiency and
low streamer probability.

A key obstacle in identifying the suitable gas mixture revolves around environmental considerations, with particular attention paid to the potential introduction of SF6, boasting a significantly higher GWP of 23,500 in contrast to HFC-134a, which has a GWP of 1,300 ~\cite{klm:IPCCRadiativeForcingMyhre2013}. Nonetheless, the research presented in Ref.~\cite{klm:glassRPCSF6} proposes a gas mixture that would replace 62\% of HFC-134a with just 2\% SF6, resulting in a noteworthy reduction in GWP emissions to the atmosphere. Regardless of the specific gas mixture employed, the collaboration is actively exploring the implementation of a gas recirculation system.

\subsubsection{Readout electronics}
\label{klm-avalanche-readout}

A corollary of the switch to avalanche-mode operation, with its much smaller
$\langle Q\rangle$, is the need for front-end preamplification of the signals as
close as feasible to their sources. The new preamplifiers would be placed at or
near the ribbon-cable connectors at the forward and backward ends of the barrel
yoke. The preamplified signals would then be routed along the existing meters-long
ribbon cables to the digitization front-end electronics located
on the magnet-yoke periphery.

The readout end of the $\phi$-measuring strips inside the detector panel are very
close to these ribbon-cable connectors. In contrast, the $z$-measuring strips are
located between $4$ and $220\cm$ from these connectors and so their signals are
propagated along interior ribbon cables to the external connectors. Any noise
pickup and/or crosstalk on these internal cables would be unavoidable.

The initial focus of our R\&D will involve measuring the existing electronic noise
levels for the $\phi$- and $z$-measuring strips when operating selected \textit{in situ}
RPCs in
avalanche mode. The signal height is expected to be approximately four to five
times smaller than that in streamer mode.

With a robust characterization of the avalanche-mode signal and noise pulses,
subsequent investigations will determine the options for the signal-processing
chain, including at least a preamplifier stage but also, perhaps, incorporating
the digitization step into a unified readout solution.
Several options are under consideration: preamplification plus continued use of
the existing binary digitizers; integrated preamplification and binary digitization;
preamplification and fine-grained pulse-height digitization (either integrated
or separate); and pramplification with waveform digitization as is done in the
present scintillator front-end readout (either integrated or separate). The
challenge in each option is to find sufficient volume in the constricted and
fully-allocated space at each end of the barrel yoke to add the preamplifiers. 

\begin{figure}[htbp]
\centering
\includegraphics[width=0.8\textwidth]{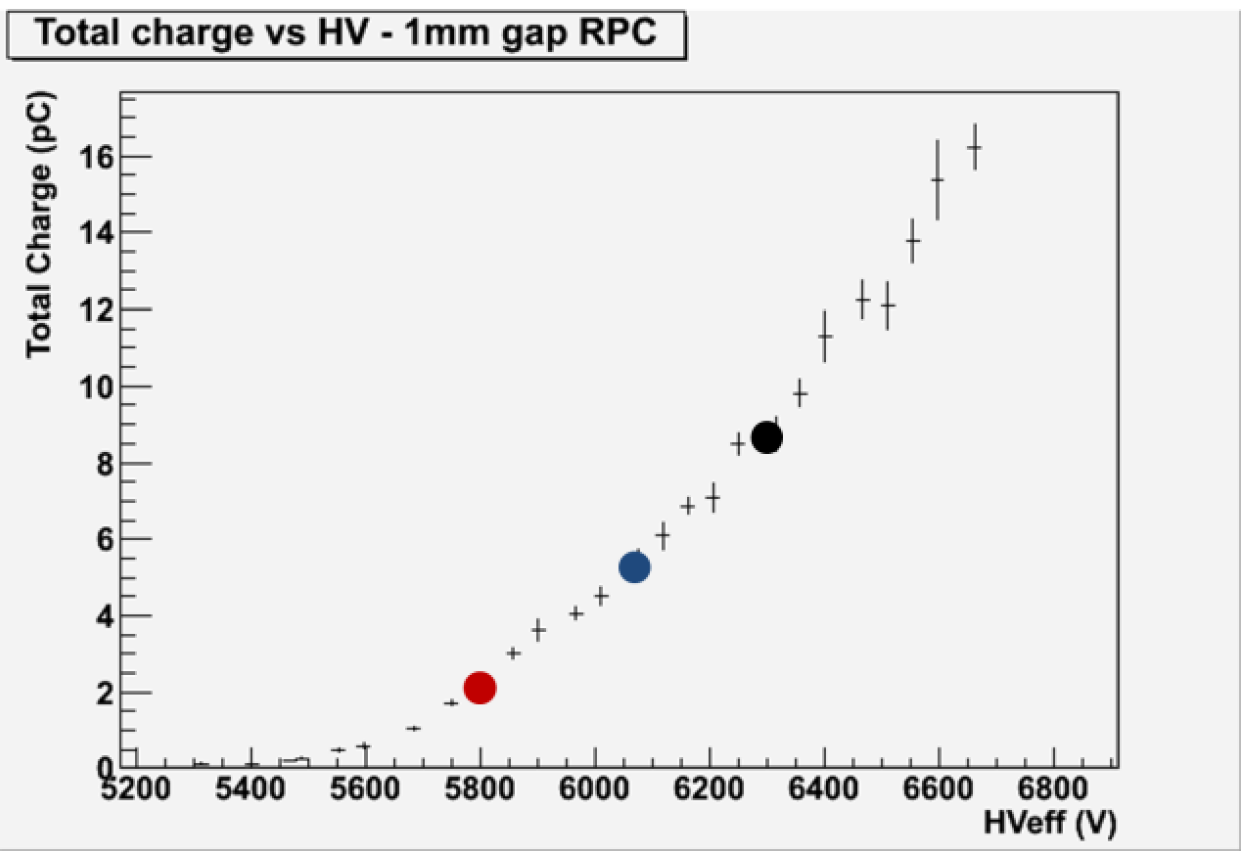}
\caption{Total Charge versus Effective High Voltage (HVeff), highlighting the point at 90\% efficiency. Depicted in the graph are different data sets: black represents No Front End (FE), blue represents Silicon Front End (Si FE), and red represents Silicon Germanium Front End (SiGe FE). \cite{klm:Cardarelli2013} }
\label{klm:RPC_Total_charge}
\end{figure}

Fig. \ref{klm:RPC_Total_charge} from the ATLAS collaboration demonstrates the 
performance of their newly designed SiGe preamplifier 
\cite{klm:Cardarelli2013}, indicating that it achieves a 95\% efficiency 
with a signal height as low as $\langle Q\rangle = 2\,{\rm pC}$, significantly
improving upon the earlier solution that required $9\,{\rm pC}$ and resulting
in a markedly higher rate capability.

The KLM RPCs exhibit a bulk resistivity approximately three orders of magnitude
higher than the ATLAS RPCs, resulting in a rate capability roughly three orders
of magnitude lower. Nevertheless, with a similar upgrade approach, we have the 
potential to increase the KLM rate capability to approximately $10\,{\rm Hz/cm}^2$, a 
level that adequately accommodates the expected background rates at design 
luminosity, even when considering the uncertainties discussed in the introduction.

\subsubsection{R\&D}

Upgrading the current RPC system of the KLM detector presents a 
significant challenge, particularly as numerous detector components are 
immovable and unalterable. Consequently, it is essential to characterize the 
existing electronic noise within the system to determine the detailed design of
a preamplifier that would operate well with the RPCs in avalanche mode. A few
spare unused full-size RPC detector panels are available for these R\&D studies,
including investigations of optimal gas mixtures with lower-voltage operating
points and evaluation of prototype pre-amplifiers. If necessary, we could use
compact radioactive sources to create a radiation environment with an incident
flux of $1{}-{}10\,{\rm Hz/cm}^2$. Nevertheless, dedicated test-beam studies
will be required for comprehensive assessments.

The RPC avalanche mode upgrade encompasses several key areas of research and development (R\&D). Firstly, the investigation into the detector itself is crucial, involving studies on various gas mixtures to validate results referenced in literature. To accomplish this, a dedicated laboratory setup using spare RPCs is essential, allowing comprehensive testing under diverse gas mixtures, bias voltages, and radiation environments. Another significant area of study focuses on the front-end electronics, as reducing the charge generated per event might render signals too weak for current digitization cards. Therefore, exploration of low-noise preamplifiers becomes necessary, inspired by successful implementations within the ATLAS Collaboration. Determining the parameters for future RPC preamplifiers is pivotal in ensuring compatibility with existing digitization cards. Additionally, investigating the detector's gas system is vital, considering the current venting of detector gas into the atmosphere. Developing a gas recirculation system, potentially inspired by practices in the ATLAS Collaboration, might be necessary based on the determined gas mixture, aiming to enhance efficiency and mitigate environmental concerns.

\subsubsection{Cost estimation and risk assessment}

At its early stages, this upgrade proposal heavily relies on substantial R\&D efforts for viability. Compared to the alternative proposal focusing on entirely new BKLM detectors, this approach primarily emphasizes upgrading the existing BKLM detector, making it notably more cost-effective. Precise cost estimations are challenging without further R\&D, but significant material costs outlined in the previous chapter, around 9.6 M\$, would be eliminated. The proposed hardware additions primarily include preamplifiers and potentially a gas recirculation system, only if it proves more cost-effective than the current venting setup. (We vent all RPC gas to the atmosphere: during beam operation, this includes 270\thinspace kg/month of HFC-134a, a greenhouse gas with a global warming potential of 1300, compared to 1 for $\rm CO_2$ and 23500 for $\rm SF_6$.) This indicates potential operational cost reduction compared to the current detector configuration, making it a promising avenue for upgrade.

For a preliminary cost estimate, it is assumed that each preamplifier costs approximately \$30, and there is an estimated need for preamplifying roughly 16,000 channels. This results in a cost estimate of around \$0.5 million.

The risks associated with this upgrade can be categorized into two parts. Firstly, there's the concern regarding the feasibility of the upgrade itself. Early-stage research and development (R\&D) can help determine the feasibility of the upgrade. In case it is not achievable, the possibility remains to continue operating with the current setup or, if there's sufficient time, switch to a full scintillator readout.

Secondly, the risk of damaging the subdetector or adjacent subdetectors exists. Access to the detector itself is minimal, reducing the likelihood of damaging adjacent subdetectors. Introducing a new gas mixture to the RPC chambers poses a potential risk, but this can be mitigated by careful consideration during the R\&D phase.

Overall, the risk of ending up with a non-functional detector after the upgrade is minimal. Additionally, the risk of damaging surrounding detectors is low since the detector remains stationary during the upgrade process.

\chapter{Software, Trigger, DAQ}
\label{sec:soft}
\section{Software, machine learning, AI}
\label{sec:Software}
\editor{T.Kuhr}

The software for the simulation, reconstruction, and analysis of the data from the upgraded detector will be an evolution of the current Belle II software, basf2~\cite{Kuhr:2018lps,basf2-zenodo}.
Nevertheless, substantial improvements are required to cope with the increased data rate, occupancy, and background level.
On the one hand this challenge has to be addressed by algorithmic improvements that find a good balance between computing and physics performance.
On the other hand new technologies provide various opportunities for software enhancements.

So far the Belle II software only runs on x86 architectures.
First attempts were made to port it to the ARM architecture which may result in a more energy efficient execution.
A large potential is expected from delegating compute-intense tasks to GPU resources.
This requires a porting of algorithms to GPU architectures, an adjustment of the software framework to run on heterogeneous computing resources, and their integration in the distributed computing system.
GPU resources are also essential for the training of large machine learning models and can facilitate their application.

Machine Learning (ML) and Artificial Intelligence (AI) are another prime example of a new technology with huge potential.
Already now AI algorithms are used at Belle~II or are under development.
ML applications at trigger level are described in Section~\ref{sec:TRG_ML}.
A prominent ML application example in the offline software is the Full Event Interpretation~\cite{Keck:2018lcd}, see also Section~\ref{sec:pp-benchmarks}.

Generative ML models are a promising technology to reduce the resource demand of simulations.
Currently a non-negligible part of storage resources is needed for an accurate simulation of beam background hits in the detector.
For this task random trigger data is taken and overlaid to simulated signal hits.
It requires DAQ bandwidth, disk space, and a distribution to MC production sites.
This can be avoided by generating the background hits on-the-fly with a generative model.

Such an approach is most beneficial for detectors with large readout windows and thus large beam background overlay event sizes.
This is currently the case for the PXD and the ECL.
It could be shown, that the effect of PXD background on the track reconstruction can be reproduced with background generated by a Generative Adversarial Network (GAN) architecture with newly developed extensions~\cite{Hashemi:2023ruu}.
GANs are also studied for the simulation of ECL waveforms.
A challenge that still has to be addressed is the generation of correlated background in different parts of the detector.

A complementary approach to the beam background is to reduce it as much as possible (see also Section~\ref{sec:BeamBackgroundsAndCountermeasures}).
This requires a good understanding of the accelerator conditions that lead to the various types backgrounds.
Such an understanding was achieved by a neural network trained on diagnostic data from SuperKEK and Belle II~\cite{Schwenker:2023bih}.
The model can be used to analyze the background composition of recorded data and to predict background levels.
The latter provides the potential to optimize the running conditions.

In general, AI has revolutionized how things are done in many areas.
Problems of classification, optimization, and generation are nowadays often tackled by AI methods.
Their potential is not yet fully exploited in particle physics in general and Belle II in particular.
And the development is still ongoing rapidly.
For example, Large Language Models (LLMs) may help to develop better physics algorithms, design more robust analyses, or write documentation and papers with fewer effort.
These are examples of areas that should be explored in the context of a Belle II Detector Upgrade.

%%%%%%%%%%%%%%%%%%%%
\section{Trigger}
\label{sec:TRG}
\editor{T.Koga}

\subsection{Introduction}

The hardware trigger system (TRG) was designed to satisfy the following requirements at the  target luminosity:
\begin{itemize}
    \item high efficiency for hadronic events from $ \Upsilon(nS) \rightarrow B\bar{B} $  and from continuum
    \item a maximum average trigger rate of 30 kHz
    \item a fixed latency of 4.5 µs
    \item a timing precision of less than 10 ns
    \item a trigger configuration that is flexible and robust
\end{itemize}

\begin{figure}[ht]
\begin{center}
\includegraphics[scale=0.5]{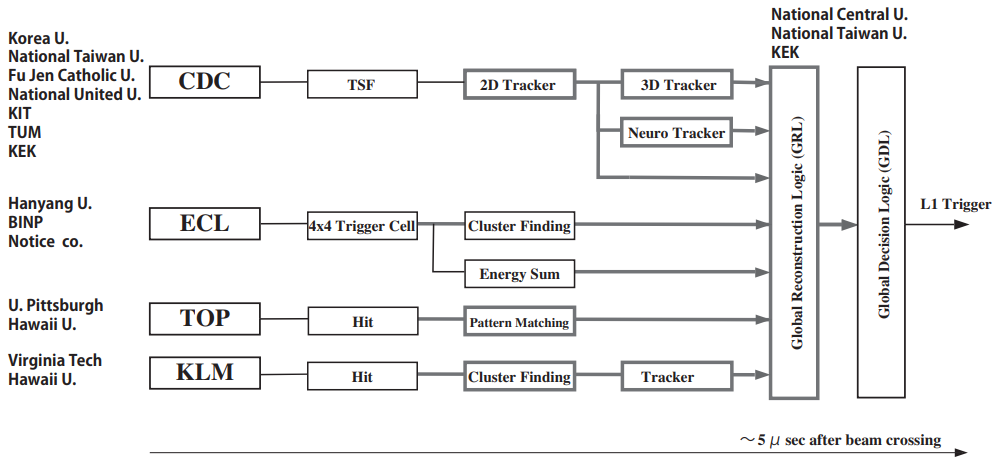}
\caption{\it Schematic view of the hardware trigger system}
\label{fig:trgview}
\end{center}
\end{figure}

Figure~\ref{fig:trgview} shows an overview of the system. The CDC and ECL take a major role to trigger charged particles and photons. The KLM triggers muons and the TOP measures event timing precisely. The information from sub detector triggers are send to the Global Reconstruction Logic (GRL) for the matching and the final trigger decision is made at the Global Decision Logic (GDL). Although TRG has been developed following the original design, further upgrades are needed to satisfy the requirements with the target luminosity, due to the higher beam originated background rate than the initial expectation. In addition, it is important to trigger the non-B physics with low multiplicity of the final state particles, such as physics with tau, dilepton, and dark/new particles searches. In order to achieve that, several upgrades are planned for both hardware and firmware with various time schedules.

%Fig. 1

\subsection{Hardware upgrade}

\subsubsection{Universal trigger board}

TRG consists of several kinds of FPGA-based electronic circuit boards. Each board has optical transceivers to communicate with other boards and the detector front-end boards. The universal trigger board (UT) has been commonly developed for all sub trigger systems to implement core trigger logic. Table~\ref{tab:trg_ut} shows a comparison of the generations 3 to 5. Upgrade of the UT is on-going to improve the resource size of the FPGA and the optical transmission rate, in order to implement more complicated trigger logic with increased information from the detector. At present, the 3rd (UT3) and 4th (UT4) generation boards are used for the physics operation. The number of UT4 boards is planned to be increased in the short and medium term. Development and production of the next generation of the UT (UT5) are scheduled in the medium and long term in 2024-2032. A high-end FPGA, Xilinx Versal, is used for UT5 to implement machine learning based logic with large DSP and AI engine. UT5 is being developed at KEK with cooperation of Belle II, ATLAS in Japan, and KEK electronics group via collider electronics forum, for not only Belle~II but also for any collider experiment.

\begin{table}[hbt]
    \caption{Specification of the universal trigger boards.}
    \centering
    \begin{tabular}{c|ccc}
        \hline 
        \hline 
        UT generation & UT3 & UT4 & UT5 \\
        \hline 
        Main FPGA (Xilinx) & Virtex6         & Virtex Ultrascale &  Versal \\
                           & XC6VHX380-565   & XCVU080-190       &                       \\
        Sub FPGA (Xilinx)  & $--$              & Artex7            &  Artex7, Zynq         \\
        \# Logic gate             &  500k & 2000k         &  8000k            \\
        Optical transmission rate &  8~Gbps  & 25~Gbps           &  58~Gbps              \\
        %RAM                    & $--$ & DDR4 & DDR4, UltraRAM  \\
        \# UT boards             &  30 & 30  & 10 \\
        Cost per a board (k\$) & 15     & 30        & 50 \\
        Time schedule          & 2014-  & 2019-2026 & 2024-2032 \\
        \hline 
        \hline 
    \end{tabular}
    \label{tab:trg_ut}
\end{table}

\subsubsection{CDCTRG upgrade with CDC front end board}

\begin{figure}
    \centering
    \includegraphics[scale=1.3]{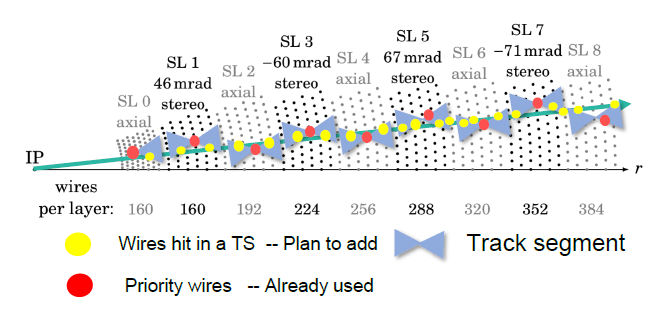}
    \caption{Wires used for present(red) and new(yellow) tracking algorithm.}
    \label{fig:trg_cdc}
\end{figure}

\begin{figure}
    \centering
    \includegraphics[scale=1.5]{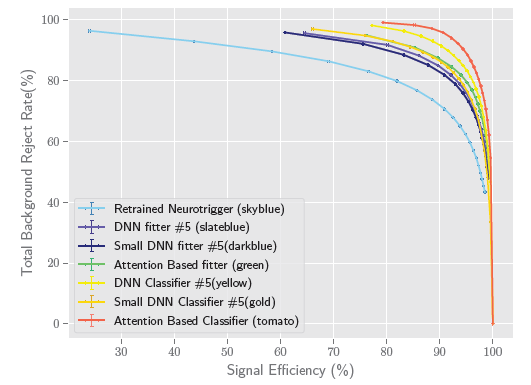}
    \caption{Expected improvements of signal efficiency and background rejection rate. Skyblue shows the performance of the current logic and the others are new logic with several machine learning architecture.}
    \label{fig:trg_cdc_z}
\end{figure}

CDC front end board and CDCTRG system will be upgraded around 2026. Bandwidth of the optical transceiver of the front end board is increased from 3Gbps to 10Gbps in order to send TDC and ADC of all wires to the trigger system, as shown in Table~\ref{tab:trg_cdc} and Fig~\ref{fig:trg_cdc}. The most upstream board of CDCTRG, the Merger board, is upgraded to UT4 to increase bandwidth of the receiver. All CDCTRG firmware will be updated with the TDC and ADC, in order to improve tracking and background rejection performances. The expected performance improvement is studied by simulation, including upgrade of tracking logic with machine learning. The vertex resolution is expected to be improved from 10~cm to 5~cm, with 50\% trigger rate reduction\cite{trg:liu}, as shown in Fig.~\ref{fig:trg_cdc_z}.

\begin{table}[hbt]
    \caption{Information send from CDC front end board to TRG.}
    \centering
    \begin{tabular}{c|ccc}
        \hline 
        \hline 
                                 & present           & new \\
        \hline 
        existence of wire hit    & 45 of 56 layers  &  all layers    \\
        hit timing (TDC)         & 9 of 56 layers   &  all layers    \\
        charge     (ADC)         & not sent         &  all layers    \\
        \hline 
        \hline 
    \end{tabular}
    \label{tab:trg_cdc}
\end{table}

\subsubsection{ECLTRG upgrade}

At present, analogue sum of 16 crystals (Trigger cell, TC) are used as a minimum unit of ECLTRG to measure energy deposit in the calorimeter. Upgrade of shaper+DSP and ECLTRG boards is considered in order to divide the analogue readout from each crystal. Background rejection performance is expected to be improved by measuring direction of the photon with the cluster shape, like ratio of the energy deposit between centered and outer crystals in a cluster. At offline analysis, 30\% background photon rejection has been achieved by using the cluster shape, and similar improvement is expected to the ECLTRG trigger rate. In addition, signal efficiency is expected to be improved by reducing pile up effects, when two daughter particles are close to each other within 20--40~cm on the ECL surface. Simulation study is on-going to evaluate the expected performance quantitatively and fix the initial design.

\subsubsection{TOPTRG upgrade}

In principle, TOP-based trigger could provide online event time with the highest precision with the high timing resolution photon sensors. However, to achieve this goal (while also making TOPTRG operate with a high efficiency) requires to implement a reliable way of estimating the location of the source of optical photons inside the quartz bars of the TOP detector. Another challenge is posed by the beam-related background which severely affects the TOP detector. To solve these problems and make TOP-based online event time useful for purposes of L1 trigger, our main focus will be in two areas: matching TOP trigger information with CDC 3D trigger and, also, taking advantage of continuing progress in the area of ML/AI, especially with the future trigger board UT5 with Versal FPGA architecture. 

\subsubsection{VXDTRG with VXD detector upgrade}

\begin{figure}
    \centering
    \includegraphics{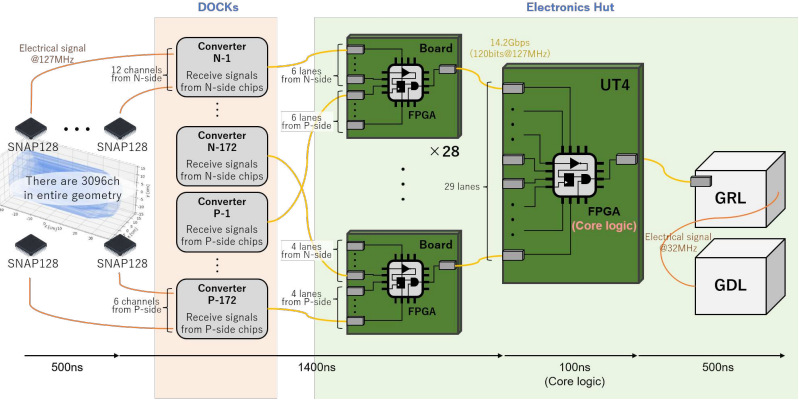}
    \caption{Over view of the VXD trigger system with a Thin Fine-Pitch SVD sensor scenario. }
    \label{fig:trg_vxd}
\end{figure}

\begin{figure}
    \centering
    \includegraphics[scale=1.5]{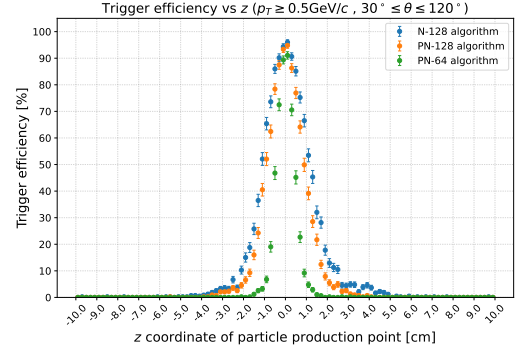}
    \caption{Expected trigger efficiency of VXDTRG along with vertex Z position of simulated particles. }
    \label{fig:trg_vxd_z}
\end{figure}

At present, PXD and SVD have no functionality to provide trigger signal. With the VXD upgrade, it might be possible to implement a vertex trigger to reject beam backgrounds originanting outside the interaction region, reducing the trigger rate. An initial evaluation of the new trigger logic performance was  done based on simulation with a Thin Fine-Pitch SVD sensor scenario~\cite{trg:shimasaki}, which is no longer an upgrade option. In that hypothesis, OR hits in multiple channels (0.6-1.0~cm area with 50-80~$\mu$m strip pitches) are sent from the front-end to the UT4 via a merger board, as shown in Fig.~\ref{fig:trg_vxd}. On the UT4, track candidates from interaction points are selected, by requiring 3 of 4 layer hits with expected hit pattern saved on FPGA with a look up table. As shown in Fig.~\ref{fig:trg_vxd_z}, 1-2~cm vertex resolution is achieved with 95\% tracking efficiency and a few percent fake trigger probability. By taking matching with CDCTRG track, trigger rate coming from beam background can be rejected more than 90\%. Similar studies are on-going for the CMOS MAPS VTX scenario, using the trigger logic under development for the Obelix chip (see Section~\ref{sec:vxd-obelix-implementation}). 

\subsection{Firmware upgrade}

In order to improve the signal efficiency and background rejection, several firmware upgrades are planned with the upgraded hardware, as shown in Table~\ref{tab:trg_firm}. Because the firmware modification can be done at any time with FPGA, the plan can be changed flexibly depending on the SuperKEKB machine condition, as a result of a physics discovery, the performance of new logic, a new idea of core logic and human resources, etc. One or two full-time workers are needed as the human resource for each upgrade component.

%
% This one should be turned around in portrait mode. with a fixed width mbox of the middle column with the long text. TO BE OPTIMIZED FOR LAYOUT 
% 
%\begin{landscape}
\begin{table}[b]
		\caption{TRG firmware upgrade plan.}
	\centering
 \scalebox{0.8}{
	\begin{tabular} {lp{0.4\linewidth}p{0.3\linewidth}lll}
	\toprule
        Component          & Feature & Improvement & Time & \#UT  \\
	\hline
        CDC cluster finder          & transmit TDC and ADC from all wires with the new CDC front end    & beamBG rejection & 2026 & 10 \\
        CDC 2Dtrack finder          & use full wire hit patterns inside clustered hit                   & increase occupancy limit & 2022 & 4 \\
        CDC 3Dtrack finder          & add stereo wires to track finding                                 & enlarge $\theta$ acceptance & 2022 & 4 \\
        CDC 3Dtrack fitter (1)      & increase the number of wires for neural net training              & beamBG rejection         & 2025 & 4 \\
        CDC 3Dtrack fitter (2)      & improve fitting algorithm with quantum annealing method            & beamBG rejection         & 2025 &  4 \\        
        Displaced vertex finder     & find track outside IP originated from long loved particle         & LLP search            & 2025 & 1 \\
        ECL waveform fitter         & improve crystal waveform fitter to get energy and timing          & resolution & 2026 & -- \\
        ECL cluster finder          & improve clustering algorithm with higher BG condition             & beamBG rejection & 2026 & 1 \\
        KLM track finder            & improve track finder with 2D information of hitting layers        & beamBG rejection & 2024 & -- \\
        TOP-CDC track matching      & include 3D information                                            & event time resolution & 2024 & -- \\
        TOPTRG ML-based algorithms          & ML-based timing measurement                  & beamBG rejection & 2026 & 2 \\
        VXD trigger                 & add VXD to TRG system with new detector and front end    & BG rejection & 2032 &  -- \\
        GRL event identification    & implement neural net based event identification algorithm         & signal efficiency  & 2025 & 1 \\
        GDL injection veto          & improve algorithm to veto beam injection BG                       & DAQ efficiency & 2024 & -- \\

	\bottomrule
	\end{tabular}
 }
    \label{tab:trg_firm}
\end{table}
%\end{landscape}

\clearpage
%%%%%%%%%%%%%%%%%%%%
\section{DAQ}
\label{sec:DAQ}
\editor{S.Yamada}

\subsection{Optically connected trigger timing distribution system}
\begin{figure}[hbt]
    \centering
    \includegraphics[width=\linewidth]{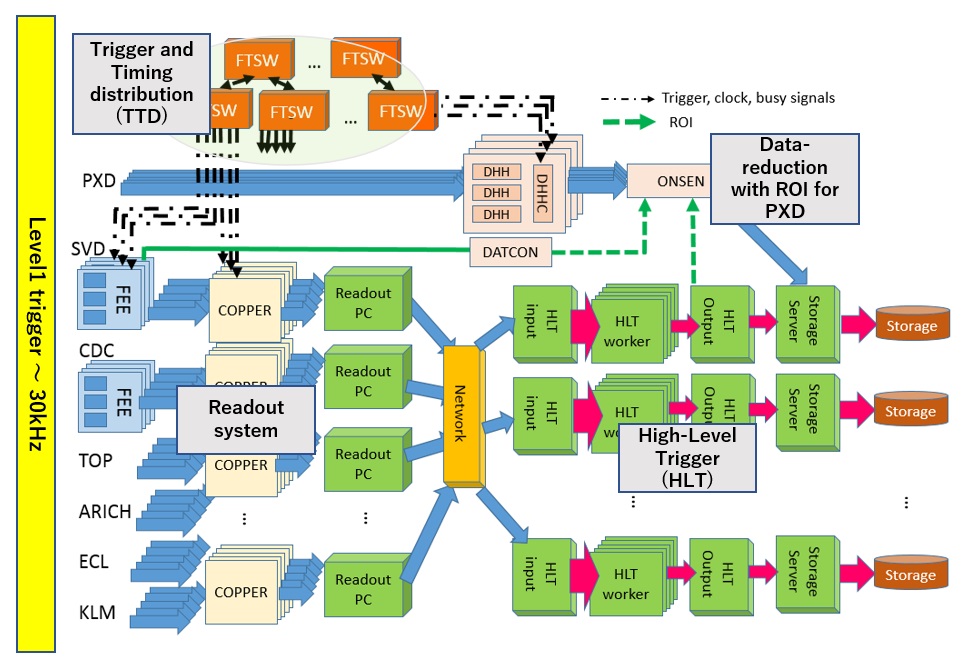}
    \caption{Schematic view of the Belle II DAQ system}
    \label{fig:belle2daq}
\end{figure}

The Belle II data acquisition system incorporates a trigger and timing distribution (TTD) system~\cite{daq:ttd}, which distributes the system clock, trigger signal, and related signals for synchronization and control to both front-end electronics and readout electronics boards. This system is designed to collect error, busy with data-processing, and other status signals from these electronics devices to monitor the data taking status and pause the trigger distribution when necessary to prevent recording corrupted data. A schematic view of the TTD system as a part of the DAQ system is shown in Fig.~\ref{fig:belle2daq}. The TTD system consists of a hierarchical structure of signal distribution modules of the two-slot-wide 6U VME form factor called FTSW (Front-end Timing Switch)~\cite{daq:ftsw}. For the communication between modules, a custom protocol called b2tt was developed and utilized. The current connections between FTSW modules and between the FTSW and front-end / readout devices are mostly made over a CAT7 network cables using the RJ45 ports of the FTSW. For longer transfer distances, optical fiber connections with a limited number of SFP transceivers are also used with a daughter card attached onto the FTSW module.
In the operation of the Belle II experiment since 2019, however, we experienced that sometimes the b2tt(Belle II trigger and timing) link via some of the CAT7 cables between FTSWs and electronics modules gets lost due to signal glitches, most likely due to external electromagnetic interference (EMI). The problem does not happen in the connection between FTSWs within the same VME crate, or in the optical connections. When the b2tt link is lost, data acquisition stops, causing a non-negligible loss in the data acquisition efficiency until the error state is reset by starting a new run

\begin{figure}[hbt]
    \centering
    \includegraphics[width=0.6\linewidth]{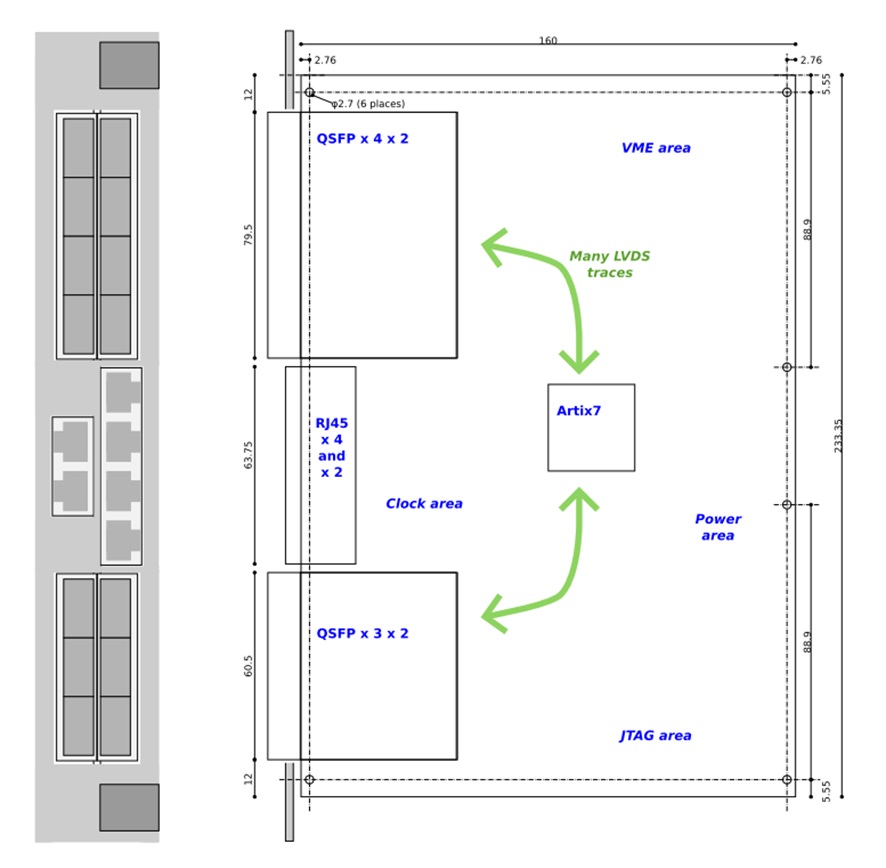}
    \caption{Conceptual drawing of the FTSW4 module}
    \label{fig:FTSW4}
\end{figure}
In order to prevent this issue and establish a more EMI-tolerant TTD system, a transition to more extensive use of the optical interconnection is planned. To increase the number of optical connections, it is essential to develop a new version of the FTSW module (FTSW4), which is capable to connect to a sufficiently large number of destinations via optical fibers. The new FTSW4 module is also designed for the trigger timing distribution to the newly designed CDC front-end boards described in Sec.~\ref{sec:CDC}, which will receive the b2tt signals over optical fibers.  The large number of the CDC boards demand the FTSW4 to have a high channel density.  Among the possible choices of the high density transceivers, we decided to adopt the QSFP transceivers, which has a moderate channel density but is highly cost-effective.  The board design we adopt is a two-slot-wide 6U VME module on which QSFP transceivers are mounted as shown in Fig.~\ref{fig:FTSW4}.  Equipped with 12 QSFP transceivers for trigger timing distribution and two more QSFP transceivers for uplink and other purposes, the FTSW4 module is capable to connect up to 24 front-end electronics or readout boards, with each connection using two pairs of optical fibers. Additionally, it has six RJ45 ports for conventional CAT7 cable connections.  The trigger timing distribution is controlled using a cost-effective Artix 7 FPGA.

The upgrade project is scheduled as follows: The first two prototype modules will be manufactured in FY2023. Following successful testing of the prototype module, mass production is set to occur in FY2024 and FY2025, aiming to produce about 40 modules, to cover the connections inside the E-hut and the connections to the newly designed CDC front-end readout boards (Table.~\ref{tab:ftsw4-plan}).

\begin{table}
    \caption{Schedule of the production of the FTSW4 modules}
   \label{tab:ftsw4-plan}
    \centering
    \begin{tabular}{cccc} \hline
         & FY2023  & FY2024  & FY2025\\ \hline
     prototype    & 2 modules &  & \\ 
     mass production    &  & 20 modules &  20 modules\\ \hline
    \end{tabular}
%    \caption{Schedule of the production of the FTSW4 modules}
%    \label{tab:ftsw4-plan}
\end{table}

\subsection{Improvement of throughput in readout system}
The functionalities of the readout system~\cite{daq:readout} are receiving data from front-end electronics boards and performing partial event-building and a basic sanity check of data. The Belle II DAQ system has a common readout interface with front-end electronics (FEE) of all sub-detectors except for PXD, which needs special data-treatment for online data-reduction. 

At the beginning of the Belle II experiment, we used a board know as COPPER~\cite{daq:copper}, on which we attached newly developed receiver cards, a CPU card, and an interface board with the TTD system. However, it became difficult to maintain the COPPER board, which was used in the previous Belle experiment, and its technology, PCI bus and Gigabit Ethernet, is somewhat obsolete. We therefore decided to replace the current readout system with state-of-the-art technology. Since electronics boards with such functionalities were already under development in other physics experiments for the basically same purpose, we decided to use the PCIe40 board~\cite{daq:pcie40}, which was originally developed for the LHCb and ALICE experiments. 

The PCIe40 board is equipped with the Intel Arria10 FPGA with 48 bi-directional serial links and 16 lanes of PCI Express 3.0 for I/O interfaces. With a much larger number of links (15) per board than in the COPPER board, which has 4, we can make a more compact readout system. We then concentrated on the development of firmware and software to make the hardware fit in with the Belle II DAQ system. The firmware of the new PCIe40-based system takes care of formatting and data checks, which have been carried out in software on the CPU board in the COPPER system. Those formatted data are read by software running on the host server of a PCIe40 board and are distributed to different HLT units for online event selection. After implementing those functionalities in the new hardware, we performed a high-trigger-rate test with dummy data and achieved the throughput of 3.4GB/s per PCIe40 board up to the readout PC.

The replacement has been completed in the LS1. Although the upgraded system can handle the 30kHz trigger rate, which is the design rate for Belle II, we will consider a further upgrade of the readout hardware to be able to handle unexpectedly high beam background levels as luminosity increases in the Run 2. One candidate for the hardware is the PCIe400 board~\cite{daq:pcie400}, an upgraded version of the PCIe40 board developed in the LHCb collaboration for the LHC-LS3 upgrade. While still in the design phase, the new board will provide enhanced capabilities, supporting higher bandwidth with a host server adopting newer technologies such as PCI Express Gen 5.0 and the latest FPGA. This will enable a higher data-processing capacity and transceiver capabilities with an increased bandwidth. Similar to the approach taken with the implementation of the PCIe40 board, our focus will be on firmware and software development to integrate it with the Belle II DAQ system.

\subsection{High-level trigger}
\subsubsection{Increase of computational power in the HLT system}
The HLT system~\cite{daq:hlt} of the Belle II DAQ system performs an online event reconstruction and issues a software trigger to reduce the number of events to be recorded on disks, which is necessary to save offline computing resources with the minimum effect on signal efficiency. The HLT system also sends region-of-interest (ROI) information to the PXD readout system, which is calculated by extrapolating a track obtained from CDC and SVD data. PXD data outside of the ROI are discarded, as most of them do not contain hits of signal tracks. For these purposes, the HLT system collects event fragments from all subdetectors except for PXD and performs online event reconstruction after the event-building~\cite{daq:eb} is done, which requires high computational power and a high network throughput. Like the readout system, the throughput requirement for the HLT system also gradually increases with the luminosity of the SuperKEKB accelerator. In addition to the luminosity, the beam background, which is difficult to precisely predict for increased luminosity in the future, can also affect the trigger rate and event size. Considering that, we adopted a staging plan to reinforce the HLT system to keep up with increasing trigger rate and event size. With this staging plan, we can also benefit from the evolution of information technology to prepare computational resources with higher performance at a later stage. 

To make the staging plan possible, the HLT system was designed to be a scalable system consisting of multiple units that can work in parallel. Adding other units will increase the total throughput of the system linearly. An HLT unit consisting of servers and network switches is stored in one rack and performs event building and online event reduction independently of other units. In the HLT unit, there are an input server, worker nodes, and an output server. The input server receives event fragments from the read-out system, builds events, and sends them to worker nodes, which perform online event reconstruction. Since the online reconstruction of an event can be done independently of other events, events that are built into the input server are distributed to different worker servers to perform online reconstruction in parallel. On each worker server, multiple reconstruction processes are running for parallel data processing to use multiple cores in the server. Event selection is performed here with reconstructed information, and events not meeting high-level trigger conditions are discarded. The output server then sends raw data with track and cluster information to a storage system, which is dedicated to recording data only from a specific HLT unit. 

From the start of the Belle II experiment, we gradually increased the number of HLT units and the number of CPU cores in each HLT unit as shown in Table~\ref{tab:hlt_cpu}. In 2019, 10 HLT units were in operation with 2880 CPU cores in total, and during the winter shutdown in 2020-2021, the total number of CPU cores increased to 4800. With this computational power, before LS1, the reconstruction software could handle data at the Level-1 trigger rate of around 15~kHz. In the LS1 period, we installed three more HLT units, and the total number of CPU cores in the HLT system reached 6212; we estimate that the HLT system can process data up to the trigger rate of 20~kHz.

\begin{table}
    \caption{Schedule of the reinforcement of HLT system}
\label{tab:hlt_cpu}
    \centering
    \begin{tabular}{cccccc} \hline
         & 2019  & 2020  & 2023  & 2024 & 2025 \\ \hline
     number of CPU cores   & 2880 & 4800 & 6212  & 6692 & 7172 \\ \hline
    \end{tabular}
\end{table}

Since the current performance does not reach the design rate of 30~kHz for the Belle II DAQ system, we will continue adding more CPUs to the HLT system. The next reinforcement plan is to add another HLT unit in 2024. After Run 2 starts, we will closely monitor the increase in trigger rates and event size; more HLT units will be added if this becomes necessary following the increase in luminosity. In addition to simply scaling up the CPU processing power, we plan to explore the implementation of heterogeneous computing in the online reconstruction within HLT. Currently, the same software used for offline analysis has been used for the online reconstruction, which is running on CPU, but offloading tracking or clustering algorithm to GPU or FPGA could be a viable option to improve the performance of HLT.

\subsubsection{Study of data-handling framework for triggerless DAQ}
A triggerless DAQ is a potential future upgrade to the Belle II DAQ framework. In this scheme, all data from front-end electronics are sent to the back-end DAQ system without the Level-1 trigger, and the event selection is performed solely by the software trigger in HLT. This upgrade could require substantial modifications in the DAQ system, the front-end electronics of each sub-detector, and the reconstruction software used by HLT. Therefore, it should be noted that this is an initial idea and has not been verified with either the detector group or the software group. 

Currently, the front-end electronics of each sub-detector receive Level-1 triggers from the TTD system to determine which data to send to the backend DAQ system. This scheme can control the throughput from the front-end electronics and reduce the workload of the back-end DAQ system. However, since the Level-1 trigger of the Belle II system is produced by an FPGA-based system with limited information, there is potential to enhance the event selection efficiency by sending all data without Level-1 trigger to the backend system to execute event selection using comprehensive detector information and a sophisticated reconstruction algorithm primarily handled by software. While the Belle II Level-1 trigger achieves nearly 100\% efficiency for a B\(\bar{B}\) event, the potential for improvement by introducing the triggerless DAQ could lie in the search for more exotic events, such as those with a displaced vertex and low multiplicity. 

\begin{figure}
    \centering
    \includegraphics[width=0.9\linewidth]{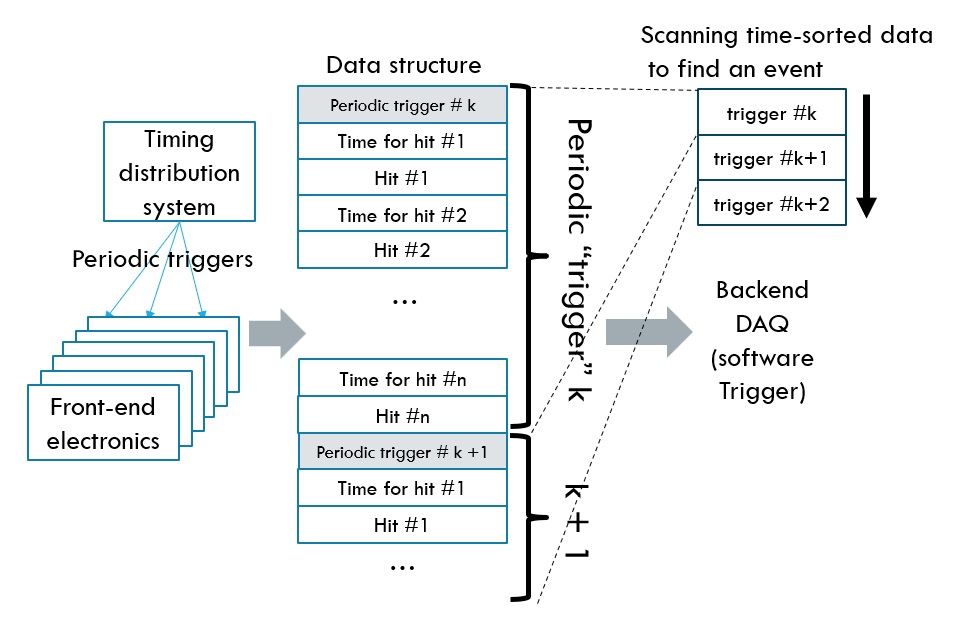}
    \caption{Data handling in triggerless DAQ}
    \label{fig:triggerless}
\end{figure}

If we adopt a triggerless DAQ scheme, besides handling the increased throughput resulting from the removal of the Level-1 trigger, which requires reinforcements in both the readout and HLT systems, a substantial upgrade in the software framework of the DAQ system is also necessary, as shown in Fig.~\ref{fig:triggerless}. Since no trigger information is available, timestamp information becomes essential for event-building without the Level-1 trigger. For this purpose, the front-end electronics must add the timing information to hits or waveform data. This information should consist of coarse timing distributed by the TTD system and fine-timing information generated internally by the front-end electronics. Subsequently, the front-end electronics needs to transmit all hits or waveform data to the back-end PC farms with the timing information. At the back-end DAQ system, software should be employed to merge data from different front-end electronics boards by using the timing information attached to the data. The software trigger could first examine a potential event time window using algorithms like track finding. Once the time window is determined, a detailed reconstruction process can take place. The selected events will subsequently be stored in a storage system for offline analysis. In this section, we have exclusively discussed the upgrade of the DAQ software framework. However, the reconstruction software developed by the Belle II software group also requires a significant upgrade for this new DAQ scheme. Detailed discussion and collaboration with the software group are necessary for the feasibility study of this aspect.

Eliminating the Level-1 trigger and sending all data to the back-end DAQ system also requires increased throughput capabilities for the front-end electronics boards of each sub-detector. With the exception of the current PXD readout system, it will be difficult to adopt the front-end electronics for other sub-detectors to the trigger-less readout. Therefore, if we proceed with the implementation of the triggerless DAQ, we should consider a staging approach, such as introducing triggerless readout first for specific sub-detectors, which has enough capability to increase throughput. In such a scenario, in addition to the normal data processing using the Level-1 trigger, additional online analysis will be conducted on data obtained from certain sub-detector data with triggerless readouts. In addition to the study of the software framework for the triggerless scheme, a feasibility study for the front-end electronics and back-end DAQ for higher throughput is ongoing.

\chapter{Beam Polarization and Chiral Belle}
\label{sec:chiral}
\editor{J.M.~Roney}
%\section{Beam Polarization and Chiral Belle}

\section{Introduction and physics motivation}
Upgrading SuperKEKB  with polarized electron
beams in the high energy ring (HER) opens a new and unique program of precision physics at a center-of-mass energy
of 10.58~GeV. We summarize here the  physics potential of this `Chiral Belle' program with a base-plan integrated luminosity of 40~ab$^{-1}$ and corresponding
technical requirements needed to achieve the goals. A brief summary of the conceptual design of
the requisite new hardware needed for the upgrade is also provided. 
These were introduced in the Snowmass White Paper
on upgrading SuperKEKB with a polarized electron beam \cite{USBelleIIGroup:2022qro}. As  data with polarized beam is also to be used for the current unpolarized physics program, it is possible to introduce and run with polarized beam while Belle~II accumulates the 50~ab$^{-1}$ sample at the design luminosity of $6\times10^{35}$cm$^{-2}$s$^{-1}$.
A separate
Conceptual Design Report on the Chiral Belle Beam Polarization Upgrade Project is being written and this section summarizes that work.

Unique and powerful sensitivities to new physics via precision neutral current measurements at 10~GeV
are enabled by upgrading SuperKEKB to have polarized electron beams. With  a  measurement of $\sin^2\theta_W$
having a precision comparable to the current world-average $Z^0$-pole value, but at 10~GeV,
Chiral Belle will be uniquely sensitive to the presence of dark sector parity violating bosons with masses below the Z$^0$.
Moreover, it would be the only facility able to probe neutral current universality relations between charm and beauty quarks
and all three charged leptons at energies below  the $Z^0$ pole and would measure ratios of neutral current couplings with
unprecedentedly high precision. As all of these measurements  are accurately  predicted in the Standard Model, any deviations would
be signatures of physics beyond the Standard Model to which only Chiral Belle would be sensitive for the foreseeable future.
The $\tau$-pairs produced with polarized beams will also provide the only means to actually measure, as opposed to setting limits on,  the
third-generation $g-2$ at an interesting precision. 
With 40~ab$^{-1}$, a measurement at the $10^{-5}$ level is possible and represents
a 1\% precision on the "Schwinger term" of the anomalous magnetic moment, which is more than 100 times more precise than current limits on the tau $g-2$ and only approach capable of achieving this precision within the next decade. 
The  40~ab$^{-1}$ Chiral Belle physics program also includes precision $\tau$ EDM measurements and QCD hadronization studies as well as
substantial improvements to the sensitivities of  lepton flavor violation searches in $\tau$ decays, and
$\tau$ Michel parameter measurements.
In the longer term, the tau anomalous magnetic moment measurement can  reach the ${\cal O}(10^{-6})$ level,
the equivalent of the muon $g-2$ anomaly scaled by $(m_{\tau}/m{_\mu})^2$ as expected in Minimal Flavor Violation scenarios. However, 
reaching ${\cal O}(10^{-6})$  will 
require substantially more statistics than the base plan of collecting  40~ab$^{-1}$ of polarized-beam data, as well as improved measurements of $m_{\tau}$ and $m_{\Upsilon(1S)}$ as well as additional theoretical work, and would represent a follow-on next stage associated for the much longer-term that includes a very high luminosity upgrade.

\section{Requirements}
\begin{figure}[htb]
    \centering
    \includegraphics[width=0.9\textwidth]{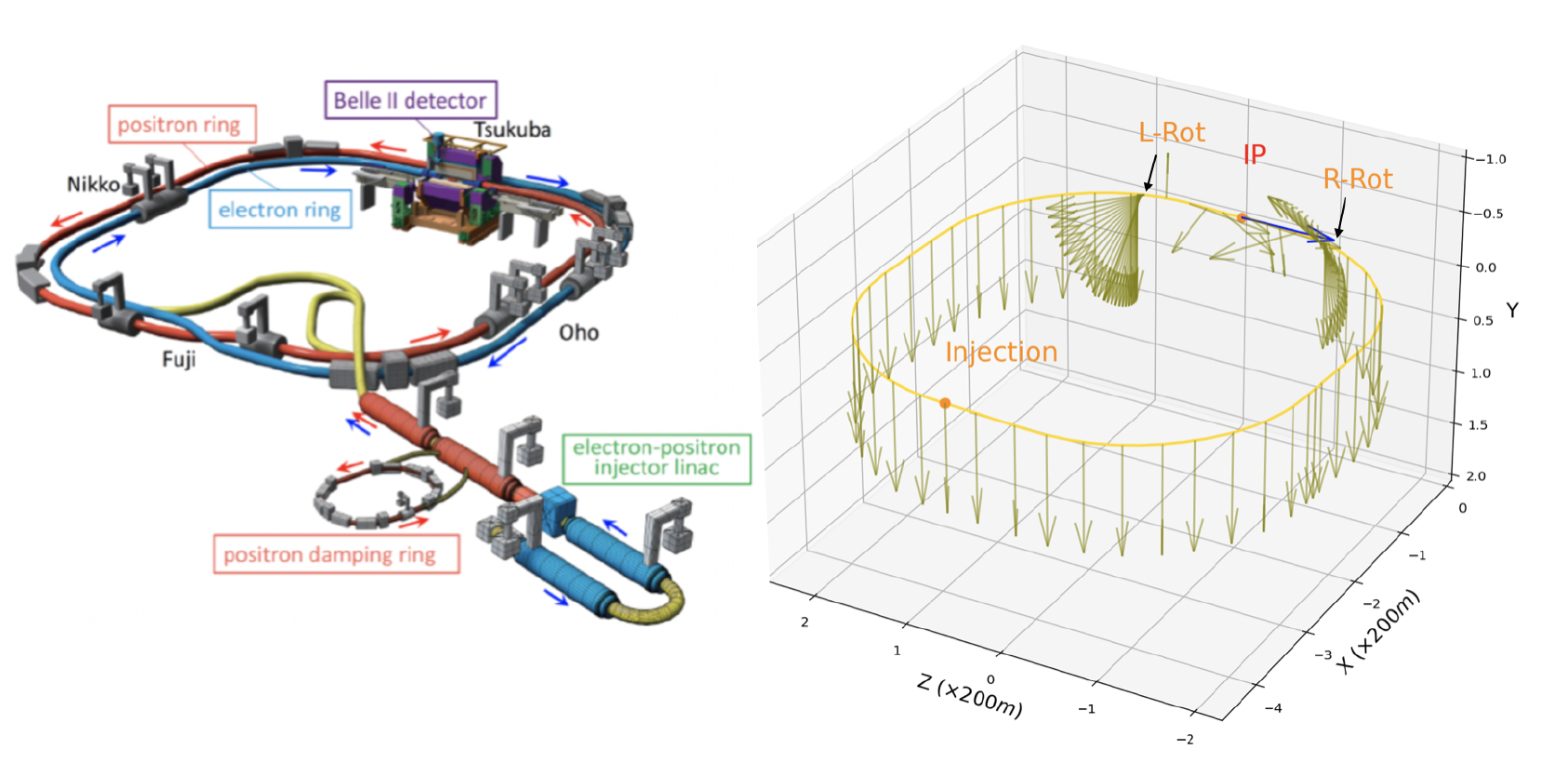}
    \caption{The spin motion of the electron  in the HER with the spin rotators installed.}
    \label{fig:HER-SpinRotation}
\end{figure}

  In order to implement $e^-$ beam polarization in the SuperKEKB HER,
  three hardware upgrades are required:\\
  1) introduction of a low-emittance polarized source that supplies SuperKEKB
  with  transversely polarized electrons with separate data sets having opposite polarization states;\\
  2) a system of spin rotator magnets that rotate the spin of the electrons in the beam to be longitudinal
  before the interaction point (IP) where the Belle~II detector is located, and then back to transversely polarized after the IP, as depicted in Figure~\ref{fig:HER-SpinRotation}; and\\
  3) a  Compton polarimeter that provides  online measurements of the beam polarization at a location between the first spin rotator 
  and the IP. A precision measurement of the polarization is also made at the IP by analysing the spin-dependent decay kinematics
  of $\tau$ leptons produced in  an  $e^+e^- \rightarrow \tau^+\tau^-$ data set, which gives Chiral Belle a unique way of controlling
  the systematic uncertainties on the beam polarization.

  The development work on a polarized source is underway and is a hardware project that has many synergies with the polarized source work currently being   undertaken for the Electron Ion Collider project at Brookhaven National Lab in the United States. The  helicity of the electrons is to be changed for different bunch trains by controlling the circular polarization of the source laser illuminating a GaAs photocathode, similar to that which was used for the SLC polarized source. The electrons circulating in the HER are required to be  transversely (i.e. vertically) polarized, as depicted in  Figure~\ref{fig:HER-SpinRotation}. 
  Since the electrons emitted from the photocathode are longitudinally polarized, there must be a spin rotator immediately after photocathode, such as that provided by a Wien filter, that ensures transversely polarized electrons are injected into the linac supplying the HER beam. Spin tracking studies from the source, through the linac and injector lattice, show that the transversely polarized state of the electrons will be preserved when injected into the HER.   Multiparticle long term tracking studies have shown that electrons injected with transverse polarization will maintain that polarization in the HER with a polarization lifetime of many hours.

\begin{figure}[htb]
    \centering
    \includegraphics[width=0.9\textwidth]{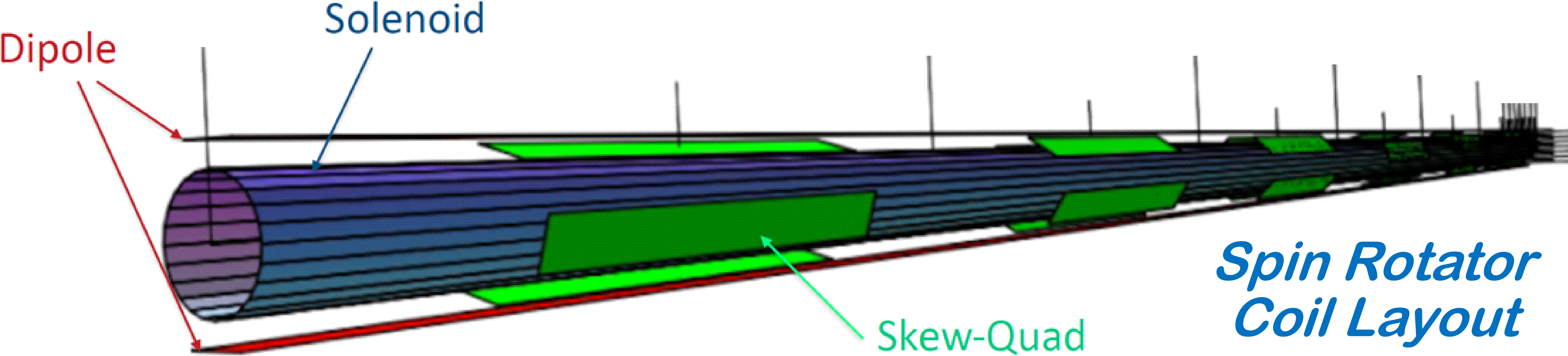}
    \caption{The Wienands concept for a compact combined function spin rotator unit with overlaid dipole, solenoid and skew-quadrupole superconducting coil fields.}
    \label{fig:spin-rotator}
\end{figure}

\begin{figure}[htb]
    \centering
    \includegraphics[width=1.0\textwidth]{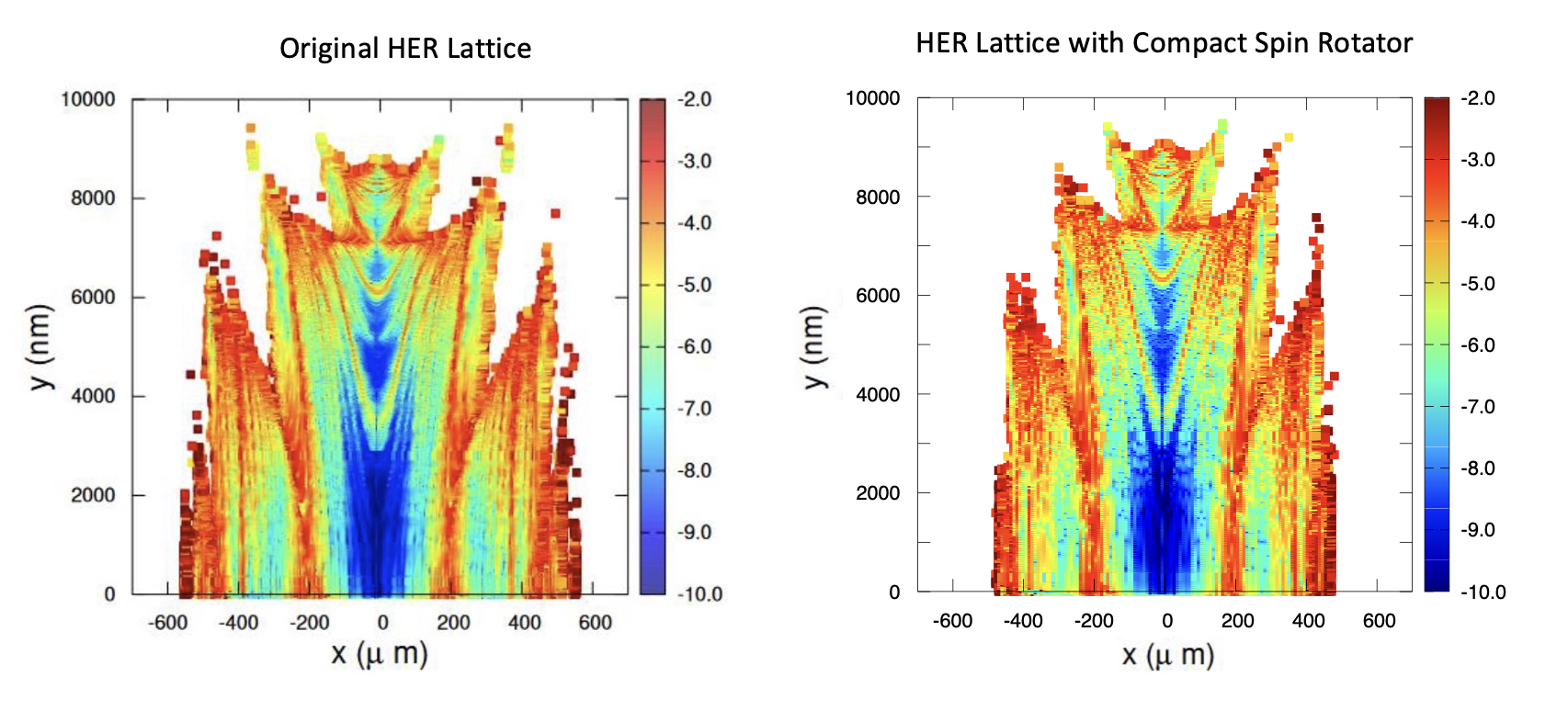}
    \caption{Frequency Map Analysis for the current HER lattice (left) and for the HER lattice with four compact combined function spin rotator magnets installed (right)  demonstrate that spin rotators can be implemented in a manner that is transparent to the rest of the lattice. This is further confirmed with long term tracking studies, as described in the text.} 
    \label{fig:fma-spinrot}
\end{figure}

  The challenging problem of providing a spin rotator in the SuperKEKB HER that is transparent to the rest of the  HER lattice
  can be solved in a manner that minimizes disruptions to the existing lattice and that easily permits the possible operation of the HER
  without any spin rotator magnet fields. The solution,  proposed  by Uli Wienands (Argonne National Laboratory),
 uses compact combined function magnets consisting of  a dipole, a solenoid and a system of six skew-quadrupole fields, as depicted in Figure~\ref{fig:spin-rotator}. It 
  can be implemented in the HER by using four compact combined function magnets, two on either side of the IP.
   These would simply  physically replace existing dipole magnets currently in the present lattice. Frequency Map Analysis dynamic  aperture studies, Figure~\ref{fig:fma-spinrot}, demonstrate the feasibility of this design.  
  Furthermore, long term tracking studies, which include radiation damping  and radiation fluctuations/quantum excitation effects, have  explored non-linear features of beam lifetime and polarization lifetime. These studies demonstrate that sufficiently long beam and polarization lifetimes can be achieved and confirm the viability of this particular spin rotator concept.

  Solutions for  implementing a Compton polarimeter have been found where the polarimeter is located in a place that minimizes any disruptions to
  the existing lattice as described in reference~\cite{Charlet:2023qvh} and can achieve polarization measurements 
  with systematic uncertainties at the 0.5\% level.
  Unique to the Chiral Belle program is the  additional means of  measuring the beam polarization at the
  IP using $\tau$ pair events with a relative systematic uncertainty of 0.4\%, as described in \cite{BaBar:2023upu},  a paper describing the technique and evaluating the statistical and systematic uncertainties using PEP-II $e^+e^-$ collision  data collected with the BaBar detector at the $\Upsilon(4S)$ center-of-mass energy. 

 \section{Implementation planning}
  The next steps in the Chiral Belle project involve completing the
  Conceptual Design Report on the Chiral Belle Beam Polarization Upgrade Project after which
  R\&D on the hardware components, including prototypes,  will converge on the Technical Design  and cost estimates. The capital costs for the polarization upgrade are expected to be substantially less than half of the annual  power costs of operating the SuperKEKB accelerator.
It can also be expected that a significant fraction of those capital costs will be provided by non-Japanese groups on Belle~II.

  A proposal to measure  the lifetime of the transverse polarization in the HER is currently being prepared, where a polarized source will inject transversely
  polarized electrons and the polarization lifetime measured using the Touschek effect, as is done in  resonant depolarization measurements. This would be
  a dedicated experiment of a few days duration and serve to validate the long term tracking studies that predict a spin lifetime that is much greater than the HER top-up rate. This Touschek-Polarization Experiment would be conducted in parallel with the other work required for the Technical Design Report.

It is  feasible  to plan for the upgrade to commence towards the end of this decade using the next long shutdown and possibly a number of summer shutdowns. In such a scenario, the polarization program could begin
while SuperKEKB completes its program of delivering 50~ab$^{-1}$ of data to Belle~II and continued beyond that program.

\chapter{Longer Term Upgrades}
\label{sec:longterm}
%%%%%%%%%%%%%%%%%%%%
\section{ARICH}
\label{sec:ARI}
\editor{R.Pestotnik}

\subsection{Introduction}
The Aerogel RICH particle identification subsystem (ARICH) located in a perpendicular 1.5~T magnetic field in front of an electromagnetic calorimeter consists of three key elements: aerogel radiator plane, photon detector plane with single photon sensors, and front-end readout electronics. Cherenkov photons emitted in the 4~cm aerogel radiator are propagated through a 16~cm  expansion gap and detected on the photon detector consisting of Hybrid Avalanche Photo-detectors (HAPD) produced by Hamamatsu \cite{ari:hapd}. 
The HAPD is a hybrid vacuum photon detector consisting of an entrance window with a bi-alkali photo-cathode with a 28\% peak quantum efficiency and four segmented avalanche photo-diodes in proximity focusing configuration. 420 HAPD sensors with 144 channels are mounted in seven concentric rings surrounding the beam pipe in the forward end-cap region of the spectrometer.
The ARICH was designed to operate up to the design luminosity of 8$\times10^{35}$~cm$^{-2}$ s$^{-1}$. Due to high background environments, its operation poses new challenges when the Belle II  operation is extended beyond the design luminosity.
   At high luminosity, the neutron background radiation and gamma radiation levels increase. They can affect the functionality of all detector components. 
 
The silica aerogel is produced in a supercritical drying process and is not affected by the irradiation \cite{ari:radaerogel}. Its performance was measured during the irradiation with a dose of 98~kGy, and no significant degradation was found. 

HAPDs, however, are sensitive to gamma and neutron irradiation. Different modifications in the APD structure have been implemented during the sensor's development. Different designs have been tested, including the insertion of a film on the active area and different thicknesses of the p+ layer. The design with an intermediate electrode, a thin p+  layer, and a film on the active area was finally selected for ARICH. The HAPD was demonstrated to work up to the design luminosity but not after its 5 $\times$  longer-term upgrade \cite{ari:iwataptep}. Also, due to its high Boron content introduced during the chip production, a Xilinx Spartan-6 FPGA, part of the front-end electronics board, is prone to single event upsets. By using a custom FPGA memory scrubbing \cite{ari:Giordano} and triple redundancy logic, we can operate the electronics in the short and mid-term. 

We have already started investigating different possible upgrade scenarios for the long-term upgrade, focusing on the possible photon detector candidates.

\subsection{Photon detector candidates}

The choice of photon detectors capable of operating in the magnetic field of 1.5~T is limited to several different technologies, which we already considered during the R\&D of the current photo-detector \cite{ari:sipmrich,ari:mcppmtrich}. The HAPD, produced by Hamamatsu Photonics, is no longer a viable option as the HAPD production line has been disassembled after the sensor production. 

\subsubsection{Silicon photo-multipliers}
The first technology is silicon photomultipliers, which have several advantages over hybrid avalanche photo-detectors. They have a very high photon detector efficiency, reaching 60\% at the peak. They are very easy to operate as they only require reverse biasing from 30 to 70 V, much less than standard or hybrid vacuum detectors. In addition, they have a very good timing resolution, which opens new possibilities for particle identification in ARICH. As semiconductor devices, they are inherently insensitive to magnetic fields.

They have, however, several drawbacks. First is the large dark count rate. Even more problematic is their sensitivity to radiation, which was a limiting factor a decade ago when selecting the photon detector candidates for the current ARICH photon detector.

The radiation damage increases the sensor currents, affects its breakdown voltage, and increases the dark count rate. The dark counts affect the measurement in two ways. First, their signals cannot be distinguished from single photon ones. As a result, they produce a high background on the detector's plane. The resulting effect on the kaon identification capabilities for different background levels can be seen in Fig.~\ref{fig:lappd}. Second, the increase of the dark count rate, proportional to the size of the device, is also responsible for the sensor baseline loss, which happens above $10^{11}$~n$_{eq}/$cm$^2$ or less for small sensitive area (1~mm$^2$) devices. The limit depends on many factors related to the SiPM design and the operation conditions. The dark count rate increases exponentially with the operating temperature, and the usability of the silicon photo-multipliers should be tested for each specific application. 
The baseline loss affects single photon counting much more than silicon photomultipliers in the multi-photon regime, e.g., in calorimetry. 
 
We are following the evolution of the technology and evaluating currently available samples. We are also working with the producers to reduce the neutron sensitivity and find the operation parameters where the damage might be under control. We will consider different engineering changes in the electric field design and sensor SPAD size \cite{ari:aidainnova}.

Several techniques can contribute to the operation of Silicon photo-multipliers: the reduction of their operating temperature, the use of timing information for background rejection, the use of a light collection system to reduce the sensor-sensitive surface, and the high-temperature annealing. We are studying the response of the silicon photomultipliers at low temperatures and developing multi-channel timing electronics to read out the sensors. 
These developments in silicon photomultipliers aim to demonstrate their capability to measure single photons and reduce the dark count rate to an acceptable level.

In the case of the RICH with the aerogel radiator where neutron fluences of up to \SI{5 e 12}{n/cm^2} are expected, the operation at lower temperatures requires the use of an additional cooling system, which introduces more material in front of the electromagnetic calorimeter impacting its performance. The effect must be carefully studied by performing detector response simulations to understand its consequences. 
% some comment should be added here, stating whether this additional material is not acceptable at all (preventing this solution) or if simulation studies will be done to understand how much this can worsen the ECL performances.

\subsubsection{Large area picosecond photo-detector}
The second possible photon detector technology we are considering for the long-term upgrade is micro-channel plate photo-multipliers (MCP-PMT). 
Compared to silicon photo-multipliers, such sensors have several drawbacks. As they are based on bi-alkali photo-cathodes and suffer from additional collection inefficiency, their photon detection efficiency is similar to the current HAPDs, much lower than that of SiPMs. As a vacuum detector with an internal amplification structure, they also suffer from the gain drop in the magnetic field and lifetime issues due to charge collection on the photocathode. In the case of Aerogel RICH, the above drawbacks can be, to a certain extent, mitigated. In the current configuration, there are gaps between the HAPD detector modules, which can be reduced during the design of the photodetector layout. Also, the gain drop in the case of the perpendicular magnetic field of ARICH is relatively small. Finally, the MCP-PMT lifetime can be extended by operating sensors at a lower gain and by the latest technological advances (atomic layer deposition treatment of the MCP structures). Compared to silicon-photomultipliers, the dark count rate of MCP-PMTs is much lower, especially after exposure to neutron irradiation; thus, considering them as a candidate for the future ARICH photon detector is very justified. 
%Here again it would be useful a comment on how these drawbacks can be managed in the future detector, and why it is anyhow worthwhile to study this solution

\begin{figure}[h]
\centering
\includegraphics[width=0.5\textwidth]{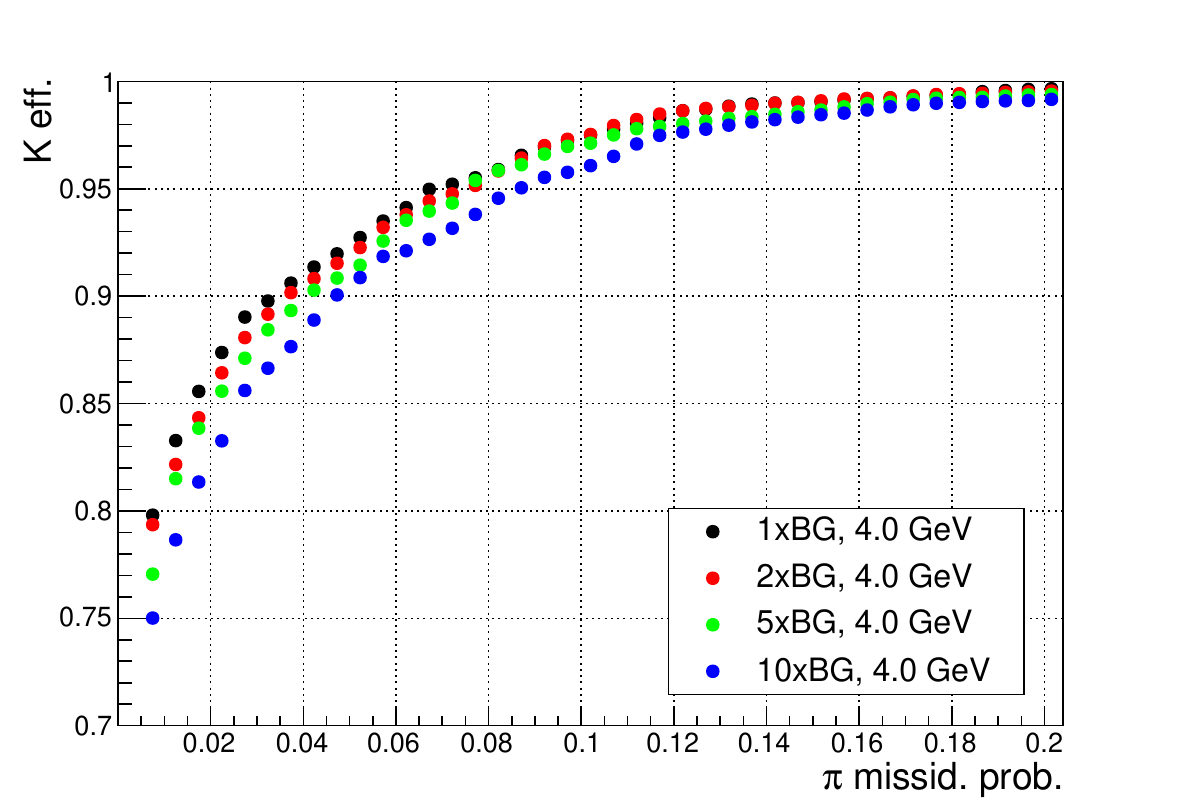}
\includegraphics[width=0.4\textwidth]{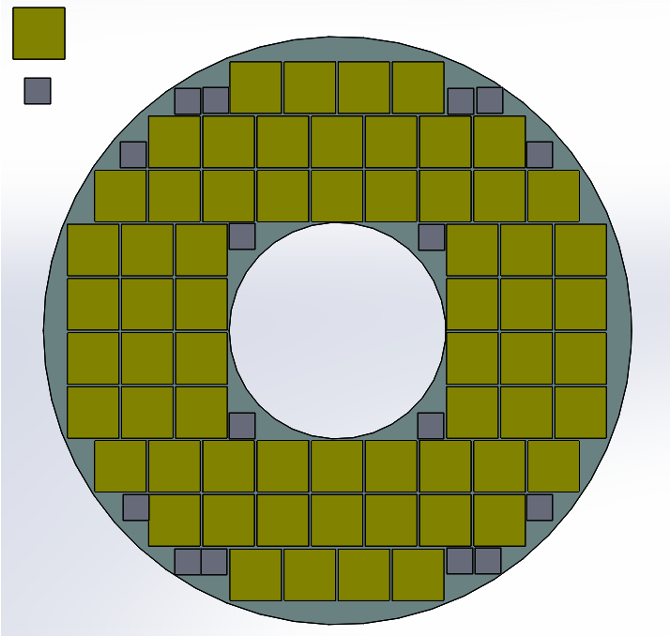}
\caption{ 
	(left) Simulated Kaon identification efficiency versus pion misidentification probability at the momentum of 4~GeV/c for various levels of nominal background. 
	Bialkali photocathode was assumed during the simulation.  Please note that these simulations were performed before the scenarios discussed in section~\ref{sec:BeamBackgroundsAndCountermeasures} were defined, and the background levels do not match exactly. Although the simulation will need to be repeated with standardized background levels, we believe the conclusions are substantially correct. 
	(right) Conceptual layout of ARICH photon detector plane populated with 20$\times$20~cm$^2$ and 10$\times$10~cm$^2$ LAPPDs.}
\label{fig:lappd}
\end{figure}

MCP-PMT production is, in general, very expensive, apart from the case of the Large Area Picosecond Photodetector (LAPPD) technology, where only inexpensive glass is used in the production process, leading to a significantly lower volume pricing. The LAPPDs, produced by Incom USA, come in two sizes,  20$\times$20~cm$^2$ and 10$\times$10~cm$^2$. The large size reduces the complexity of the photon detector. Unfortunately, it would introduce additional inactive areas on the photon detector. A conceptual design layout of the ARICH photon detector plane equipped with LAPPDs is shown in the right plot of Fig.~\ref{fig:lappd}. A version of the devices with capacitively coupled electrodes enables the design of custom readout granularities. We are optimizing the size of the pads and studying possible charge sharing between several pixels, which may impact the operation when signals from photons start to overlap.

%If I understand correctly this is a solution to implement the 10x10cm^2 LAPPD. This should be clarified at the beginning of the sentence: "To allow the use od more segmented sensors...."

\subsection{Sensor read-out electronics}
Both possible sensors require reading out the signals in a narrow time window of several ns concerning the trigger signal, requiring optimized internal design and high integration with the read-out electronics at the back side of the sensor. We are studying two potential front-end chip options. The first is based on upgrading the current ASIC used for Hybrid avalanche photo-diodes \cite{ari:asic}. The chip has been designed; the first samples are currently under functional tests. This option allows the use of a proven technology; however, it is not yet demonstrated how it will handle high dark count rates. The second option uses novel FastIC ASIC, featuring low noise fast timing capabilities, developed in collaboration with CERN and the University of Barcelona \cite{ari:fastic}. As the new ASIC will be used in the coming upgrades of the LHC experiments, the synergies in common development (e.g., with LHCb RICH) are expected to reduce the development costs and commissioning time.

\subsection{Radiator}
A study has already been launched to estimate the possible gains of using more transparent aerogel tiles and possible higher refractive index. The aerogel production technology has been transferred from Chiba University to a local company. We are evaluating the quality of aerogel after the technology transfer by measuring the characteristics of new samples.

\subsection{Potentially interested community in Belle II}
The following collaborators expressed their interest in the ARICH hardware upgrade activities: KEK, Tokyo Metropolitan University, and Niigata University, all from Japan; the University of Naples, Italy; and Jožef Stefan Institute, Slovenia.

\subsection{Cost and schedule}
The current Aerogel RICH can operate efficiently until the design luminosity is reached. Therefore, all the sub-detector changes considered above are projected for the long-term upgrade. Preliminary cost estimates are given for two photodetector options: silicon photomultipliers with 250,000 1~mm$^2$ channels and 64 20$\times$20~cm$^2$ and 16  10$\times$10~cm$^2$ LAPPDs.   The estimates are based on the current understanding of the technologies under development. 
\begin{table}[t]
\caption{Preliminary cost estimates in kUSD for photon detector upgrade based on SiPM and LAPPD. These include the costs for the associate R\&D.}
\vspace{-0.6cm}
\begin{center}
\begin{tabular}{|l| c| c|}
    \hline
    {\bf Option} & {\bf SiPM  } & {\bf LAPPD}  \\
    \hline
    {\bf Description} & {\bf k\$ } & {\bf k\$}  \\
    \hline
    \hline
     \multicolumn{3}{|l|}{\it Photo-sensor} \\
    \hline
    Research and development  & 500 & 500 \\
    \hline
    Production &  1.500 & 2.300 \\
     \hline
    \multicolumn{3}{|l|}{\it Electronics} \\
    \hline
    ASIC including TDC & 700 & 700  \\
    \hline
    Front-end board & 200 & 200  \\
    \hline
    Merger boards & 200 & 200 \\
    \hline
    \multicolumn{3}{|l|}{\it Services and detector infrastructure} \\
    \hline
    Photon detector mechanical frame & 200 & 200\\
    \hline
    Cooling system & 300 & re-use current system\\
    \hline
    Power supplies & 50 & re-use current system\\
    \hline
    \multicolumn{3}{|l|}{\it Radiator} \\
    \hline
    
    Aerogel upgrade & 700 & 700\\
    
    \hline \hline
    Total & 4.350 & 4.800\\ 
    \hline
    Contingency & \multicolumn{2}{|l|}{\it +15\% }  \\
    \hline
    
    {\bf Grand total} & {\bf 5000}  & {\bf 5520}  \\
    \hline
\end{tabular}
\end{center}
\label {tab:ARICost}
\end{table}
%%%%%%%%%%%%%%%%%%%%
\clearpage
\section{ECL}
\label{sec:ECL}
\editor{C.Cecchi}

This section contains a summary of the upgrade of the Belle II Electromagnetic Calorimeter (ECL). The current detector limitations are due to insensitivity to the incident photon direction and to possible deterioration of the energy resolution due to the signals pileup that can be produced under severe background conditions. Some upgrade options have been discussed in order to cope with the high occupancy and background rate and to maintain good particle reconstruction performances.

\subsection{Anticipated performance limitations}
\begin{figure}[htb]
    \centering
    \includegraphics[width=0.47\textwidth]{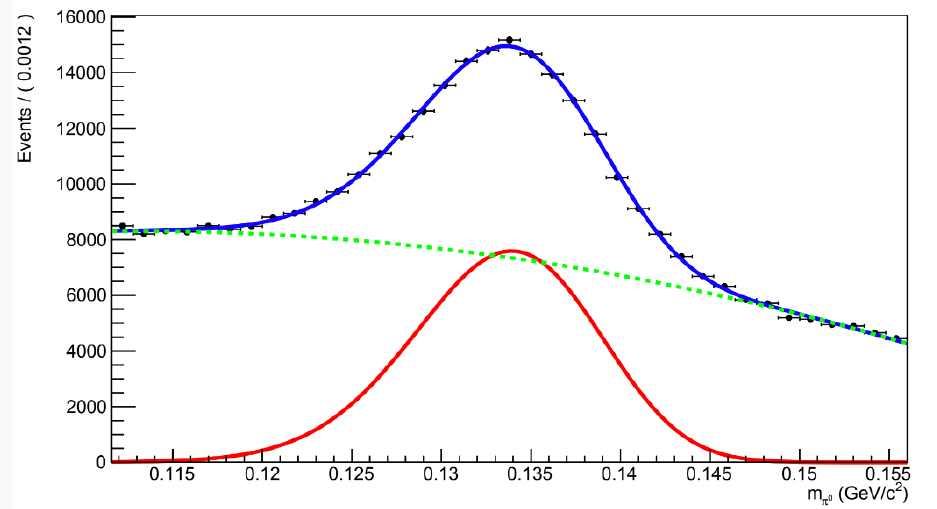}
    \includegraphics[width=0.45\textwidth]{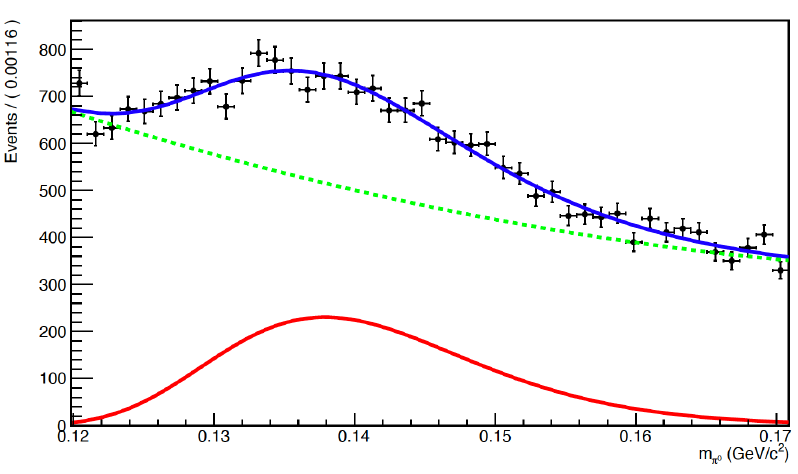}
    \caption{Reconstructed $\pi^0$ mass for the early phase-3 luminosity with BGx1 background simulation (left) and for the ultimate phase-3 BGx1 background simulation (right). The signal peak has been fitted with a Novosibirsk function while the background is described by a Chebichev distribution.}
    \label{fig:pizero-mass-fit}
\end{figure}

Since CsI(Tl) crystals scintillation decay time is relatively long, $\sim 1~\mathrm{\mu}s$, the increase of beam background would results in a larger pileup noise that would deteriorate the detector energy resolution. This effect is particularly relevant for neutral particles with respect to electron and positrons, as they are measured only in the ECL, without taking advantage of a good precise momentum measurement performed in the tracking detector. Simulated beam background has been superimposed on generic Monte Carlo (MC) $B\bar B$ events, and studies on the $\pi^0$ mass resolution have been done for the early phase-3 luminosity ($L = 1 \times 10^{34}\mbox{cm}^{-2}\mbox{s}^{-1}$) with nominal, ×2 and ×5 beam background as well as the ultimate phase-3 luminosity ($L = 8 \times 10^{35}\mbox{cm}^{-2}\mbox{s}^{-1}$) with nominal beam background cases. The $\gamma \gamma$ invariant mass spectrum is fitted with a Novosibirsk and a Chebichev functions for the $\pi^0$ signal and background, respectively. Resultant distributions are shown in Figure \ref{fig:pizero-mass-fit}, early phase-3 luminosity with nominal background (left) and ultimate phase-3 luminosity (right). In the early phase-3 case, the $\pi^0$ mass resolution and reconstruction efficiency are found to be 6 MeV and 30$\%$, while they become 9 MeV and 15$\%$ in phase-3 case. The ultimate phase-3 environment with higher background deteriorates the matching of the ECL deposit with the true photon because of the higher combinatorial in the number of ECL clusters. This could represent a strong limiting factor for physics analysis with rare B decays where high efficiency is required.

\begin{figure}
\centering
%\vspace{-15pt}
\includegraphics[width=0.6\textwidth]{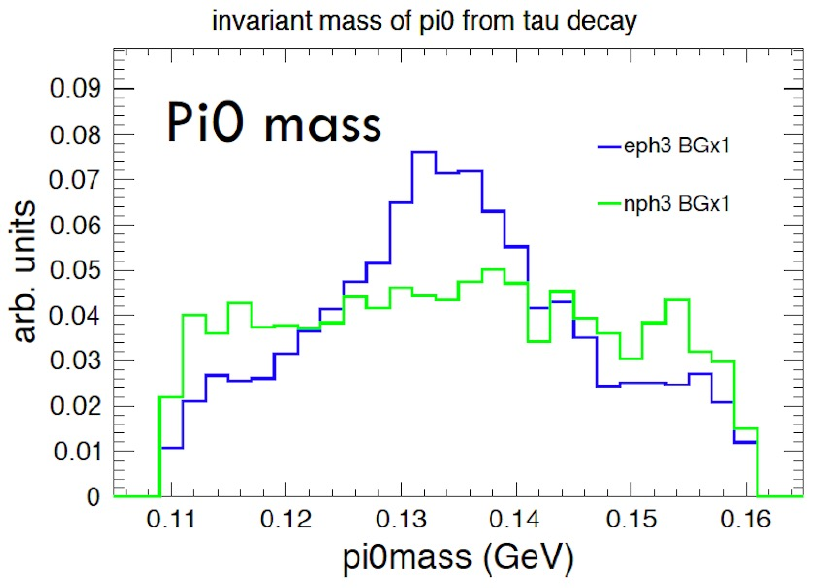}
\caption{$\pi^0$ mass distribution in $B\rightarrow \tau \nu$ events for ultimate phase-3 luminosity (green) compared to the early phase-3 luminosity with nominal BGX1 background distribution (blue).}
\label{fig:pizero-mass-nominal}
% \vspace{-15pt}
 %\vspace{1pt}
\end{figure}

In the $B\rightarrow \tau \nu$ analysis, one of the $B$ mesons in the event is fully reconstructed and the rest of the event is required to have $\tau$ daughters. Since substantial fraction of $\tau$ decay modes contain $\pi^0$, a good reconstruction of its invariant mass is crucial. The $\gamma \gamma$ invariant mass distribution is shown in Figure \ref{fig:pizero-mass-nominal}. In Phase 3, the $\pi^0$ signal peak is unseen, thus indicating a serious limit in reconstructing fundamental physics objects. These simulations were performed before the scenarios discussed in section~\ref{sec:BeamBackgroundsAndCountermeasures} were defined  and the background levels do not match exactly. Although the simulation will need to be repeated with standardized background levels, we believe the conclusions are substantially correct. 

%% FF - removed wrapfigure because of formatting issues
%\begin{wrapfigure}{r}{0.5\textwidth}
%\end{wrapfigure}

\begin{figure}[htb]
\centering
\includegraphics[width=0.45\textwidth]{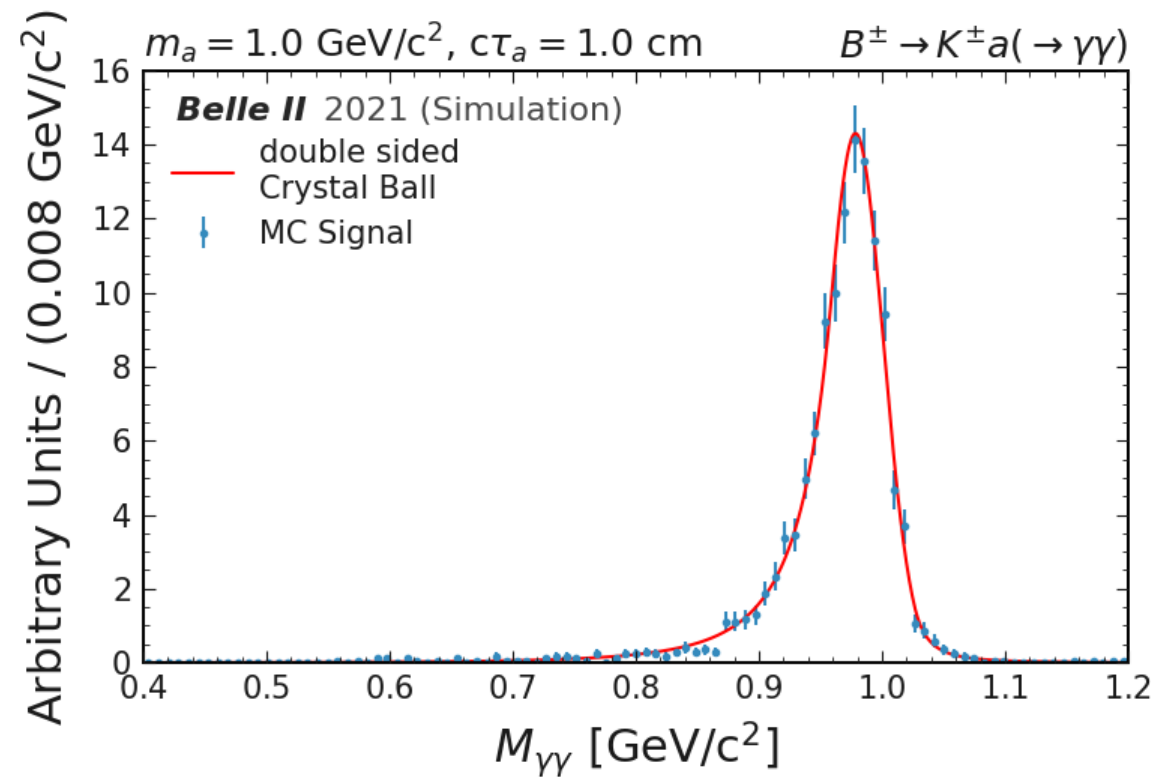}
\includegraphics[width=0.45\textwidth]{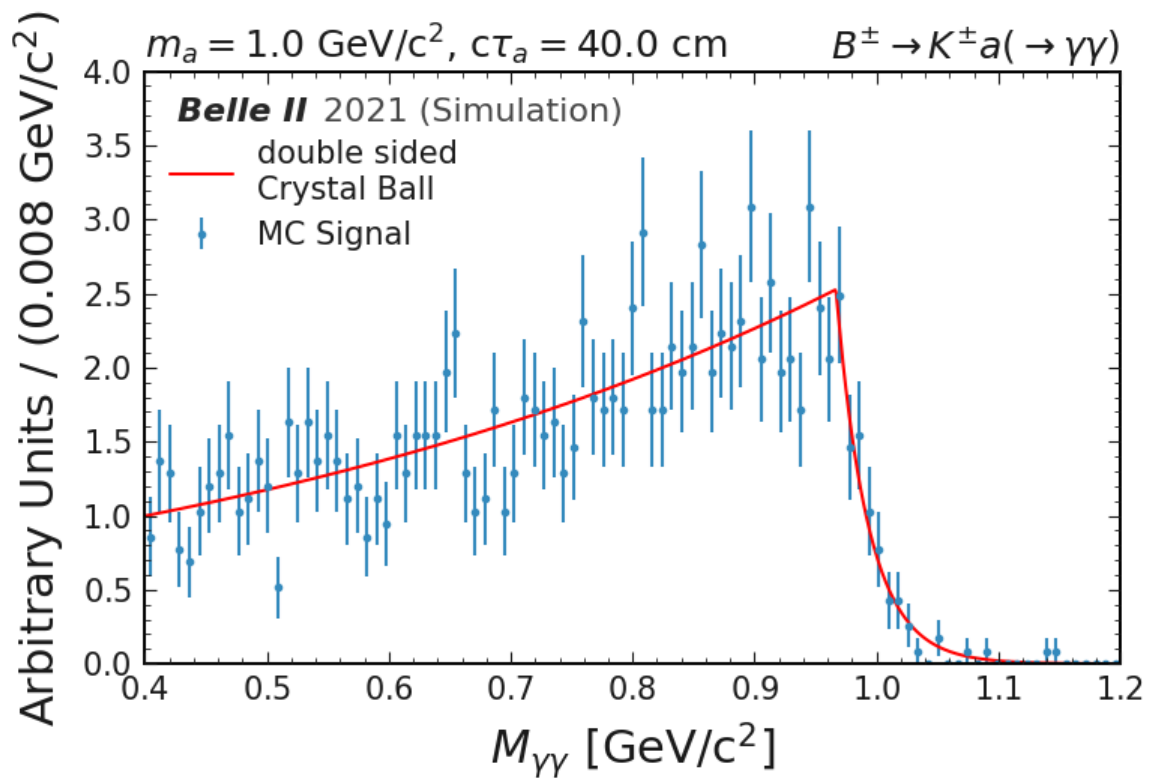}
\caption{The reconstructed photon pair's invariant mass distribution in $B \to Ka$, $a \to \gamma \gamma$ MC sample for the axion-like-particle (ALP), $a$ mass $m_a=1$ GeV/$c^2$, lifetime $c\tau=1$ cm (left) and 40 cm (right) cases.}
\label{fig:LonglivedALP}
\end{figure}

To search for dark-sector particles in Belle II data, importance to search for
a long-lived particle is increasing. A very interesting process it is the production of an axion-like-particle (ALP), ${a}$, in the B decay of $B \rightarrow Ka$. Such ALP is likely to be a long-lived particle decaying into a photon pair. Since current ECL configuration has very little sensitivity to incident photon direction whose reconstruction relies only on subtle shower shape difference, its mass determination for the long-lived ALP can only be done at the edge of the reconstructed invariant mass distribution when its lifetime ($\tau$) is several tens of cm in $c\tau$ , as shown in Figure \ref{fig:LonglivedALP}. Adding an extra device sensitive to incident photon direction is expected to significantly increase the mass determination power for such physics
case.

\subsection{Upgrade hypothesis}
\subsubsection{Replacement of CsI(Tl) crystals with pure CsI scintillator counters}
In order to eliminate pile up noise, faster scintillation crystal gives a primary solution. Pure CsI is suitable for this purpose because its fast light component decay time is about 30 ns, a factor of 40 faster than CsI(Tl). However pure CsI light output is about 20 times smaller than CsI(Tl) and its peak emission of 310 nm is in the near-UV region where photosensors are less sensitive. That is why the use of CsI crystals for the Belle II calorimetry is very challenging and requires the identification of performing photodetectors in the near-UV region. Front end electronics should provide a fast integration and a high amplification for the signal readout. To match this condition, avalanche photodiode (APD) readout based on the Hamamatsu APD S8664-55 \cite{ecl:hama2} and S8664-1010 \cite{ecl:hama1} has been studied. The S8664 series APD is usually operated at the gain $\times 50$, using it at higher gain of $\times 100$ is found to be still stably operational. Attaching to the photosensor a wavelength shifter (WLS) to emit about 500 nm wavelength light is also effective to get larger signal pulse by a factor 2$\sim$3 \cite{Jin:2016mic}. The best solution is then to combine these two approaches by placing 4 Hamamatsu APD S8664-55 placed on the edge of the crystal with the photodetector coupled to a WLS.
Simulation studies have been carried out to compare energy resolution of the current detector with a pure CsI calorimeter and it turned out that to have a resolution of the order of 5$\%$ at 100 MeV energy we should have an Equivalent Noise Energy (ENC) not higher than 1 MeV. The described configuration with pure CsI crystal connected to a WLS and 4 APD's has shown to satisfy the requirement on the ENC at sub MeV level. 

\subsubsection{A preshower detector in front of the ECL}

%% FF - removed wrapfigure because of formatting issues
%\begin{wrapfigure}{R}{0.5\textwidth}
\begin{figure}
\centering
%\vspace{-18pt}
\includegraphics[width=0.4\textwidth]{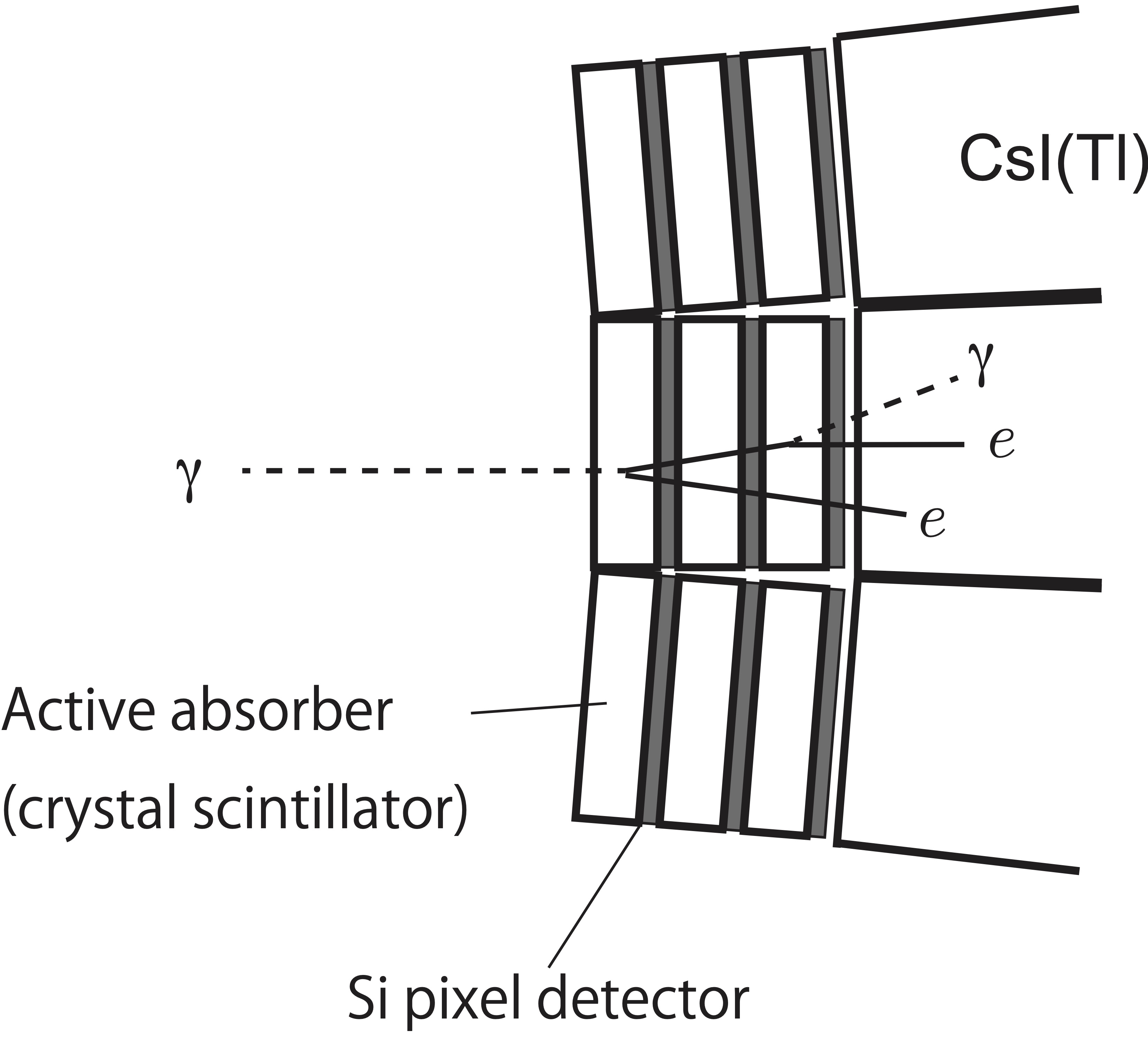}
\caption{Concept of the preshower detector. It is placed just in front of ECL and detects shower position and energy deposition, thus being sensitive to the photon incident direction.}
\label{fig:concept}
% \vspace{-18pt}
% \vspace{1pt}
\end{figure}
%\end{wrapfigure}

A second option considered for the ECL upgrade is a small detector system just in front of the ECL acting as a preshower detector as shown in Fig.\ref{fig:concept}. 
That has a sensitivity for photon incident direction, thanks to an excellent incident position resolution. Superior timing resolution can be featured. One layer consists of an active absorber with 1 radiation length thick crystal scintillator and a Si pixel detector to detect electron and positron passage when producing an electromagnetic shower. 
A GEANT4 based simulation with BGO/LYSO as the active absorber material and 1 mm$^2$ silicon pixels shows that the incident direction can be reconstructed for 80\% of the 1 GeV photons because they produce hits on two or more Si pixel detector layers. The position and angular resolutions are found to be 1.9 mm and 0.08 rad., respectively, for normal incidence. For the long-lived ALP decay to a photon pair with $m_a=2$ GeV/$c^2$ and 40 cm flight distance, the photons from IP can be separated from those produced by the ALP decay by about 2$\sigma$. We also see that significant portion of events' energy deposition is a few hundred MeV, thus the absorber part must be active with a reasonably good energy resolution. As for introduction of such a preshower detector, (1) it can act as a shield to stop significant fraction of beam background soft photons before hitting into the main ECL, thus mitigating pile up noise effect, (2) longitudinal sampling may help to distinguish photons and neutral hadrons by their different shower development characteristics, such possibility is worth to pursuit in addition to the already implemented pulse shape discrimination (PSD)~\cite{Longo:2020zqt} technique to improve particle identification capability inside ECL, (3) it is interesting to identify physics cases due to better photon incident position resolution, timing resolution and capability to reconstruct incident photon direction, and (4) geometrical activity match between the preshower and the main ECL also may help to reduce backgrounds.

\subsubsection{Complementing the PiN diode photosensors with avalanche photodiodes}

\begin{figure}[ht]
    \centering
    \includegraphics[scale=0.5]{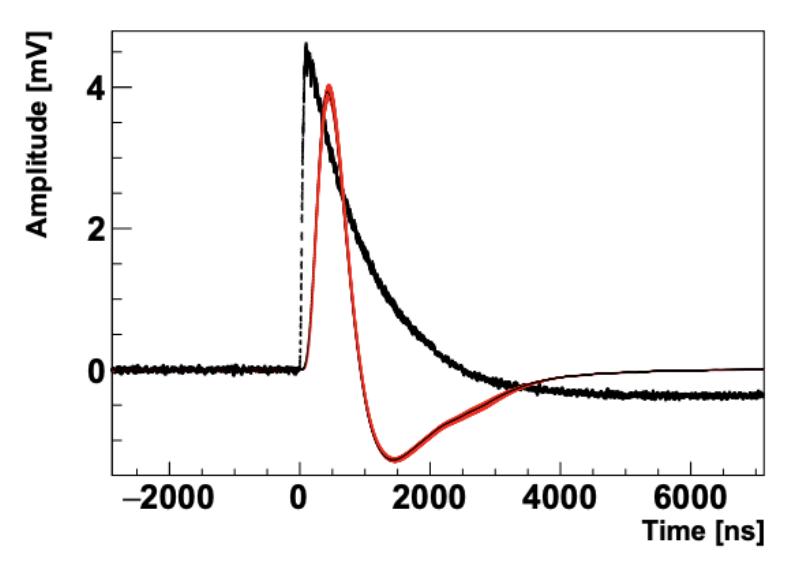}\includegraphics[scale=0.5]{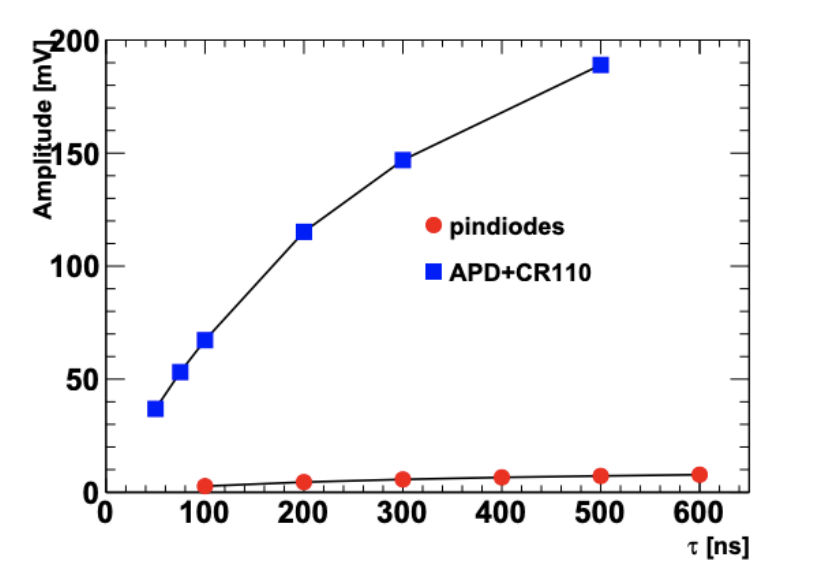}
    \caption{Signal amplitude for the APD plus transimpedence preamplifier configuration, the red curve represents the filtered waveform with a $\tau$ of 200 ns (left). Signal amplitude after shaping in the case of APD plus CR110 preamplifier (right). Notice the difference in the timescale on the horizontal axis for the two plots.}
    \label{fig:signal-CsI(Tl)-APD-TZA-CR110}
\end{figure}

In this section the possibility of complementing the Belle photodiodes readout with
two high-gain avalanche photodiodes (APDs) for the readout of the CsI(Tl) crystals will be described. The APDs were readout with either transimpedence amplifiers (TZA) [13] or with the charge-integrating CR-110 amplifiers.
The APD+TZA signal is shown in Figure \ref{fig:signal-CsI(Tl)-APD-TZA-CR110} (left) while the amplitude, after the $CR-(RC)^4$ filter, from APDs and the charge-integrating CR-110 is shown on the right side of the same Figure.

\begin{figure}[ht]
    \centering
    \includegraphics[scale=0.5]{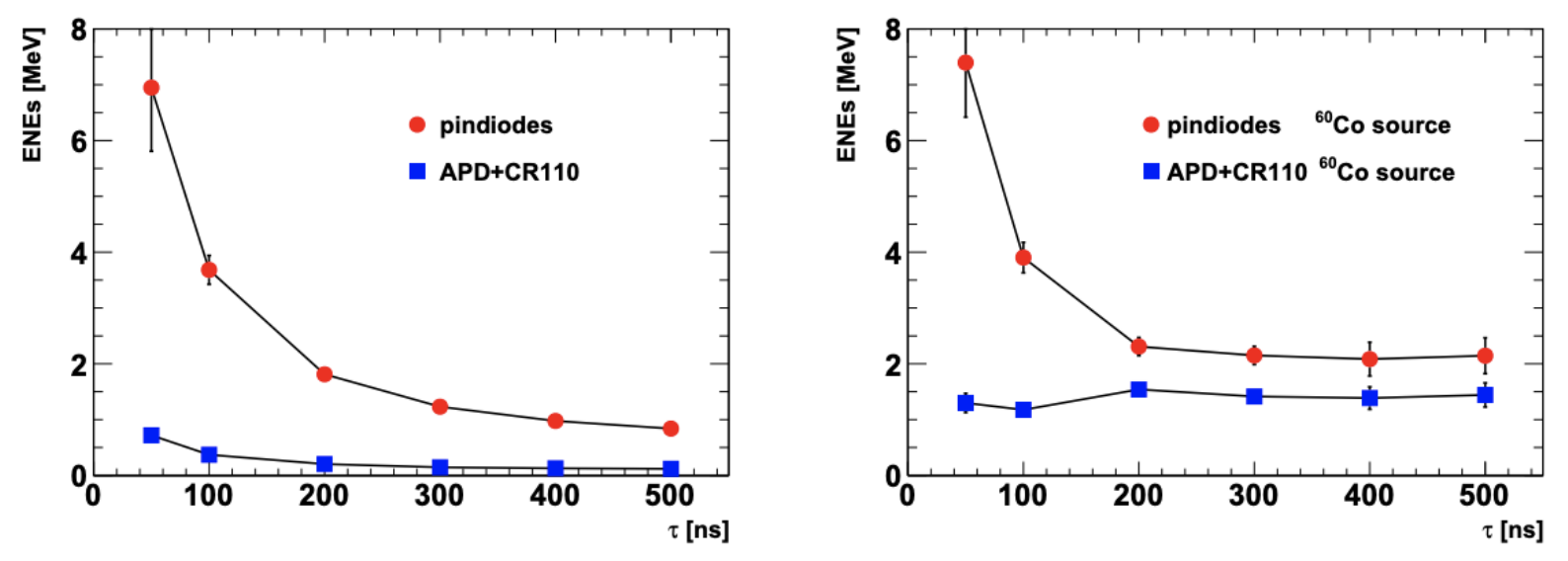}
    \caption{Equivalent Noise Energy as a function of the shaping time comparing pin diodes with APD with CR110 amplifier without (left) and with (right) $^{60}$Co source.}
    \label{fig:ENE-CsI(Tl)-APD-CR110}
\end{figure}

The Equivalent Noise Energy (the input-referred noise calculated from the signal to noise ratio at the shaper output, indicated as ENE) of the sum of two APDs with CR-110 readout, compared to the pin diodes on the same data run, is shown in Figure \ref{fig:ENE-CsI(Tl)-APD-CR110} with and without background overlay. 

 The large amplitude signal grant to APD's a very small electronic noise and a reduced pile-up noise with respect to the pin diodes.

 \subsubsection{Complementing the PiN diode photosensors with Silicon Photomultipliers (SiPM)}

\begin{figure}[ht]
    \centering
    \includegraphics[scale=0.8]{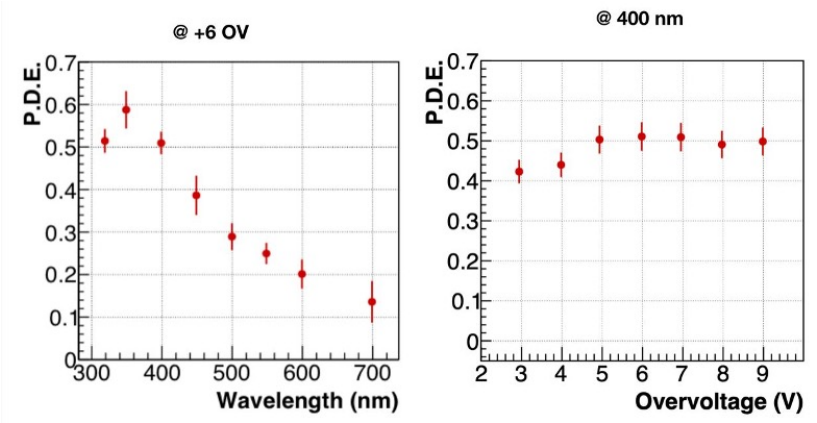}
    \caption{Photodetection efficiency (PDE) as a function of the wavelength (left). Photo detection efficiency (PDE) as a function of the overvoltage (right).}
    \label{fig:PDE}
\end{figure}

 To control the pile-up in condition of high background it is possible to use a high gain photodetector to perform an accurate measurement of the timing exploiting the large amount of light emitted by the CsI(Tl) crystals.
SiPM’s are a very good option thanks to their fast response, high gain and simplicity in use. Some preliminary studies have been done trying to understand the actual possibility of using these photodetectors for timing measurement, keeping the PiN diodes for the signal amplitude measurement. To start the R$\&$D with SiPMs as the readout system for CsI(Tl), crystals from Belle and a prototype SiPM by FBK (Fondazione Bruno Kessler) ~\cite{ecl:fbk}, developed for the CTA (Cherenkov Telescope Array) experiment, have been used. This prototype SiPM is not optimized to match the emission peak (550 nm) of the light emitted by CsI(Tl) counter, but it is good enough for our test presenting 25$\%$ of photo detection efficiency (PDE) at the above mentioned wavelength. Figure \ref{fig:PDE} (left) shows the PDE as a function of the wavelength while on the right panel the PDE as a function of the overvoltage is shown.

\begin{figure}[ht]
    \centering
    \includegraphics[scale=0.45]{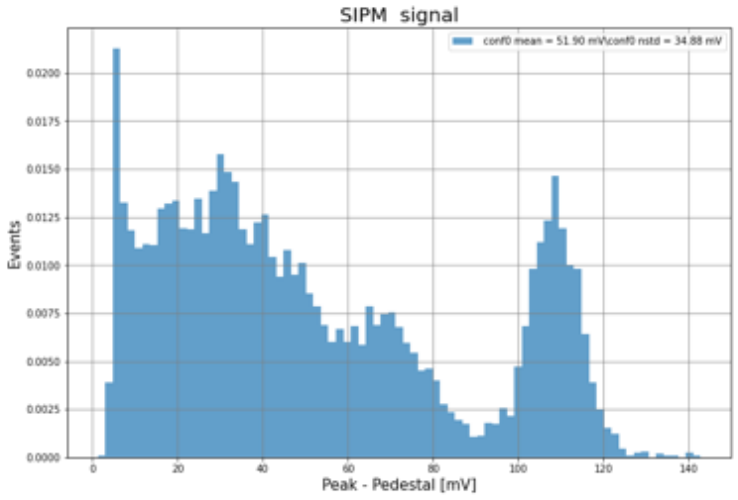}
    \includegraphics[scale=0.48]{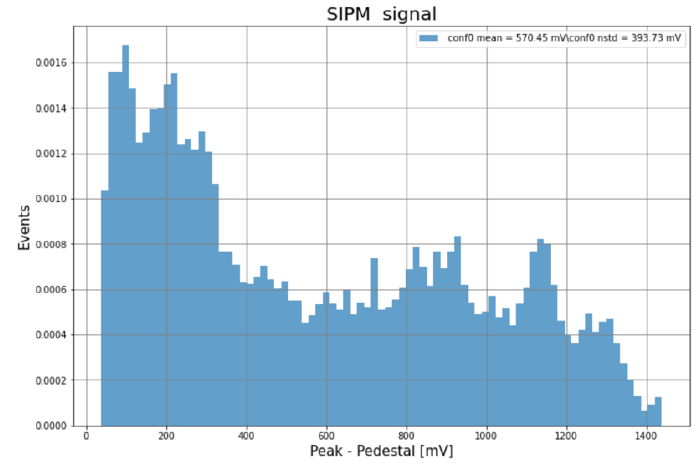}
    \caption{$^{137}$Cs spectrum measured with the FBK SiPM (left). $^{60}$Co spectrum measured with the FBK SiPM (right).}
    \label{fig:sources-spectrum}
\end{figure}

The SiPM has an active area of 6x6 mm$^2$ (effective area 80$\%$ of the active region), breakdown voltage of 26.5 V and a nominal gain at breakdown of $1.1 \times 10^6$ electrons/p.e. which increases linearly with the overvoltage ($2-3 \times 10^6$ at +5 OV). The signal produced by the SiPM has been readout with a board from AdvanSid ~\cite{ecl:advansid} which contains a transimpedance amplifier with output impedance of Z=500~$\Omega$ corresponding to a gain of a factor 10$^3$. The integral output signal is proportional to the detected charge. To calibrate the system of CsI(Tl) plus SiPM the response to radioactive sources signal has been studied with a $^{137}$Cs source emitting a photon at 700 KeV and with a $^{60}$Co source emitting two photons of energy 1.17 MeV and 1.33 MeV. The calibration has been done with a sample of LaBr crystal which has high light output (70000 photons/MeV), a very short decay time (20 ns) and the emission peak at about 360 nm which perfectly matches the region of the highest PDE of the FBK SiPM. In Figure \ref{fig:sources-spectrum} the measured spectrum for the two sources, on the left $^{137}$Cs and on the right $^{60}$Co, is shown. The second peak of the $^{60}$Co spectrum is fainter than the first one, as expected, and in both distributions we can observe the contribution of the Compton edge.

\begin{figure}[ht]
    \centering
    \includegraphics[scale=0.8]{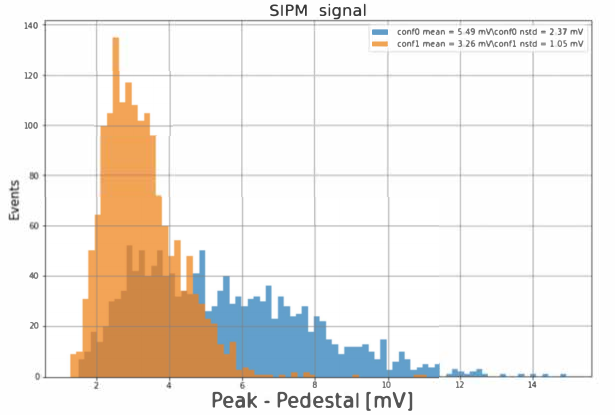}  
    \caption{Peak amplitude measured with the FBK SiPM with (blue) and without (orange) Cs$_{137}$ source. }
    \label{fig:CsI(Tl)-w/wo-source}
\end{figure}

Using the measurements of the integrated charge performed with this setup, the gain of the Advansid board has been derived to be 2$ \times 10^5$, 1.3$\times 10^6$ and 3$\times 10^6$ for 27, 29 and 31 bias voltages respectively. To evaluate the impact of the noise coming from the SiPM plus the Advansid board a measurement with and without radioactive sources on a Belle CsI(Tl) crystal has been performed. Figure \ref{fig:CsI(Tl)-w/wo-source} shows the signal peak amplitude with (blue distribution) and without source (orange distribution). The distribution of the signal of the CsI(Tl), when the Cs$_{137}$ source is present, is much larger than the one observed with the LaBr crystal, furthermore the noise (orange distribution) is not well separated from the signal preventing the possibility of  subtracting it in order to extract the signal. 

Others SiPM's better matching the CsI(Tl) emission spectrum as well as a customized low noise front end board will be tested to improve the separation between signal and noise. 
 
\subsection{Cost and schedule}

An estimation of the main cost for the Pure CsI crystals with APD readout + WLS option is reported in Table \ref{tab:CsIPure-cost}.

\begin{table}[htb]
    \centering
    \begin{tabular}{|l|c|c|} \hline \hline 
                           & Cost/unit (k\euro  ) & Total cost (M\euro)\\ \hline
         Crystals (8736)   & 4.9                 & 42.5  \\
         APD's (4/crystal) & 0.1                 & 3.5 \\
         WLS   (1/crystal) & 0.05                & 0.4        \\
         FE    (4/crystal) & 0.3                 & 2.6  \\ \hline \hline 
         TOTAL             &                     & 49 \\ \hline
         
    \end{tabular}
    \caption{Estimated cost for the Pure CsI crystal calorimeter with APD + WLS readout.}
    \label{tab:CsIPure-cost}
\end{table}

A rough estimation of the main cost for the pre-shower option has been evaluated (see Table \ref{tab:pre-shower-cost}) considering the following dimensions:
each endcap has 4.4 m$^2$ (Outer radius 1.25 m, Inner radius 0.42 m),
the Barrel Inner cylinder surface is 11.7 m$^2$ which results in 20 m$^2$ in total.
\begin{table}[htb]
    \centering
    \begin{tabular}{|l|c|c|} \hline \hline 
                    & Cost/unit (k\euro) & Total cost(M\euro)\\ \hline
Si sensors          & 100/m$^2$                & 2  \\
LYSO                & 600/m$^2$                  & 12 \\
Photosensor (3 layers, 2pc, 8736 crystal) & 100 & 5.2       \\
Readout                  &                 & 2  \\ 
Mechanics                  &                 & 2  \\ \hline \hline 
TOTAL                &                     & 23.2 \\ \hline
         
    \end{tabular}
    \caption{Estimated cost for the pre-shower calorimeter.}
    \label{tab:pre-shower-cost}
\end{table}

Finally, an estimation of the main cost for the CsI(Tl) crystals with LAAPD readout option is reported in Table \ref{tab:CsI(Tl)-cost}.

\begin{table}[htb]
    \centering
    \begin{tabular}{|l|c|c|} \hline \hline 
                           & Cost/unit (k\euro  ) & Total cost (M\euro)\\ \hline
         APD's (2/crystal) & 0.5                 & 4.4 \\
         %WLS   (1/crystal) & 0.05                & 0.4        \\
         FE    (2/crystal) & 0.2                 & 1.7  \\ 
         ECL extraction    &                     & 2.0 \\ \hline \hline 
         TOTAL             &                     & 8.1 \\ \hline
         
    \end{tabular}
    \caption{Estimated cost for the CsI(Tl) crystal calorimeter with APD readout.}
    \label{tab:CsI(Tl)-cost}
\end{table}

At the moment, the cost for the replacement of the crystals and for the pre-shower options do not appear to be jusitfied. However, the idea of complementing the readout photosensors with SiPM looks interesting and economically feasible.

\subsection{Potentially interested community in Belle II}
The following collaborators expressed their interest in the hardware upgrade activities: Budker Institute of Nuclear Physics (BINP), Russia, KEK, Japan, Nara Women's University, Japan, University of Naples and INFN-NA, Italy, University of Perugia and INFN-PG, Italy.

\printbibliography

\end{document}